\documentclass[iop]{emulateapj}
\usepackage[utf8]{inputenc}
\usepackage{float,hyperref,xcolor}
\usepackage{longtable}

\newcommand\fbol{$\mathcal{F}_{bol}$}
\newcommand\rstar{$R_*$}

\newcommand\msun{$M_\odot$}
\newcommand\delR{$\Delta R_*$}
\newcommand\teff{$T_{\rm eff}$}
\newcommand\teffs{$T_{\rm eff}$s}

\newcommand\BV{$B-V$}
\newcommand\VI{$V-I_{c}$}
\newcommand\VK{$V-K_{\rm s}$}
\newcommand\dance{\textit{DANCe}}

\newcommand\logg{$\log g$}

\defcitealias{casagrande10}{C10}

\begin{document}

\title{A Measurement of Radius Inflation in the Pleiades and its Relation to Rotation and Lithium Depletion}
\author{Garrett Somers\altaffilmark{1} and Keivan G.\ Stassun\altaffilmark{1,2}}
\altaffiltext{1}{Department of Physics \& Astronomy, Vanderbilt University, 6301 Stevenson Center Ln., Nashville, TN 37235, USA}
\altaffiltext{2}{Department of Physics, Fisk University, 1000 17th Ave.\ N., Nashville, TN  37208, USA}

\begin{abstract}
    Precise measurements of eclipsing binary parameters and statistical studies of young clusters have suggested that some magnetically active low-mass dwarfs possess radii inflated by $\sim$5--15\% relative to theoretical expectations. If true, this effect should be pronounced in young open clusters, due to the rapid rotation and strong magnetic activity of their most extreme members. We explore this possibility by determining empirical radii for 83 members of the nearby Pleiades open cluster, using spectral energy distribution fitting to establish \fbol\ with a typical accuracy of $\approx$3\% together with color and spectro-photometric indices to determine \teff. We find several Pleiades members with radii inflated above radius-\teff\ models from state-of-the-art calculations, and apparent dispersions in radii for the K-dwarfs of the cluster. Moreover, we demonstrate that this putative radius inflation correlates strongly with rotation rate, consistent with inflation of young stars by magnetic activity and/or starspots. We argue that this signal is not a consequence of starspot-induced color anomalies, binarity, or depth effects in the cluster, employing {\it Gaia} DR1 distances as a check. Finally, we consider the lithium abundances of these stars, demonstrating a triple correlation between rotation rate, radius inflation, and enhanced lithium abundance. Our result---already significant to $\sim$99.99\% confidence---provides strong support for a magnetic origin of the inflated radii and lithium dispersion observed in young, low-mass stars.
\end{abstract}

\section{Introduction}\label{sec:intro}
    
    Precise knowledge of the fundamental parameters of stars is of considerable importance to understanding both their exact nature, their exo-planets, and the timescales of diverse astrophysical phenomenon such as star formation and circum-stellar disk evaporation. Standard stellar theory has been extremely successful at predicting stellar properties throughout the wide and varied life cycles of stars, but their fidelity with regard to the radii of low mass ($M < 1$\msun) stars has been called into question by a consistent trend emerging from direct measurements: young, active stars appear to have radii that are larger than standard predictions by $\sim 5-15$\%. This phenomenon has been claimed in eclipsing binaries \citep[e.g][]{popper97,torres02,lopez-morales05}, statistical studies of open clusters \citep[e.g.][]{jackson16}, on both sides of the fully-convective boundary of $0.35$~\msun\ \citep[e.g.][]{clausen09,Stassun:2012}, and on both the pre-main sequence and main sequence \citep[e.g.][]{torres10,feiden12,torres14,Stassun:2014}. Additionally, corresponding anomalies in the \teffs\ of the afflicted stars have been noted in several instances \citep[e.g.][]{stassun06,dupuy16}. While the underlying cause remains controversial, it has been shown that in some cases the degree of radius inflation appears correlated with the strength of surface magnetic activity \citep[e.g.][]{torres06,lopez-morales07,morales08,Stassun:2012}. Newer generations of theoretical models have begun to consider such effects, and ongoing research is studying their impact on the fundamental parameters, abundances, and evolutionary timescales of young stars \citep[e.g.][]{mullan01,chabrier07,macdonald10,feiden13,feiden14,jackson14a,jackson14b,Somers:2014,Somers:2015b,Somers:2015a}.
    Results have been particularly promising for young stars: the well-known lithium-rotation correlation in pre- and zero-age main sequence clusters \citep[e.g.][]{soderblom93,messina2016,bouvier2016} is a direct prediction of an activity-radius connection on the pre-main sequence \citep[e.g.][]{ventura98,Somers:2015b,Somers:2015a,jeffries2016}.
    
    If an activity-radius connection truly exists, young ($\lesssim 200$~Myr) main sequence open clusters present a valuable test bed to uncover the nature of the correlation. Young clusters generally contain members with extraordinarily rapid rotation, as the magnetic stellar winds which efficiently break stars on the main sequence have not yet significantly depleted the initial stellar angular momentum \citep[e.g.][]{pinsonneault89,gallet2015,somers16b}. Furthermore, young clusters host large dispersions in rotation rate at fixed mass due to the diversity of rotative initial conditions in star-forming regions, and consequently show a range of activity levels and starspot coverage from moderate to extreme \citep[e.g.][]{soderblom93,odell95,gallet2015,fang16}. Such clusters have been the target of several recent statistical studies examining the fundamental parameters of stars \citep{littlefair11,cottaar14,jackson14a,jackson14b,jackson16,lanzafame2016}. In several cases, these studies have concluded both that the average stellar radius is larger compared to standard predictions, and that dispersions likely exist in radius at fixed \teff.

    In particular, \citet{jackson14a} recently examined the well-studied open cluster, the Pleiades, for signs of radius inflation. By measuring the projected radii of a large sample of its members, and statistically analyzing their results, they found that the average Pleiad radius is $\sim 10$\% larger than theoretical predictions below $1$~\msun. The Pleiades is an advantageous laboratory for such experiments, given its proximity \citep[$d \sim 136$~pc,][]{melis14}, near-solar composition of [Fe/H]~$\sim 0.03$ \citep{soderblom09}, and young age of 125~Myr \citep{stauffer1998}. Furthermore, its members more massive than $\sim 0.6$~\msun\ have reached the main sequence, meaning that the harrowing uncertainties dogging pre-main sequence models can be avoided in comparisons with theory \citep[e.g.][]{Stassun:2014}.
    
    In this paper, we apply a distinct method to this same open cluster to search for corroborating signs of radius inflation and dispersion. Our method involves fitting the broadband spectral energy distributions (SEDs) of individual members to establish their bolometric fluxes (\fbol), and combining this result with the known distance of the cluster and an estimate of the \teff, to calculate the stellar radius. This approach is attractive because, in principle, individual stars can be tested for radius inflation, and correlations with non-standard physical effects like activity and rotation can be explored. We devote considerable discussion to the accuracy of our \teffs, as the active and spotted nature of young stars complicates simple extrapolation from photometry. In the end, we find a clear connection between rotation rate and apparent radius inflation above a putative \teff-radius relation.
    
    The paper is organized as follows. In $\S$\ref{sec:methods}, we describe our sample selection, our methods for deriving \fbol\ and \teff\ for our stars, and our procedure for deriving the corresponding radii. In $\S$\ref{sec:results}, we present these radii and compare them with predictions from theoretical models, looking particular at the influence of rotation on the agreement. In $\S$\ref{sec:discussion}, discuss the possible contaminating influence of starspots, binaries, and extinction, and describe the relationship between inflated radii and lithium abundance in this cluster. Finally, we summarize our conclusions in $\S$\ref{sec:summary}.

\section{Methods}\label{sec:methods}

    The familiar astrophysical formulation of the Stefan-Boltzmann law defines the effective temperature of a star as a function of radius and luminosity: $L = 4 \pi R_*^2 \sigma_{\rm SB} T_{\rm eff}^4$. By inverting this equation, the radius of a star can be directly calculated if the \teff\ and luminosity can be determined with fidelity. This is the fundamental method we employ for this study. In this section, we describe how we infer these two properties, and how the associated uncertainties propagate into radius errors. We first select a sample of Pleiads to study in $\S$\ref{sec:selection}, requiring accurate BVK$_{\rm s}$ photometry, spectro-photometric \teffs\ from the \dance\ collaboration \citep{bouy15,barrado16}, and literature lithium abundances. We then discuss our \teff\ derivations in $\S$\ref{sec:teffs}, describe the broadband photometric data employed to derive the bolometric flux in $\S$\ref{sec:photo}, detail our spectral energy distribution fitting technique in $\S$\ref{sec:fitting}, and execute the radius derivations in $\S$\ref{sec:radii}.

\subsection{Sample selection}\label{sec:selection}

    We began with the newly-assembled catalog of literature lithium measurements reported by \citet{barrado16}, who accepted only stars with membership probability $> 0.75$. A small number of Pleiads lie behind an HI cloud, and are thus far more extincted than the rest \citep[e.g.][]{gordon84}; for simplicity, we discard these members. We then cross correlated this sample with the famous $UBV$ photometric catalog of \citet{johnson58}, who produced homogeneous photometry for a large sample of Pleiades members. We further selected stars with measured rotation rates, either from the HATNet collaboration \citep{hartman10}, or the recent analysis of \textit{K2} data \citep{rebull16a,rebull16b,stauffer16}, preferring the later for joint detections. Finally, we queried {\tt VizieR\footnote{\url{http://vizier.u-strasbg.fr/}}} to obtain $K_{\rm s}$-band magnitudes from the \textit{2MASS} catalog \citep{cutri03}, which detected every star in our reduced sample. These criteria produced a total of 83 high-probability cluster members, stretching from early-F to late-K type. We refer to these stars as our ``sample''. Basic parameters for our sample are listed in Table~\ref{tab1}.
    
    For the analysis of $\S$\ref{sec:results}, we exclude known binaries (see $\S$\ref{sec:binaries}), and ignore stars warmer than 6250~K, the approximate temperature where the rotative properties of stars change due to their vanishingly small surface convection zones \citep[e.g.][]{kraft67}.

\subsection{Effective Temperatures}\label{sec:teffs}

    We next establish \teff\ values for our sample. Accurate \teffs\ are crucial for this study, as their errors represent the dominant contribution to the uncertainties in our derived radii. This can be understood by considering the equation relating luminosity, radius, and effective temperature. Re-arranging this equation shows that $R \propto T_{\rm eff}^2 L^{1/2}$, implying that fractional radius errors go like twice the fractional \teff\ errors, but only as half the fractional luminosity errors. The measured \teff\ also sets which theoretical radius value to be used as comparison, and also contributes to errors in the bolometric flux derivation ($\S$\ref{sec:bolofluxes}). 

    There is a well-known problem in the Pleiades related to the colors of the cluster's K-dwarfs. Going back at least to \citet{herbig62}, and continuing through the studies of \citet{jones72}, \citet{stauffer80}, \citet{mermilliod92}, \citet{stauffer03}, and as recently as \citet{covey16}, it has been shown that the Pleiades K-dwarfs are substantially bluer than expected from a calibrated zero-age main sequence. This is generally attributed to the presence of starspots on the photospheres of the young stars (see the discussion in \citealt{stauffer03}), whose accompanying plages emit substantial short wavelength radiation, thus altering the relative fractions of photons within the $B$ and $V$ band-passes. \citet{stauffer03} noted a similar but opposite-sign effect for \VK, which produces a color-magnitude diagram (CMD) too red compared to a less magnetically active cluster (in their case, the $\sim 600$~Myr old Praesepe), presumably a result of the cool starspot surfaces themselves. They finally note that, perhaps as a result of balance between the cool spots and the hotter photosphere/plage regions, spotted stars generally lie in their ``expected'' locations in a \VI\ CMD.
    
    To better understand this phenomenon, \citet{Somers:2015a} constructed evolutionary models of spotted stars and investigated the consequence for different photometric bands. For a 50\% spotted star with an 80\% temperature contrast between the hot and cool regions, appropriate for the inferred properties of Pleiades stars, they found that the stars would appear bluer in \BV, redder in \VK, and approximately unchanged in \VI, in good agreement with the empirical Pleiades results. However, they noted that the \teff\ of a spotted star is always lower than an equivalent star without spots. Consequently, one might expect the true \teff\ of a spotted star to be somewhat cooler than implied by \VI\ and perhaps somewhat warmer or cooler than implied by \VK\ band depending on the properties of the spots, but far cooler than implied by \BV.
    
    With these considerations, we believe \VI\ photometry is the ideal photometric index for deriving temperatures for this study, because a \VI-derived \teff\ may be interpreted as a close upper limit to the true temperature. This is advantageous as it results in a \textit{lower limit} on the true radius, useful for detecting radius inflation. However, reliable $I_C$-band photometry does not exist for the majority of our sample. This is remarkable, given the extensive scrutiny placed on the Pleiades for over a century. The recent and exquisite $I_C$-band data of Kamai et al. (2014) overlaps with only two of our Li-rotation-selected stars, and although there is greater overlap with the comprehensive catalog of Stauffer et al. (2007), many of their reported $I_C$ band values are of mixed origin and uncertain calibration.
    
    \begin{figure*}[!ht]
    \centering
    \includegraphics[width=6.0in]{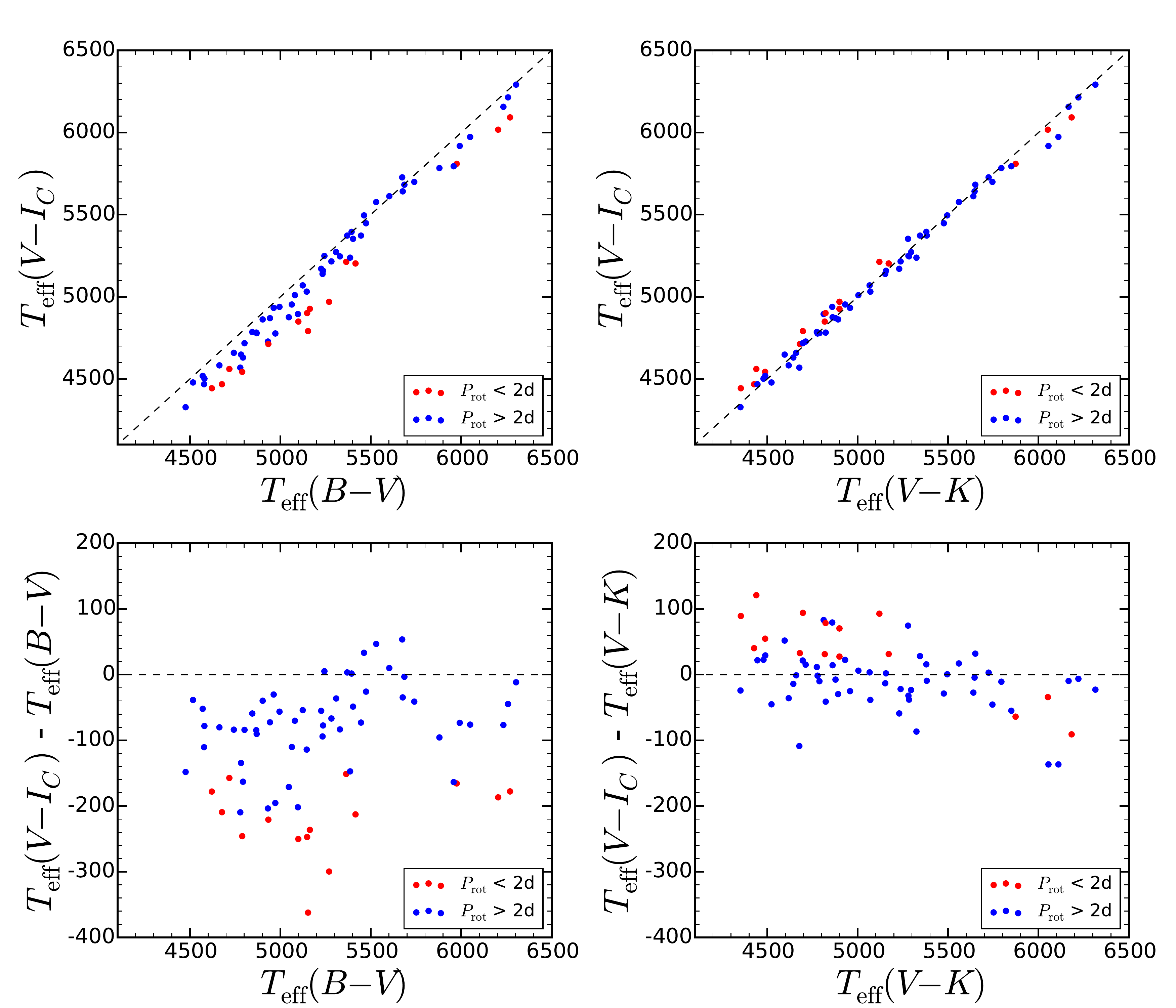}
    \caption{Comparisons between \teff\ values derived with different color combinations, from the data of Kamai et al (2014). \textit{Top left:} \BV\ and \VI\ \teffs\ are compared with one another. Each point represents a single star, grouped into fastest (red) and slower (blue) rotators. \BV\ values are consistently hotter than \VI\ values, and the problem is exacerbated for the fastest rotating stars. \textit{Bottom left:} The difference between the two \teff\ derivations as a function of \BV\ temperature. The agreement is extremely poor overall, demonstrating the unreliability of \BV\ temperatures of spotted stars. \textit{Top right:} Same as top left, for \VK\ and \VI. The agreement is generally much better between these bands. \textit{Bottom right:} Same as bottom left, for \VK\ and \VI. Essentially no systematic offset exists for the slow rotators, and only a mild one for the fastest rotators. This figure attests to similarity of \teffs\ derived from \VI\ and \VK\ photometry of spotted stars.}
    \label{fig:teffcomp}
    \end{figure*}
    
    Instead, we elected to proceed with \teffs\ derived from \VK\ photometry. While the above discussion suggests that \VK\ photometry produces \teffs\ of uncertain accuracy, we find that for Pleiades stars, \teffs\ derived from \VK\ and \VI\ photometry are quite similar. As we expect these \teffs\ to approximately bracket the possible \teff\ range, their similarity is an encouraging sign that our \teffs\ are reasonable estimates.

As a demonstration, we obtained the $BVI_C$ data set of \citet{kamai14}, and cross correlated their members with both {\it 2MASS} and the joint HATNet/\textit{K2} rotation data set ($\S$\ref{sec:selection}), to obtain $K_S$ photometry and rotation rates. We assume a reddening of $E(B-V)$ = 0.04 \citep{an07} and a selective extinction of $R_{\rm V}$ = 3.1 which, with a standard \citet{Cardelli:1989} reddening law, gives $E(V-I_C)$ = 0.06 and $E(V-K_S)$ = 0.11. We then derived \teffs\ from the infrared flux method (IRFM) calibration of \citet[][C10 hereafter]{casagrande10}, and examined the resulting \teffs\ in Fig. \ref{fig:teffcomp}. The left column compares \BV\ and \VI\ \teffs\ for stars with rotation periods slower than 2~days (blue points), and faster than 2~days (red points), in both absolute (top) and differential (bottom) terms. Considering first the left panels, \BV\ photometry implies substantially higher \teffs\ than does \VI\ for nearly every star, with the fastest rotating objects showing the greatest departure. Significantly, the dispersion reaches several hundred Kelvin between 4500 and 5500K. This illustrates the adverse impact of magnetic activity on \BV\ photometry. By contrast, the right compares \VI\ and \VK\ colors, showing good agreement for the slowest rotators, and a far weaker systematic offset amongst the fastest rotators (65~K on average), below $\sim 5800$~K. This agreement may be somewhat better than previously thought, perhaps as a consequence of the weaker shift in \teff\ for a given change in \VK\ when compared to \teff\ shift for an equal change in \BV. The similarity of these two temperatures, in the context of expectations from the \citet{Somers:2015a} models, gives us confidence in the general reliability of our \teffs. We consider how large the systematic errors bars might be in $\S$\ref{sec:starspots}.

    As a further test, we fit for the offset between \VI\ and \VK-derived \teffs\ in the combined \citet{kamai14} and \citet{hartman10} sample as a function of rotation rate, and ``corrected'' our \VK\ \teffs\ to the new scale. To do this, we subdivided our stars into a ``hot'' and ``cool'' sample, taking 5500~K as the dividing point. This is because the hotter and cooler stars show opposite-sign relationships with rotation (see section 3.4 in \citealt{kamai14}). We then fit second order polynomials to the difference between the \VI\ and \VK\ \teffs\ as a function of the later. Finally, we applied these offsets to our derived \VK\ \teffs\ to obtain estimated \VI\ \teffs. Performing the analysis outlined in this paper with these \VI\ values does not influence our results or conclusions in any substantial way, increasing our confidence in our results. 

    Given these justifications, we proceed with \teffs\ from the \VK\ data of the combined {\it 2MASS} and \citet{johnson58} photometric catalogs using the color-\teff\ relations of \citetalias{casagrande10}. For this conversion, we adopt a cluster [Fe/H] = 0.03 \citep{soderblom09}. Propagated errors result from formal uncertainties on the colors, the systematic offsets between the true \teff\ scale at the \citetalias{casagrande10} scale (quoted at 25~K), and uncertainties on the reddening and metallicity. These are likely lower limits on the errors, as the departure from pristine surfaces will somewhat affect the calibration. Our values are shown in Table~\ref{tab2}.

    As a check on the validity of our results, we perform in parallel the forthcoming analysis with \teffs\ derived from an independent method. As a comparison sample, we extract \teff\ values from the recent \dance\ analysis of Pleiades members with literature Li values \citep{barrado16}. These values were obtained by fitting single-temperature spectral models to the SEDs of their targets, and minimizing the chi-square residual. The authors quote a typical error of 125~K, which we adopt in every case. These values are listed in Table~\ref{tab3}. In what follows, we report results with both \VK\ and \dance\ \teffs.

\subsection{Bolometric Fluxes}\label{sec:bolofluxes}

    Next, we derive bolometric fluxes by considering the full SED of our target stars. In this section, we discuss first the provenance of our SED data, and second our measurement procedure.

\subsubsection{Broadband photometric data from the literature}\label{sec:photo}
    In order to systematize and simplify our procedures, we opted to assemble for each star the available broadband photometry from the following large, all-sky catalogs (listed here in approximate order by wavelength coverage) via the {\tt VizieR} query service: 
\begin{itemize}
    \item {\it GALEX} All-sky Imaging Survey (AIS): FUV and NUV at $\approx$0.15 \micron\ and $\approx$0.22 \micron, respectively. 
    \item Catalog of Homogeneous Means in the $UBV$ System for bright stars from \citet{Mermilliod:2006}: Johnson $UBV$ bands ($\approx$0.35--0.55 \micron).
    \item {\it Tycho-2\/}: Tycho $B$ ($B_{\rm T}$) and Tycho $V$ ($V_{\rm T}$) bands ($\approx$0.42 \micron\ and $\approx$0.54 \micron, respectively). 
    \item Str\"omgren Photometric Catalog by \citet{Paunzen:2015}: Str\"omgren $uvby$ bands ($\approx$0.34--0.55 \micron).
    \item AAVSO Photometric All-Sky Survey (APASS) DR6 (obtained from the UCAC-4 catalog): Johnson $BV$ and SDSS $gri$ bands ($\approx$0.45--0.75 \micron). 
    \item Two-Micron All-Sky Survey (2MASS): $JHK_S$ bands ($\approx$1.2--2.2 \micron). 
    \item All-WISE: {\it WISE1--4} bands ($\approx$3.5--22 \micron). 
\end{itemize}

    We found $BV$, $JHK_S$, and {\it WISE1--3} photometry---spanning a wavelength range $\approx$0.4--10~\micron---for nearly all of the stars in our study sample. Most of the stars also have {\it WISE4\/} photometry, and many of the stars also have Str\"omgren and/or {\it GALEX\/} photometry, thus extending the wavelength coverage to $\approx$0.15--22~\micron. We adopted the reported measurement uncertainties unless they were less than 0.01 mag, in which case we assumed an uncertainty of 0.01 mag. In addition, to account for an artifact in the Kurucz atmospheres at 10~\micron, we artificially inflated the {\it WISE3\/} uncertainty to 0.1 mag unless the reported uncertainty was already larger than 0.1 mag. The assembled SEDs are presented in Appendix \ref{sec:sed_appendix}. 

\subsubsection{Spectral energy distribution fitting\label{sec:fitting}} 

    We followed the SED fitting procedures described in \citet{Stassun:2016}.
    Briefly, the observed SEDs were fitted with standard stellar atmosphere models from \citet{Kurucz:2013}. 
    As summarized in Tables~\ref{tab2}-\ref{tab3}, for each star we have \teff, and we assume a main-sequence \logg\ and solar metallicity. We interpolated in the model grid to obtain the appropriate model atmosphere for each star in units of emergent flux. To redden the SED model, we adopted the interstellar extinction law of \citet{Cardelli:1989}. We then fitted the atmosphere model to the flux measurements to minimize $\chi^2$ by varying just two fit parameters: extinction ($A_{\rm V}$) and overall normalization (effectively the ratio of the stellar radius to the distance, $R_\star^2 / d^2$). (The adopted stellar \teff\ also has an associated uncertainty; this is handled in a later step via the propagation of errors through the stellar angular radius, $\Theta$; see Section \ref{sec:radii}.) We allowed the fit to vary $A_{\rm V}$ within the generally accepted reddening range of $E(B-V) = 0.04 \pm 0.01$ \citep[e.g.,][]{an07}.  The best-fit model SED with extinction is shown for each star in Appendix \ref{sec:sed_appendix}, and the reduced $\chi^2$ values ($\chi_\nu^2$) are given in Tables~\ref{tab2}-\ref{tab3}. 

    The primary quantity of interest for each star is \fbol, which we obtained via direct summation of the fitted SED, {\it without} extinction, over all wavelengths. The formal uncertainty in \fbol\ was determined according to the standard criterion of $\Delta\chi^2 = 2.30$ for the case of two fitted parameters \citep[e.g.,][]{Press:1992}, where we first re-normalized the $\chi^2$ of the fits such that $\chi_\nu^2 = 1$ for the best fit model. Because $\chi_\nu^2$ is in almost all cases greater than 1 (see Tables~\ref{tab2}-\ref{tab3}), this $\chi^2$ renormalization is equivalent to inflating the photometric measurement errors by a constant factor and results in a more conservative final uncertainty in \fbol\ according to the $\Delta\chi^2$ criterion. 
    
\begin{figure}[!t]
    \centering
    \includegraphics[width=3.5in]{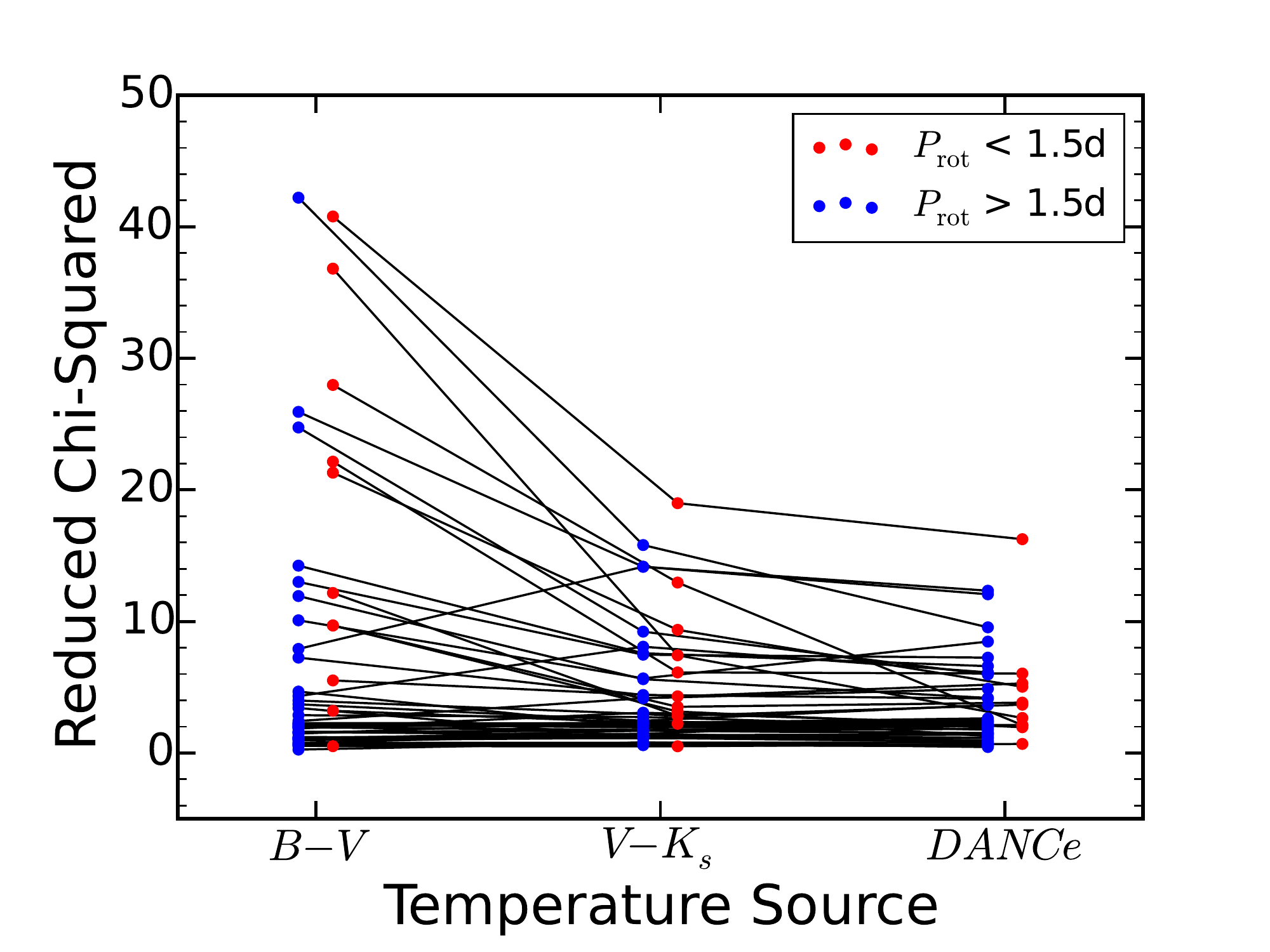}
    \caption{Comparison of the $\chi_\nu^2$ values arising during our SED fitting procedure ($\S$\ref{sec:fitting}), for the rapid (red) and slow (blue) members of our sample. Rapid and slow samples have been slightly offset for visibility. \BV\ temperatures perform the worst on the whole, producing the largest number of extremely poor fits ($\chi_\nu < 20$), whereas \VK\ and \dance\ temperatures produce far better. This attests to the poorness of \BV\ \teffs\ in spotted stars, and the superiority of the \VK\ and \dance\ values.}
    \label{fig:chisquare}
    \end{figure}

    Fig.~\ref{fig:chisquare} illustrates the resulting $\chi_\nu^2$ values resulting from the three \teff\ scales discussed in $\S$\ref{sec:teffs}, distinguishing between the faster (red) and slower (blue) rotating subsets of our sample. The \BV\ temperatures produce many extremely poor fits, with nearly half the rapid rotators falling above $\chi_\nu^2$ = 20. By contrast, these outliers generally produce acceptable fits when the \VK\ and \dance\ scales are employed, showing little difference in the $\chi_\nu^2$ distribution of the slow and rapid rotators. This serves as further justification for our choice of \teff\ scale.
    
    
\subsection{Radius Derivation and Errors\label{sec:radii}}

    Finally, the stellar radius can be derived by combining \fbol\ and \teff\ into an angular radius $\Theta$, and multiplying by the distance to the star, $d_*$. $\Theta$ is related to our derived quantities by Eq. \ref{frt},
    
    \begin{equation}\label{frt}
    \mathcal{F}_{bol} = \Theta^2 \sigma_{\rm SB} T_{\rm eff}^4,
    \Theta = \sigma_{\rm SB}^{-0.5} \mathcal{F}_{bol}^{0.5} T_{\rm eff}^{-2},
    \end{equation}
    
    \noindent where $\sigma_{\rm SB}$ is the Stefan-Boltzmann constant. Contributions to the error budget of \rstar\ thus come from random and systematic errors on \teff, uncertainties in \fbol, and uncertainties in the distances of each individual star. We adopt for our purposes a distance of 134 $\pm$ 3 pc, based on \citet{soderblom05}, where the uncertainty can be thought of as reflecting the depth of the cluster. As we are searching for differential signals, the distance we adopt has very little influence on our results. However, differences in the distances to individual Pleiads can, and does, influence our answer. The recent {\it Gaia} data release 1 contained distances to several members of our sample, allowing us to verify that no significant distance outliers exist. We discuss this issue in $\S$\ref{sec:nonmembers}. Our final radii are listed in Tables~\ref{tab2} and \ref{tab3}. The formal errors on the radii are typically 3-5\%, though this does not account for potential offsets from the fundamental \teff\ scale (see $\S$\ref{sec:starspots}).
    
\section{Results}\label{sec:results}

\subsection{Angular Radii}\label{sec:angradii}

\begin{figure}[!t]
    \centering
    \includegraphics[width=\linewidth]{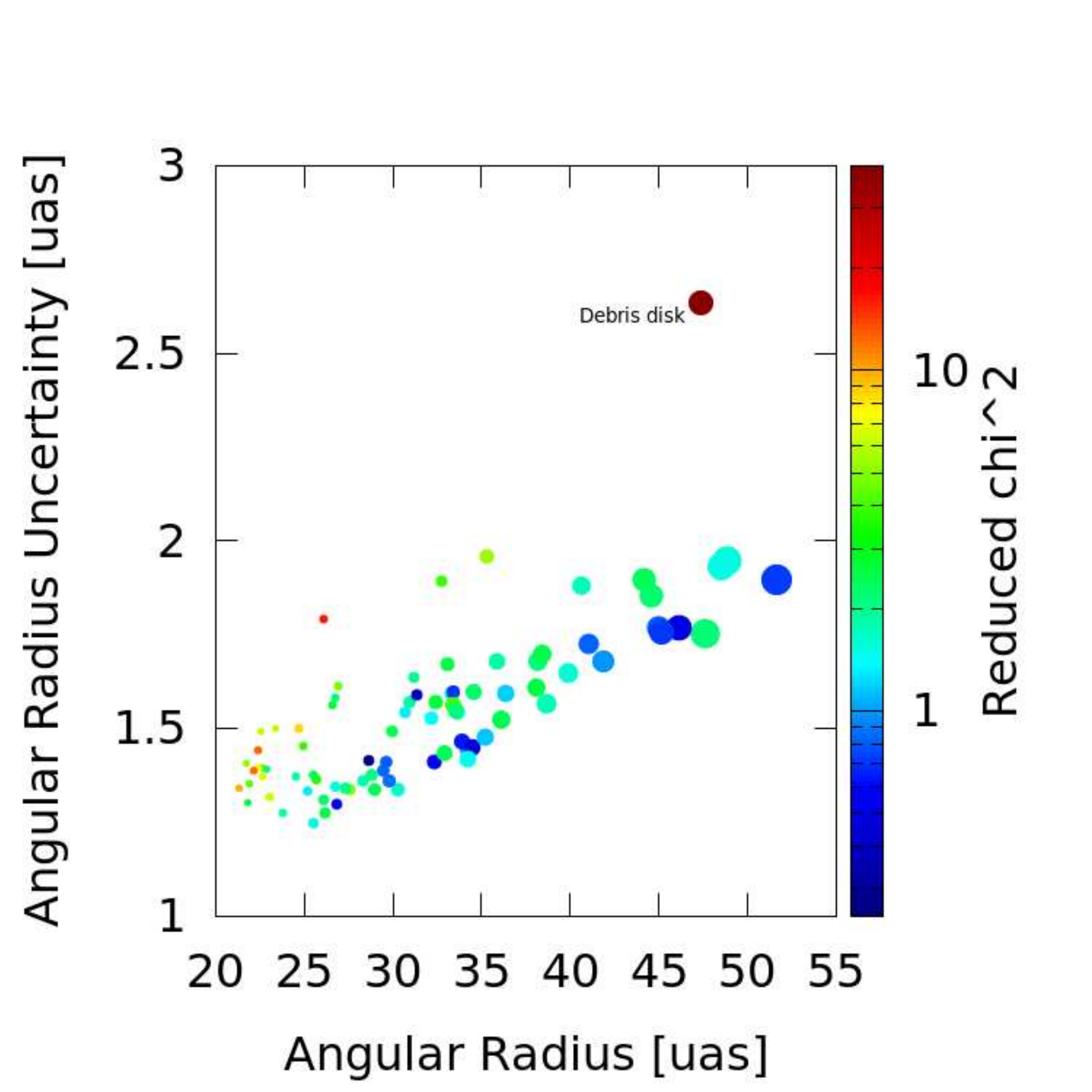}
    \caption{Stellar angular radii, $\Theta$ and uncertainties (in units of $\mu$as) newly derived in this work. Color represents the $\chi_\nu^2$ of the SED fit. Symbol size is proportional to \fbol. One star with a debris disk and $\chi_\nu^2 > 20$ is labeled and excluded from analysis (see the text).}
    \label{fig:theta}
    \end{figure}

    A fundamental product of this work is a measurement of the angular radius, $\Theta$, of each star in our sample, determined empirically from the measured \fbol\ and \teff. We report the resulting $\Theta$ values and their uncertainties in Tables~\ref{tab2}-\ref{tab3}. Figure~\ref{fig:theta} presents the distributions of $\Theta$ and their uncertainties in relation to the $\chi_\nu^2$ of the SED fits and the \fbol\ that result from the SEDs. For simplicity, this figure shows only the \dance\ \teff\ sample. The median precision achieved on \fbol\ is $\approx$3\% which, together with the typical uncertainty on \teff\ of 2--3\%, results in a median precision on $\Theta$ for this sample of 1.5~$\mu$as, or $\approx$5\%, dominated by the uncertainty on \teff. The typical uncertainly for the \VK\ derivation is closer to $\approx$3\%, due to the smaller formal \teff\ errors.

    We note that the poor-fit outlier in Fig.~\ref{fig:theta} is HII 1132. This star hosts a well-known debris disk \citep[e.g][]{spangler01}, and the resulting IR-excess is evidently the cause of the poor fit (Fig. \ref{fig:seds_7}). However, given its \teff\ of 6400-6500~K, it is already excluded from the forthcoming analysis, and thus does not present a problem.

\subsection{Stellar Radii}\label{sec:infradii}

    \begin{figure*}[!ht]
    \centering
    \includegraphics[width=6.0in]{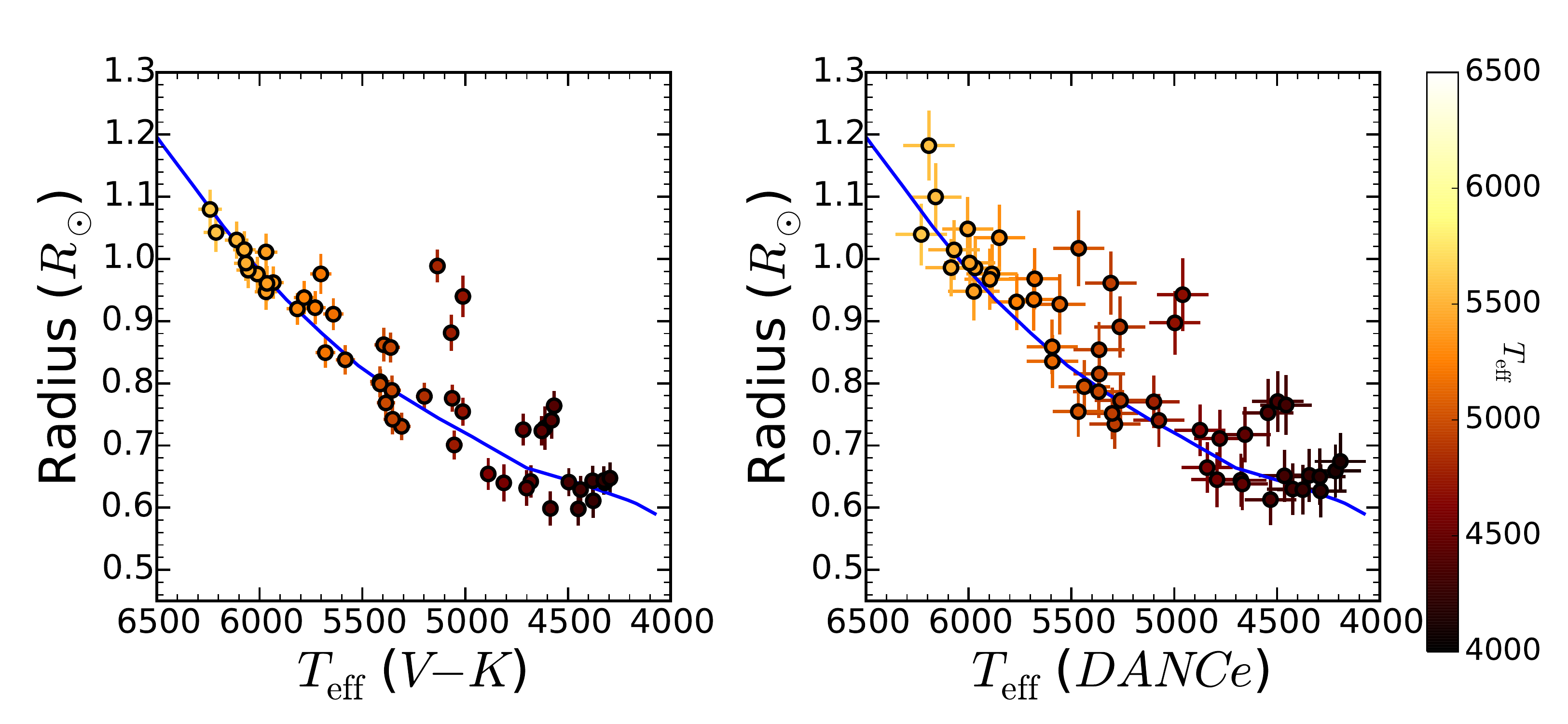}
    \caption{A comparison between the radii of our stars, derived from \VK\ (left) and \dance\ (right) temperatures according to the procedure outlined in $\S$\ref{sec:methods}, and the stellar isochrones (blue line) of Bressan et al (2012). There is little evidence for a radius dispersion above $\sim 5700$~K, but significant scatter below this value for both \teff\ proxies.}
    \label{fig:teffradius}
    \end{figure*}

    Using these angular sizes, we next derive the radii of the Pleiads using the method describe in $\S$\ref{sec:radii}, and compare them to the \teff-radius relation from the Padova isochrones\footnote{http://stev.oapd.inaf.it/cgi-bin/cmd} \citep{bressan12,chen2014}, calculated for Pleiades age and metallicity. These isochrones, like all standard evolutionary tracks, are designed to model stars with negligible magnetic and rotative effects, and have accordingly been calibrated on the relatively quiet, inactive Sun. This procedure generally involves setting the free parameters of the stellar model, namely the convective mixing length and the solar helium abundance, to values which reproduce observed properties of the Sun, like its radius and luminosity (see \citealt{bressan12} for details). Once calibrated, these models make specific predictions for stellar properties as a function of mass, composition, and age. To verify the reliability of these models, we compared a 1~Gyr, solar metallicity Padova isochrone to the catalogs of single, main-sequence stars with interferometric radii from \citet{boyajian2012} and \citet{boyajian2013}. We found that despite the age, metallicity, evolutionary state, and parallax accuracy variance in the Boyajian sample, the isochrone traced well the lower envelope of stars in the \teff-Radius plane, demonstrating the generic reliability of the Padova calibration. In this context, discrepancies between the radii of young, active stars and isochrone predictions could be interpreted as signatures of radius inflation driven by rapid rotation, magnetic activity, and starspots, particularly when radius dispersion is present at fixed \teff.
    
    Our comparison appears in Fig. \ref{fig:teffradius}. The left panel shows the \teffs\ and radii derived from \VK, and the right shows the \dance\ temperatures and resulting radii. For both temperature scales, the hottest stars (\teff$> 6000$~K) cluster near the isochrone, and within the errors show little sign of dispersion. This suggests that, at least among the hottest stars, our derived radii are reasonably accurate, and that Pleiads in the mass range $\sim 1.1-1.2$\msun\ are generally un-inflated. For stars cooler than $\sim 5700$~K, the picture changes. While many of these stars still straddle the \teff-radius relationship predicted by Padova, several lie significantly above this line, and indeed at larger radii than other stars of equivalent \teff. 

    It is notable that the onset of scatter in stellar radii sets in around the same \teff\ as in the inflated radii measurements of \citet{jackson14a}. These authors combined spot modulation periods and {\it v}~sin~{\it i} values to measure the average Pleiad radius from projected stellar radii. They found that the average radius is $\sim 10$\% larger than stellar models for approximately 1\msun\ and below. This mass corresponds to $\sim 5700$~K in the Padova isochrones determined for Pleiades parameters, and it is indeed around this temperature at which star-to-star scatter first becomes evident in our derivations, demonstrating good agreement between the methods. This \teff\ also corresponds to the onset of extraordinary surface activity among the fastest rotators in the cluster \citep[e.g.][]{micela99}. This suggests a connection between activity and the anomalous derived radii, as suggested by many authors (see $\S$\ref{sec:intro}). If physical, this could be a consequence of the greater rapidity of rotation among the fastest K-dwarfs when compared to G- and F-dwarfs in the cluster, or could relate to the changing magnetic properties as convection zones deepen. 

\subsection{Radius Inflation and its Relation to Rotation}\label{sec:radiusrotation}

    \begin{figure*}[!ht]
    \centering
    \includegraphics[width=6.0in]{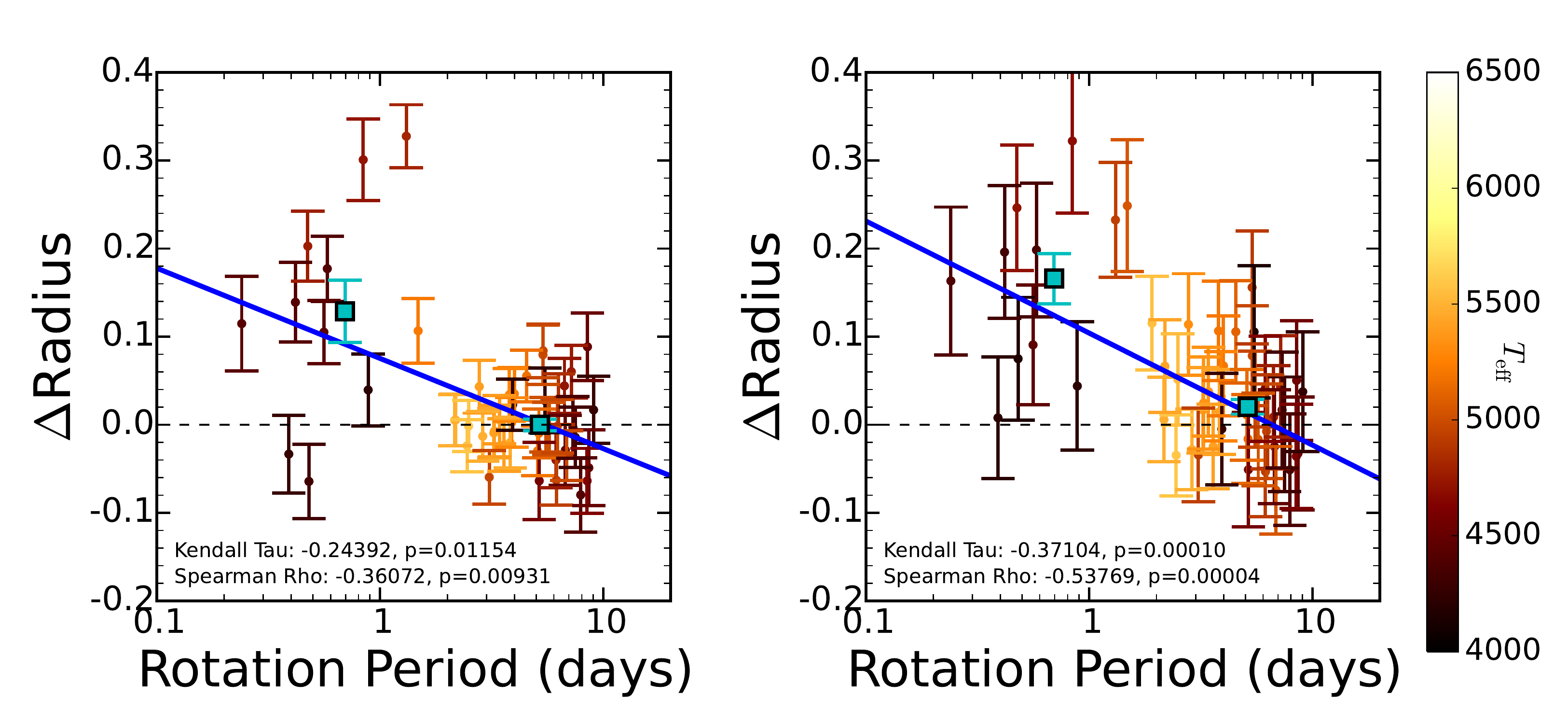}
    \caption{A comparison between the rotation period of stars in our sample and their fractional height above the \citet{bressan12} isochrone for \VK\ and \dance\ temperatures. In both panels, stars rotating slower than 1.5~days show good agreement with predictions, but faster rotating stars are systematically large by on average 10-20\%. The purple squares shows the average \delR\ among the slower and faster stars, divided at 1.5~days. The trend is statistically significant in both bins according to Kendall's $\tau$ and Spearman's $\rho$ coefficients. This suggests that rapid rotation drives radius inflation, perhaps through the influence of correlated starspots and magnetic activity.}
    \label{fig:rotdradius}
    \end{figure*}

    Given this apparent connection with magnetic activity, the exquisite Pleiades rotation data provided by HATNet \citep{hartman10} and {\it K2} \citep{rebull16a,rebull16b,stauffer16} offer an interesting comparison point. To explore this, we first calculated the fractional height above the isochrone of each data point from Fig. \ref{fig:teffradius}, using Eq. \ref{eqn:delRad}.

    \begin{equation}\label{eqn:delRad}
    \Delta R_* = \frac{R_* - R_{\rm isoc}}{R_*},
    \end{equation}
    
    where $R_{\rm isoc}$ is the radius predicted by the \citet{bressan12} isochrones at the calculated \teff\ of each stars. We then compared each value to the spot-modulation rotation periods collected in Table \ref{tab1}. The results appear in Fig. \ref{fig:rotdradius}. For stars rotating at slower than 1.5~day periods, the values cluster nicely around \delR\ = 0, with an average value (cyan square) consistent with standard expectations. The scatter around \delR\ = 0 is consistent with Poisson noise, with an R.M.S. equal to the size of the error bars on the data points. On the other hand, the faster rotating stars show a clear systematic trend towards larger \delR, with values ranging from $\sim$0-30\% larger than the Padova prediction. The average calculated radius inflation for those with faster than 2~day rotation periods is $12.3 \pm 3.3$\% for the \VK\ \teffs, and $15.6 \pm 2.8$\% for the \dance\ \teffs, showing good agreement between our \teff\ metrics. The significance of the anti-correlation is tested with both Kendall' $\tau$ and Spearman's $\rho$ tests, showing strong correlations in both panels (blue lines).
    
    These conclusions are unaltered if we transform the \VK\ temperatures into \VI\ temperatures as described in $\S$\ref{sec:teffs}, and redo the analysis. We find in that case an average \delR\ among the rapid rotators of $10.0 \pm 3.3$\%, with essentially no change in the p-values of the significance tests. The slightly lower average \delR\ reflects the fact that \VI\ photometry gives marginally higher \teffs\ for rapidly rotating Pleiades, by about 65~K ($\S$\ref{sec:teffs}). We also re-derived \delR\ values using the isochrones of \citet{baraffe15}, and with isochrones derived with our own YREC evolution code (see \citealt{Somers:2015b} for details), to determine whether the offset could be due to our choice of comparison models. We find that the overall normalization is affected by isochrone choice, largely as a consequence of the adopted solar abundance, but in each case the rapid rotator average remained larger by $\sim 10$\% compared to the slow rotators, and the significance of the \delR\ correlation with rotation rate was unchanged.
    
    \begin{figure*}[!ht]
    \centering
    \includegraphics[width=6.0in]{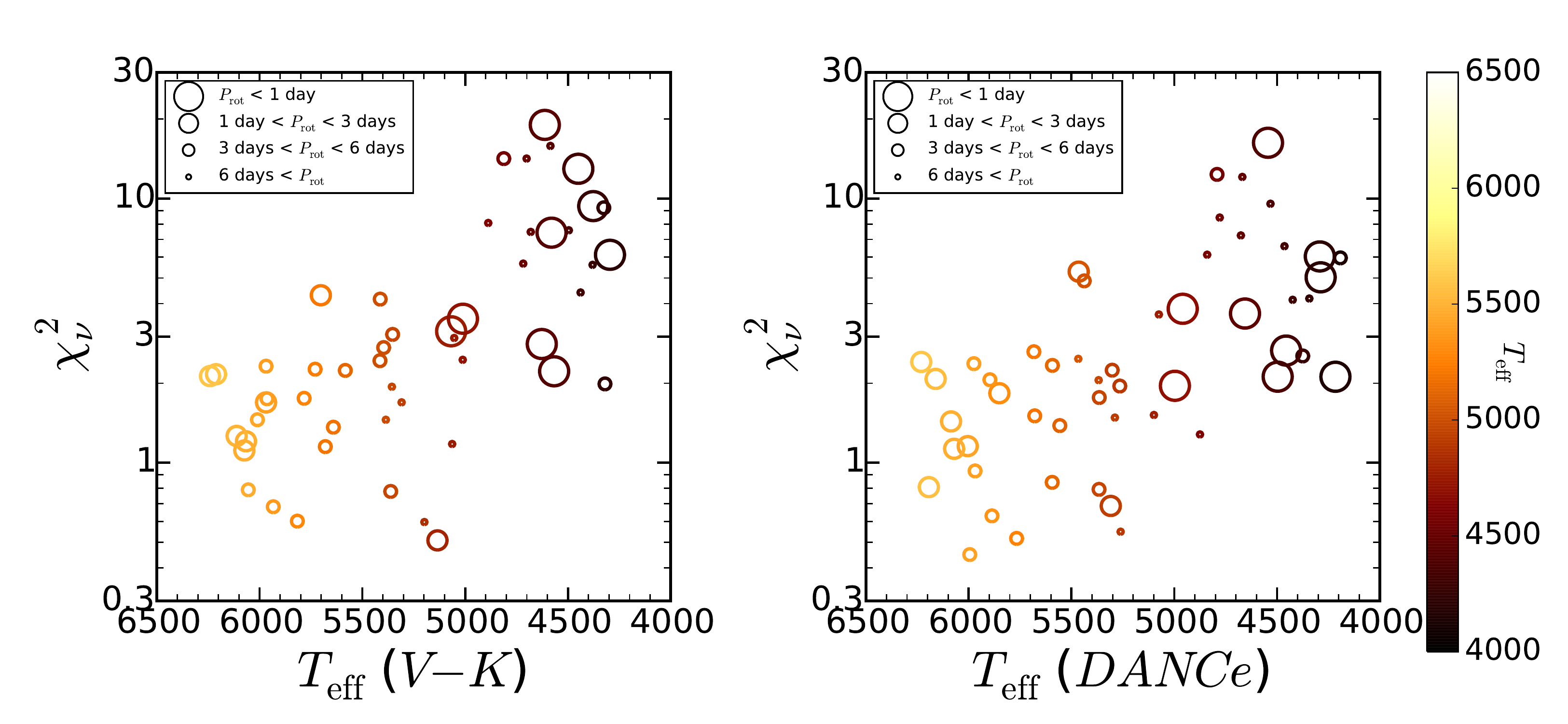}
    \caption{The $\chi_{\nu}^2$ residuals for our SED fits, plotted as a function of temperature for our \VK\ and \dance\ \teffs. The size of the circles reflects rotation rate, as indicated by the key. There is little connection between rotation rate and SED residuals, though a clear connection exists between \teff\ and the average goodness-of-fit.}
    \label{fig:chiteffrot}
    \end{figure*}
    
    To ensure that this result does not stem from systematically poor fitting of the rapid-rotator SEDs, we plot in Fig. \ref{fig:chiteffrot} the $\chi_{\nu}^2$ residuals of each \fbol\ calculation as a function of \teff. The size of each point reflects its rotation rate, given by the inset key. There exists a clear trend of poorer average fits towards lower temperatures. Above $\sim 5000$~K, the stars show generally good fits, but cooler objects are significantly worse.
    Whether this is a consequence of generic deficiencies in the atmosphere models or related to intrinsic properties of the SEDs of young, cool stars is not clear, and deserves attention in future work. However, there is little evidence that the fastest rotators have worse \fbol\ fits, and consequently does not influence our results due to the differential nature of the radius inflation signal.

    It is interesting that the rapid rotators do not follow a clear inflationary trend with rotation rate, but instead scatter between consistent with expectations and $\sim 30$\% inflation. This could simply be due to observational errors, as the standard deviation about the mean for the fast rotators in the \dance\ sample is comparable to the error bar size. This is not the case for the \VK\ values, but these error bars may be under-estimates as the presence of starspots adds additional uncertainty. If indeed there is scatter above a given rotation threshold, this behavior would mirror additional properties of saturated stars, such as their spread in activity \citep{pizzolato03,wright11,argiroffi16}, spot modulation amplitudes \citep{covey16}, and inferred starspot coverage \citep{fang16}. It is also notable that among the inflated stars, none are hotter than $\sim 5700$~K. This may reflect the fact that higher mass stars are far more likely to have spun down to the slow sequence, leaving a paucity of extremely rapid rotators among the G-dwarfs by Pleiades age.

\section{Discussion}\label{sec:discussion}

\subsection{Potential Sources of False Signals}

    While a truly inflated radius would be the most interesting conclusion of our investigation, it is important to consider potential sources of false positives for over-inflated stars. In this section, we discuss the possibilities of systematic \teff\ offsets due to starspots, contamination by binaries, and problems with differential extinction. 

\subsubsection{Starspots and Plages}\label{sec:starspots}

    Perhaps the most pressing concern is the distortion of the stellar SED by starspots and plages. The presence of cool regions on the surface has long been known to alter the shape of the stellar spectrum, thus leading to anomalous colors relative to quiet, inactive stars (see $\S$\ref{sec:teffs}). When using a standard color-\teff\ transformation, this can result in a systematic \teff\ offset which depends on the level of spot coverage, the spot temperature distribution, and even the phase of the stellar rotation. In $\S$\ref{sec:teffs}, we discussed in detail our efforts to minimize the impact of these effects, including performing our analysis with two distinct \teff\ scales, and comparing our resulting values with more reliable \VI\ photometry. However, a strong underlying systematic \teff\ shift among the fastest rotators could plausible still impact our results. In this section, we discuss how large such an offset must be in order to produce a spurious rotation-radius correlation, and estimate whether such an offset is plausible.
    
    For \teff\ errors to mimic a systematic increase in radius of $\sim 12$\% among rapid rotators, our derived temperatures for these must be systematically low by $\sim 6$\%, but accurate for stars rotating at slower than 1.5~d periods. Given the typical \teff\ of our inflated targets of 4750~K, we calculate that this offset must be approximately 230~K on average. This is a rather extreme shift, and in fact approaches the difference in \teff\ derived between \BV\ and \VK\ photometric bands for the individual rapid rotators (Fig. \ref{fig:teffcomp}). As \BV\ \teffs\ are thought to be the least accurate ($\S$\ref{sec:teffs}), and our rapid rotator \VK\ temperatures only differed from the \VI\ scale by $\sim 65$~K, this seems an implausibly large shift based on distorted colors. Even taking an average of \teffs\ derived from \BV\ and \VK, and redoing the analysis of $\S$\ref{sec:results} still produces an average radius inflation of $\sim 7$~\% among rapid rotators, and given the stronger reliability of \VK\ temperatures, one would expect this to represent a lower limit on the true strength of the rotation-\delR\ correlation.
    
    However, a systematic offset of the required magnitude ($\sim 230$~K) could conceivably arise from a substantial offset in the color-\teff\ relations from \citet{casagrande10} in the regime of highly active stars. This offset would have to affect the most rapidly rotating stars to a much greater degree than those with periods less than 1.5~days, as our inflationary signal is ultimately a differential sign rather than an absolute sign. We cannot exclude this possibility at present, but we can determine whether the \citet{casagrande10} scale is uniquely subject to a systematic of this nature. To do this, we re-derived our \VK\ temperatures using the empirically calibrated isochrones of \citet{an07}, and compare them to our adopted values. We find that the new scale produces somewhat cooler \teffs\ for the lower mass stars and somewhat warmer \teffs\ for the higher mass stars, a directionality which would actually {\it enhance} the observed radius-rotation correlation in Fig. \ref{fig:rotdradius}. On the whole, the average offset between the two scales for our sample is $6 \pm 52$~K, showing overall good agreement. A similar exercise with the empirical isochrones of young stars determined by \citet{pecaut13} again produced commensurate results, with average offsets of $-17 \pm 6$~K and $-4 \pm 41$~K, using their pre-main sequence and main sequence isochrones, respectively. As the \citet{pecaut13} tracks were calibrated on active spotted stars, this agreement greatly strengthens our belief in the accuracy of the IRFM-derived \teffs\ we have used. Finally, we note that the good agreement between the \dance\ temperatures and our color-derived temperatures would be surprising if systematic offsets afflicted our color temperatures.
    
    On balance, it is our judgement that starspots have a small impact on our derived \teffs, perhaps as much as 100~K from the \VK\ values, but do not distort our results enough to produce a spurious rotation-radius correlation. However, our analysis would benefit greatly from advances in understanding the spectral energy distributions of rapidly rotating, heavily spotted stars.
    
\subsubsection{Binaries}\label{sec:binaries}

    Another false-positive source are binaries, which can impact our radius derivations in two ways. First, for near-equal-mass binaries, \fbol\ will be significantly increased with little change in inferred \teff. A doubling in \fbol\ corresponds to a $\sim 40$\% increase in \rstar, thus appearing as an inflated object. Second, for binaries with a significantly lower mass companion, \fbol\ will be only marginally affected, but the long wave-length emission will be enhanced by the lower mass companion peaking in the near-IR, leading to a lower inferred temperature. Consequently, a larger \rstar\ will be inferred, as a lower \teff\ star demands a large emitting surface at fixed \fbol.
    
    While these would be extremely problematic in an unfiltered sample, the \dance\ stars used in our selection already attempted to exclude known binaries. To supplement this initial cut, we searched the literature for additional information on binarity, drawing from \citet{mermilliod92}, \citet{bouvier97}, \citet{sestito05}, \citet{margheim07}, and \citet{hartman10} some of whom collate previous binary catalogs. We find an additional 22 objects out of our sample of 83 for which one or more of these publications indicates possible binarity. Each of these stars have been excluded from the analysis plots of $\S$\ref{sec:results}. We thus expect the contamination fraction of binaries, or at least of known binaries, to be small. Moreover, as the inflated stars described in $\S$\ref{sec:radiusrotation} were all among the most rapidly rotating, it would require an extraordinary cosmic conspiracy were the signal a product of unknown binaries.
    
\begin{figure}[!t]
    \centering
    \includegraphics[width=3.0in]{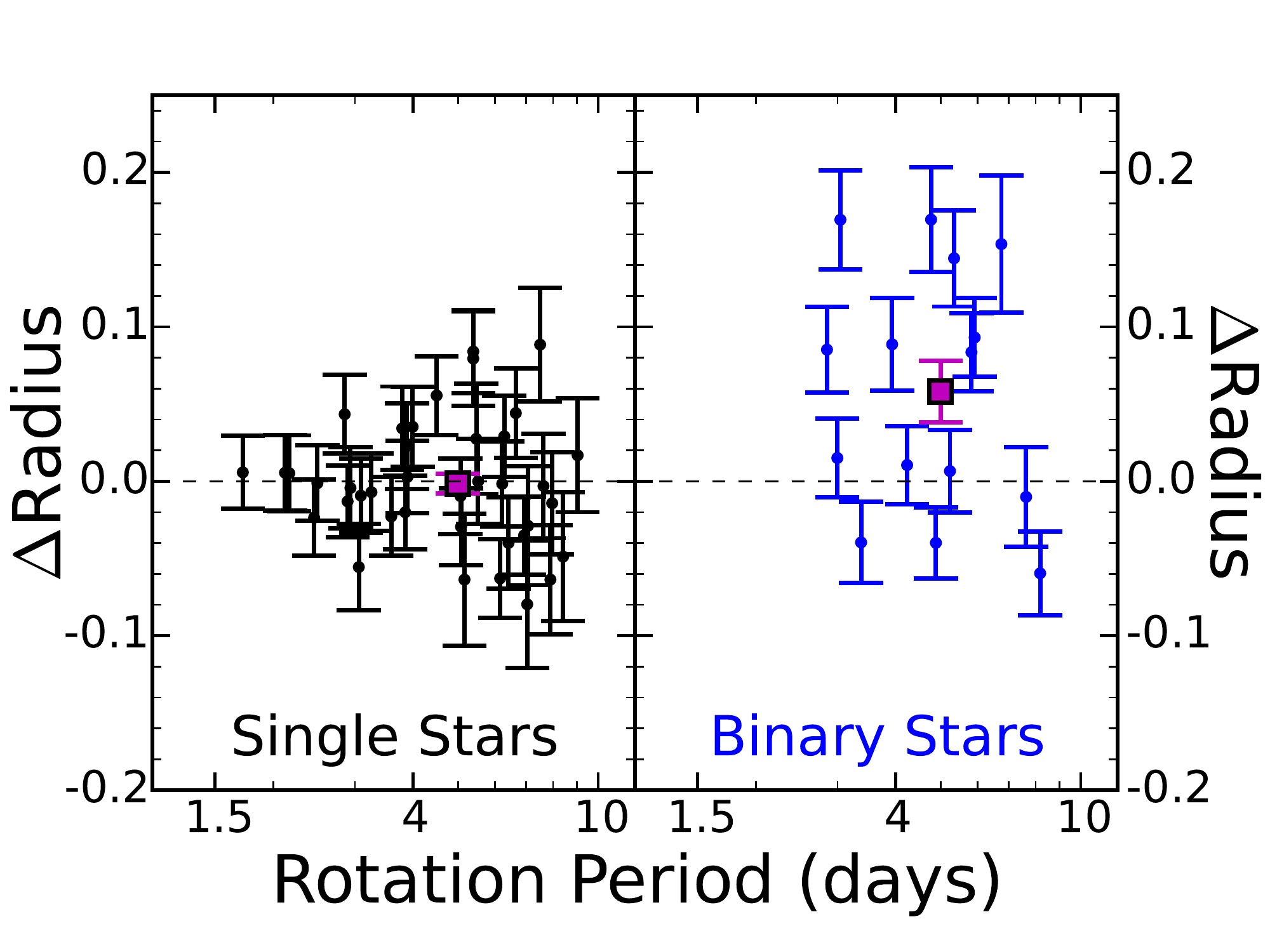}
    \caption{Illustration of the impact of binarity on \delR, as defined in Eq. \ref{eqn:delRad}. We show slowly rotating single stars from our sample (left) and binary stars in our sample (right). The binaries have a larger average \delR\ by $\sim 6$~\%, demonstrating the statistical impact of enhanced bolometric flux at fixed \teff\ on our calculation. These binaries have been excluded from the analysis of $\S$\ref{sec:results}. The dissimilarity between the two plots suggests that we have filtered binaries out with reasonable completeness.}
    \label{fig:binaries}
    \end{figure}

    In the interest of exploring the typical contaminating influence binaries could have on our results, we compare in Fig. \ref{fig:binaries} the derived radii of the slowly rotating binaries to the slowly rotating single stars in our sample. Compared to the single stars, the binaries show a similar \delR\ floor, but a much higher ceiling as a consequence of the increased bolometric flux due to the blended light of the companions. The average is about 6\% higher, showing the likely magnitude of this effect. Visually, it is clear that binaries can produce quite large spurious \delR\ values, and thus stronger verification of the single-star-nature of our putative inflated stars would be valuable. However, it also appears that there is little clear contamination of the slow-rotator sample with binaries, which may indicate that our binary exclusion is largely complete.

\subsubsection{Reddening}\label{sec:reddening}

    Bad measures of extinction could also affect the \teff\ and \fbol\ determinations of individual objects. In fact, several Pleiads are known to sit behind an area of enhanced extinction, due to an HI cloud occupying a portion of the cluster foreground (e.g. \citealt{gordon84}). For simplicity, we have excluded these stars from our analysis, and adopt the canonical $E(B-V)$ = 0.04 (e.g. \citealt{an07}) for the remainder. We do not anticipate that this complicates our results, and again it would be extraordinary were the rapid rotators preferentially distorted by differential reddening.
    
\subsubsection{Distance errors}\label{sec:nonmembers}

    A final way that a star might appear spuriously inflated is if its true distance is substantially less than we have assumed ($134 \pm 3$~pc). A closer star, at fixed bolometric luminosity and \teff, obviously requires a lower radius. Given the intrinsic depth of the cluster, it is expected that some stars indeed lie closer than our lower bound of 131~pc, and thus may require some adjustment to their inferred radius. As \fbol~$\propto d_*^2$ and also \fbol~$\propto R_*^2$, a given fractional distance error converts directly into an equivalent fractional radius error. Thus, a three-sigma outlier from the cluster center could induce as much as a 7\% increase in inferred radius. Non-member interlopers would also appear as abnormally inflated objects. Each star in our sample has a membership probability $> 85$\%, with most $\geq 99$\%, so the membership of our sample is generally secure. However, it is conceivable that a few of our targets are interloper non-members.
    
    \begin{figure*}[!t]
    \centering
    \includegraphics[width=6.0in]{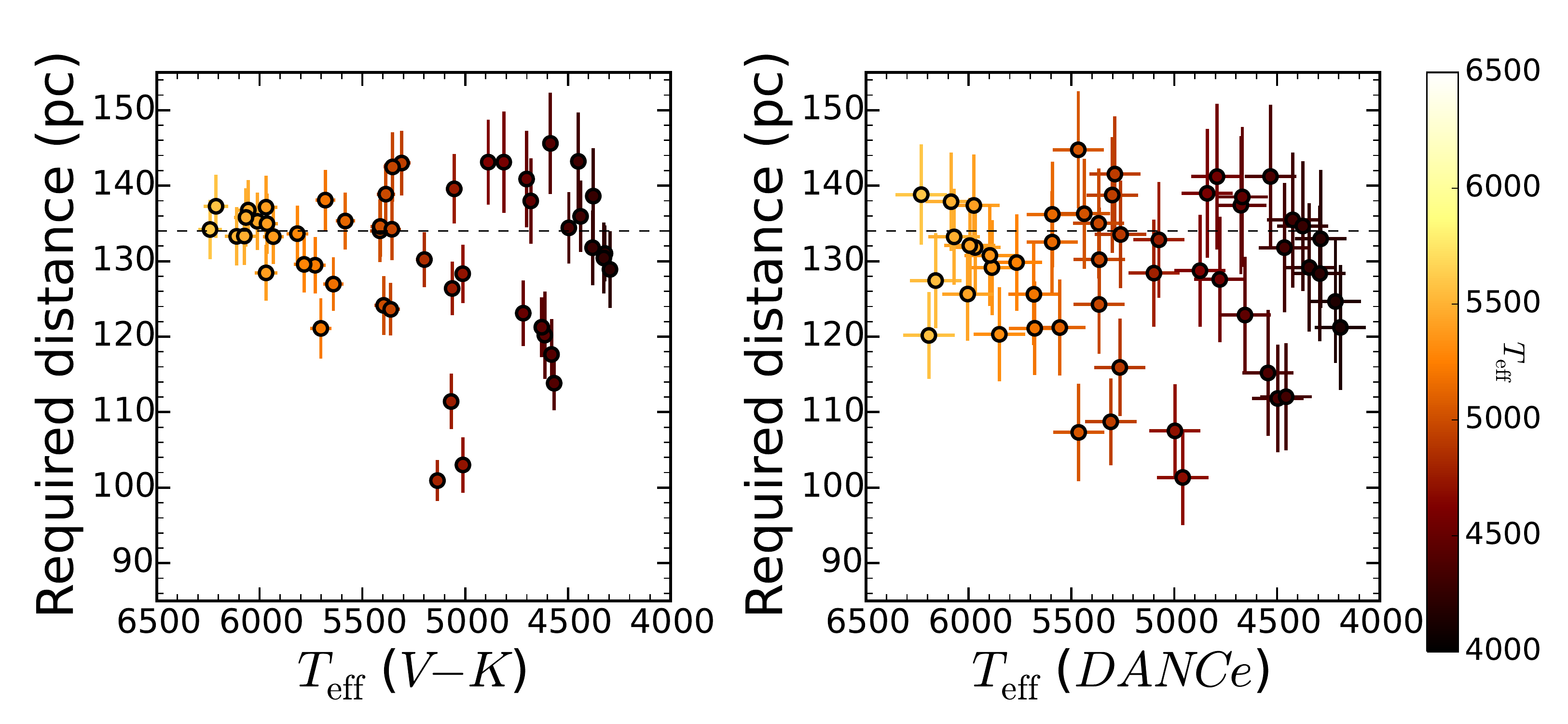}
    \caption{The required distance for each star to lie on the Padova isochrone, given its measured \fbol\ and \teff. The large required distances from the center of the cluster ($\sim 134$~pc) indicates that in order to not be over-luminous, they must either be non-member interlopers, or members that are currently quite distance from the cluster center.}
    \label{fig:recdist}
    \end{figure*}

    Given our measured \fbol\ and \teff\ for each star, we can easily determine the distance it must lie at in order to have a normal radius, as defined by the Padova isochrones we have adopted in this paper. Fig. \ref{fig:recdist} shows this calculation for each of our sample stars. Many of the warmest members cluster around the putative Pleiades distance of 134~pc, but as we move to cooler temperatures, more and more scatter far from the mean. In particular, three stars around 5000~K would have to be much closer than the cluster center in order to appear as bright as they are without radius inflation.
    
    \begin{figure}[!]
    \centering
    \includegraphics[width=3.0in]{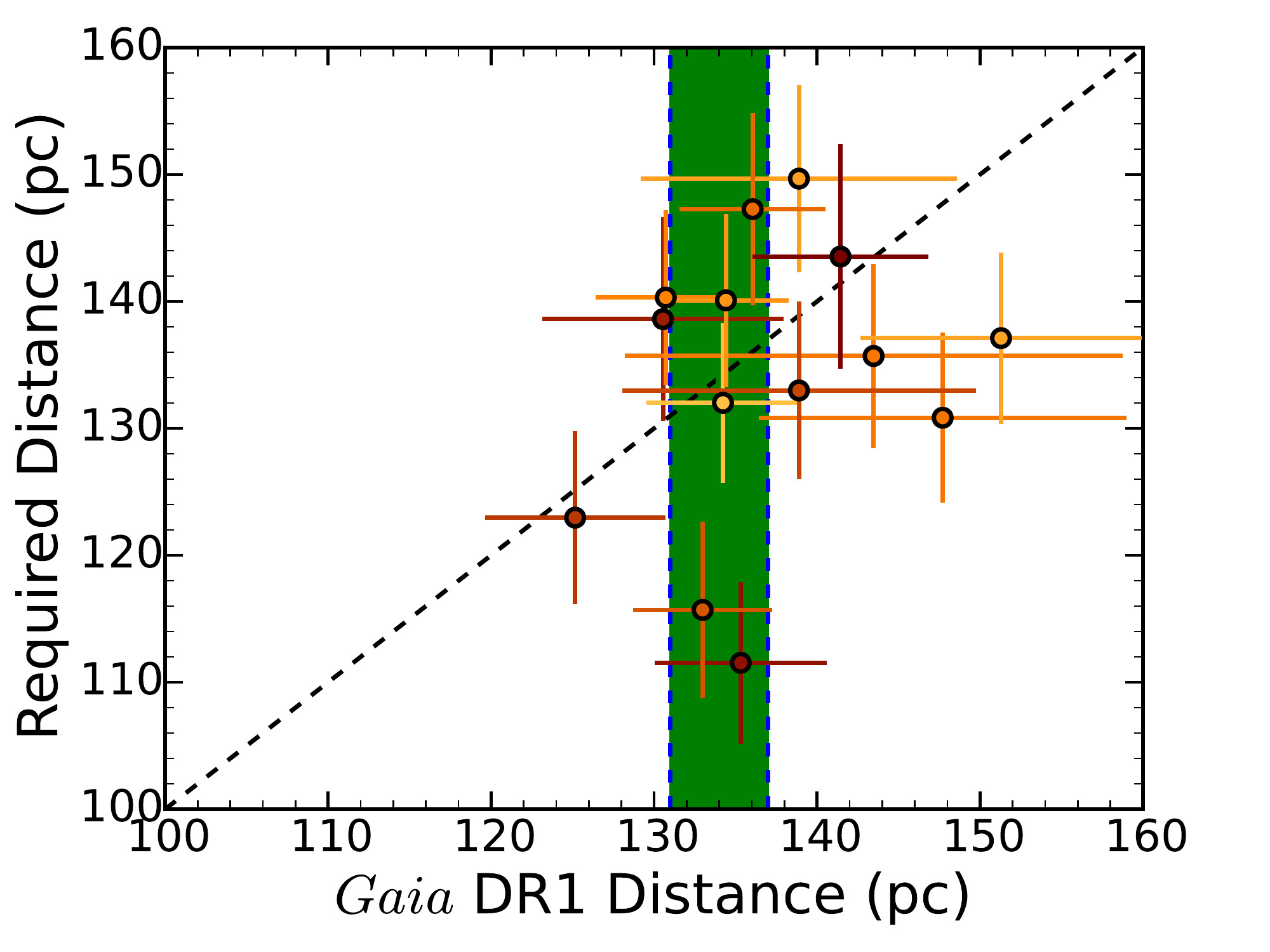}
    \caption{A comparison between the required distances from Fig. \ref{fig:recdist} and the true distances from {\it Gaia} DR1, for those members of our sample present in the first data release. The stars generally cluster around the true distance to the cluster and not along the one-to-one line, suggesting that distances effects are not responsible for the radius dispersion we find.}
    \label{fig:gaia}
    \end{figure}
    
    Fortunately, \textit{Gaia} has just unveiled its first data release, which includes parallaxes for nearly half of our sample, albeit the brighter half where the putative radius dispersion is not as pronounced. We can use these values to judge whether the most inflated stars could plausibly be interlopers. We compare our required distances to the measured {\it Gaia} distances in Fig. \ref{fig:gaia}. We find that these stars lie at an average distance of $133 \pm 8$~pc, nicely consistent with our adopted distance to the center of the cluster of $134 \pm 3$~pc (green shaded region), indicating that the DR1 Pleiades distances agree well with previous calculations. While many of these stars required distances under 120~pc to agree with the Padova isochrone predictions, none are found to be closer than 125~pc. In particular, two of the three highly inflated stars mentioned above lie right at the adopted Pleiades distance in the {\it Gaia} determination, whereas they could only be explained with a standard radius if they lay at $\sim 105$~pc. This strongly demonstrates that interlopers and depth effects cannot account for the radius spread we have inferred. Future {\it Gaia} data releases will permit similar exercises for the entire sample, and allow us to confirm or refute the inflated nature of the most extreme objects in the cluster.
    
\subsection{Connection with Lithium Abundances}\label{sec:lithium}

    \begin{figure*}[!ht]
    \centering
    \includegraphics[width=6.0in]{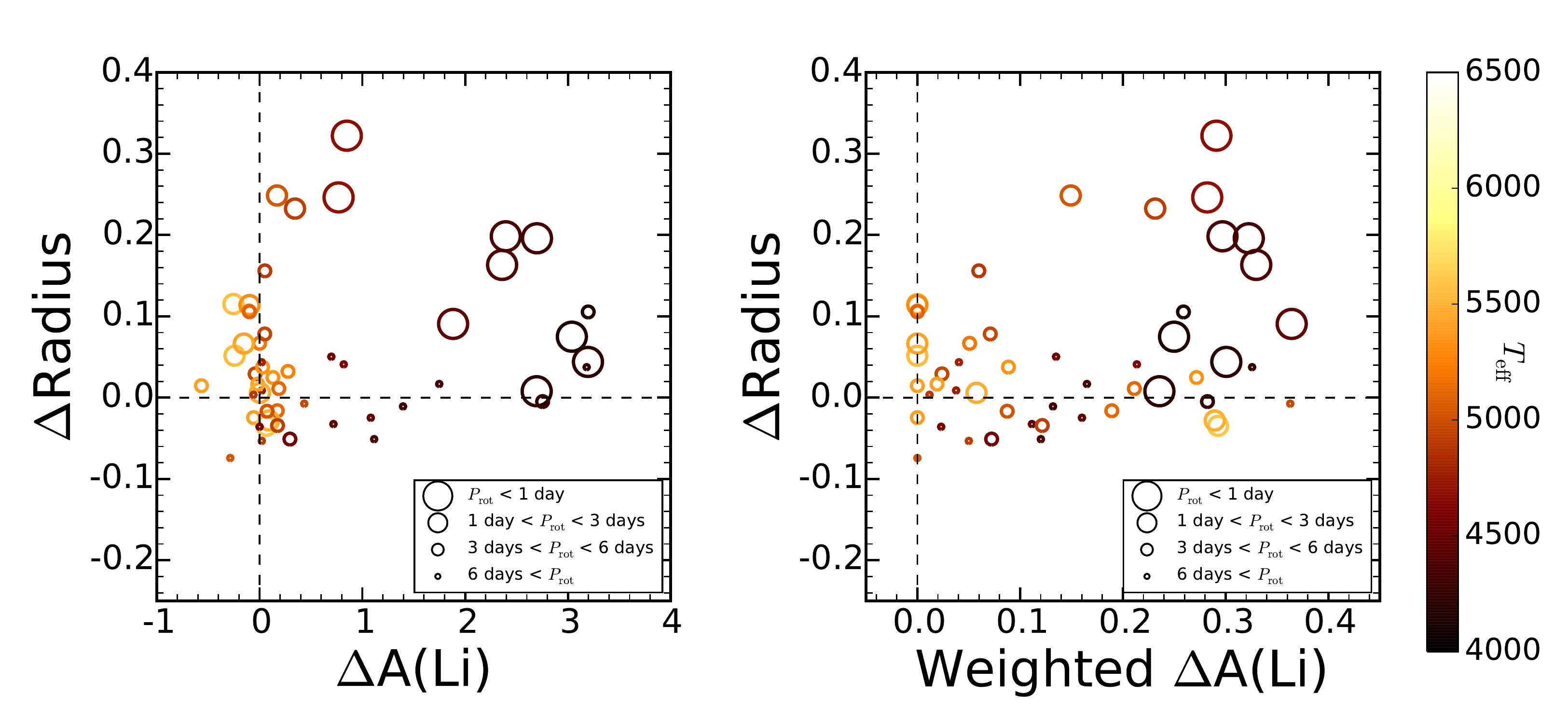}
    \caption{\textit{Left:} Comparison between the height above \teff-radius relation of \citet{bressan12}, and the height above the standard stellar model lithium predictions of \citep{Somers:2015a}, with rotation rate indicated by point size. Slower rotating stars, and warm stars, tend to cluster near the origin, whereas the rapidly rotating, cooler stars tend to extend to the upper right. This suggests a triple correlation between rotation, radius, and lithium abundance. \textit{Right:} Same as left panel, but the x-axis new reflects how inflated each star must have been on the pre-main sequence in order to possess its current abundance. This panel solidifies the tripartite correlation.}
    \label{fig:lithium}
    \end{figure*}

    There is a well known correlation between rotation rate and lithium abundance in the Pleiades \citep[e.g.][]{soderblom93}. The correlation is in the sense that at a given \teff, the most rapidly rotating stars tend to be the richest in lithium. Moreover, equivalent correlations have been found in younger clusters such as the $\sim 5$~Myr NGC 2264 and the $\sim 24$~Myr $\beta$~Pictoris \citep{bouvier2016,messina2016}, suggesting that the Li dispersion sets in during the early pre-main sequence. This picture contradicts standard stellar predictions \citep[e.g.][]{iben65}, which anticipate no dispersion in lithium at fixed \teff\ in a cluster like the Pleiades. Explanations for the origin of this rotationally correlated dispersion in lithium include core-envelope de-coupling influencing the rate of rotational mixing \citep{bouvier08}, accretion altering the central temperatures of proto-stars \citep{baraffe10}, and magnetic activity inducing radius inflation during the pre-main sequence, thus suppressing the central temperatures of stars and slowing the rate of Li destruction \citep{ventura98,Somers:2014,Somers:2015a,Somers:2015b}. 
    
    Given our detection of a correlation between rotation rate and radius, it is interesting to consider whether the \delR\ we have determined in the Pleiades also correlate with lithium. To this end, we draw lithium abundances for our sample from \citet{barrado16}. In order to quantify how Li-rich each star is, we use the lithium depletion models of \citet{Somers:2015a} to determine the difference in A(Li) between the measured value of each star and the values predicted for its \teff. We refer to this metric as $\Delta$A(Li), as it is in some ways analogous to our \delR\ metric. We compare these values to our \delR\ values in the left panel of Fig. \ref{fig:lithium}. Here, the size of the circles reflects the rotation rate, with the largest circles being the most rapidly spinning. The general trend in this panel is for smaller circles to cluster around the origin, and larger circles to be either at high \delR, high $\Delta$A(Li), or both. At each value of \delR, the fastest rotating stars have the largest $\Delta$A(Li) values, confirming that the three quantities correlate with one another.
    
    However, we note that there is a strong mass dependence in the largest possible $\Delta$A(Li). This is because lower mass stars tend to deplete more lithium, thus producing a larger dynamic range with their Li-rich counterparts (see figure 7 in \citealt{Somers:2015a}). To attempt to remove this effect, and thus get a cleaner comparison of lithium and radius anomalies across the full range of \teff, we create a new metric which weights $\Delta$A(Li) by the width of the lithium dispersion at each \teff. This amounts to determining how much radius inflation is required on the pre-main sequence in order to produce the current day abundance of each star (see \citealt{somers16a} for a discussion of this method). We compare these values to \delR\ in the right panel of Fig. \ref{fig:lithium}. Here,the abcissa values correspond to the predicted fraction of the surface covered in starspots, based on the \teff\ and A(Li) values for each star \citep{Somers:2015a}. With this metric, essentially all of the rapidly rotating stars now show large lithium anomalies, and (as in $\S$\ref{sec:radiusrotation}) show a range of radius anomalies between 0--30\%. It is clear that there is no one-to-one correspondence between rotation rate, radius, and lithium abundance, but it is clear that the most rapidly rotating, lithium rich, cool stars are also the most inflated. 

\section{Summary and Conclusions}\label{sec:summary}

    Previous studies have reported anomalously large radii among low-mass stars in the Pleiades and other young clusters, using ensemble averages of projected rotation velocity measurements. Here, we have measured the radii of several Pleiades cluster members via the Stefan-Boltzmann law, combining (i) \teffs\ determined through color and spectro-photometric techniques, (ii) bolometric fluxes determined by summing the observed spectral energy distributions, and (iii) the known cluster distance. Our sample specifically includes stars with previously determined rotation periods and lithium abundances. We compare our radius measurements to literature isochrones, calibrated on older, inactive stars, and find that in many cases the Pleiades radii can be larger by 10--20\% compared to expectations. We further show that this over-radius correlates with rapid rotation at greater than 99.99\% confidence, strongly suggesting a magnetic origin. We discuss whether this radius-rotation correlation could be a spurious artifact brought on by poorly calculated radii due to the distorted SEDs of rapidly rotating stars. A very large systematic offset in \teff, afflicting only stars which rotate with periods shorter than 1.5~days, would be required to reconcile the rapidly rotating Pleiads with the model expectations. However, our quantitative measures of SED distortion find that this is principally a function of \teff, not rotation, with the coolest stars tending to show modestly distorted SEDs that might be better fit by a two-temperature model. We conclude that the most likely explanation is magnetically-driven radius inflation amongst the most rapidly rotating Pleiads.

\acknowledgments

G.E.S.\ acknowledges the support of the Vanderbilt Office of the Provost through the Vanderbilt Initiative in Data-intensive Astrophysics (VIDA) fellowship. K.G.S.\ acknowledges partial support through NSF PAARE grant AST-1358862. This work has made use of the Filtergraph data visualization service \citep{Burger:2013} at \url{filtergraph.vanderbilt.edu}.

\clearpage 
\LongTables
\begin{deluxetable}{cccccccccc}
\tabletypesize{\scriptsize}
\tablecaption{Basic Pleiades Data \label{tab1}}
\tablewidth{0pt}
\tablehead{
\colhead{HII}
&\colhead{Tycho-2}
&\colhead{\textit{2MASS}}
&\colhead{R.A.}
&\colhead{Dec.}
&\colhead{$V_{\rm mag}$}
&\colhead{$B-V$}
&\colhead{$V-K_s$}
&\colhead{$P_{\rm rot}$}
&\colhead{Binary?}
} 
\startdata
0025 & 1803-478-1  & 03425511+2429350 & 55.729626 & +24.493065 & 9.470  & 0.48$\pm$0.02 & 1.207$\pm$0.029 & 1.41       & N \\ 
0034 & 1803-400-1  & 03430293+2440110 & 55.762230 & +24.669737 & 12.030 & 0.94$\pm$0.02 & 2.303$\pm$0.028 & 6.69       & N \\ 
0097 &     ...     & 03432662+2459395 & 55.860922 & +24.994333 & 12.500 & 1.08$\pm$0.02 & 2.705$\pm$0.030 & 6.74947635 & Y \\ 
0120 & 1799-118-1  & 03433195+2340266 & 55.883139 & +23.674074 & 10.790 & 0.70$\pm$0.02 & 1.687$\pm$0.029 & 3.99       & N \\ 
0129 & 1799-1268-1 & 03433440+2345429 & 55.893373 & +23.761917 & 11.470 & 0.88$\pm$0.02 & 2.100$\pm$0.029 & 5.44       & Y \\ 
0152 & 1799-780-1  & 03433772+2332096 & 55.907204 & +23.536011 & 10.750 & 0.70$\pm$0.02 & 1.626$\pm$0.029 & 3.88781400 & N \\ 
0164 & 1799-1037-1 & 03434286+2335412 & 55.928608 & +23.594801 & 9.540  & 0.48$\pm$0.02 & 1.307$\pm$0.033 & ....       & Y \\ 
0174 & 1803-8-1    & 03434833+2500157 & 55.951377 & +25.004387 & 11.620 & 0.85$\pm$0.02 & 2.246$\pm$0.028 & 0.47429706 & N \\ 
0193 & 1799-816-1  & 03435070+2414508 & 55.961274 & +24.247465 & 11.290 & 0.81$\pm$0.02 & 1.946$\pm$0.028 & 5.36       & N \\ 
0250 & 1803-818-1  & 03440424+2459233 & 56.017681 & +24.989822 & 10.680 & 0.68$\pm$0.02 & 1.619$\pm$0.029 & 4.23218914 & Y \\ 
0253 & 1803-282-1  & 03440353+2430151 & 56.014746 & +24.504206 & 10.660 & 0.68$\pm$0.02 & 1.707$\pm$0.029 & 1.48       & N \\ 
0263 &     ...     & 03440484+2416318 & 56.020167 & +24.275501 & 11.540 & 0.88$\pm$0.02 & 2.154$\pm$0.029 & 4.68       & Y \\ 
0293 & 1803-812-1  & 03441391+2446457 & 56.057980 & +24.779388 & 10.800 & 0.70$\pm$0.02 & 1.738$\pm$0.030 & 4.03       & Y \\ 
0345 &     ...     & 03442627+2435229 & 56.109485 & +24.589705 & 11.570 & 0.84$\pm$0.02 & 2.304$\pm$0.028 & 0.84       & N \\ 
0380 &     ...     & 03443741+2508160 & 56.155898 & +25.137793 & 13.330 & 1.21$\pm$0.02 & 3.091$\pm$0.031 & 9.03288129 & N \\ 
0405 & 1803-542-1  & 03444075+2449067 & 56.169800 & +24.818542 & 9.830  & 0.54$\pm$0.02 & 1.317$\pm$0.031 & 1.91       & N \\ 
0430 &     ...     & 03444398+2413523 & 56.183251 & +24.231215 & 11.400 & 0.85$\pm$0.02 & 1.931$\pm$0.030 & 5.51856881 & N \\ 
0470 & 1799-1157-1 & 03445123+2316082 & 56.213483 & +23.268969 & 8.950  & 0.39$\pm$0.02 & 0.966$\pm$0.039 & ....       & Y \\ 
0489 &     ...     & 03445639+2425574 & 56.234988 & +24.432623 & 10.390 & 0.63$\pm$0.02 & 1.524$\pm$0.028 & 2.78       & N \\ 
0514 & 1803-1061-1 & 03450400+2515282 & 56.266691 & +25.257841 & 10.690 & 0.69$\pm$0.02 & 1.649$\pm$0.028 & 3.79041574 & N \\ 
0530 & 1799-170-1  & 03450528+2342097 & 56.272009 & +23.702707 & 8.950  & 0.39$\pm$0.02 & 0.950$\pm$0.030 & ....       & Y \\ 
0627 & 1803-388-1  & 03452412+2453095 & 56.350531 & +24.885977 & 9.680  & 0.50$\pm$0.02 & 1.261$\pm$0.031 & 1.81       & N \\ 
0636 &     ...     & 03452219+2328182 & 56.342478 & +23.471731 & 12.480 & 1.06$\pm$0.02 & 2.632$\pm$0.027 & 8.48       & N \\ 
0708 &     ...     & 03453539+2404595 & 56.397484 & +24.083210 & 10.130 & 0.62$\pm$0.02 & 1.568$\pm$0.029 & 1.02       & Y \\ 
0746 &     ...     & 03454184+2425534 & 56.424374 & +24.431513 & 11.270 & 0.92$\pm$0.02 & 1.872$\pm$0.030 & 5.23230299 & Y \\ 
0879 &     ...     & 03460649+2434027 & 56.527062 & +24.567419 & 12.790 & 1.07$\pm$0.02 & 2.677$\pm$0.028 & 6.75       & N \\ 
0882 &     ...     & 03460412+2324199 & 56.517177 & +23.405537 & 12.660 & 1.07$\pm$0.02 & 2.824$\pm$0.027 & 0.58       & N \\ 
0883 &     ...     & 03460689+2433461 & 56.528742 & +24.562807 & 13.050 & 1.08$\pm$0.02 & 2.800$\pm$0.031 & 7.91       & N \\ 
0916 &     ...     & 03461174+2437203 & 56.548937 & +24.622314 & 11.710 & 0.87$\pm$0.02 & 2.165$\pm$0.028 & 5.98       & Y \\ 
0923 & 1800-1917-1 & 03461005+2320240 & 56.541875 & +23.340015 & 10.120 & 0.62$\pm$0.02 & 1.461$\pm$0.026 & 1.42       & Y \\ 
0996 &     ...     & 03462267+2434126 & 56.594480 & +24.570177 & 10.420 & 0.64$\pm$0.02 & 1.497$\pm$0.026 & 3.23       & N \\ 
1015 & 1804-2366-1 & 03462735+2508080 & 56.613966 & +25.135563 & 10.540 & 0.65$\pm$0.02 & 1.547$\pm$0.026 & 3.42       & N \\ 
1032 &     ...     & 03462841+2426021 & 56.618376 & +24.433918 & 11.340 & 0.86$\pm$0.02 & 2.181$\pm$0.026 & 1.31       & N \\ 
1095 &     ...     & 03463777+2444517 & 56.657390 & +24.747698 & 11.920 & 0.88$\pm$0.02 & 2.251$\pm$0.028 & 7.18       & N \\ 
1110 &     ...     & 03463889+2431132 & 56.662043 & +24.520344 & 13.290 & 1.19$\pm$0.02 & 3.004$\pm$0.027 & 7.48       & N \\ 
1122 &     ...     & 03463932+2406116 & 56.663857 & +24.103243 & 9.290  & 0.46$\pm$0.02 & 1.099$\pm$0.025 & 0.87       & Y \\ 
1124 &     ...     & 03463938+2401468 & 56.664103 & +24.029688 & 12.120 & 0.92$\pm$0.02 & 2.261$\pm$0.029 & 6.13       & N \\ 
1132 & 1800-1574-1 & 03463839+2255112 & 56.659985 & +22.919804 & 9.420  & 0.49$\pm$0.02 & 1.267$\pm$0.031 & 6.84       & N \\ 
1139 & 1800-1672-1 & 03463999+2306373 & 56.666655 & +23.110371 & 9.380  & 0.48$\pm$0.02 & 1.137$\pm$0.032 & 1.82       & Y \\ 
1182 & 1800-1774-1 & 03464706+2254525 & 56.696086 & +22.914593 & 10.460 & 0.64$\pm$0.02 & 1.532$\pm$0.028 & 3.00       & Y \\ 
1200 & 1800-1627-1 & 03465053+2314211 & 56.710572 & +23.239195 & 9.900  & 0.54$\pm$0.02 & 1.354$\pm$0.029 & ....       & N \\ 
1207 & 1804-2205-1 & 03465491+2447468 & 56.728828 & +24.796349 & 10.470 & 0.62$\pm$0.02 & 1.493$\pm$0.028 & 3.37       & Y \\ 
1215 & 1800-1616-1 & 03465373+2335009 & 56.723910 & +23.583599 & 10.520 & 0.65$\pm$0.02 & 1.524$\pm$0.029 & 3.59       & N \\ 
1220 &     ...     & 03465326+2252513 & 56.721920 & +22.880943 & 11.740 & 0.88$\pm$0.02 & 2.021$\pm$0.028 & 6.15       & N \\ 
1298 &     ...     & 03470678+2342546 & 56.778252 & +23.715170 & 12.180 & 1.02$\pm$0.02 & 2.341$\pm$0.028 & 8.48       & Y \\ 
1305 &     ...     & 03470734+2313349 & 56.780599 & +23.226362 & 13.460 & 1.18$\pm$0.02 & 3.096$\pm$0.028 & 0.39       & N \\ 
1309 & 1800-1935-1 & 03471005+2416360 & 56.791894 & +24.276672 & 9.460  & 0.47$\pm$0.02 & 1.178$\pm$0.032 & 0.78       & N \\ 
1332 &     ...     & 03471352+2342515 & 56.806374 & +23.714310 & 12.410 & 1.04$\pm$0.02 & 2.401$\pm$0.028 & 7.85       & Y \\ 
1454 &     ...     & 03473367+2441032 & 56.890319 & +24.684223 & 12.780 & 1.16$\pm$0.02 & 2.652$\pm$0.029 & 8.60       & N \\ 
1514 &     ...     & 03474044+2421525 & 56.918539 & +24.364594 & 10.480 & 0.64$\pm$0.02 & 1.527$\pm$0.028 & 3.25       & N \\ 
1531 &     ...     & 03474143+2358190 & 56.922657 & +23.971945 & 13.300 & 1.15$\pm$0.02 & 2.988$\pm$0.029 & 0.48       & N \\ 
1593 & 1800-2170-1 & 03474811+2313053 & 56.950462 & +23.218140 & 11.120 & 0.75$\pm$0.02 & 1.796$\pm$0.027 & 5.14       & N \\ 
1613 &     ...     & 03475252+2356286 & 56.968837 & +23.941282 & 9.880  & 0.54$\pm$0.02 & 1.306$\pm$0.028 & 1.40       & N \\ 
1776 &     ...     & 03481769+2502523 & 57.073718 & +25.047886 & 10.910 & 0.72$\pm$0.02 & 1.752$\pm$0.028 & 4.53       & N \\ 
1794 &     ...     & 03481712+2353253 & 57.071342 & +23.890385 & 10.360 & 0.64$\pm$0.02 & 1.469$\pm$0.029 & 3.82       & N \\ 
1797 &     ...     & 03481691+2338125 & 57.070459 & +23.636806 & 10.110 & 0.56$\pm$0.02 & 1.374$\pm$0.029 & 2.45       & N \\ 
1856 &     ...     & 03482616+2402544 & 57.109023 & +24.048445 & 10.020 & 0.56$\pm$0.02 & 1.357$\pm$0.027 & 2.49       & N \\ 
1883 &     ...     & 03482802+2318027 & 57.116787 & +23.300774 & 12.600 & 1.06$\pm$0.02 & 2.764$\pm$0.027 & 0.24       & N \\ 
1924 &     ...     & 03483451+2326053 & 57.143806 & +23.434818 & 10.330 & 0.62$\pm$0.02 & 1.462$\pm$0.029 & 2.88       & N \\ 
2016 &     ...     & 03484542+2320201 & 57.189275 & +23.338938 & 13.520 & 1.22$\pm$0.02 & 3.185$\pm$0.027 & 3.92       & N \\ 
2034 &     ...     & 03484932+2358383 & 57.205510 & +23.977320 & 12.570 & 0.99$\pm$0.02 & 2.578$\pm$0.028 & 0.55       & Y \\ 
2106 &     ...     & 03485848+2312044 & 57.243692 & +23.201229 & 11.530 & 0.86$\pm$0.02 & 2.153$\pm$0.027 & 6.01       & Y \\ 
2126 &     ...     & 03490232+2315088 & 57.259705 & +23.252468 & 11.640 & 0.86$\pm$0.02 & 1.983$\pm$0.028 & 3.08       & N \\ 
2244 &     ...     & 03492059+2446360 & 57.335794 & +24.776670 & 12.670 & 1.04$\pm$0.02 & 2.745$\pm$0.030 & 0.56       & N \\ 
2284 &     ...     & 03492405+2350214 & 57.350217 & +23.839287 & 11.350 & 0.78$\pm$0.02 & 1.972$\pm$0.028 & 5.59       & Y \\ 
2311 &     ...     & 03492873+2342440 & 57.369749 & +23.712238 & 11.360 & 0.83$\pm$0.02 & 1.932$\pm$0.030 & 5.71       & N \\ 
2341 &     ...     & 03493312+2347435 & 57.388005 & +23.795427 & 10.870 & 0.71$\pm$0.02 & 1.656$\pm$0.029 & 4.94       & Y \\ 
2345 & 1800-727-1  & 03493272+2322494 & 57.386341 & +23.380413 & 9.100  & 0.44$\pm$0.02 & 1.064$\pm$0.031 & 0.24       & N \\ 
2366 &     ...     & 03493653+2417460 & 57.402244 & +24.296135 & 11.530 & 0.82$\pm$0.02 & 1.980$\pm$0.028 & 6.21783952 & N \\ 
2462 &     ...     & 03495035+2342202 & 57.459828 & +23.705631 & 11.550 & 0.83$\pm$0.02 & 1.955$\pm$0.028 & 6.85       & N \\ 
2506 & 1800-471-1  & 03495648+2313071 & 57.485363 & +23.218641 & 10.230 & 0.60$\pm$0.02 & 1.434$\pm$0.028 & 2.18       & N \\ 
2644 &     ...     & 03502089+2428003 & 57.587080 & +24.466755 & 11.030 & 0.73$\pm$0.02 & 1.723$\pm$0.027 & 5.07153391 & N \\ 
2665 & 1800-669-1  & 03502130+2305470 & 57.588751 & +23.096392 & 11.360 & 0.83$\pm$0.02 & 1.975$\pm$0.029 & 5.37       & N \\ 
2741 &     ...     & 03503457+2430281 & 57.644052 & +24.507822 & 12.600 & 1.00$\pm$0.02 & 2.520$\pm$0.027 & 5.15798051 & N \\ 
2786 &     ...     & 03504007+2355590 & 57.666983 & +23.933065 & 10.310 & 0.60$\pm$0.02 & 1.457$\pm$0.028 & 2.16       & N \\ 
2870 & 1800-586-1  & 03505143+2319447 & 57.714333 & +23.329100 & 12.450 & 1.07$\pm$0.02 & 2.435$\pm$0.029 & 8.45       & N \\ 
2880 &     ...     & 03505508+2411508 & 57.729525 & +24.197466 & 11.750 & 0.86$\pm$0.02 & 2.121$\pm$0.027 & 6.28238456 & N \\ 
3019 &     ...     & 03512440+2405147 & 57.851675 & +24.087425 & 13.450 & 1.19$\pm$0.02 & 3.176$\pm$0.030 & 5.46870223 & N \\ 
3031 &  1804-67-1  & 03512722+2431072 & 57.863420 & +24.518692 & 8.830  & 0.38$\pm$0.02 & 0.953$\pm$0.028 & ....       & N \\ 
3063 &     ...     & 03512993+2353572 & 57.874730 & +23.899225 & 13.520 & 1.17$\pm$0.02 & 3.224$\pm$0.030 & 0.88393824 & N \\ 
3163 &     ...     & 03515338+2423132 & 57.972446 & +24.387018 & 12.690 & 0.98$\pm$0.02 & 2.807$\pm$0.030 & 0.41747738 & N \\ 
3179 & 1800-1415-1 & 03515685+2354070 & 57.986888 & +23.901966 & 10.040 & 0.56$\pm$0.02 & 1.408$\pm$0.027 & 4.85456989 & Y \\ 
3187 &     ...     & 03515733+2320219 & 57.988895 & +23.339439 & 13.120 & 1.16$\pm$0.02 & 2.922$\pm$0.029 & 7.30       & N
\enddata
\tablecomments {$V_{mag}$ and $B-V$ data from \citet{johnson58}, $K_s$ data from \citet{cutri03}. Rotation periods with two decimal places are from the {\it K2} Pleiades campaign \citep{rebull16a,rebull16b,stauffer16}, and those with more decimals places are from the HATNet campaign \citep{hartman10}. Sources of binary info are described in $\S$\ref{sec:binaries}.}

\end{deluxetable}
\clearpage 
\begin{deluxetable}{ccccccc}
\tabletypesize{\scriptsize}
\tablecaption{Derived Stellar Properties ($V-K_s$) \label{tab2}}
\tablewidth{0pt}
\tablehead{
\colhead{HII}
&\colhead{$T_{\rm eff}$}
&\colhead{$\mathcal{F}_{bol}$}
&\colhead{$\chi_{\nu}^2$}
&\colhead{Angular Diameter}
&\colhead{Radius}
&\colhead{$\Delta$Radius}\\
\colhead{}
&\colhead{}
&\colhead{$\times 10^{-10}$}
&\colhead{}
&\colhead{$\times 10^{-2}$}
&\colhead{}
&\colhead{}\\
\colhead{}
&\colhead{(K)}
&\colhead{(erg cm$^{-2}$ s$^{-1}$)}
&\colhead{}
&\colhead{(mas)}
&\colhead{($R_{\odot})$}
&\colhead{(\%)}
} 
\startdata
0025 & 6513$^{+55}_{-68}$ & 47.38$_{-1.71}^{+0.59}$ & 2.18 & 8.89$_{-0.23}^{+0.18}$ & 1.280$_{-0.044}^{+0.038}$ & 6.4$^{+3.7}_{-3.2}$ \\
0034 & 5011$^{+27}_{-34}$ & 5.77$_{-0.15}^{+0.21}$ & 2.45 & 5.24$_{-0.09}^{+0.12}$ & 0.754$_{-0.022}^{+0.024}$ & 4.4$^{+3.0}_{-3.3}$ \\
0097 & 4658$^{+23}_{-28}$ & 4.36$_{-0.28}^{+0.26}$ & 7.05 & 5.27$_{-0.18}^{+0.17}$ & 0.759$_{-0.031}^{+0.030}$ & 15.4$^{+4.7}_{-4.5}$ \\
0120 & 5729$^{+40}_{-49}$ & 14.71$_{-0.28}^{+0.29}$ & 2.26 & 6.40$_{-0.12}^{+0.12}$ & 0.922$_{-0.027}^{+0.027}$ & 3.5$^{+3.0}_{-3.0}$ \\
0129 & 5220$^{+31}_{-38}$ & 9.06$_{-0.27}^{+0.28}$ & 1.99 & 6.05$_{-0.12}^{+0.12}$ & 0.871$_{-0.026}^{+0.026}$ & 14.4$^{+3.4}_{-3.5}$ \\
0152 & 5816$^{+41}_{-51}$ & 15.55$_{-0.15}^{+0.15}$ & 0.60 & 6.39$_{-0.11}^{+0.11}$ & 0.920$_{-0.026}^{+0.026}$ & 0.3$^{+2.8}_{-2.8}$ \\
0164 & 6328$^{+57}_{-69}$ & 43.69$_{-1.10}^{+1.14}$ & 0.87 & 9.04$_{-0.21}^{+0.22}$ & 1.302$_{-0.042}^{+0.043}$ & 16.2$^{+3.8}_{-3.8}$ \\
0174 & 5068$^{+27}_{-34}$ & 8.23$_{-0.33}^{+0.35}$ & 3.14 & 6.12$_{-0.14}^{+0.15}$ & 0.881$_{-0.029}^{+0.029}$ & 20.3$^{+3.9}_{-4.0}$ \\
0193 & 5396$^{+32}_{-41}$ & 10.12$_{-0.29}^{+0.40}$ & 2.73 & 5.99$_{-0.12}^{+0.14}$ & 0.862$_{-0.026}^{+0.028}$ & 7.9$^{+3.2}_{-3.5}$ \\
0250 & 5826$^{+41}_{-51}$ & 16.01$_{-0.31}^{+0.32}$ & 1.17 & 6.46$_{-0.12}^{+0.12}$ & 0.930$_{-0.027}^{+0.027}$ & 1.0$^{+2.9}_{-3.0}$ \\
0253 & 5702$^{+39}_{-48}$ & 16.17$_{-0.49}^{+0.70}$ & 4.30 & 6.78$_{-0.15}^{+0.18}$ & 0.976$_{-0.030}^{+0.034}$ & 10.7$^{+3.5}_{-3.8}$ \\
0263 & 5163$^{+30}_{-37}$ & 8.78$_{-0.26}^{+0.37}$ & 1.57 & 6.09$_{-0.12}^{+0.15}$ & 0.877$_{-0.026}^{+0.029}$ & 16.9$^{+3.5}_{-3.9}$ \\
0293 & 5659$^{+40}_{-49}$ & 14.75$_{-0.58}^{+0.30}$ & 1.86 & 6.57$_{-0.17}^{+0.12}$ & 0.946$_{-0.032}^{+0.028}$ & 8.9$^{+3.7}_{-3.2}$ \\
0345 & 5010$^{+27}_{-34}$ & 8.94$_{-0.48}^{+0.41}$ & 3.51 & 6.53$_{-0.19}^{+0.17}$ & 0.940$_{-0.035}^{+0.032}$ & 30.1$^{+4.8}_{-4.5}$ \\
0380 & 4380$^{+20}_{-24}$ & 2.45$_{-0.12}^{+0.15}$ & 5.61 & 4.47$_{-0.12}^{+0.15}$ & 0.643$_{-0.023}^{+0.026}$ & 1.7$^{+3.6}_{-4.1}$ \\
0405 & 6310$^{+54}_{-66}$ & 34.17$_{-0.00}^{+0.39}$ & 0.85 & 8.04$_{-0.15}^{+0.16}$ & 1.158$_{-0.034}^{+0.035}$ & 4.1$^{+3.1}_{-3.1}$ \\
0430 & 5414$^{+36}_{-44}$ & 8.89$_{-0.31}^{+0.24}$ & 2.43 & 5.57$_{-0.13}^{+0.11}$ & 0.803$_{-0.026}^{+0.024}$ & -0.0$^{+3.2}_{-3.0}$ \\
0470 & 7018$^{+87}_{-101}$ & 74.86$_{-1.86}^{+0.96}$ & 1.95 & 9.62$_{-0.28}^{+0.27}$ & 1.386$_{-0.051}^{+0.049}$ & 2.9$^{+3.8}_{-3.7}$ \\
0489 & 5968$^{+43}_{-54}$ & 20.85$_{-0.43}^{+0.22}$ & 1.69 & 7.02$_{-0.14}^{+0.12}$ & 1.011$_{-0.030}^{+0.029}$ & 4.3$^{+3.1}_{-2.9}$ \\
0514 & 5783$^{+39}_{-49}$ & 15.81$_{-0.31}^{+0.32}$ & 1.75 & 6.51$_{-0.12}^{+0.12}$ & 0.938$_{-0.027}^{+0.027}$ & 3.4$^{+3.0}_{-3.0}$ \\
0530 & 7055$^{+68}_{-84}$ & 75.16$_{-0.93}^{+0.95}$ & 1.51 & 9.54$_{-0.21}^{+0.21}$ & 1.374$_{-0.044}^{+0.044}$ & 2.4$^{+3.2}_{-3.3}$ \\
0627 & 6412$^{+57}_{-69}$ & 39.54$_{-0.91}^{+0.47}$ & 0.95 & 8.38$_{-0.19}^{+0.17}$ & 1.206$_{-0.039}^{+0.037}$ & 4.3$^{+3.3}_{-3.2}$ \\
0636 & 4717$^{+21}_{-27}$ & 4.19$_{-0.18}^{+0.24}$ & 5.67 & 5.04$_{-0.12}^{+0.15}$ & 0.725$_{-0.024}^{+0.027}$ & 8.8$^{+3.6}_{-4.1}$ \\
0708 & 5901$^{+43}_{-53}$ & 26.88$_{-0.96}^{+0.67}$ & 1.65 & 8.15$_{-0.20}^{+0.17}$ & 1.174$_{-0.039}^{+0.036}$ & 24.2$^{+4.1}_{-3.8}$ \\
0746 & 5486$^{+37}_{-45}$ & 9.91$_{-0.26}^{+0.28}$ & 2.38 & 5.73$_{-0.12}^{+0.12}$ & 0.825$_{-0.025}^{+0.025}$ & 0.7$^{+3.0}_{-3.1}$ \\
0879 & 4680$^{+22}_{-28}$ & 3.18$_{-0.20}^{+0.22}$ & 7.47 & 4.46$_{-0.15}^{+0.16}$ & 0.642$_{-0.025}^{+0.027}$ & -2.9$^{+3.9}_{-4.1}$ \\
0882 & 4567$^{+20}_{-25}$ & 4.08$_{-0.16}^{+0.17}$ & 2.22 & 5.30$_{-0.11}^{+0.12}$ & 0.764$_{-0.024}^{+0.024}$ & 17.7$^{+3.7}_{-3.7}$ \\
0883 & 4585$^{+23}_{-28}$ & 2.55$_{-0.19}^{+0.21}$ & 15.80 & 4.16$_{-0.16}^{+0.18}$ & 0.599$_{-0.026}^{+0.029}$ & -8.0$^{+4.1}_{-4.4}$ \\
0916 & 5151$^{+29}_{-36}$ & 7.56$_{-0.07}^{+0.07}$ & 0.31 & 5.68$_{-0.08}^{+0.08}$ & 0.817$_{-0.021}^{+0.021}$ & 9.3$^{+2.9}_{-2.9}$ \\
0923 & 6067$^{+41}_{-53}$ & 26.01$_{-0.57}^{+0.59}$ & 2.70 & 7.59$_{-0.15}^{+0.15}$ & 1.093$_{-0.032}^{+0.032}$ & 8.5$^{+3.2}_{-3.2}$ \\
0996 & 6010$^{+39}_{-51}$ & 19.96$_{-0.39}^{+0.20}$ & 1.45 & 6.78$_{-0.12}^{+0.11}$ & 0.976$_{-0.028}^{+0.027}$ & -0.9$^{+2.9}_{-2.7}$ \\
1015 & 5933$^{+38}_{-49}$ & 18.42$_{-0.18}^{+0.18}$ & 0.68 & 6.68$_{-0.10}^{+0.11}$ & 0.962$_{-0.026}^{+0.026}$ & 0.6$^{+2.8}_{-2.8}$ \\
1032 & 5134$^{+26}_{-34}$ & 10.92$_{-0.12}^{+0.26}$ & 0.51 & 6.87$_{-0.09}^{+0.11}$ & 0.989$_{-0.026}^{+0.028}$ & 32.8$^{+3.4}_{-3.7}$ \\
1095 & 5063$^{+28}_{-35}$ & 6.36$_{-0.17}^{+0.12}$ & 1.18 & 5.39$_{-0.10}^{+0.08}$ & 0.776$_{-0.022}^{+0.021}$ & 6.0$^{+3.1}_{-2.9}$ \\
1110 & 4438$^{+18}_{-23}$ & 2.47$_{-0.12}^{+0.13}$ & 4.41 & 4.37$_{-0.11}^{+0.12}$ & 0.629$_{-0.022}^{+0.022}$ & -1.4$^{+3.4}_{-3.5}$ \\
1122 & 6728$^{+51}_{-67}$ & 55.22$_{-0.00}^{+0.68}$ & 0.97 & 8.99$_{-0.16}^{+0.17}$ & 1.295$_{-0.037}^{+0.038}$ & 0.0$^{+2.8}_{-2.9}$ \\
1124 & 5053$^{+28}_{-35}$ & 5.14$_{-0.21}^{+0.22}$ & 2.97 & 4.87$_{-0.12}^{+0.12}$ & 0.701$_{-0.023}^{+0.023}$ & -4.0$^{+3.1}_{-3.2}$ \\
1132 & 6401$^{+56}_{-69}$ & 52.38$_{-4.65}^{+3.75}$ & 36.26 & 9.68$_{-0.47}^{+0.40}$ & 1.394$_{-0.075}^{+0.065}$ & 20.9$^{+6.5}_{-5.6}$ \\
1139 & 6651$^{+63}_{-77}$ & 51.28$_{-0.62}^{+0.63}$ & 0.91 & 8.87$_{-0.20}^{+0.20}$ & 1.277$_{-0.040}^{+0.040}$ & 1.1$^{+3.2}_{-3.2}$ \\
1182 & 5956$^{+43}_{-53}$ & 19.39$_{-0.39}^{+0.40}$ & 1.86 & 6.80$_{-0.13}^{+0.13}$ & 0.979$_{-0.029}^{+0.029}$ & 1.5$^{+3.0}_{-3.0}$ \\
1200 & 6246$^{+49}_{-61}$ & 31.51$_{-0.35}^{+0.72}$ & 1.76 & 7.88$_{-0.15}^{+0.17}$ & 1.135$_{-0.033}^{+0.035}$ & 4.8$^{+3.1}_{-3.2}$ \\
1207 & 6016$^{+44}_{-55}$ & 18.93$_{-0.71}^{+0.37}$ & 3.86 & 6.58$_{-0.17}^{+0.13}$ & 0.948$_{-0.032}^{+0.028}$ & -4.0$^{+3.2}_{-2.8}$ \\
1215 & 5968$^{+44}_{-55}$ & 18.29$_{-0.35}^{+0.55}$ & 2.32 & 6.58$_{-0.13}^{+0.15}$ & 0.947$_{-0.028}^{+0.030}$ & -2.3$^{+2.9}_{-3.1}$ \\
1220 & 5309$^{+32}_{-39}$ & 6.82$_{-0.17}^{+0.23}$ & 1.69 & 5.07$_{-0.09}^{+0.11}$ & 0.731$_{-0.021}^{+0.023}$ & -6.3$^{+2.7}_{-2.9}$ \\
1298 & 4975$^{+26}_{-32}$ & 4.94$_{-0.24}^{+0.22}$ & 5.16 & 4.92$_{-0.13}^{+0.12}$ & 0.709$_{-0.025}^{+0.024}$ & -1.0$^{+3.5}_{-3.3}$ \\
1305 & 4376$^{+18}_{-22}$ & 2.20$_{-0.17}^{+0.17}$ & 9.36 & 4.24$_{-0.17}^{+0.17}$ & 0.611$_{-0.028}^{+0.028}$ & -3.3$^{+4.4}_{-4.4}$ \\
1309 & 6569$^{+61}_{-74}$ & 47.93$_{-1.70}^{+1.18}$ & 2.29 & 8.79$_{-0.24}^{+0.21}$ & 1.265$_{-0.045}^{+0.042}$ & 3.1$^{+3.6}_{-3.4}$ \\
1332 & 4918$^{+25}_{-31}$ & 4.13$_{-0.16}^{+0.14}$ & 3.71 & 4.60$_{-0.10}^{+0.09}$ & 0.663$_{-0.021}^{+0.020}$ & -6.0$^{+3.0}_{-2.8}$ \\
1454 & 4700$^{+23}_{-28}$ & 3.13$_{-0.23}^{+0.24}$ & 14.16 & 4.38$_{-0.17}^{+0.17}$ & 0.631$_{-0.028}^{+0.029}$ & -4.9$^{+4.3}_{-4.3}$ \\
1514 & 5963$^{+42}_{-53}$ & 18.76$_{-0.54}^{+0.19}$ & 1.75 & 6.67$_{-0.14}^{+0.11}$ & 0.961$_{-0.030}^{+0.027}$ & -0.7$^{+3.1}_{-2.8}$ \\
1531 & 4449$^{+19}_{-24}$ & 2.25$_{-0.15}^{+0.19}$ & 12.95 & 4.15$_{-0.15}^{+0.18}$ & 0.598$_{-0.025}^{+0.029}$ & -6.4$^{+3.9}_{-4.5}$ \\
1593 & 5583$^{+34}_{-44}$ & 10.96$_{-0.20}^{+0.20}$ & 2.24 & 5.82$_{-0.10}^{+0.10}$ & 0.838$_{-0.023}^{+0.023}$ & -1.0$^{+2.8}_{-2.8}$ \\
1613 & 6330$^{+48}_{-61}$ & 31.80$_{-0.68}^{+0.35}$ & 1.89 & 7.71$_{-0.16}^{+0.14}$ & 1.110$_{-0.034}^{+0.032}$ & -1.0$^{+3.0}_{-2.9}$ \\
1776 & 5641$^{+36}_{-46}$ & 13.51$_{-0.26}^{+0.13}$ & 1.36 & 6.33$_{-0.11}^{+0.10}$ & 0.911$_{-0.026}^{+0.025}$ & 5.5$^{+3.0}_{-2.9}$ \\
1794 & 6054$^{+46}_{-57}$ & 20.82$_{-0.21}^{+0.42}$ & 0.79 & 6.82$_{-0.12}^{+0.14}$ & 0.982$_{-0.028}^{+0.029}$ & -2.0$^{+2.8}_{-2.9}$ \\
1797 & 6211$^{+49}_{-60}$ & 26.01$_{-0.53}^{+0.55}$ & 2.16 & 7.24$_{-0.15}^{+0.15}$ & 1.043$_{-0.032}^{+0.032}$ & -2.4$^{+3.0}_{-3.0}$ \\
1856 & 6241$^{+46}_{-58}$ & 28.41$_{-0.60}^{+0.31}$ & 2.13 & 7.50$_{-0.15}^{+0.13}$ & 1.080$_{-0.032}^{+0.031}$ & -0.1$^{+3.0}_{-2.8}$ \\
1883 & 4612$^{+20}_{-26}$ & 3.85$_{-0.34}^{+0.30}$ & 18.99 & 5.06$_{-0.23}^{+0.21}$ & 0.728$_{-0.036}^{+0.034}$ & 11.5$^{+5.6}_{-5.2}$ \\
1924 & 6066$^{+46}_{-57}$ & 21.46$_{-0.22}^{+0.22}$ & 1.21 & 6.90$_{-0.12}^{+0.12}$ & 0.993$_{-0.028}^{+0.028}$ & -1.3$^{+2.8}_{-2.8}$ \\
2016 & 4319$^{+16}_{-21}$ & 2.29$_{-0.08}^{+0.06}$ & 1.99 & 4.44$_{-0.08}^{+0.07}$ & 0.640$_{-0.019}^{+0.018}$ & 2.3$^{+3.0}_{-2.8}$ \\
2034 & 4762$^{+23}_{-29}$ & 3.94$_{-0.13}^{+0.17}$ & 2.89 & 4.79$_{-0.10}^{+0.12}$ & 0.690$_{-0.021}^{+0.023}$ & 2.4$^{+3.1}_{-3.4}$ \\
2106 & 5164$^{+28}_{-35}$ & 8.79$_{-0.26}^{+0.27}$ & 1.43 & 6.09$_{-0.12}^{+0.12}$ & 0.877$_{-0.026}^{+0.026}$ & 16.9$^{+3.5}_{-3.5}$ \\
2126 & 5352$^{+31}_{-40}$ & 7.26$_{-0.24}^{+0.31}$ & 3.06 & 5.15$_{-0.11}^{+0.13}$ & 0.742$_{-0.023}^{+0.025}$ & -6.0$^{+2.9}_{-3.2}$ \\
2244 & 4627$^{+22}_{-28}$ & 3.86$_{-0.14}^{+0.18}$ & 2.81 & 5.02$_{-0.11}^{+0.13}$ & 0.723$_{-0.022}^{+0.025}$ & 10.5$^{+3.4}_{-3.8}$ \\
2284 & 5365$^{+32}_{-40}$ & 9.80$_{-0.09}^{+0.09}$ & 0.52 & 5.96$_{-0.09}^{+0.09}$ & 0.858$_{-0.023}^{+0.023}$ & 8.4$^{+2.9}_{-2.9}$ \\
2311 & 5413$^{+35}_{-43}$ & 8.80$_{-0.23}^{+0.24}$ & 4.16 & 5.55$_{-0.11}^{+0.11}$ & 0.799$_{-0.024}^{+0.024}$ & -0.4$^{+3.0}_{-3.0}$ \\
2341 & 5773$^{+40}_{-50}$ & 13.31$_{-0.12}^{+0.25}$ & 1.94 & 6.00$_{-0.10}^{+0.11}$ & 0.864$_{-0.024}^{+0.025}$ & -4.4$^{+2.7}_{-2.8}$ \\
2345 & 6802$^{+65}_{-80}$ & 64.81$_{-0.82}^{+0.84}$ & 1.62 & 9.53$_{-0.21}^{+0.21}$ & 1.373$_{-0.043}^{+0.043}$ & 4.1$^{+3.3}_{-3.3}$ \\
2366 & 5356$^{+31}_{-40}$ & 8.22$_{-0.28}^{+0.22}$ & 1.94 & 5.48$_{-0.12}^{+0.10}$ & 0.789$_{-0.025}^{+0.023}$ & -0.2$^{+3.1}_{-2.9}$ \\
2462 & 5385$^{+32}_{-40}$ & 7.98$_{-0.20}^{+0.21}$ & 1.45 & 5.34$_{-0.10}^{+0.10}$ & 0.768$_{-0.022}^{+0.023}$ & -3.5$^{+2.8}_{-2.8}$ \\
2506 & 6111$^{+44}_{-56}$ & 23.80$_{-0.49}^{+0.25}$ & 1.26 & 7.16$_{-0.14}^{+0.12}$ & 1.030$_{-0.031}^{+0.029}$ & 0.5$^{+3.0}_{-2.8}$ \\
2644 & 5680$^{+36}_{-46}$ & 12.06$_{-0.32}^{+0.22}$ & 1.15 & 5.90$_{-0.12}^{+0.10}$ & 0.849$_{-0.025}^{+0.024}$ & -3.0$^{+2.9}_{-2.7}$ \\
2665 & 5362$^{+33}_{-41}$ & 9.77$_{-0.19}^{+0.19}$ & 0.78 & 5.96$_{-0.10}^{+0.10}$ & 0.858$_{-0.024}^{+0.024}$ & 8.4$^{+3.0}_{-3.1}$ \\
2741 & 4811$^{+23}_{-29}$ & 3.53$_{-0.26}^{+0.30}$ & 14.15 & 4.44$_{-0.17}^{+0.19}$ & 0.640$_{-0.029}^{+0.031}$ & -6.4$^{+4.2}_{-4.6}$ \\
2786 & 6074$^{+44}_{-55}$ & 22.53$_{-0.46}^{+0.23}$ & 1.11 & 7.05$_{-0.14}^{+0.12}$ & 1.015$_{-0.030}^{+0.029}$ & 0.5$^{+3.0}_{-2.8}$ \\
2870 & 4887$^{+26}_{-32}$ & 3.92$_{-0.21}^{+0.26}$ & 8.08 & 4.54$_{-0.13}^{+0.16}$ & 0.654$_{-0.024}^{+0.027}$ & -6.4$^{+3.4}_{-3.9}$ \\
2880 & 5198$^{+28}_{-36}$ & 7.12$_{-0.19}^{+0.13}$ & 0.60 & 5.41$_{-0.10}^{+0.08}$ & 0.779$_{-0.022}^{+0.021}$ & 2.9$^{+3.0}_{-2.8}$ \\
3019 & 4325$^{+18}_{-22}$ & 2.33$_{-0.12}^{+0.13}$ & 9.22 & 4.47$_{-0.12}^{+0.13}$ & 0.644$_{-0.023}^{+0.024}$ & 2.7$^{+3.6}_{-3.8}$ \\
3031 & 7048$^{+64}_{-81}$ & 86.19$_{-1.15}^{+0.00}$ & 0.68 & 10.24$_{-0.22}^{+0.21}$ & 1.474$_{-0.046}^{+0.045}$ & 9.7$^{+3.4}_{-3.3}$ \\
3063 & 4295$^{+18}_{-22}$ & 2.29$_{-0.14}^{+0.15}$ & 6.12 & 4.50$_{-0.14}^{+0.15}$ & 0.647$_{-0.025}^{+0.026}$ & 4.0$^{+4.0}_{-4.2}$ \\
3163 & 4579$^{+22}_{-27}$ & 3.87$_{-0.25}^{+0.23}$ & 7.42 & 5.14$_{-0.17}^{+0.16}$ & 0.740$_{-0.030}^{+0.029}$ & 13.9$^{+4.6}_{-4.4}$ \\
3179 & 6154$^{+44}_{-56}$ & 28.14$_{-0.92}^{+0.32}$ & 2.34 & 7.67$_{-0.18}^{+0.13}$ & 1.105$_{-0.036}^{+0.031}$ & 5.9$^{+3.4}_{-3.0}$ \\
3187 & 4495$^{+20}_{-25}$ & 2.70$_{-0.13}^{+0.14}$ & 7.58 & 4.45$_{-0.12}^{+0.13}$ & 0.641$_{-0.022}^{+0.023}$ & -0.3$^{+3.4}_{-3.6}$ \\
\enddata
\tablecomments {Derived parameters for our \VK\ \teffs\ ($\S$\ref{sec:methods}). Bolometric flux values are in units of $10^{-10}$ erg cm$^{-2}$ s$^{-1}$, and angular diameters in units of $10^{-2}$ milliseconds of arc.}

\end{deluxetable}
\clearpage 
\begin{deluxetable}{ccccccc}
\tabletypesize{\scriptsize}
\tablecaption{Derived Stellar Properties ($DANCe$) \label{tab3}}
\tablewidth{0pt}
\tablehead{
\colhead{HII}
&\colhead{$T_{\rm eff}$}
&\colhead{$\mathcal{F}_{bol}$}
&\colhead{$\chi_{\nu}^2$}
&\colhead{Angular Diameter}
&\colhead{Radius}
&\colhead{$\Delta$Radius}\\
\colhead{}
&\colhead{}
&\colhead{$\times 10^{-10}$}
&\colhead{}
&\colhead{$\times 10^{-2}$}
&\colhead{}
&\colhead{}\\
\colhead{}
&\colhead{(K)}
&\colhead{(erg cm$^{-2}$ s$^{-1}$)}
&\colhead{}
&\colhead{(mas)}
&\colhead{($R_{\odot})$}
&\colhead{(\%)}
} 
\startdata
0025 & 6481$^{+125}_{-125}$ & 46.67$_{-1.69}^{+1.18}$ & 2.20 & 8.91$_{-0.38}^{+0.36}$ & 1.283$_{-0.062}^{+0.059}$ & 8.0$^{+5.2}_{-5.0}$ \\
0034 & 5075$^{+125}_{-125}$ & 5.84$_{-0.20}^{+0.26}$ & 3.64 & 5.14$_{-0.27}^{+0.28}$ & 0.740$_{-0.042}^{+0.043}$ & 0.9$^{+5.7}_{-5.9}$ \\
0097 & 4586$^{+125}_{-125}$ & 4.26$_{-0.20}^{+0.22}$ & 4.73 & 5.38$_{-0.32}^{+0.32}$ & 0.775$_{-0.049}^{+0.050}$ & 19.1$^{+7.6}_{-7.7}$ \\
0120 & 5683$^{+125}_{-125}$ & 14.64$_{-0.71}^{+0.45}$ & 2.63 & 6.49$_{-0.33}^{+0.30}$ & 0.935$_{-0.051}^{+0.048}$ & 6.7$^{+5.9}_{-5.5}$ \\
0129 & 5253$^{+125}_{-125}$ & 9.10$_{-0.26}^{+0.28}$ & 2.46 & 5.99$_{-0.30}^{+0.30}$ & 0.862$_{-0.047}^{+0.047}$ & 12.3$^{+6.1}_{-6.1}$ \\
0152 & 5767$^{+125}_{-125}$ & 15.40$_{-0.15}^{+0.15}$ & 0.52 & 6.46$_{-0.28}^{+0.28}$ & 0.931$_{-0.046}^{+0.046}$ & 3.2$^{+5.1}_{-5.1}$ \\
0164 & 6405$^{+125}_{-125}$ & 45.49$_{-0.57}^{+0.00}$ & 0.79 & 9.01$_{-0.36}^{+0.35}$ & 1.297$_{-0.059}^{+0.058}$ & 12.4$^{+5.1}_{-5.1}$ \\
0174 & 4997$^{+125}_{-125}$ & 8.06$_{-0.25}^{+0.27}$ & 1.95 & 6.23$_{-0.33}^{+0.33}$ & 0.897$_{-0.051}^{+0.051}$ & 24.6$^{+7.1}_{-7.1}$ \\
0193 & 5265$^{+125}_{-125}$ & 9.79$_{-0.39}^{+0.31}$ & 1.95 & 6.19$_{-0.32}^{+0.31}$ & 0.891$_{-0.050}^{+0.049}$ & 15.6$^{+6.5}_{-6.3}$ \\
0250 & 5655$^{+125}_{-125}$ & 15.14$_{-0.61}^{+0.48}$ & 1.26 & 6.67$_{-0.32}^{+0.31}$ & 0.960$_{-0.051}^{+0.050}$ & 10.7$^{+5.9}_{-5.8}$ \\
0253 & 5465$^{+125}_{-125}$ & 14.82$_{-0.80}^{+1.05}$ & 5.29 & 7.06$_{-0.37}^{+0.41}$ & 1.017$_{-0.059}^{+0.063}$ & 24.9$^{+7.2}_{-7.8}$ \\
0263 & 5007$^{+125}_{-125}$ & 8.25$_{-0.09}^{+0.18}$ & 0.41 & 6.28$_{-0.32}^{+0.32}$ & 0.904$_{-0.050}^{+0.050}$ & 25.2$^{+6.9}_{-7.0}$ \\
0293 & 5565$^{+125}_{-125}$ & 13.98$_{-0.70}^{+0.60}$ & 2.52 & 6.61$_{-0.34}^{+0.33}$ & 0.952$_{-0.053}^{+0.052}$ & 13.3$^{+6.3}_{-6.2}$ \\
0345 & 4959$^{+125}_{-125}$ & 8.63$_{-0.47}^{+0.51}$ & 3.82 & 6.54$_{-0.37}^{+0.38}$ & 0.942$_{-0.058}^{+0.059}$ & 32.2$^{+8.1}_{-8.3}$ \\
0380 & 4343$^{+125}_{-125}$ & 2.43$_{-0.10}^{+0.11}$ & 4.18 & 4.53$_{-0.28}^{+0.28}$ & 0.652$_{-0.043}^{+0.043}$ & 3.7$^{+6.8}_{-6.8}$ \\
0405 & 6194$^{+125}_{-125}$ & 33.05$_{-0.77}^{+0.80}$ & 0.81 & 8.21$_{-0.34}^{+0.35}$ & 1.182$_{-0.056}^{+0.056}$ & 11.5$^{+5.3}_{-5.3}$ \\
0430 & 5365$^{+125}_{-125}$ & 8.84$_{-0.24}^{+0.16}$ & 1.76 & 5.66$_{-0.27}^{+0.27}$ & 0.815$_{-0.044}^{+0.043}$ & 2.9$^{+5.5}_{-5.4}$ \\
0470 & 7048$^{+125}_{-125}$ & 74.70$_{-1.83}^{+0.94}$ & 2.15 & 9.53$_{-0.36}^{+0.34}$ & 1.373$_{-0.060}^{+0.058}$ & 2.2$^{+4.5}_{-4.3}$ \\
0489 & 5851$^{+125}_{-125}$ & 20.14$_{-0.84}^{+0.67}$ & 1.83 & 7.18$_{-0.34}^{+0.33}$ & 1.034$_{-0.054}^{+0.053}$ & 11.4$^{+5.9}_{-5.7}$ \\
0514 & 5679$^{+125}_{-125}$ & 15.66$_{-0.48}^{+0.33}$ & 1.50 & 6.72$_{-0.31}^{+0.30}$ & 0.968$_{-0.050}^{+0.049}$ & 10.6$^{+5.7}_{-5.6}$ \\
0530 & 6918$^{+125}_{-125}$ & 72.92$_{-1.86}^{+2.93}$ & 1.53 & 9.77$_{-0.37}^{+0.40}$ & 1.408$_{-0.062}^{+0.066}$ & 4.8$^{+4.7}_{-4.9}$ \\
0627 & 6413$^{+125}_{-125}$ & 39.56$_{-0.91}^{+0.47}$ & 0.95 & 8.38$_{-0.34}^{+0.33}$ & 1.207$_{-0.056}^{+0.055}$ & 4.2$^{+4.8}_{-4.7}$ \\
0636 & 4779$^{+125}_{-125}$ & 4.24$_{-0.25}^{+0.27}$ & 8.46 & 4.94$_{-0.30}^{+0.30}$ & 0.711$_{-0.046}^{+0.047}$ & 5.0$^{+6.7}_{-6.9}$ \\
0708 & 5882$^{+125}_{-125}$ & 26.39$_{-0.94}^{+0.99}$ & 1.72 & 8.13$_{-0.37}^{+0.38}$ & 1.171$_{-0.060}^{+0.060}$ & 24.8$^{+6.4}_{-6.4}$ \\
0746 & 5465$^{+125}_{-125}$ & 9.88$_{-0.27}^{+0.28}$ & 2.06 & 5.77$_{-0.28}^{+0.28}$ & 0.830$_{-0.044}^{+0.044}$ & 1.9$^{+5.4}_{-5.4}$ \\
0879 & 4676$^{+125}_{-125}$ & 3.19$_{-0.22}^{+0.19}$ & 7.24 & 4.47$_{-0.28}^{+0.27}$ & 0.644$_{-0.044}^{+0.042}$ & -2.5$^{+6.6}_{-6.4}$ \\
0882 & 4497$^{+125}_{-125}$ & 3.90$_{-0.15}^{+0.16}$ & 2.12 & 5.35$_{-0.32}^{+0.32}$ & 0.771$_{-0.049}^{+0.049}$ & 19.8$^{+7.6}_{-7.6}$ \\
0883 & 4532$^{+125}_{-125}$ & 2.55$_{-0.14}^{+0.17}$ & 9.55 & 4.26$_{-0.26}^{+0.27}$ & 0.613$_{-0.040}^{+0.042}$ & -5.1$^{+6.2}_{-6.5}$ \\
0916 & 5117$^{+125}_{-125}$ & 7.49$_{-0.14}^{+0.07}$ & 0.27 & 5.73$_{-0.28}^{+0.28}$ & 0.825$_{-0.045}^{+0.044}$ & 11.2$^{+6.1}_{-6.0}$ \\
0923 & 6051$^{+125}_{-125}$ & 25.90$_{-0.29}^{+0.59}$ & 2.51 & 7.61$_{-0.32}^{+0.33}$ & 1.096$_{-0.052}^{+0.053}$ & 9.6$^{+5.2}_{-5.3}$ \\
0996 & 5969$^{+125}_{-125}$ & 19.81$_{-0.20}^{+0.20}$ & 0.93 & 6.84$_{-0.29}^{+0.29}$ & 0.986$_{-0.047}^{+0.047}$ & 1.7$^{+4.9}_{-4.9}$ \\
1015 & 5887$^{+125}_{-125}$ & 18.37$_{-0.37}^{+0.19}$ & 0.63 & 6.78$_{-0.30}^{+0.29}$ & 0.976$_{-0.048}^{+0.047}$ & 3.8$^{+5.1}_{-5.0}$ \\
1032 & 5309$^{+125}_{-125}$ & 11.79$_{-0.25}^{+0.13}$ & 0.69 & 6.67$_{-0.32}^{+0.32}$ & 0.961$_{-0.051}^{+0.050}$ & 23.3$^{+6.6}_{-6.5}$ \\
1095 & 5099$^{+125}_{-125}$ & 6.44$_{-0.17}^{+0.12}$ & 1.52 & 5.35$_{-0.27}^{+0.27}$ & 0.770$_{-0.043}^{+0.042}$ & 4.3$^{+5.8}_{-5.7}$ \\
1110 & 4424$^{+125}_{-125}$ & 2.44$_{-0.12}^{+0.13}$ & 4.13 & 4.37$_{-0.27}^{+0.27}$ & 0.630$_{-0.041}^{+0.042}$ & -1.1$^{+6.5}_{-6.5}$ \\
1122 & 6622$^{+125}_{-125}$ & 54.60$_{-0.68}^{+0.70}$ & 0.49 & 9.23$_{-0.35}^{+0.35}$ & 1.329$_{-0.059}^{+0.059}$ & 6.3$^{+4.7}_{-4.7}$ \\
1124 & 4875$^{+125}_{-125}$ & 4.76$_{-0.12}^{+0.13}$ & 1.28 & 5.03$_{-0.27}^{+0.27}$ & 0.725$_{-0.042}^{+0.042}$ & 4.1$^{+6.0}_{-6.0}$ \\
1132 & 6449$^{+125}_{-125}$ & 51.68$_{-3.87}^{+4.37}$ & 38.32 & 9.47$_{-0.51}^{+0.54}$ & 1.364$_{-0.080}^{+0.084}$ & 16.2$^{+6.8}_{-7.2}$ \\
1139 & 6600$^{+125}_{-125}$ & 51.63$_{-1.26}^{+0.00}$ & 0.70 & 9.04$_{-0.36}^{+0.34}$ & 1.301$_{-0.059}^{+0.057}$ & 4.8$^{+4.8}_{-4.6}$ \\
1182 & 5879$^{+125}_{-125}$ & 19.00$_{-0.76}^{+0.60}$ & 2.24 & 6.91$_{-0.32}^{+0.31}$ & 0.995$_{-0.052}^{+0.050}$ & 6.1$^{+5.5}_{-5.4}$ \\
1200 & 6207$^{+125}_{-125}$ & 31.53$_{-0.71}^{+0.37}$ & 1.58 & 7.98$_{-0.33}^{+0.32}$ & 1.150$_{-0.055}^{+0.053}$ & 7.9$^{+5.1}_{-5.0}$ \\
1207 & 5977$^{+125}_{-125}$ & 18.98$_{-1.08}^{+0.38}$ & 3.91 & 6.68$_{-0.34}^{+0.29}$ & 0.962$_{-0.053}^{+0.047}$ & -1.1$^{+5.5}_{-4.8}$ \\
1215 & 5975$^{+125}_{-125}$ & 18.39$_{-0.53}^{+0.36}$ & 2.37 & 6.58$_{-0.29}^{+0.28}$ & 0.948$_{-0.047}^{+0.046}$ & -2.5$^{+4.8}_{-4.7}$ \\
1220 & 5289$^{+125}_{-125}$ & 6.78$_{-0.17}^{+0.17}$ & 1.48 & 5.10$_{-0.25}^{+0.25}$ & 0.734$_{-0.039}^{+0.040}$ & -5.3$^{+5.1}_{-5.1}$ \\
1298 & 4846$^{+125}_{-125}$ & 4.77$_{-0.13}^{+0.18}$ & 2.46 & 5.09$_{-0.27}^{+0.28}$ & 0.734$_{-0.042}^{+0.043}$ & 6.3$^{+6.1}_{-6.3}$ \\
1305 & 4288$^{+125}_{-125}$ & 2.13$_{-0.12}^{+0.11}$ & 5.03 & 4.35$_{-0.28}^{+0.28}$ & 0.627$_{-0.043}^{+0.043}$ & 0.8$^{+6.9}_{-6.8}$ \\
1309 & 6467$^{+125}_{-125}$ & 45.57$_{-1.64}^{+1.74}$ & 2.33 & 8.84$_{-0.38}^{+0.38}$ & 1.273$_{-0.061}^{+0.062}$ & 7.8$^{+5.2}_{-5.2}$ \\
1332 & 4800$^{+125}_{-125}$ & 3.99$_{-0.10}^{+0.10}$ & 1.75 & 4.75$_{-0.25}^{+0.25}$ & 0.684$_{-0.040}^{+0.040}$ & 0.4$^{+5.8}_{-5.8}$ \\
1454 & 4669$^{+125}_{-125}$ & 3.11$_{-0.19}^{+0.21}$ & 12.06 & 4.43$_{-0.27}^{+0.28}$ & 0.638$_{-0.042}^{+0.043}$ & -3.3$^{+6.4}_{-6.5}$ \\
1514 & 5897$^{+125}_{-125}$ & 18.18$_{-0.54}^{+0.75}$ & 2.06 & 6.72$_{-0.30}^{+0.32}$ & 0.967$_{-0.049}^{+0.051}$ & 2.5$^{+5.1}_{-5.4}$ \\
1531 & 4216$^{+125}_{-125}$ & 2.20$_{-0.06}^{+0.06}$ & 2.12 & 4.58$_{-0.28}^{+0.28}$ & 0.659$_{-0.043}^{+0.043}$ & 7.5$^{+7.0}_{-7.0}$ \\
1593 & 5593$^{+125}_{-125}$ & 10.97$_{-0.29}^{+0.20}$ & 2.33 & 5.80$_{-0.27}^{+0.26}$ & 0.835$_{-0.043}^{+0.042}$ & -1.6$^{+5.1}_{-5.0}$ \\
1613 & 6316$^{+125}_{-125}$ & 31.74$_{-0.69}^{+0.35}$ & 1.71 & 7.74$_{-0.32}^{+0.31}$ & 1.114$_{-0.052}^{+0.051}$ & -0.1$^{+4.7}_{-4.6}$ \\
1776 & 5557$^{+125}_{-125}$ & 13.17$_{-0.39}^{+0.41}$ & 1.38 & 6.44$_{-0.30}^{+0.31}$ & 0.927$_{-0.049}^{+0.049}$ & 10.6$^{+5.8}_{-5.8}$ \\
1794 & 5995$^{+125}_{-125}$ & 20.48$_{-0.21}^{+0.21}$ & 0.45 & 6.90$_{-0.29}^{+0.29}$ & 0.993$_{-0.047}^{+0.047}$ & 1.5$^{+4.8}_{-4.8}$ \\
1797 & 6231$^{+125}_{-125}$ & 26.16$_{-0.79}^{+0.55}$ & 2.41 & 7.22$_{-0.31}^{+0.30}$ & 1.039$_{-0.050}^{+0.049}$ & -3.5$^{+4.7}_{-4.6}$ \\
1856 & 6161$^{+125}_{-125}$ & 27.99$_{-1.20}^{+0.63}$ & 2.08 & 7.64$_{-0.35}^{+0.32}$ & 1.100$_{-0.056}^{+0.052}$ & 5.2$^{+5.4}_{-5.0}$ \\
1883 & 4544$^{+125}_{-125}$ & 3.88$_{-0.35}^{+0.28}$ & 16.25 & 5.23$_{-0.37}^{+0.34}$ & 0.753$_{-0.056}^{+0.052}$ & 16.3$^{+8.7}_{-8.1}$ \\
1924 & 6086$^{+125}_{-125}$ & 21.43$_{-0.21}^{+0.22}$ & 1.43 & 6.85$_{-0.28}^{+0.28}$ & 0.986$_{-0.046}^{+0.046}$ & -2.8$^{+4.6}_{-4.6}$ \\
2016 & 4375$^{+125}_{-125}$ & 2.33$_{-0.08}^{+0.08}$ & 2.53 & 4.37$_{-0.26}^{+0.26}$ & 0.629$_{-0.040}^{+0.040}$ & -0.5$^{+6.3}_{-6.3}$ \\
2034 & 4669$^{+125}_{-125}$ & 3.81$_{-0.10}^{+0.14}$ & 1.79 & 4.91$_{-0.27}^{+0.28}$ & 0.706$_{-0.042}^{+0.043}$ & 7.1$^{+6.4}_{-6.5}$ \\
2106 & 5140$^{+125}_{-125}$ & 8.74$_{-0.26}^{+0.18}$ & 1.30 & 6.13$_{-0.31}^{+0.30}$ & 0.883$_{-0.049}^{+0.048}$ & 18.4$^{+6.6}_{-6.5}$ \\
2126 & 5302$^{+125}_{-125}$ & 7.17$_{-0.24}^{+0.25}$ & 2.24 & 5.22$_{-0.26}^{+0.26}$ & 0.752$_{-0.041}^{+0.041}$ & -3.4$^{+5.3}_{-5.3}$ \\
2244 & 4656$^{+125}_{-125}$ & 3.89$_{-0.17}^{+0.18}$ & 3.67 & 4.98$_{-0.29}^{+0.29}$ & 0.718$_{-0.045}^{+0.045}$ & 9.1$^{+6.8}_{-6.8}$ \\
2284 & 5418$^{+125}_{-125}$ & 9.97$_{-0.19}^{+0.19}$ & 0.82 & 5.89$_{-0.28}^{+0.28}$ & 0.848$_{-0.044}^{+0.044}$ & 5.6$^{+5.5}_{-5.5}$ \\
2311 & 5438$^{+125}_{-125}$ & 8.87$_{-0.30}^{+0.24}$ & 4.88 & 5.52$_{-0.27}^{+0.26}$ & 0.795$_{-0.043}^{+0.042}$ & -1.7$^{+5.3}_{-5.2}$ \\
2341 & 5750$^{+125}_{-125}$ & 13.37$_{-0.12}^{+0.25}$ & 1.58 & 6.06$_{-0.26}^{+0.27}$ & 0.872$_{-0.043}^{+0.043}$ & -2.7$^{+4.8}_{-4.8}$ \\
2345 & 6726$^{+125}_{-125}$ & 64.14$_{-2.46}^{+0.85}$ & 1.48 & 9.70$_{-0.41}^{+0.37}$ & 1.397$_{-0.066}^{+0.061}$ & 8.0$^{+5.1}_{-4.7}$ \\
2366 & 5368$^{+125}_{-125}$ & 8.25$_{-0.28}^{+0.22}$ & 2.05 & 5.46$_{-0.27}^{+0.26}$ & 0.787$_{-0.043}^{+0.042}$ & -0.7$^{+5.4}_{-5.3}$ \\
2462 & 5467$^{+125}_{-125}$ & 8.17$_{-0.27}^{+0.28}$ & 2.47 & 5.24$_{-0.25}^{+0.26}$ & 0.755$_{-0.040}^{+0.041}$ & -7.4$^{+4.9}_{-5.0}$ \\
2506 & 6005$^{+125}_{-125}$ & 22.96$_{-0.72}^{+0.50}$ & 1.16 & 7.28$_{-0.32}^{+0.31}$ & 1.048$_{-0.052}^{+0.051}$ & 6.7$^{+5.3}_{-5.2}$ \\
2644 & 5594$^{+125}_{-125}$ & 11.60$_{-0.21}^{+0.22}$ & 0.84 & 5.96$_{-0.27}^{+0.27}$ & 0.859$_{-0.044}^{+0.044}$ & 1.1$^{+5.1}_{-5.1}$ \\
2665 & 5366$^{+125}_{-125}$ & 9.71$_{-0.18}^{+0.19}$ & 0.79 & 5.93$_{-0.28}^{+0.28}$ & 0.854$_{-0.045}^{+0.045}$ & 7.8$^{+5.7}_{-5.7}$ \\
2741 & 4792$^{+125}_{-125}$ & 3.52$_{-0.26}^{+0.27}$ & 12.33 & 4.48$_{-0.29}^{+0.29}$ & 0.645$_{-0.044}^{+0.044}$ & -5.1$^{+6.5}_{-6.5}$ \\
2786 & 6071$^{+125}_{-125}$ & 22.46$_{-0.46}^{+0.23}$ & 1.13 & 7.04$_{-0.30}^{+0.29}$ & 1.014$_{-0.049}^{+0.048}$ & 0.6$^{+4.8}_{-4.7}$ \\
2870 & 4840$^{+125}_{-125}$ & 3.89$_{-0.18}^{+0.20}$ & 6.12 & 4.61$_{-0.26}^{+0.27}$ & 0.664$_{-0.041}^{+0.041}$ & -3.6$^{+5.9}_{-6.0}$ \\
2880 & 5261$^{+125}_{-125}$ & 7.34$_{-0.13}^{+0.13}$ & 0.55 & 5.36$_{-0.26}^{+0.26}$ & 0.772$_{-0.041}^{+0.041}$ & 0.3$^{+5.3}_{-5.3}$ \\
3019 & 4192$^{+125}_{-125}$ & 2.26$_{-0.12}^{+0.09}$ & 5.96 & 4.68$_{-0.31}^{+0.29}$ & 0.674$_{-0.047}^{+0.045}$ & 10.5$^{+7.7}_{-7.4}$ \\
3031 & 6935$^{+125}_{-125}$ & 82.20$_{-1.11}^{+1.14}$ & 0.71 & 10.33$_{-0.38}^{+0.38}$ & 1.487$_{-0.064}^{+0.064}$ & 10.6$^{+4.8}_{-4.8}$ \\
3063 & 4292$^{+125}_{-125}$ & 2.30$_{-0.14}^{+0.15}$ & 6.03 & 4.51$_{-0.30}^{+0.30}$ & 0.650$_{-0.045}^{+0.046}$ & 4.4$^{+7.2}_{-7.4}$ \\
3163 & 4456$^{+125}_{-125}$ & 3.71$_{-0.15}^{+0.11}$ & 2.66 & 5.31$_{-0.32}^{+0.31}$ & 0.765$_{-0.049}^{+0.048}$ & 19.6$^{+7.6}_{-7.5}$ \\
3179 & 6120$^{+125}_{-125}$ & 27.59$_{-0.90}^{+0.95}$ & 2.48 & 7.68$_{-0.34}^{+0.34}$ & 1.106$_{-0.055}^{+0.055}$ & 7.6$^{+5.3}_{-5.3}$ \\
3187 & 4464$^{+125}_{-125}$ & 2.70$_{-0.13}^{+0.12}$ & 6.60 & 4.52$_{-0.28}^{+0.27}$ & 0.651$_{-0.042}^{+0.042}$ & 1.7$^{+6.6}_{-6.5}$ \\
\enddata
\tablecomments {Derived parameters for the \dance\ \teffs\ ($\S$\ref{sec:methods}). Bolometric flux values are in units of $10^{-10}$ erg cm$^{-2}$ s$^{-1}$, and angular diameters in units of $10^{-2}$ milliseconds of arc.}

\end{deluxetable}

\clearpage
\appendix

\section{Spectral Energy Distribution Measurements and Fits}\label{sec:sed_appendix} 

\begin{figure}[!ht]
\centering
\includegraphics[trim=70 70 70 50,clip,width=\linewidth]{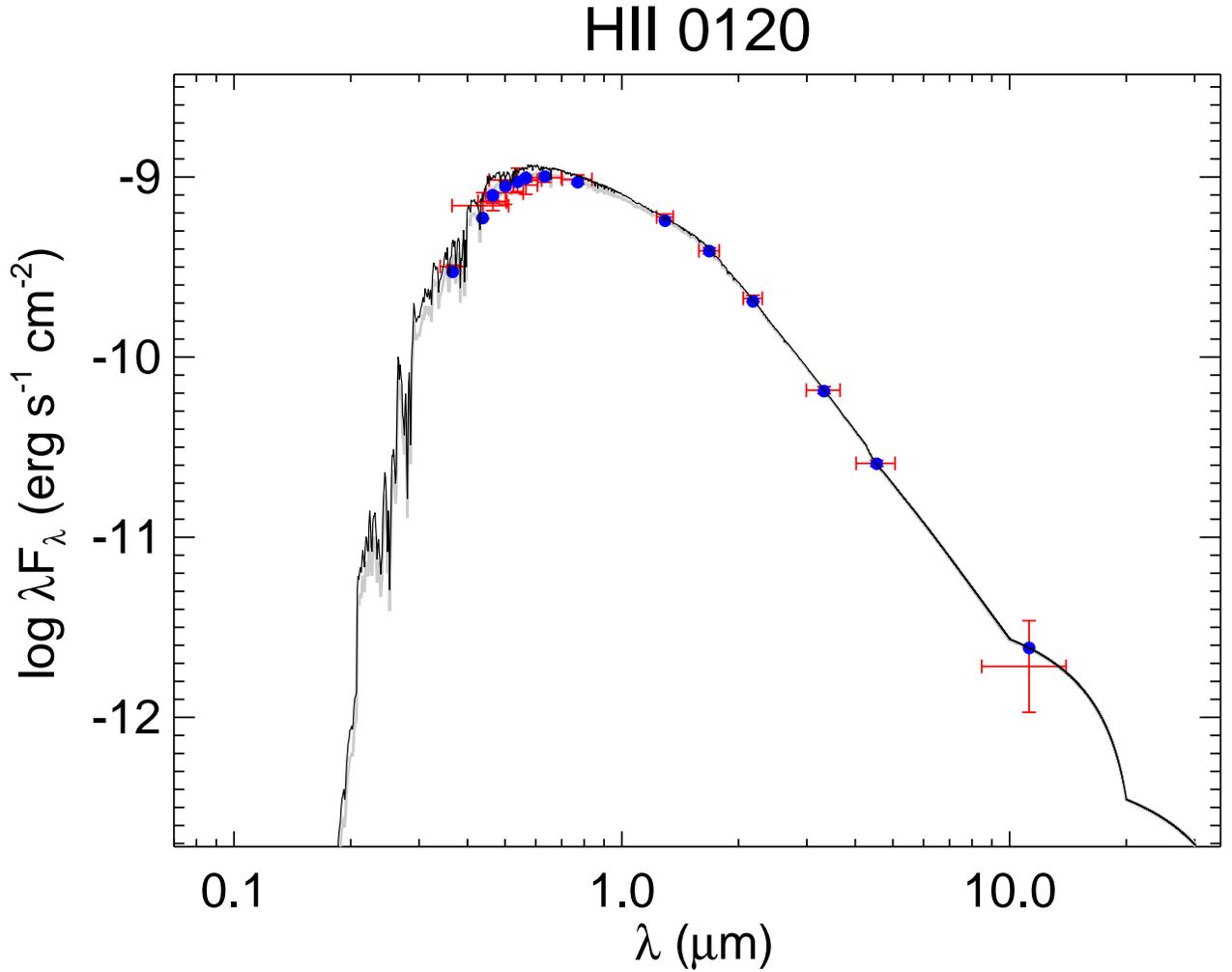}
\caption{HII~0120 is shown as example of the figure set. Each panel in the Figure Set is labeled at top by the HII number, and shows the observed fluxes (in units of erg cm$^{-2}$ s$^{-1}$) versus wavelength (in \micron) as red error bars, where the vertical error bar represents the uncertainty in the measurement and the horizontal ``error" bar represents the effective width of the passband. Also in each figure is the fitted SED model including extinction (light gray curve), on which is shown the model passband fluxes as blue dots. 
The corresponding un-extincted SED model is also shown (dark black curve); the reported \fbol\ is the sum over all wavelengths of this un-extincted model (see the text). The full figure set is displayed in Figures \ref{fig:seds_1}--\ref{fig:seds_14}.}
\label{fig:seds}
\end{figure}

In Figure Set \ref{fig:seds} we present the observed and fitted spectral energy distributions of the stars in our study sample (Tables~\ref{tab2}-\ref{tab3}). 

\begin{figure}[H]
  \centering
  \includegraphics[trim=60 60 60 60,clip,width=0.49\linewidth]{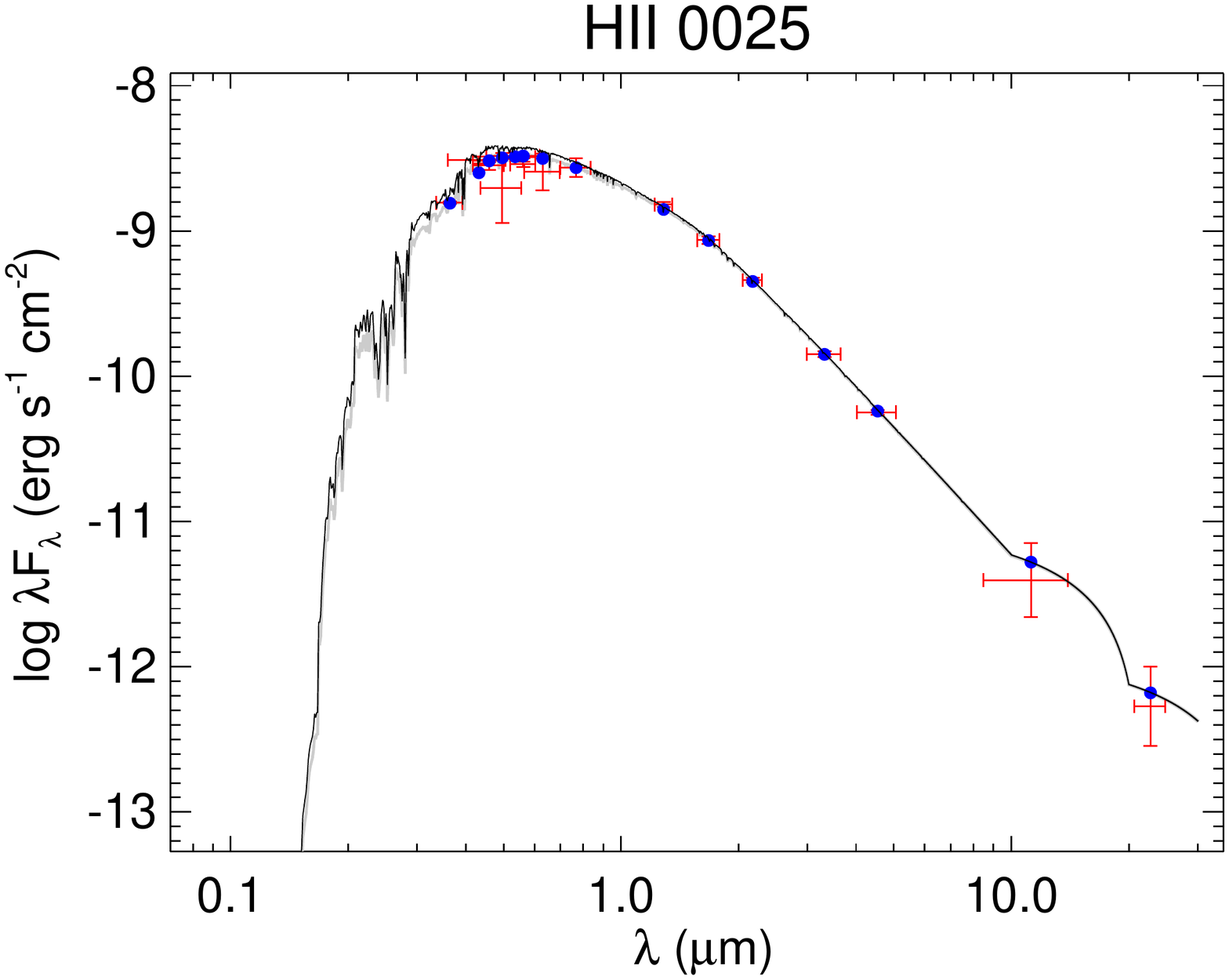}
  \includegraphics[trim=60 60 60 60,clip,width=0.49\linewidth]{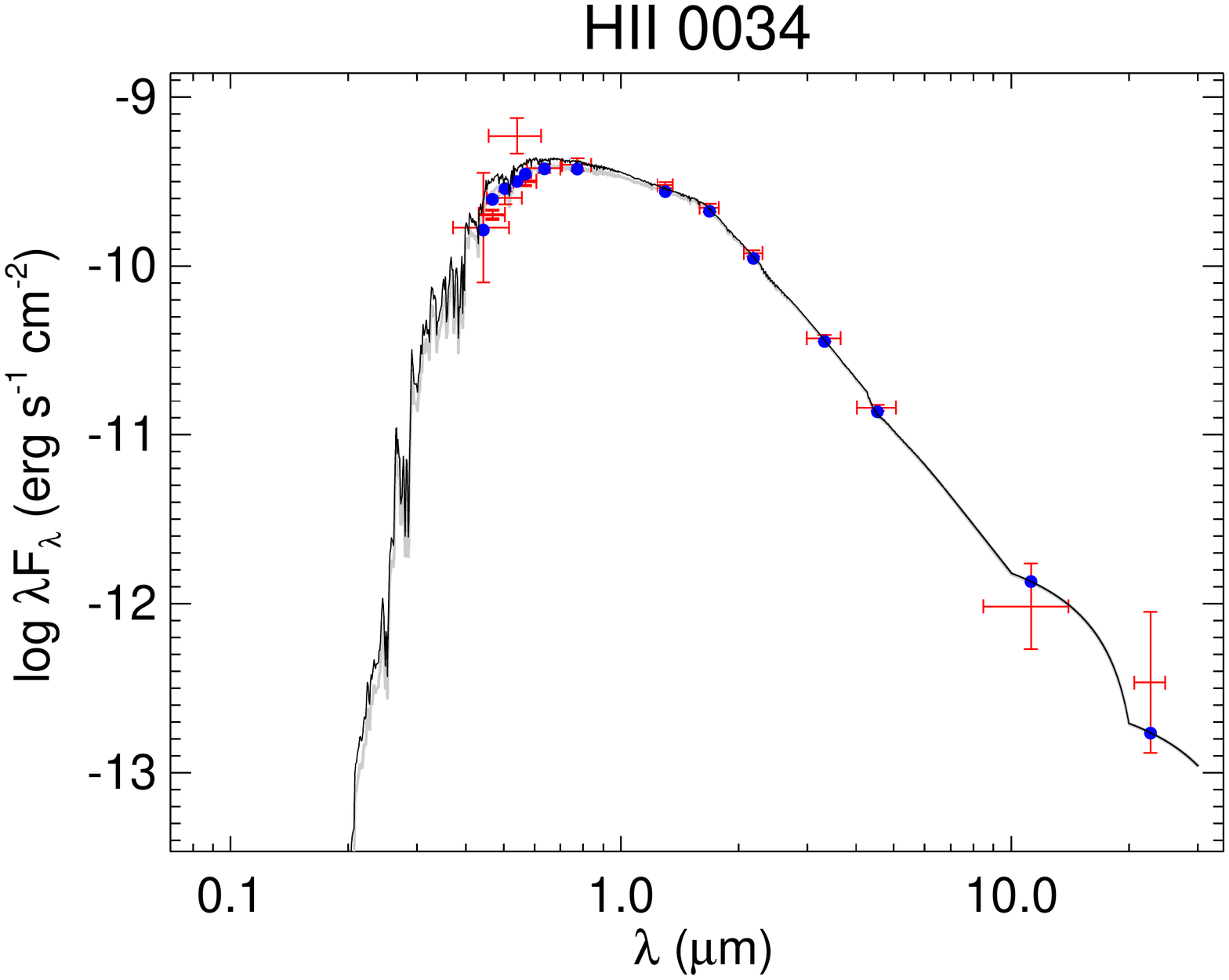}
  \includegraphics[trim=60 60 60 60,clip,width=0.49\linewidth]{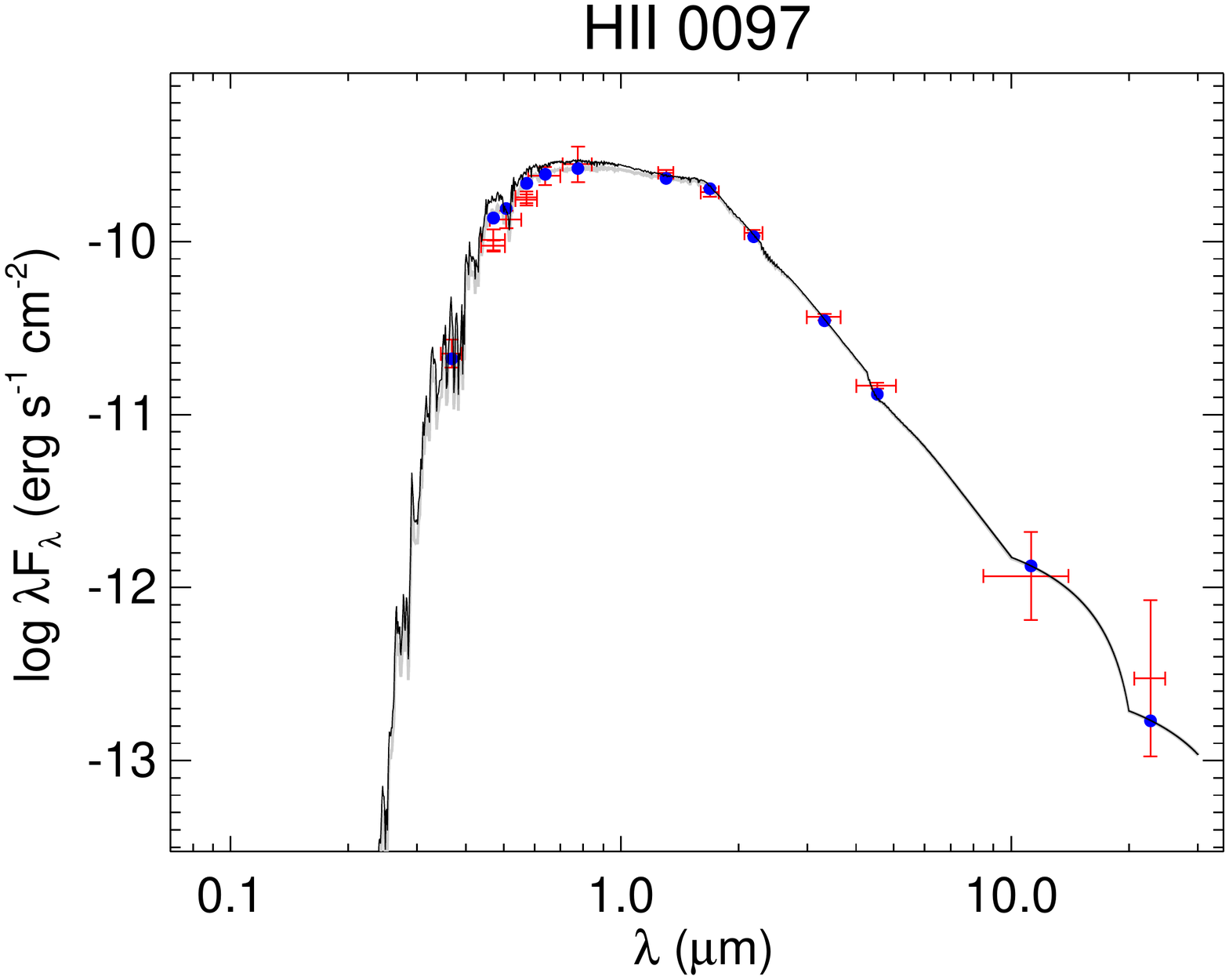}
  \includegraphics[trim=60 60 60 60,clip,width=0.49\linewidth]{pl_0120.pdf}
  \includegraphics[trim=60 60 60 60,clip,width=0.49\linewidth]{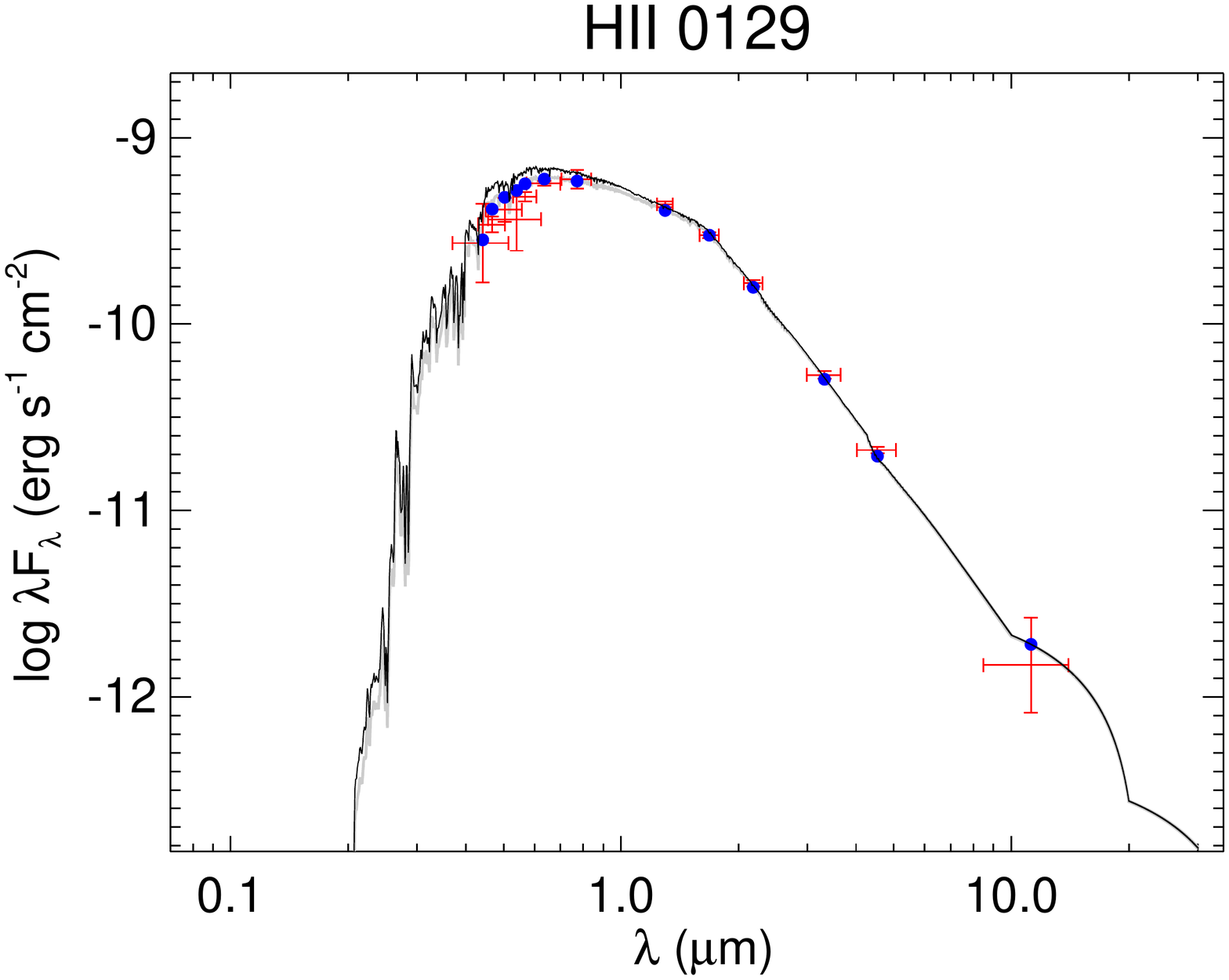}
  \includegraphics[trim=60 60 60 60,clip,width=0.49\linewidth]{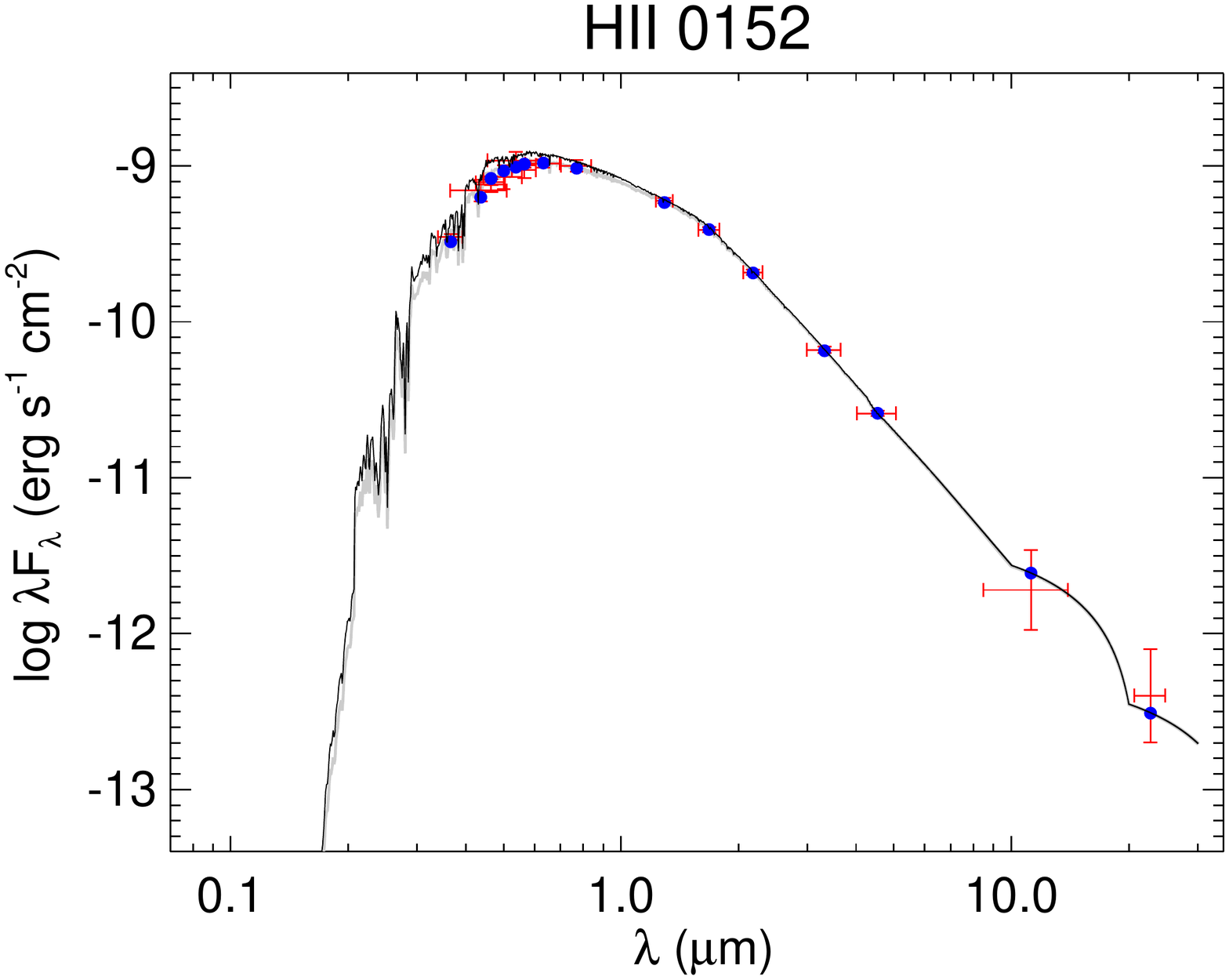}
  \caption{All labels, lines, symbols, and colors as in Figure \ref{fig:seds}.}
  \label{fig:seds_1}
\end{figure}

\begin{figure}[H]
  \centering
  \includegraphics[trim=60 60 60 60,clip,width=0.49\linewidth]{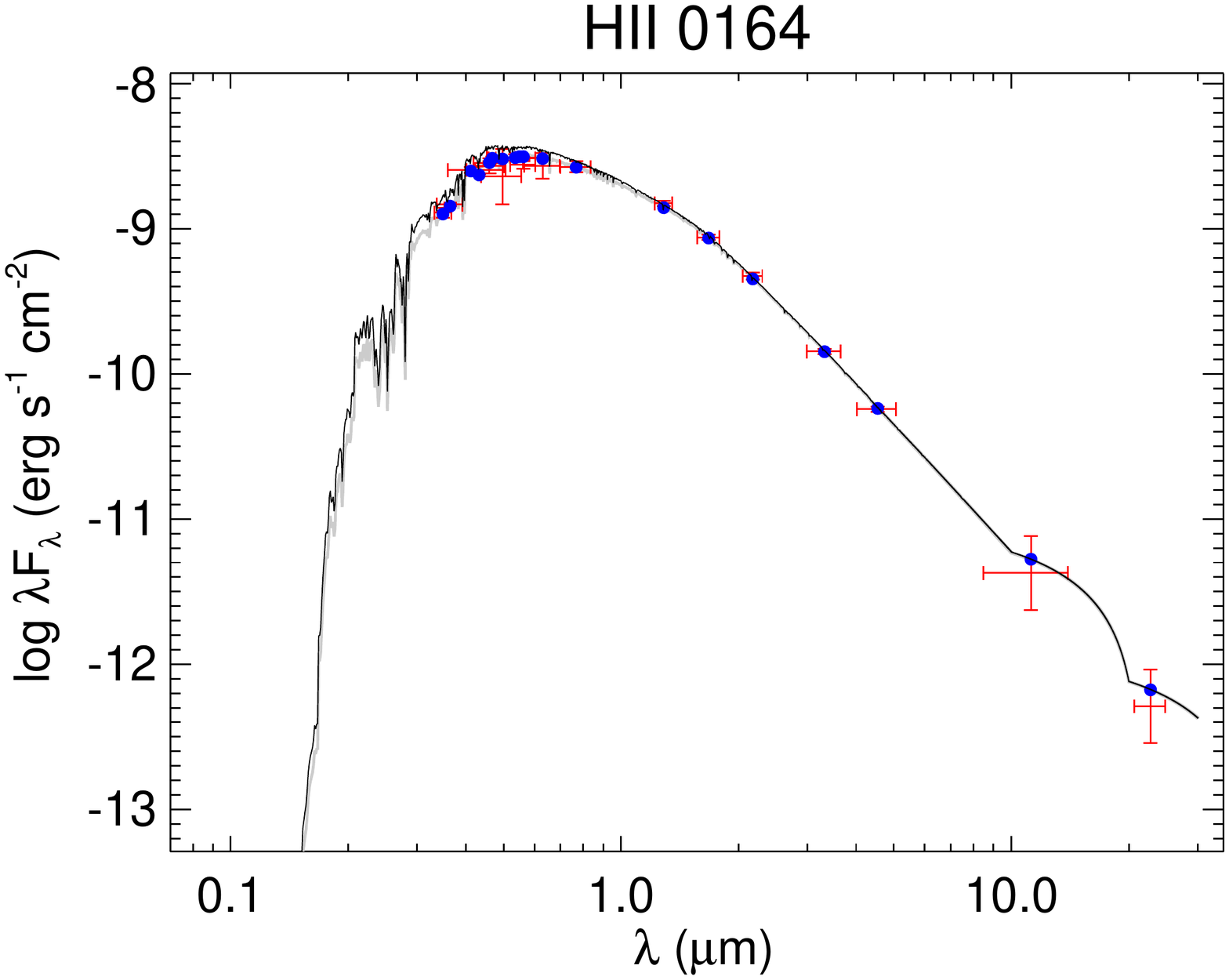}
  \includegraphics[trim=60 60 60 60,clip,width=0.49\linewidth]{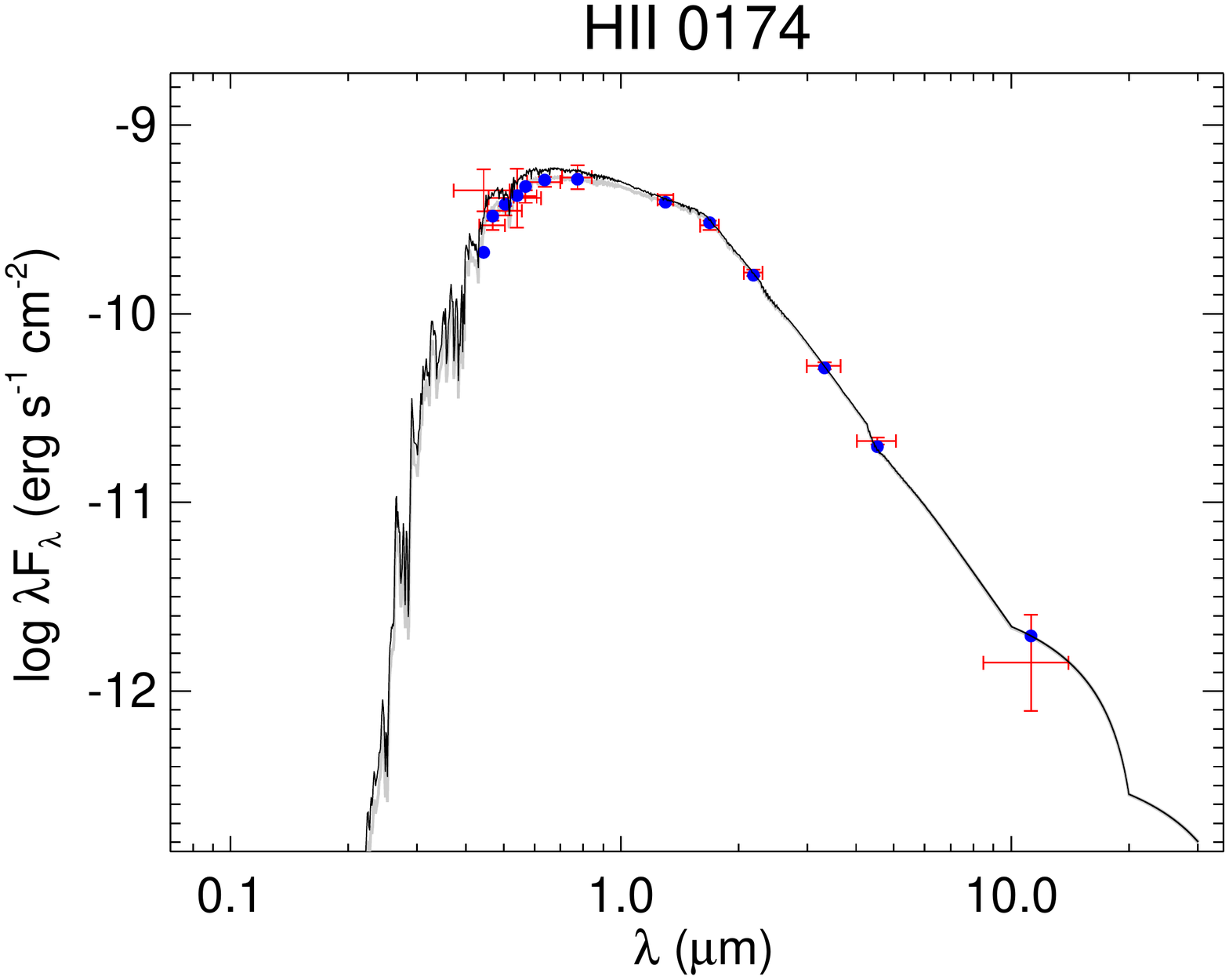}
  \includegraphics[trim=60 60 60 60,clip,width=0.49\linewidth]{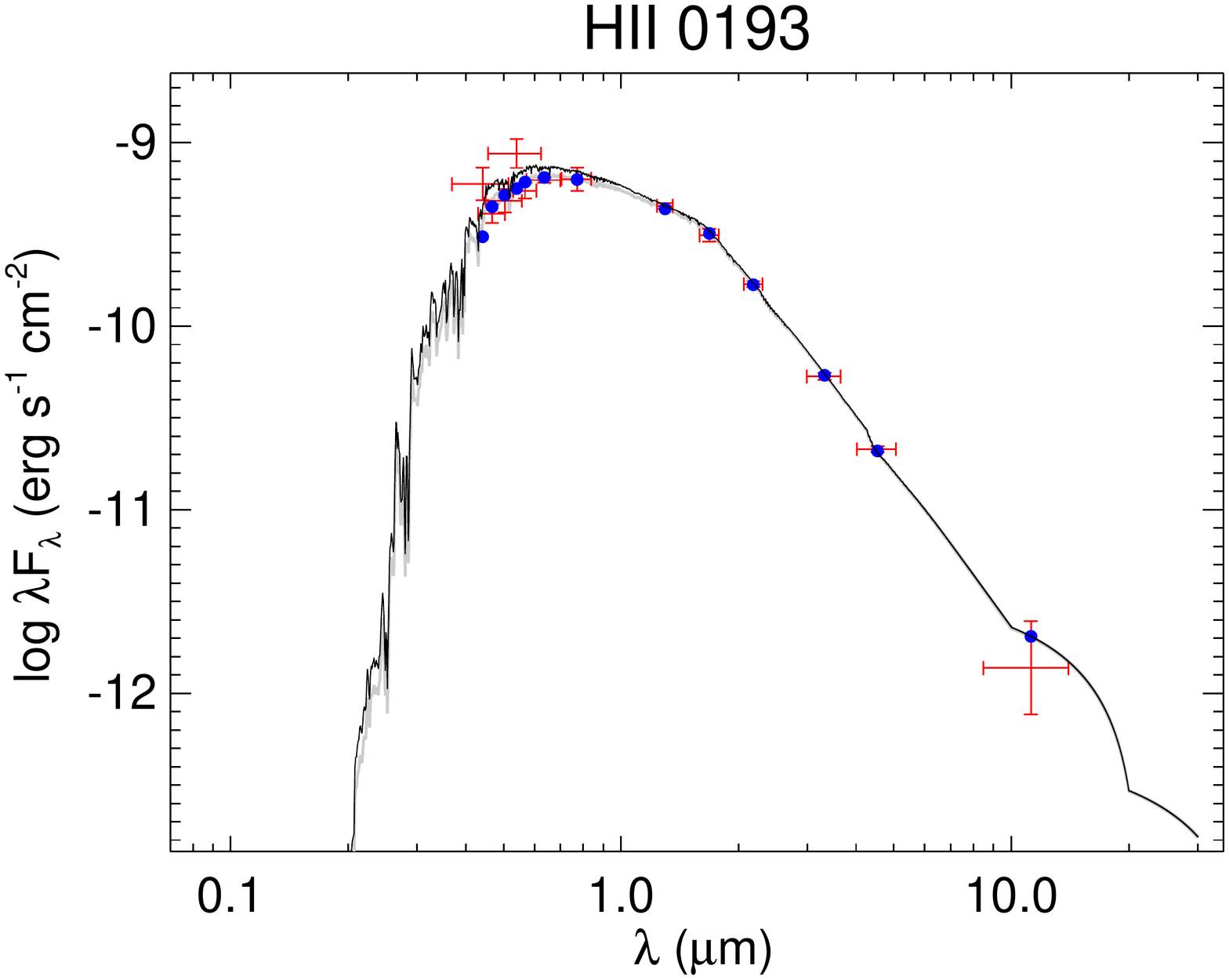}
  \includegraphics[trim=60 60 60 60,clip,width=0.49\linewidth]{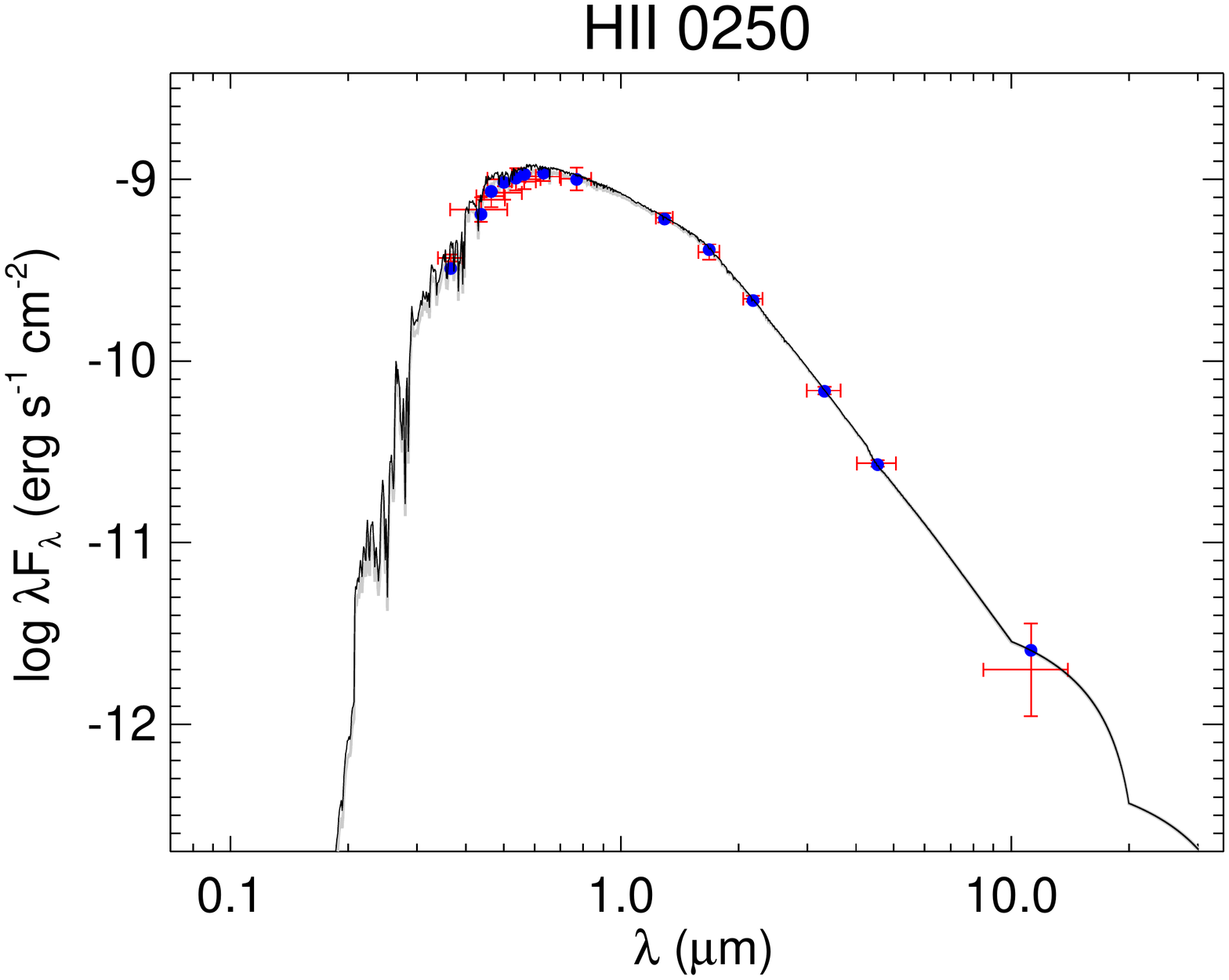}
  \includegraphics[trim=60 60 60 60,clip,width=0.49\linewidth]{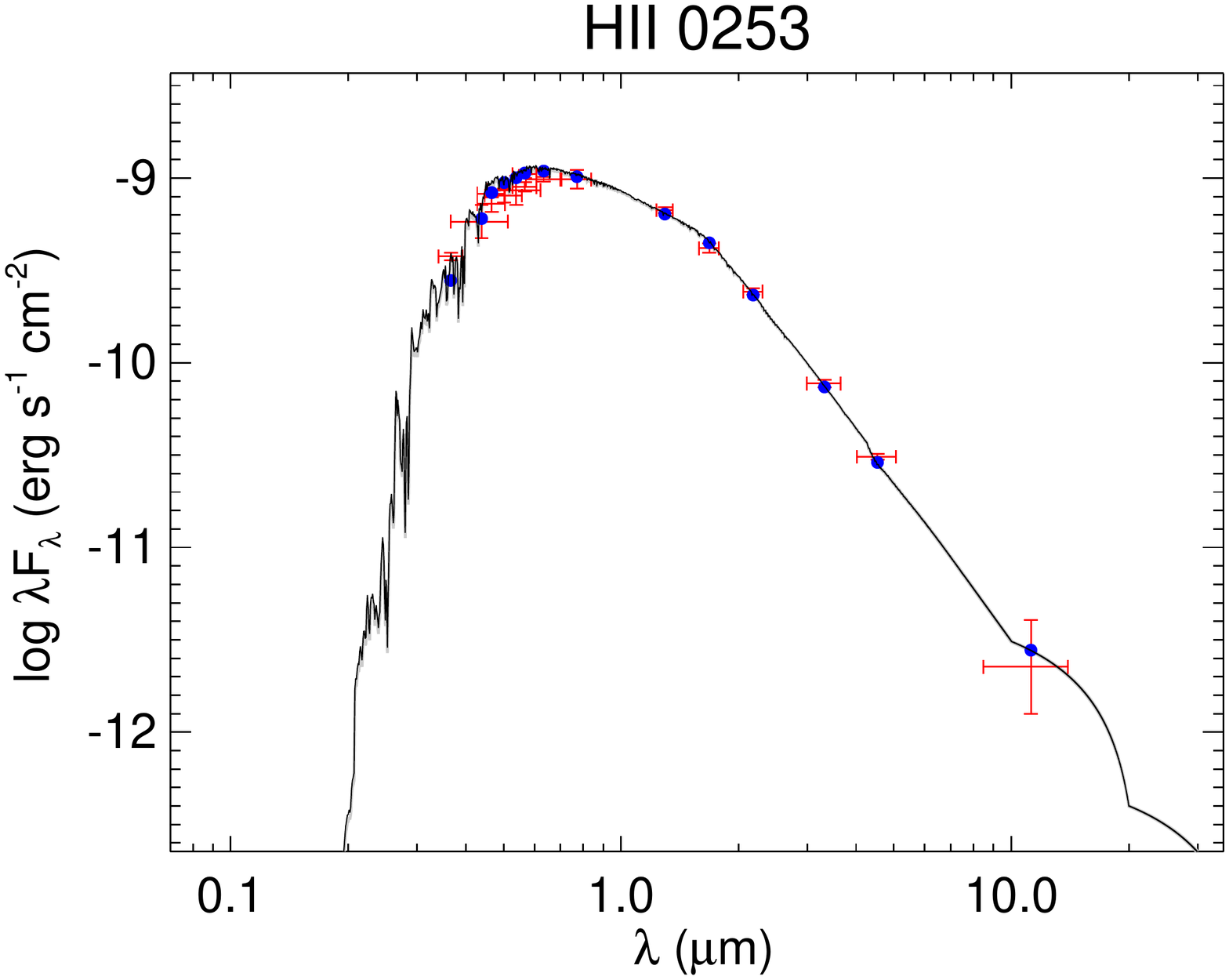}
  \includegraphics[trim=60 60 60 60,clip,width=0.49\linewidth]{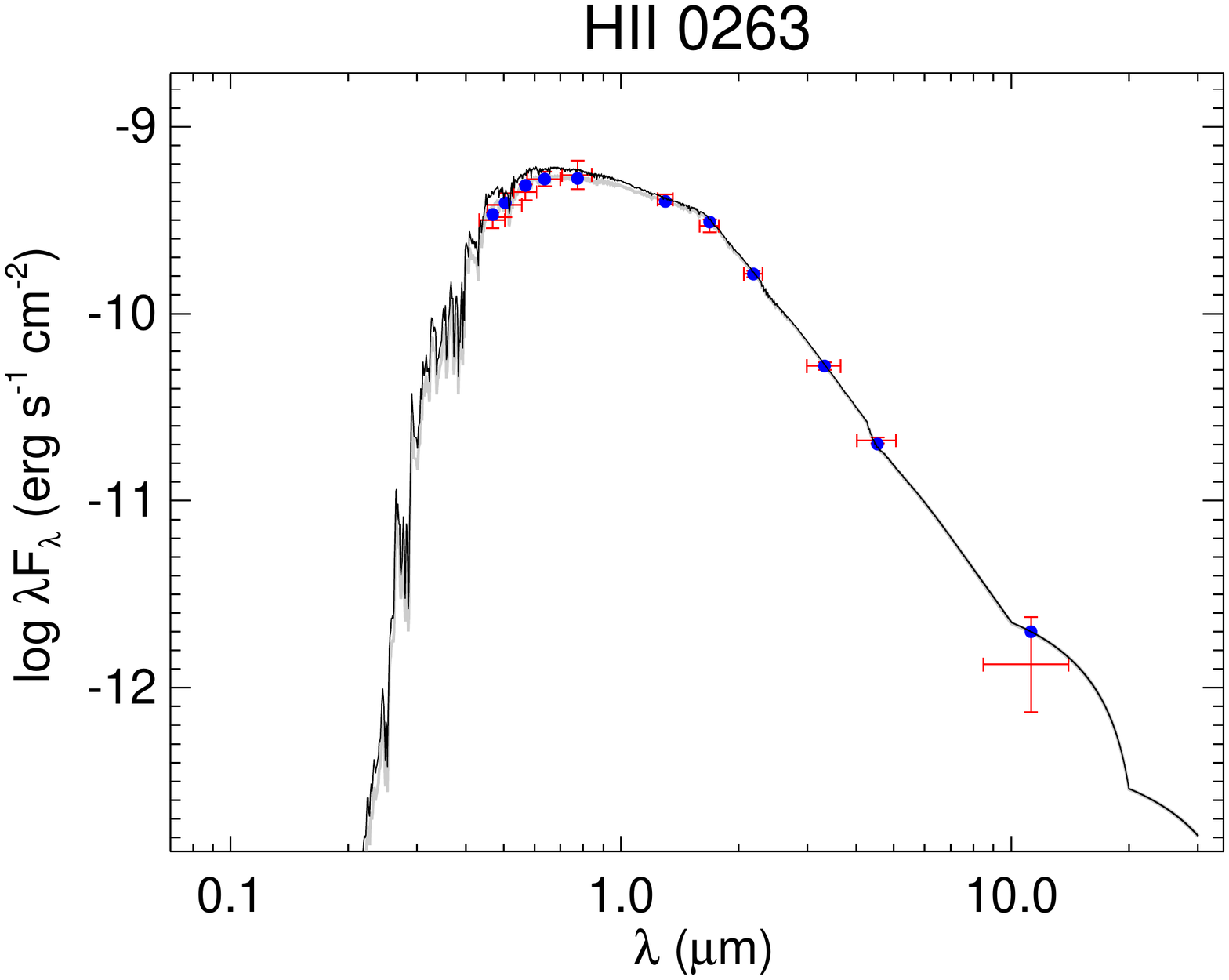}
  \caption{All labels, lines, symbols, and colors as in Figure \ref{fig:seds}.}
  \label{fig:seds_2}
\end{figure}

\begin{figure}[H]
  \centering
  \includegraphics[trim=60 60 60 60,clip,width=0.49\linewidth]{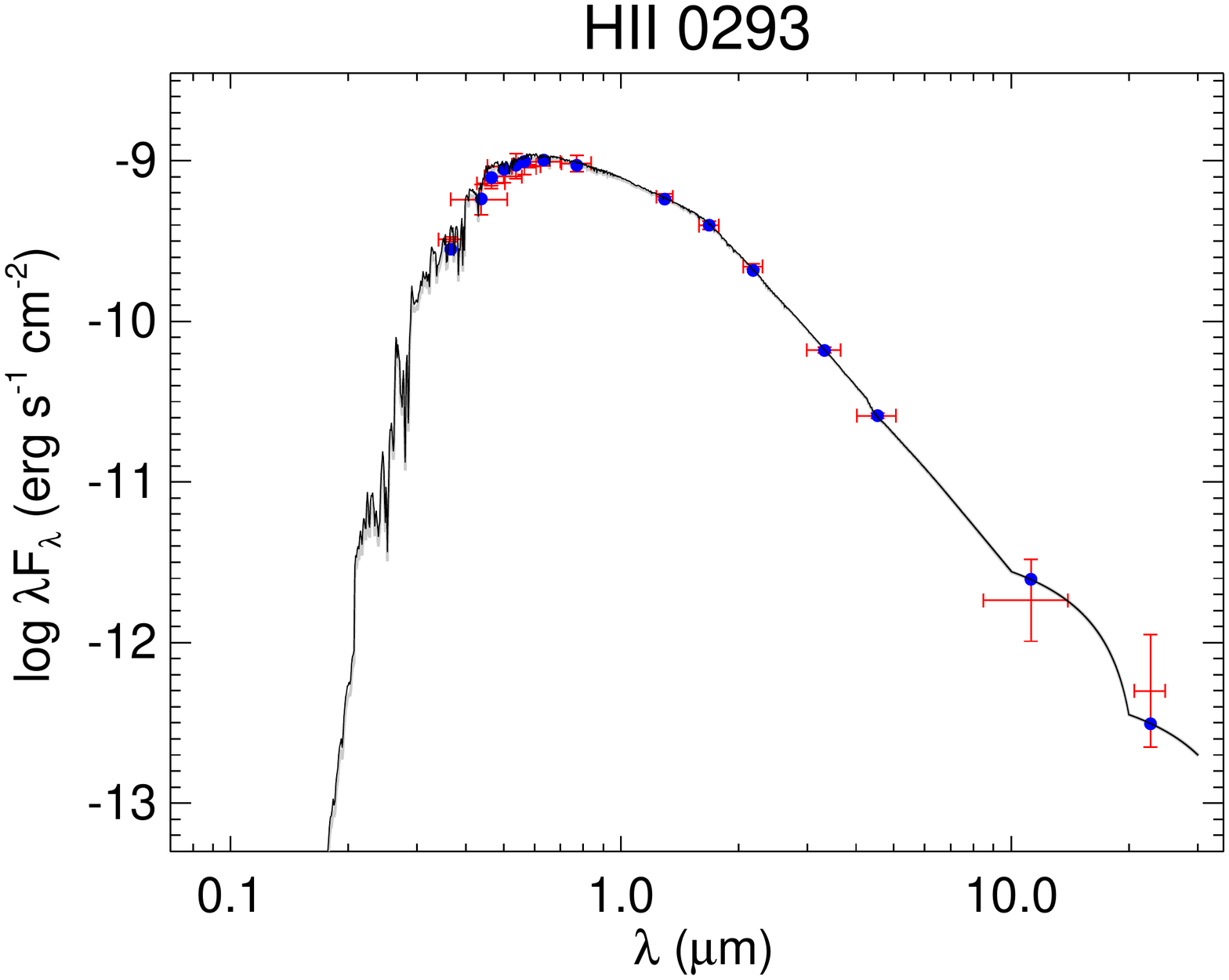}
  \includegraphics[trim=60 60 60 60,clip,width=0.49\linewidth]{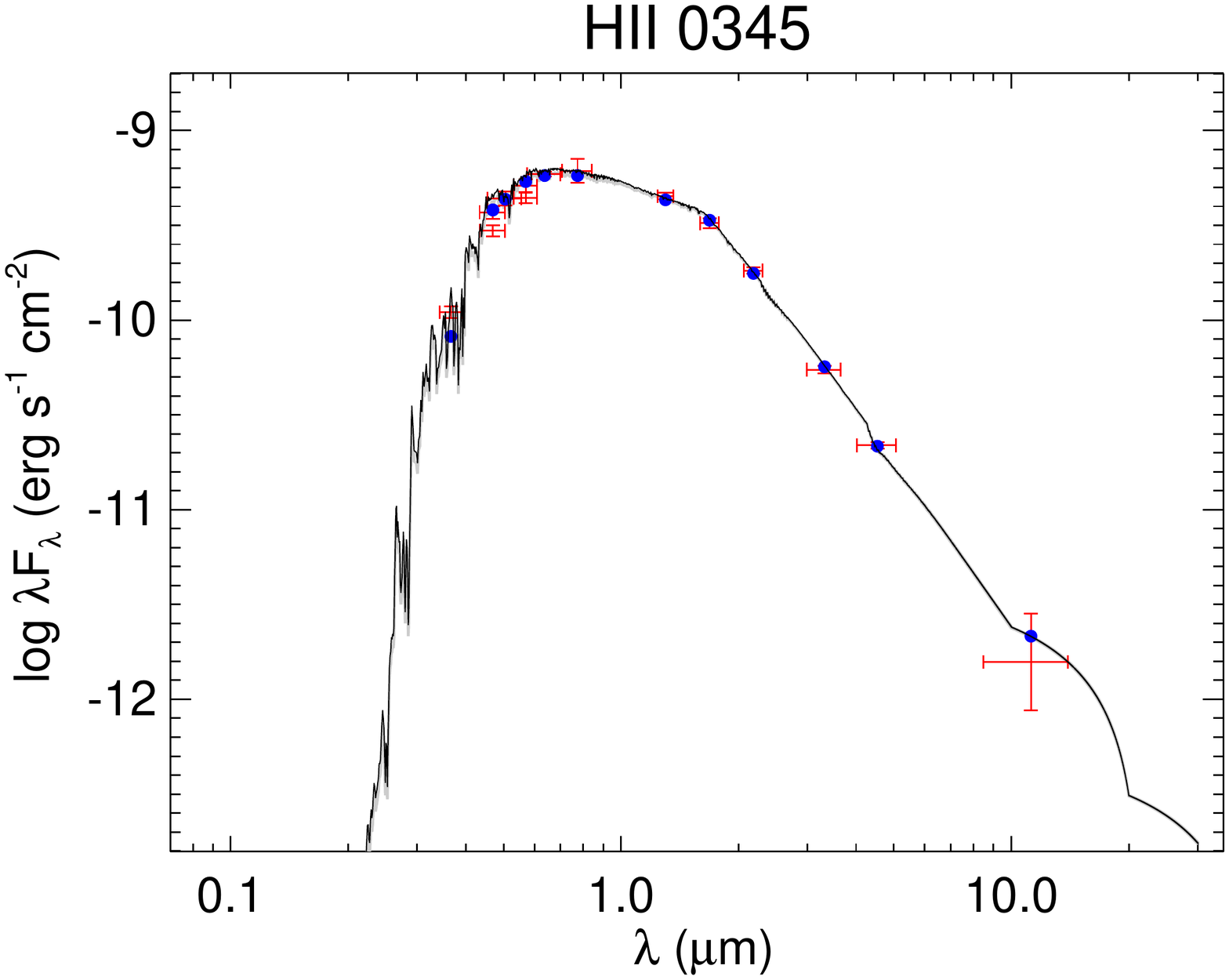}
  \includegraphics[trim=60 60 60 60,clip,width=0.49\linewidth]{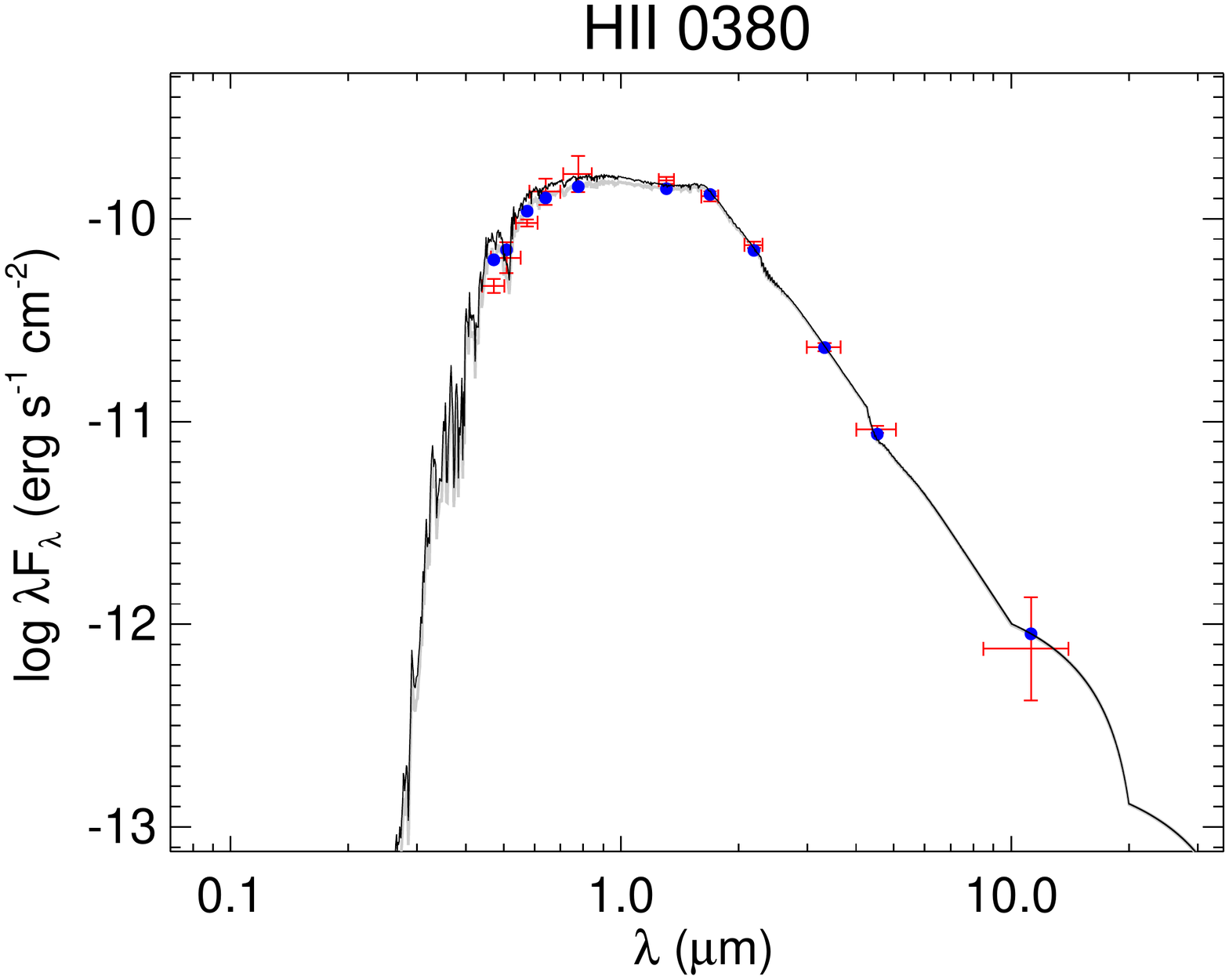}
  \includegraphics[trim=60 60 60 60,clip,width=0.49\linewidth]{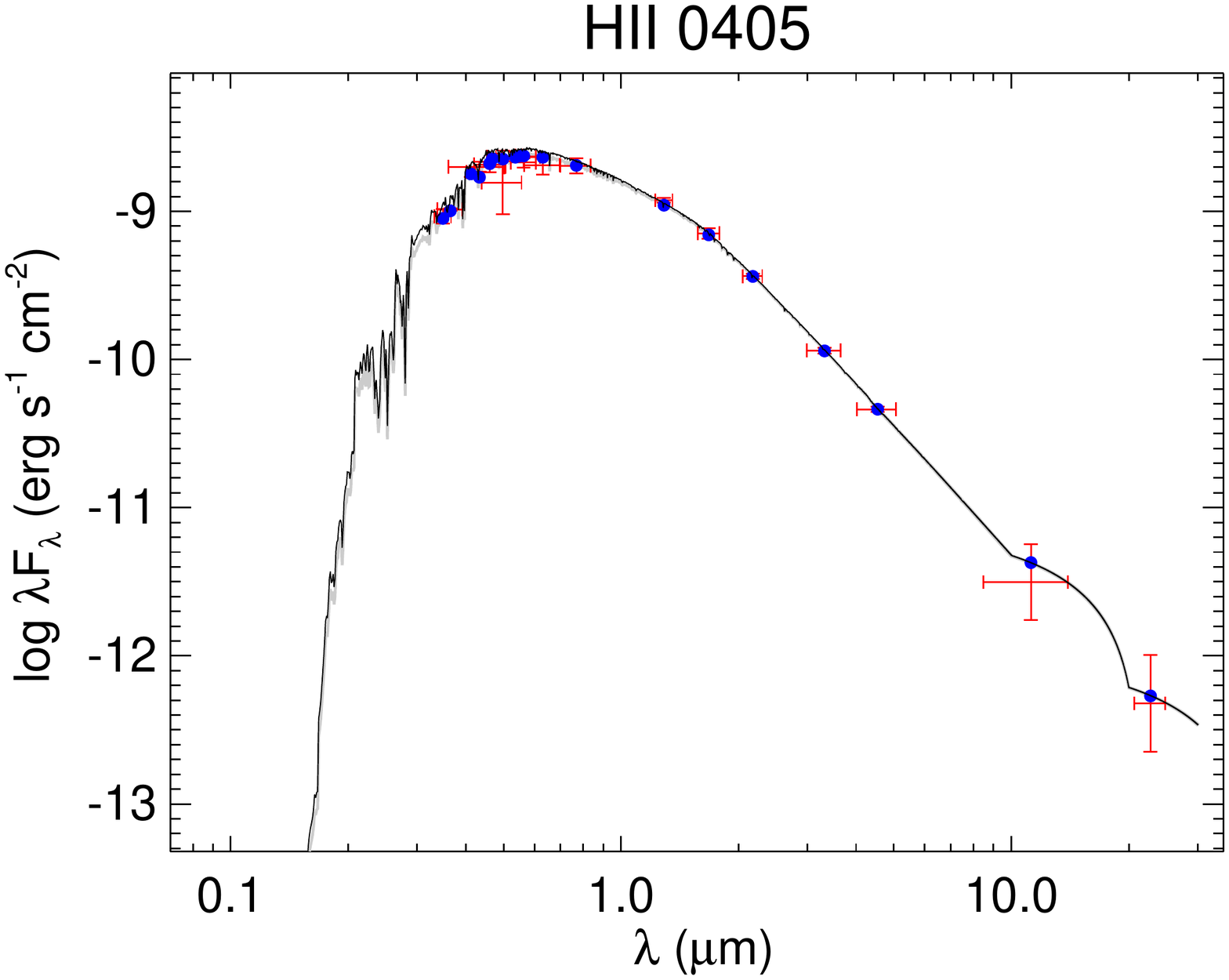}
  \includegraphics[trim=60 60 60 60,clip,width=0.49\linewidth]{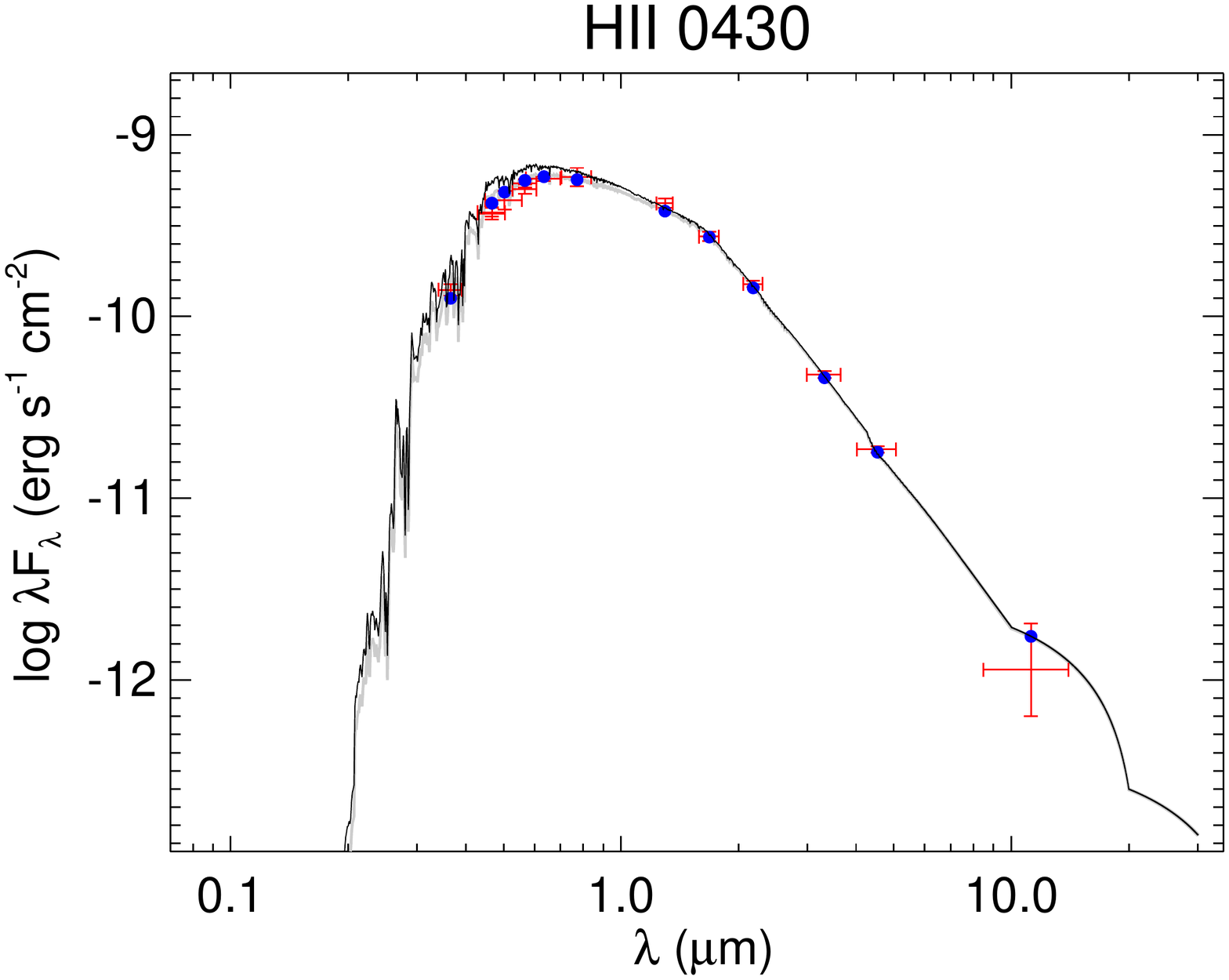}
  \includegraphics[trim=60 60 60 60,clip,width=0.49\linewidth]{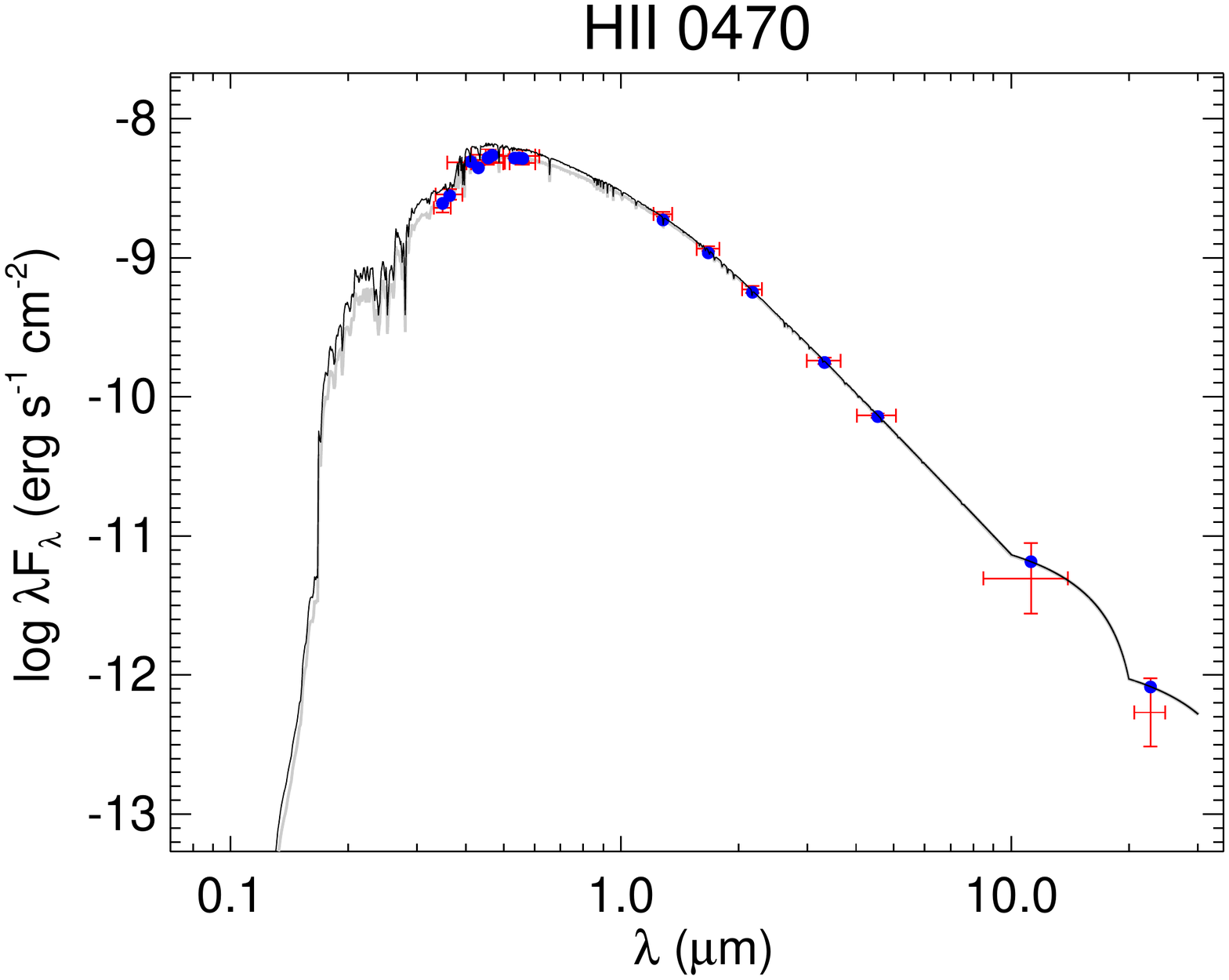}
  \caption{All labels, lines, symbols, and colors as in Figure \ref{fig:seds}.}
  \label{fig:seds_3}
\end{figure}

\begin{figure}[H]
  \centering
  \includegraphics[trim=60 60 60 60,clip,width=0.49\linewidth]{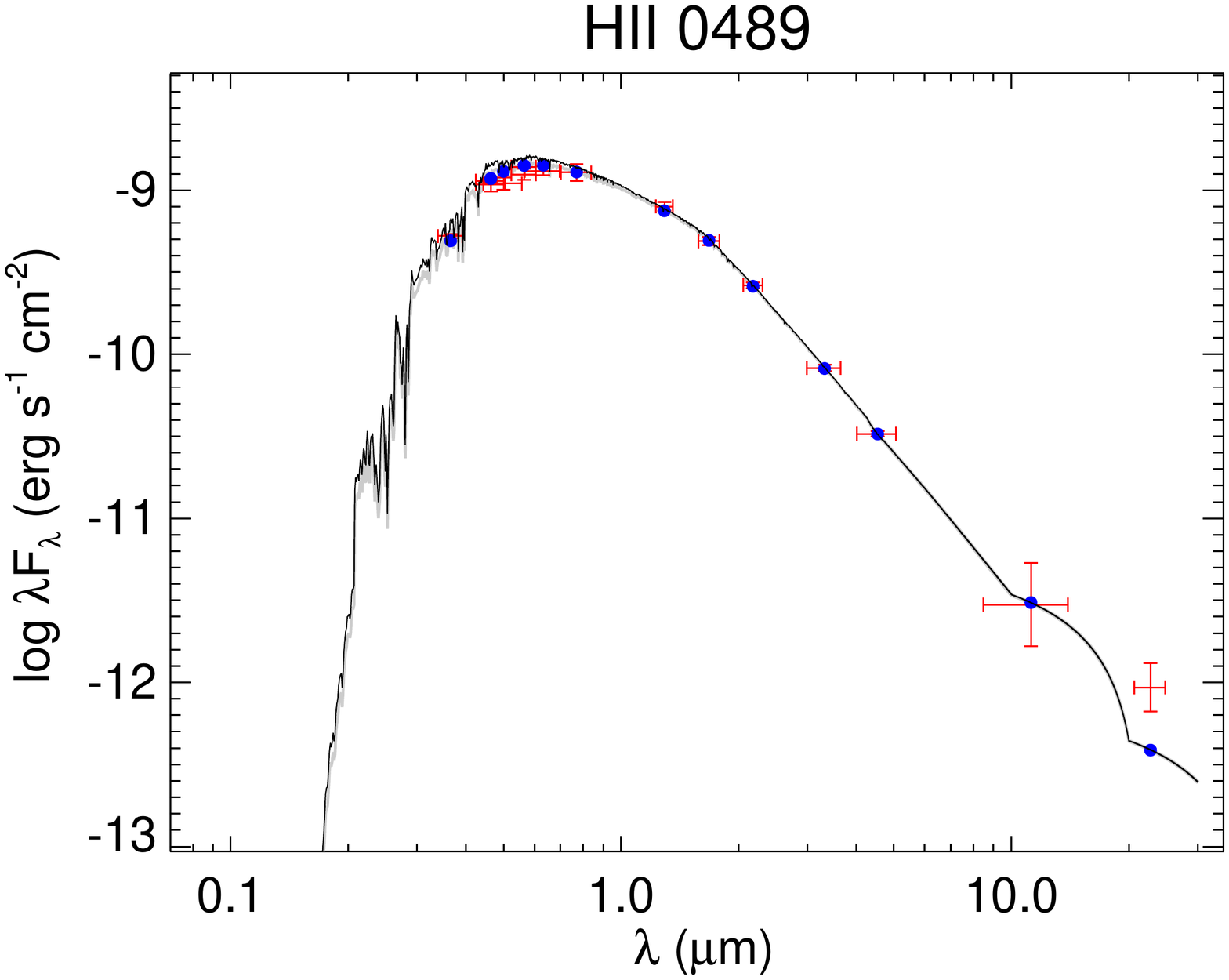}
  \includegraphics[trim=60 60 60 60,clip,width=0.49\linewidth]{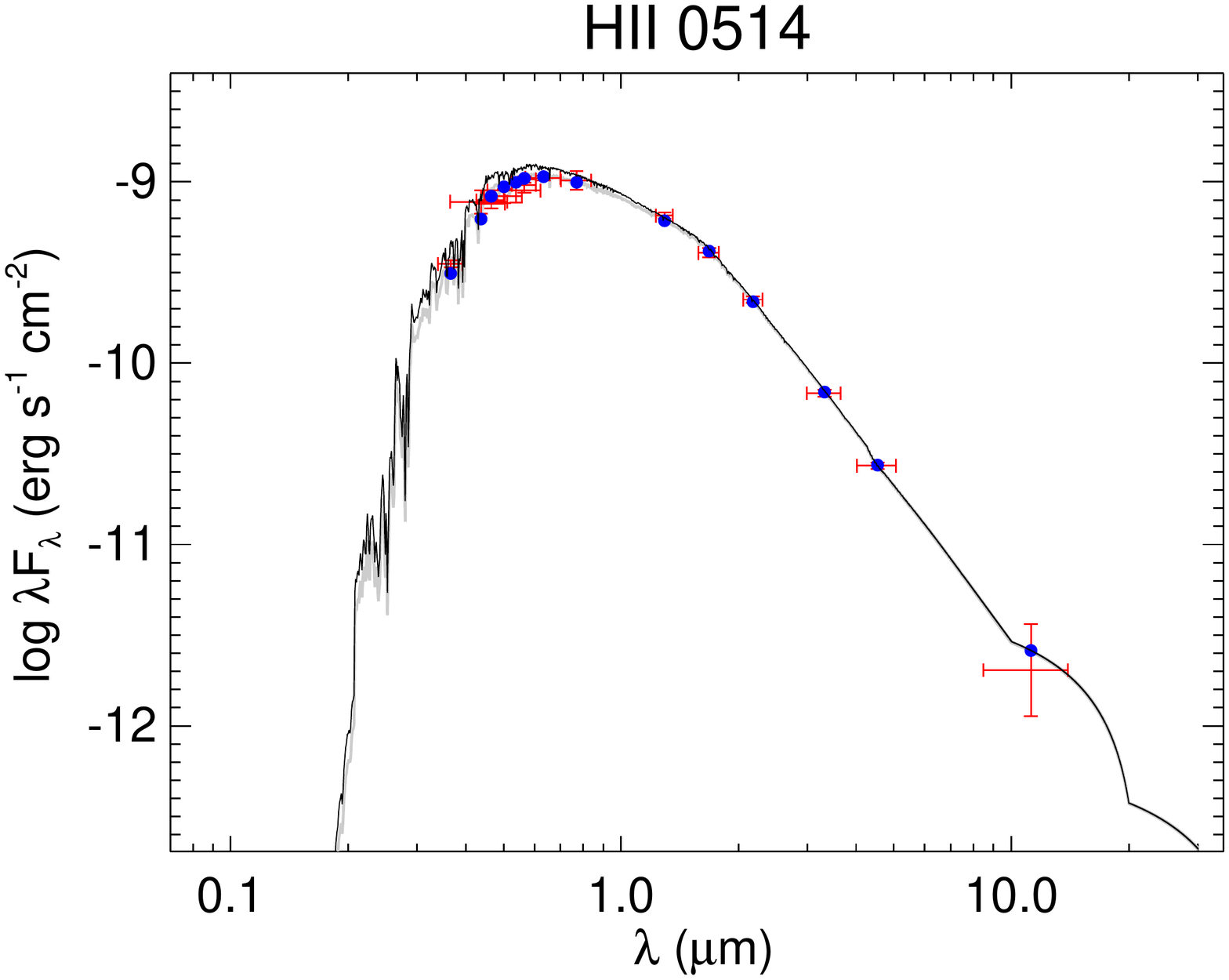}
  \includegraphics[trim=60 60 60 60,clip,width=0.49\linewidth]{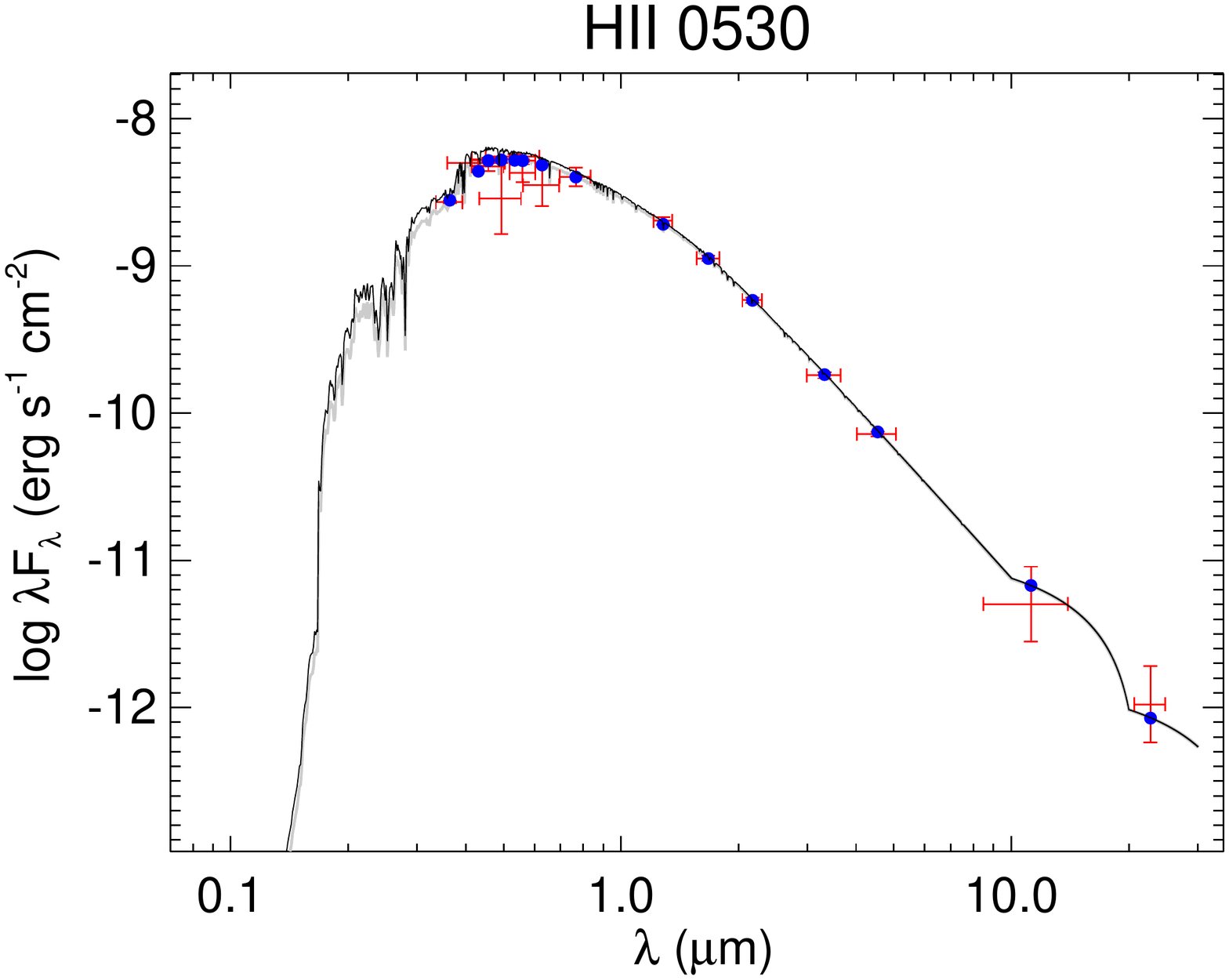}
  \includegraphics[trim=60 60 60 60,clip,width=0.49\linewidth]{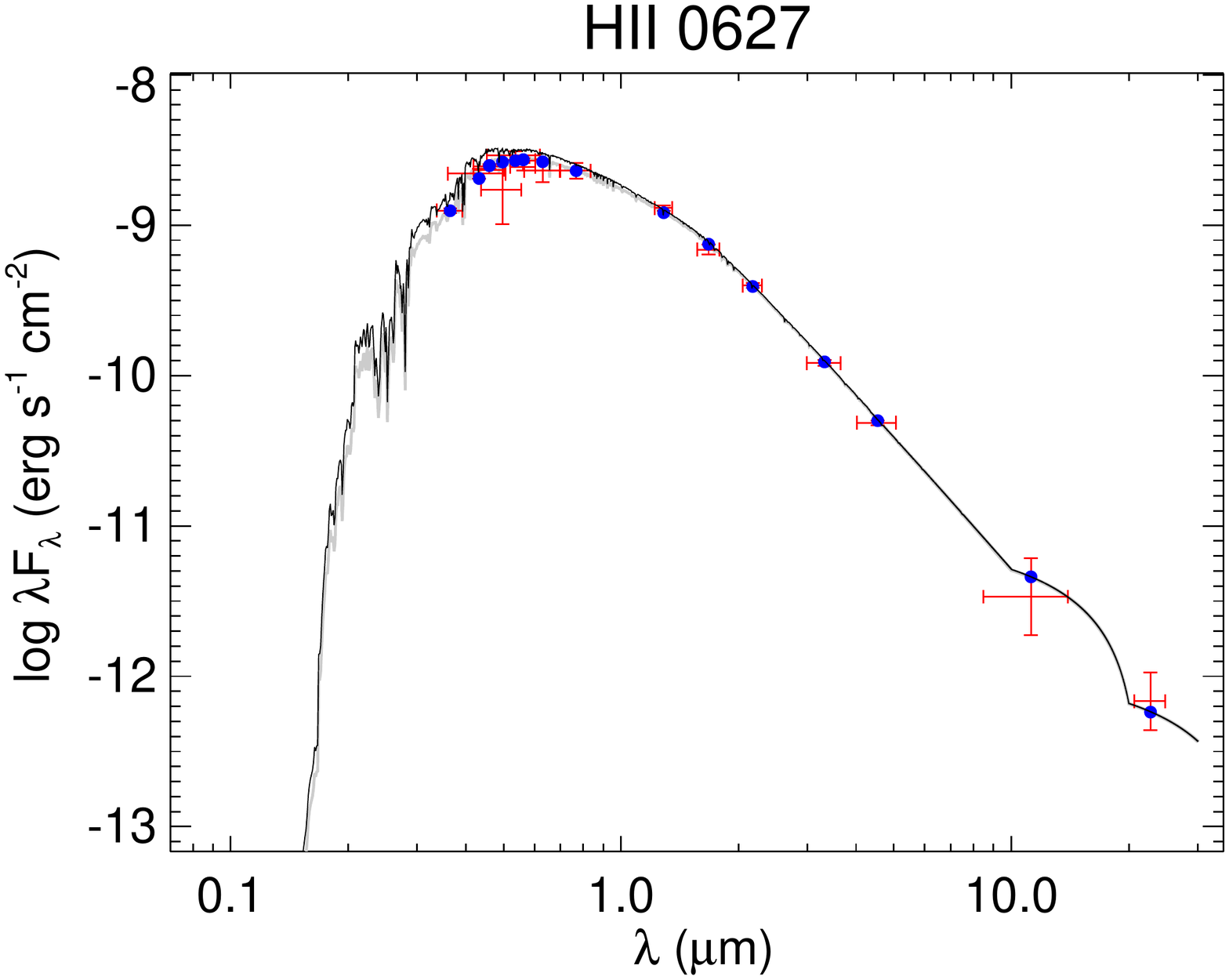}
  \includegraphics[trim=60 60 60 60,clip,width=0.49\linewidth]{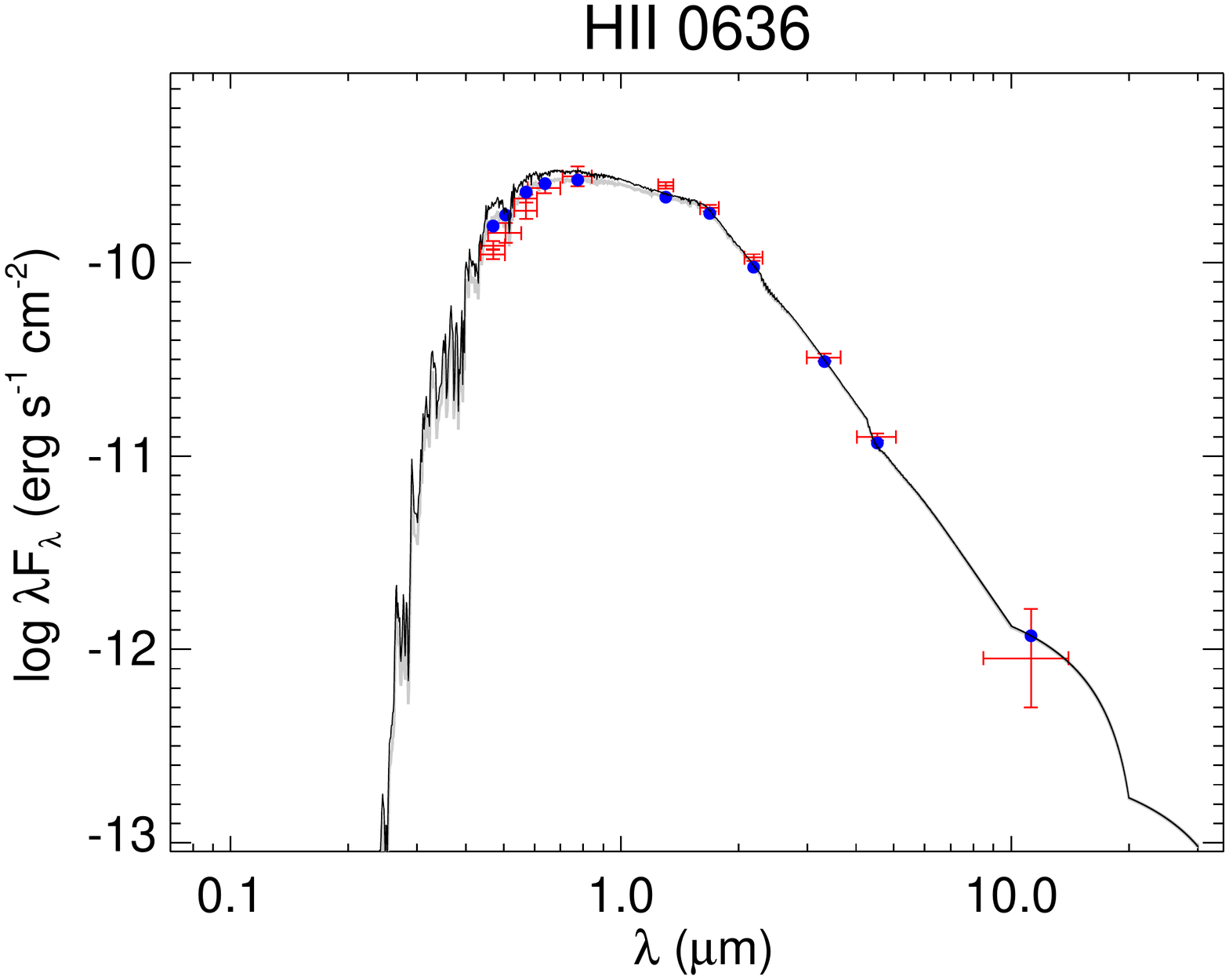}
  \includegraphics[trim=60 60 60 60,clip,width=0.49\linewidth]{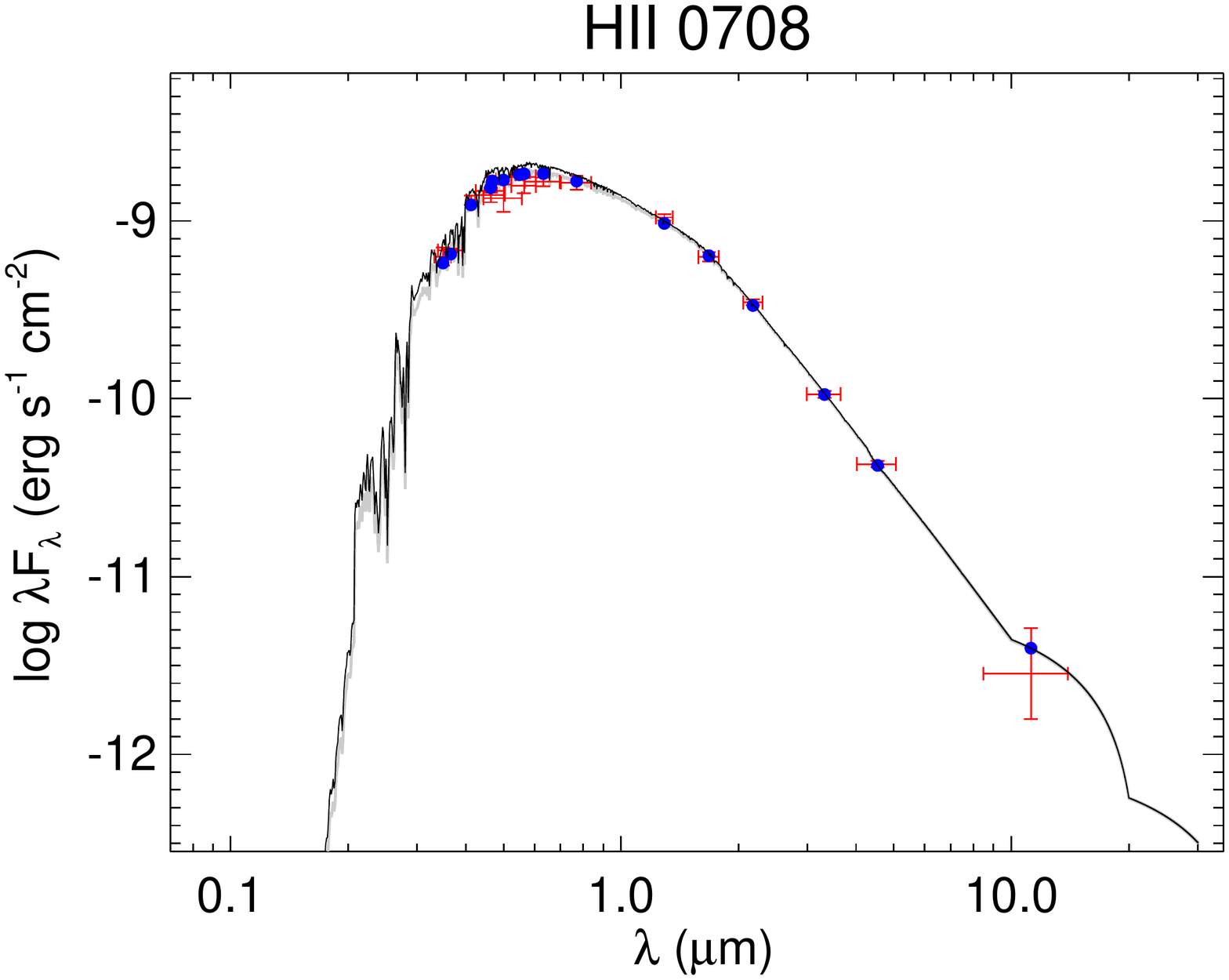}
  \caption{All labels, lines, symbols, and colors as in Figure \ref{fig:seds}.}
  \label{fig:seds_4}
\end{figure}

\begin{figure}[H]
  \centering
  \includegraphics[trim=60 60 60 60,clip,width=0.49\linewidth]{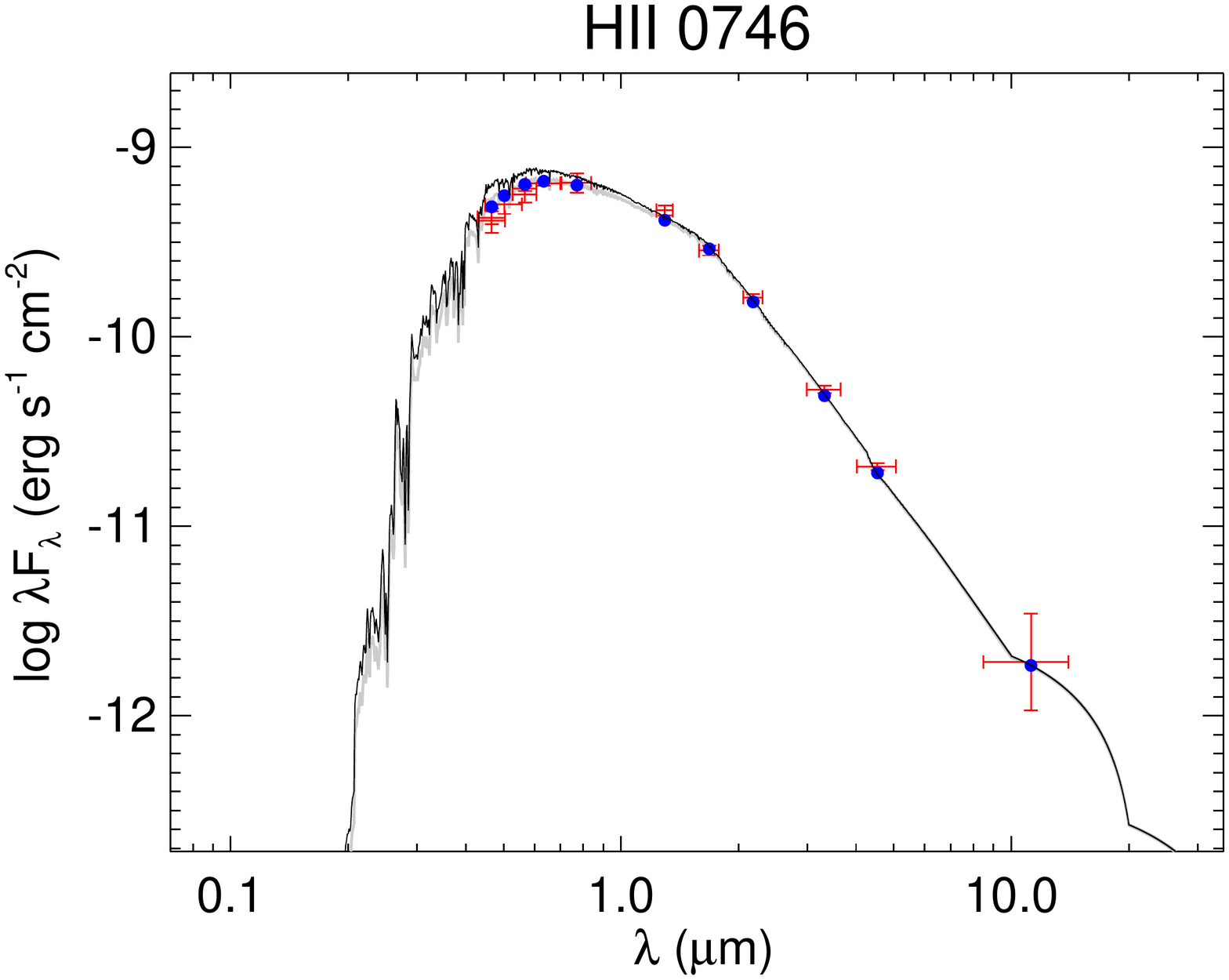}
  \includegraphics[trim=60 60 60 60,clip,width=0.49\linewidth]{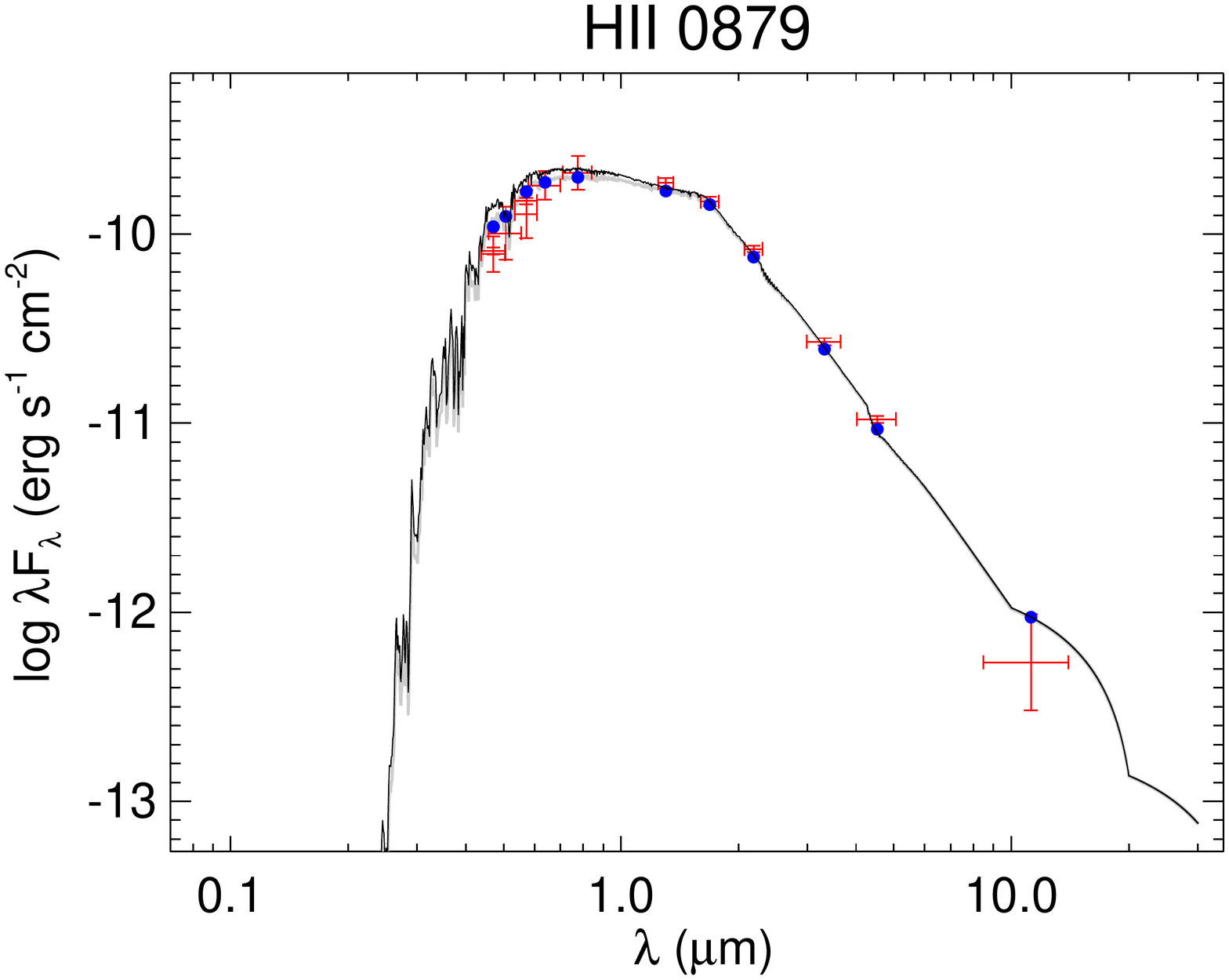}
  \includegraphics[trim=60 60 60 60,clip,width=0.49\linewidth]{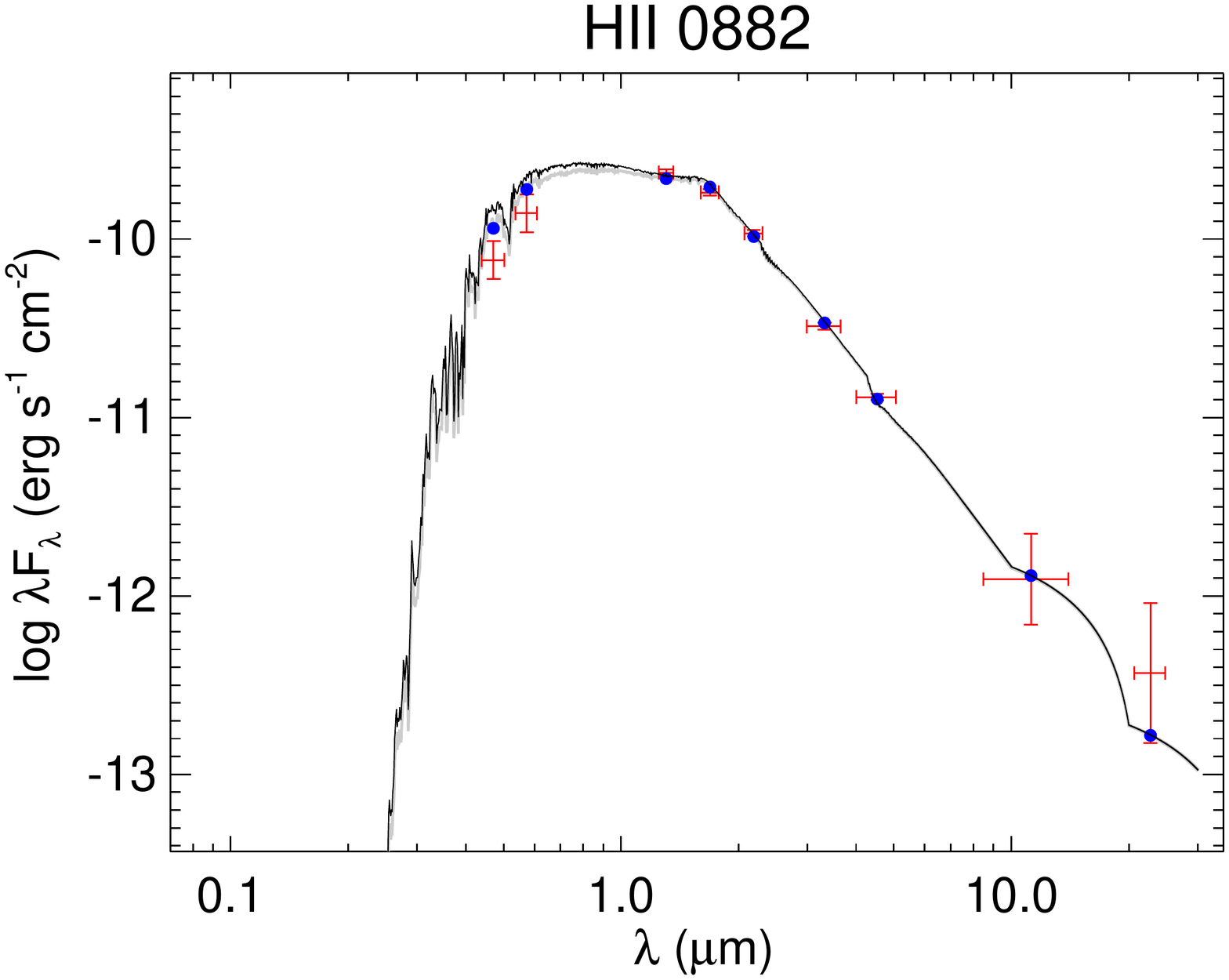}
  \includegraphics[trim=60 60 60 60,clip,width=0.49\linewidth]{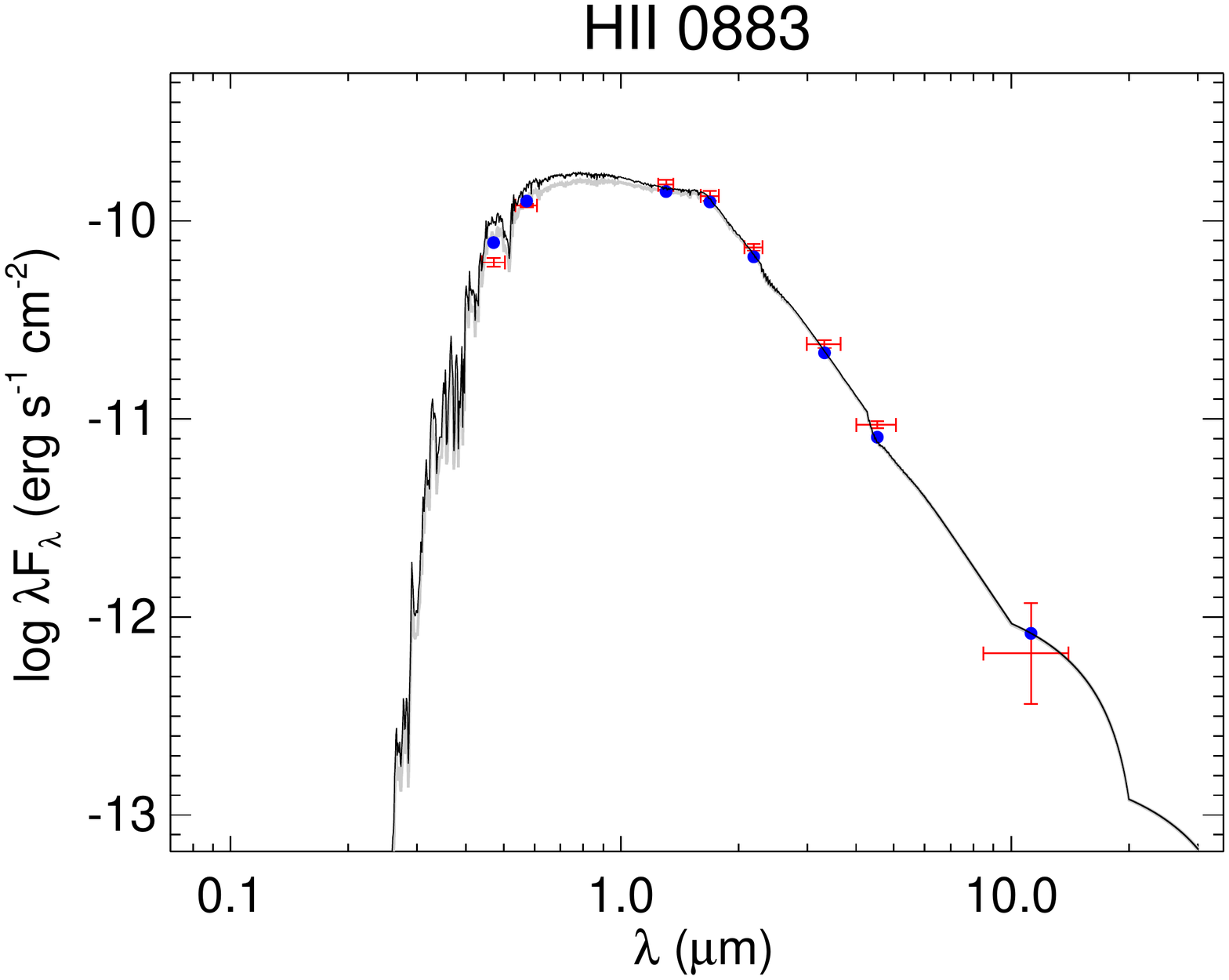}
  \includegraphics[trim=60 60 60 60,clip,width=0.49\linewidth]{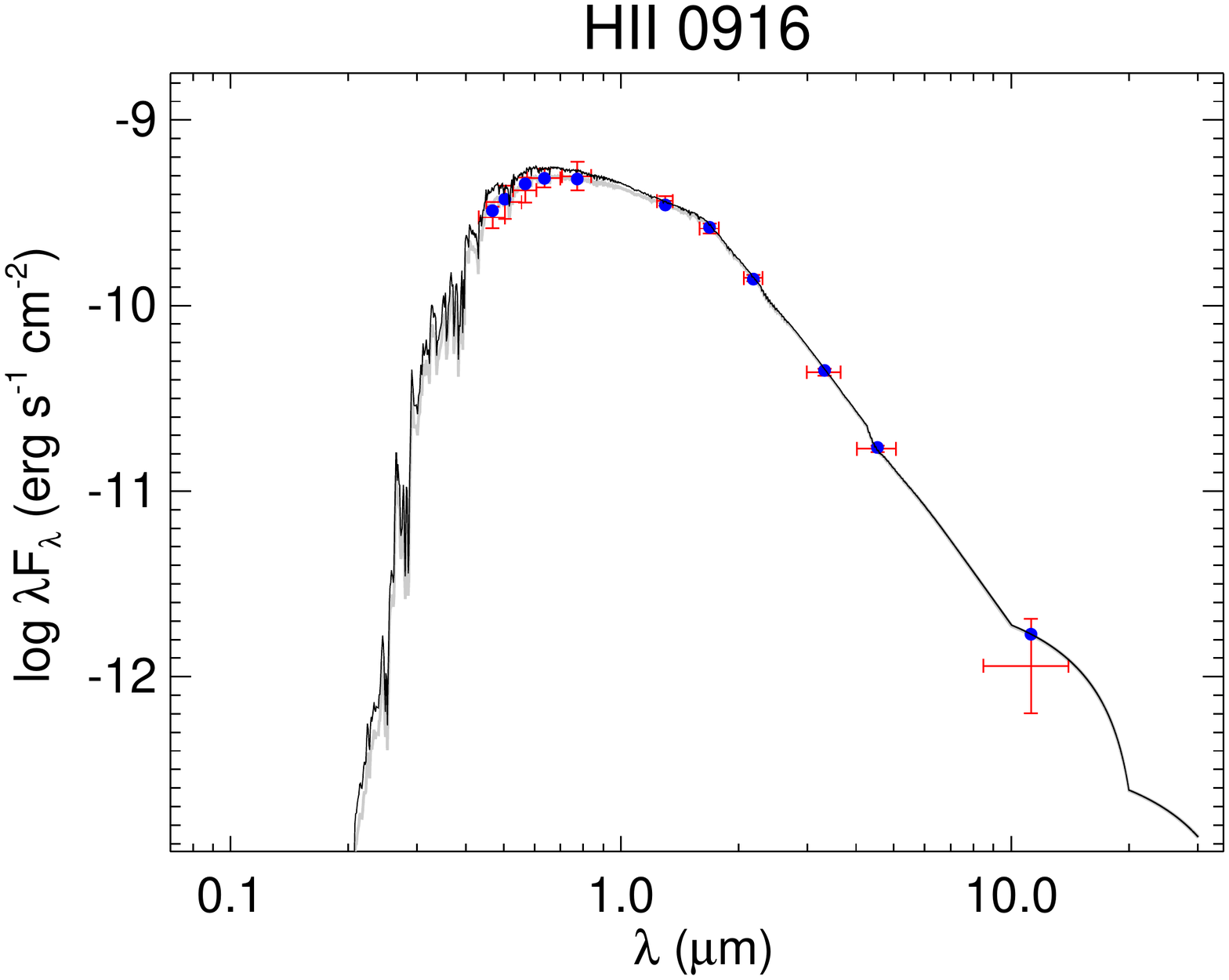}
  \includegraphics[trim=60 60 60 60,clip,width=0.49\linewidth]{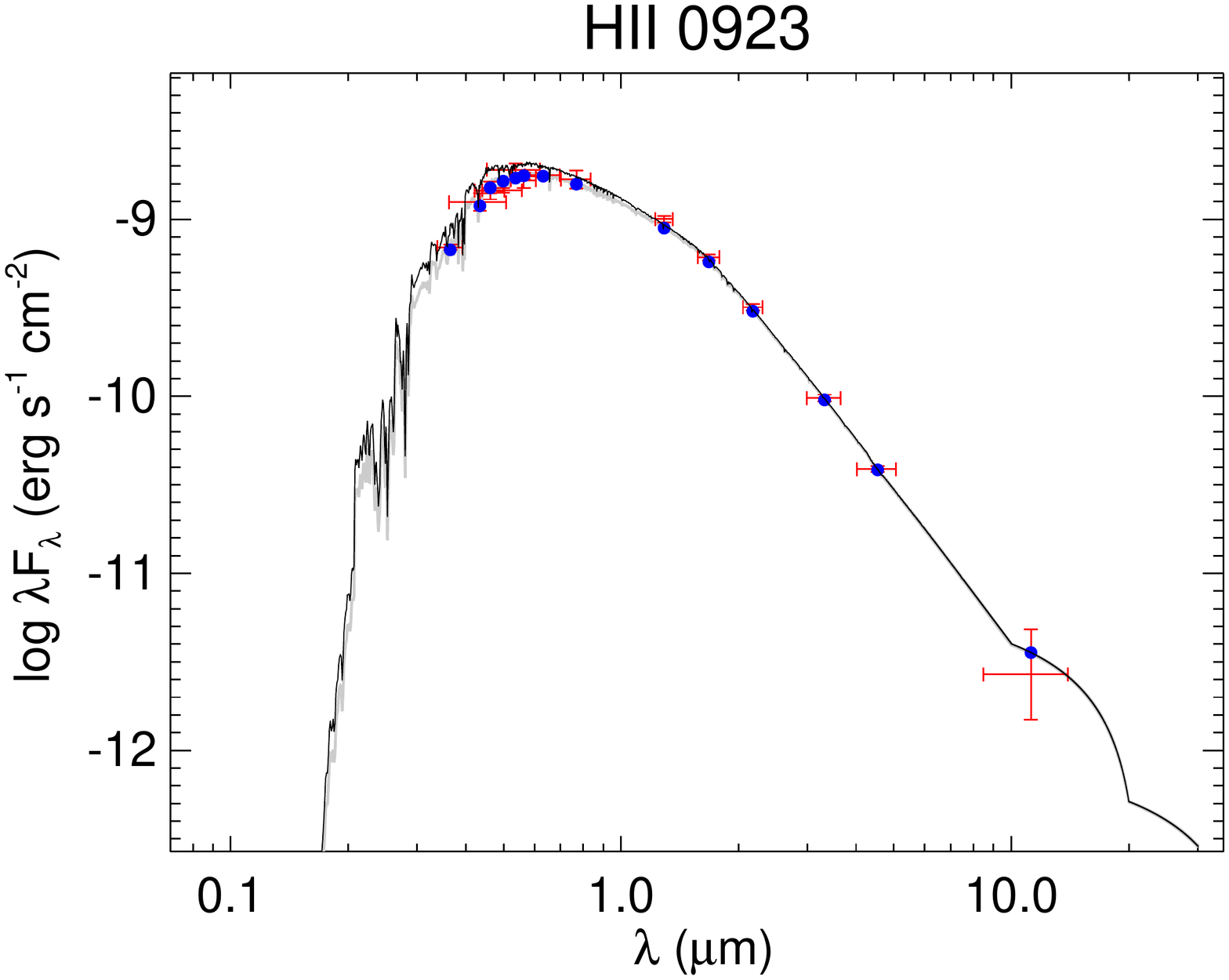}
  \caption{All labels, lines, symbols, and colors as in Figure \ref{fig:seds}.}
  \label{fig:seds_5}
\end{figure}

\begin{figure}[H]
  \centering
  \includegraphics[trim=60 60 60 60,clip,width=0.49\linewidth]{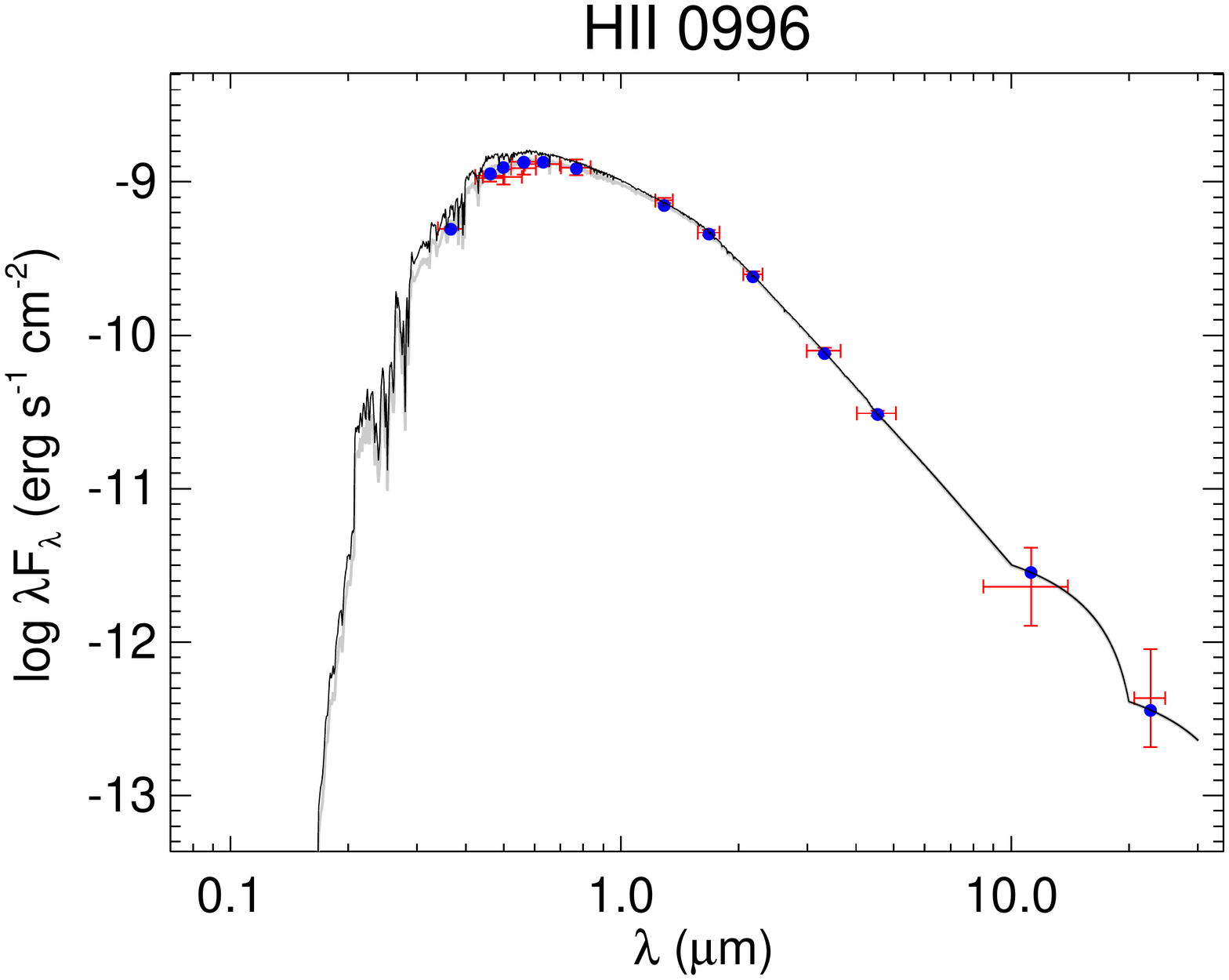}
  \includegraphics[trim=60 60 60 60,clip,width=0.49\linewidth]{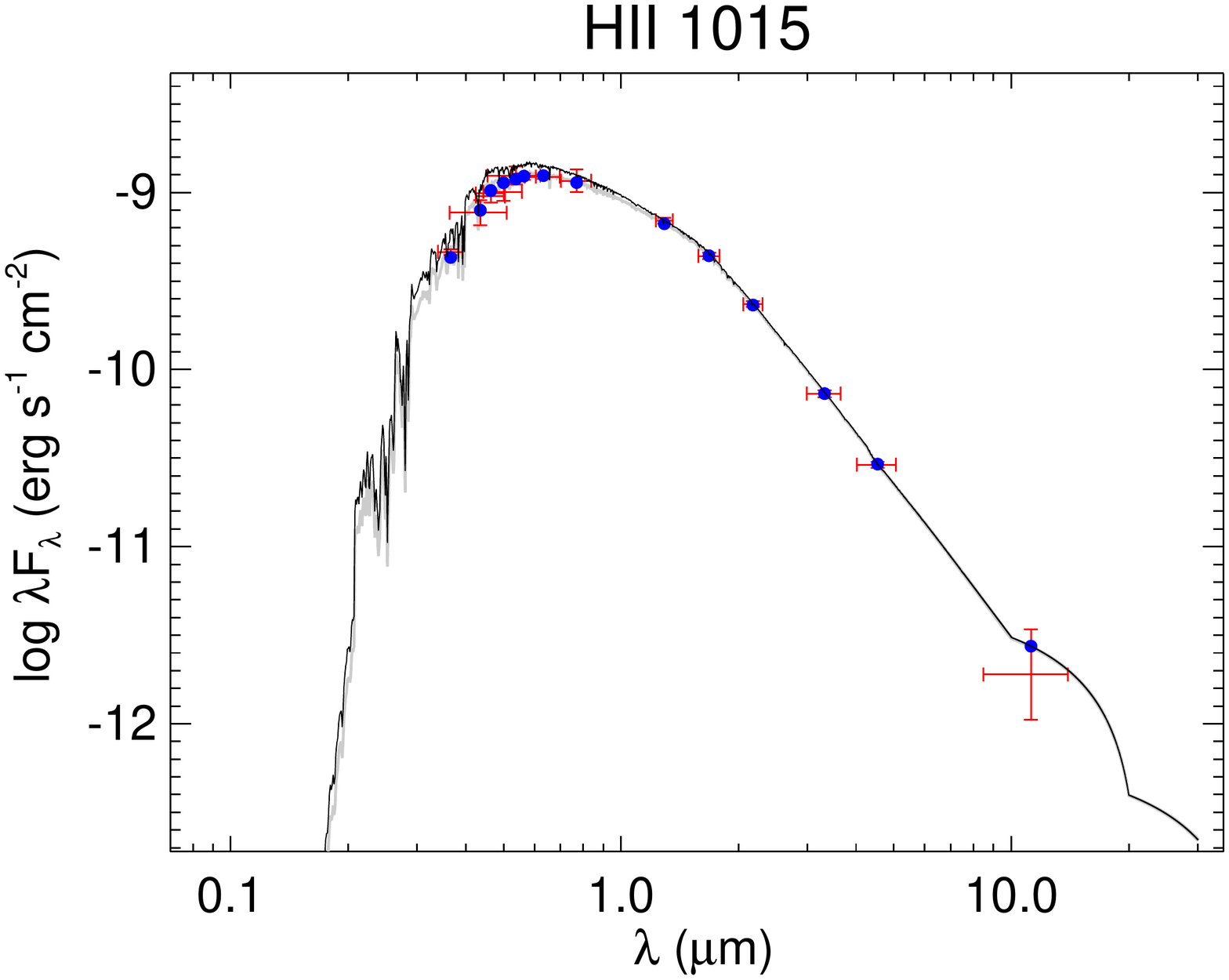}
  \includegraphics[trim=60 60 60 60,clip,width=0.49\linewidth]{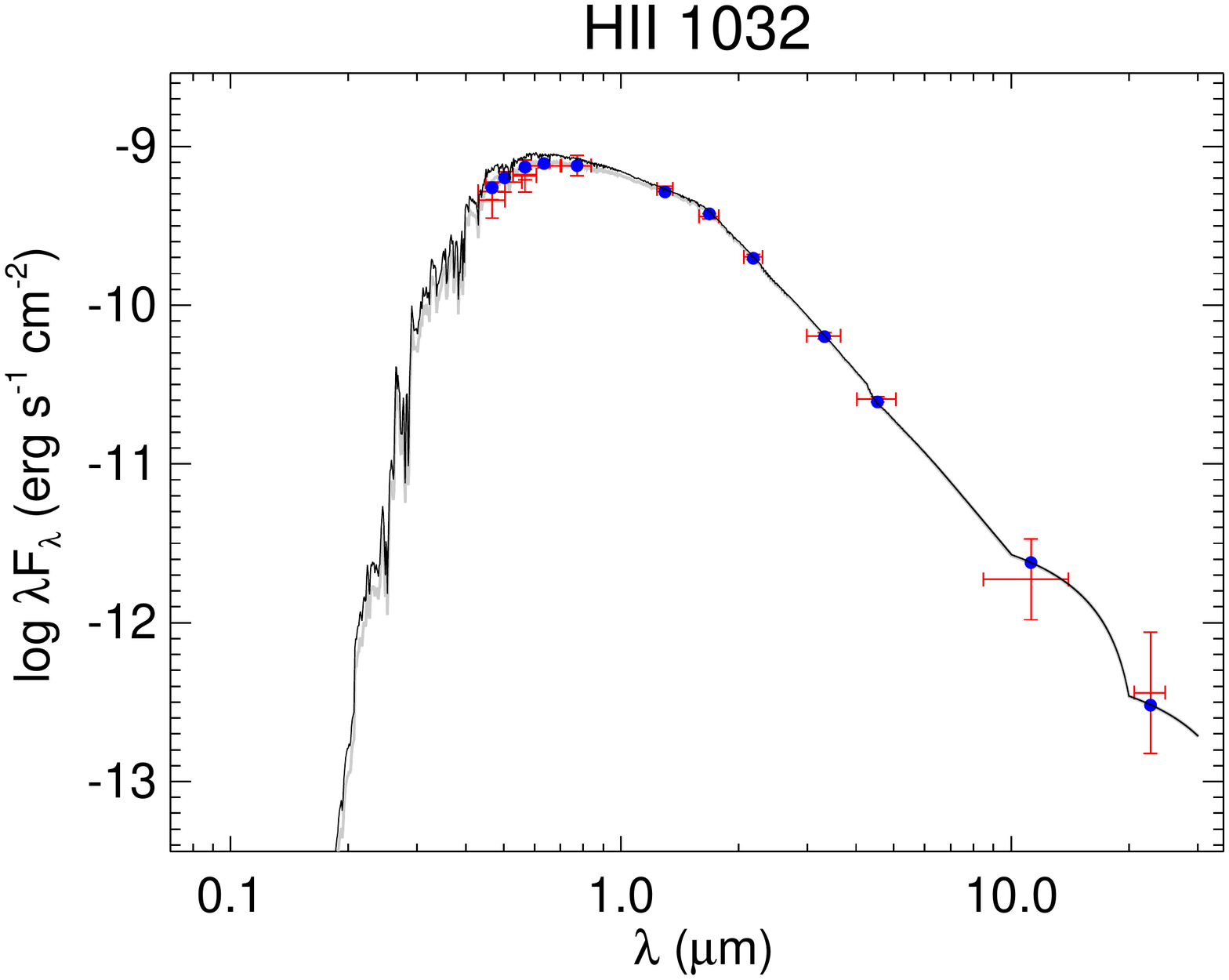}
  \includegraphics[trim=60 60 60 60,clip,width=0.49\linewidth]{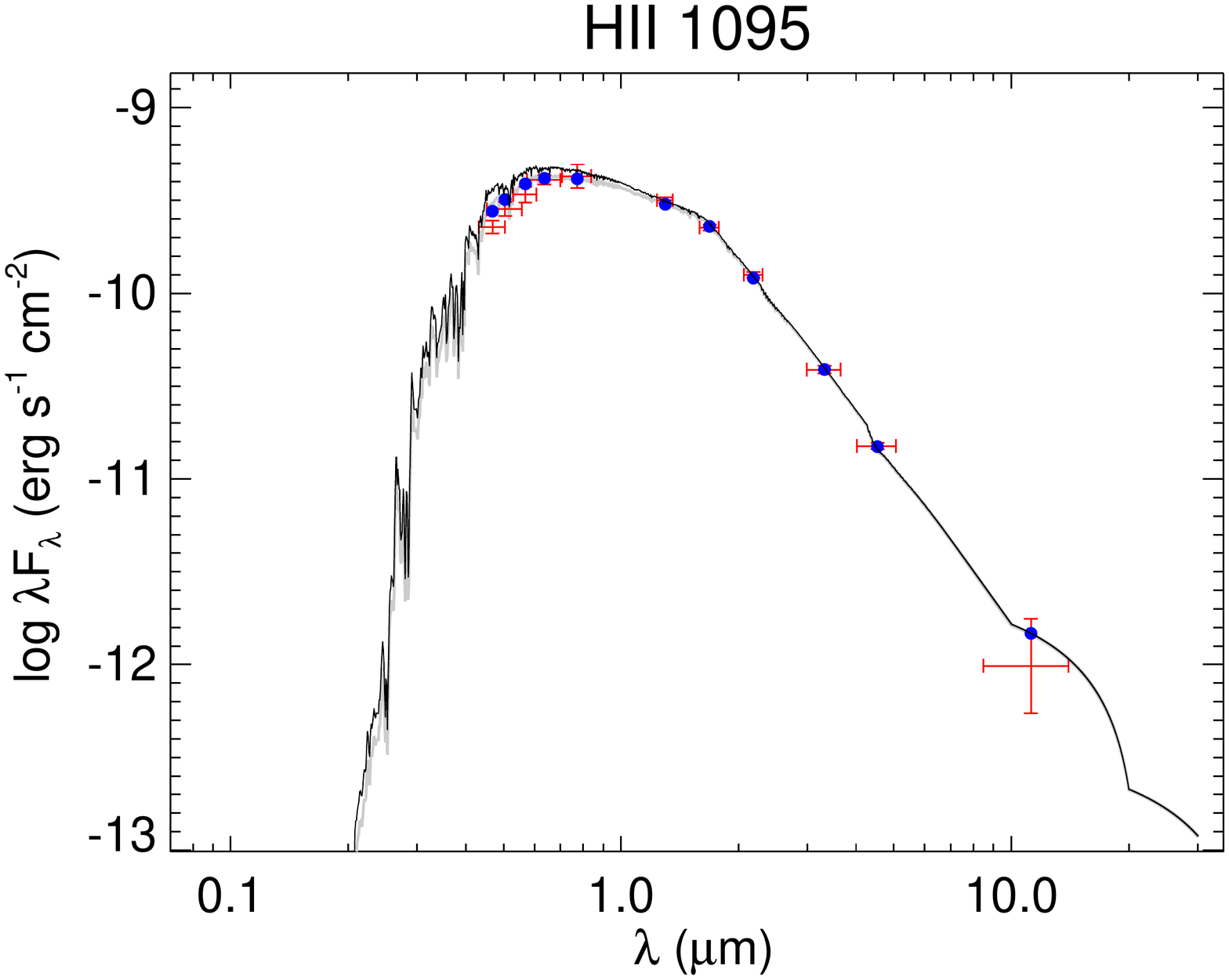}
  \includegraphics[trim=60 60 60 60,clip,width=0.49\linewidth]{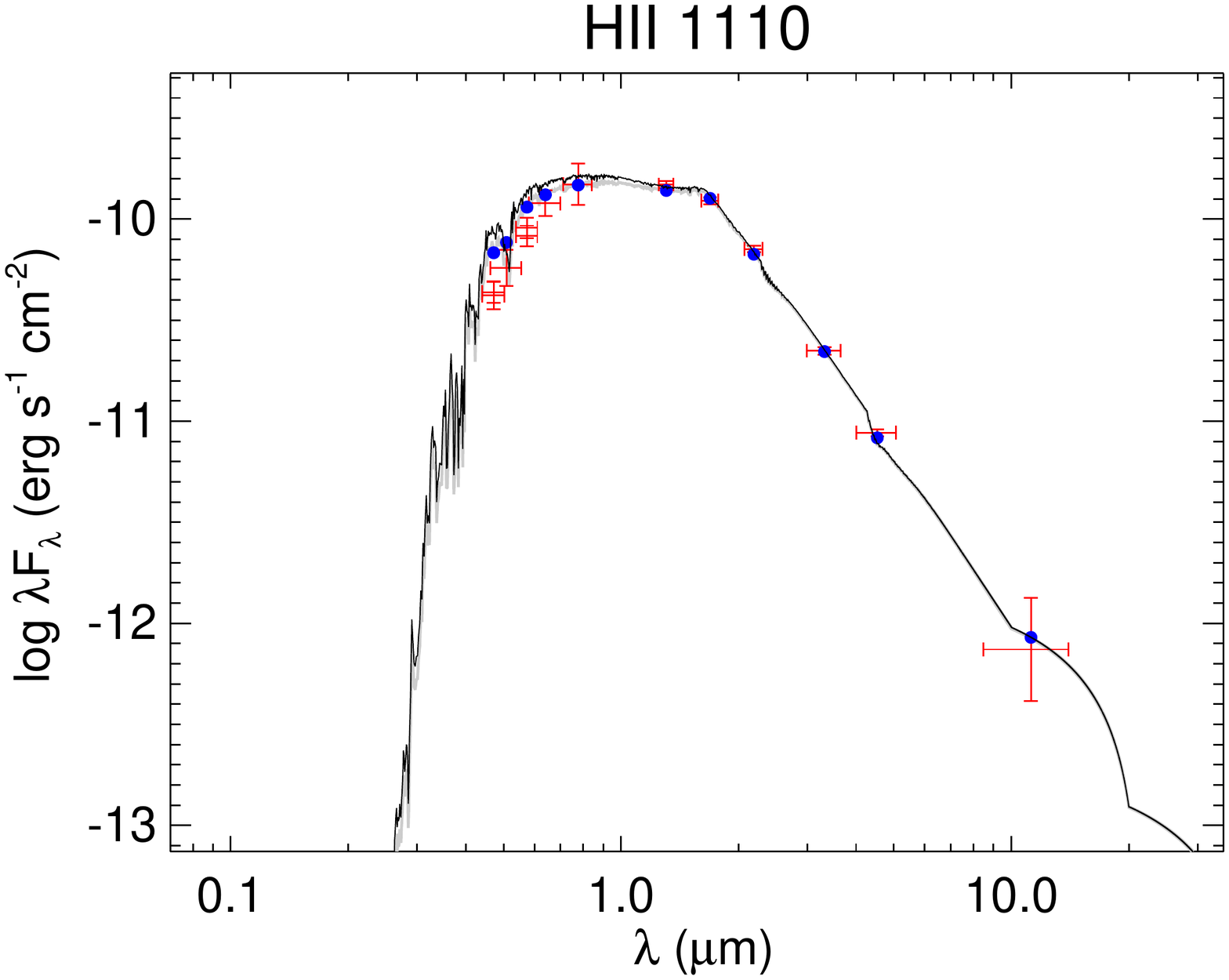}
  \includegraphics[trim=60 60 60 60,clip,width=0.49\linewidth]{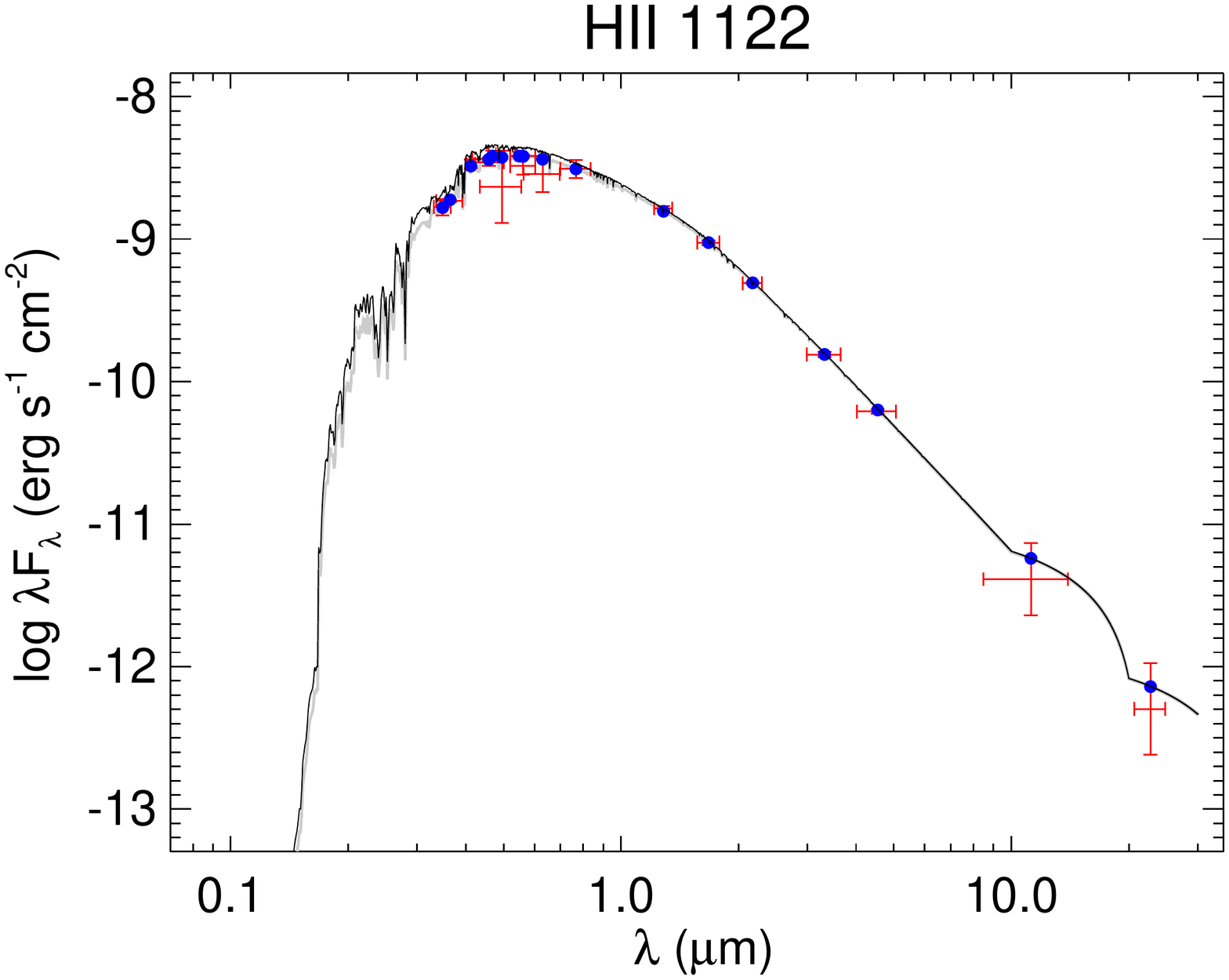}
  \caption{All labels, lines, symbols, and colors as in Figure \ref{fig:seds}.}
  \label{fig:seds_6}
\end{figure}

\begin{figure}[H]
  \centering
  \includegraphics[trim=60 60 60 60,clip,width=0.49\linewidth]{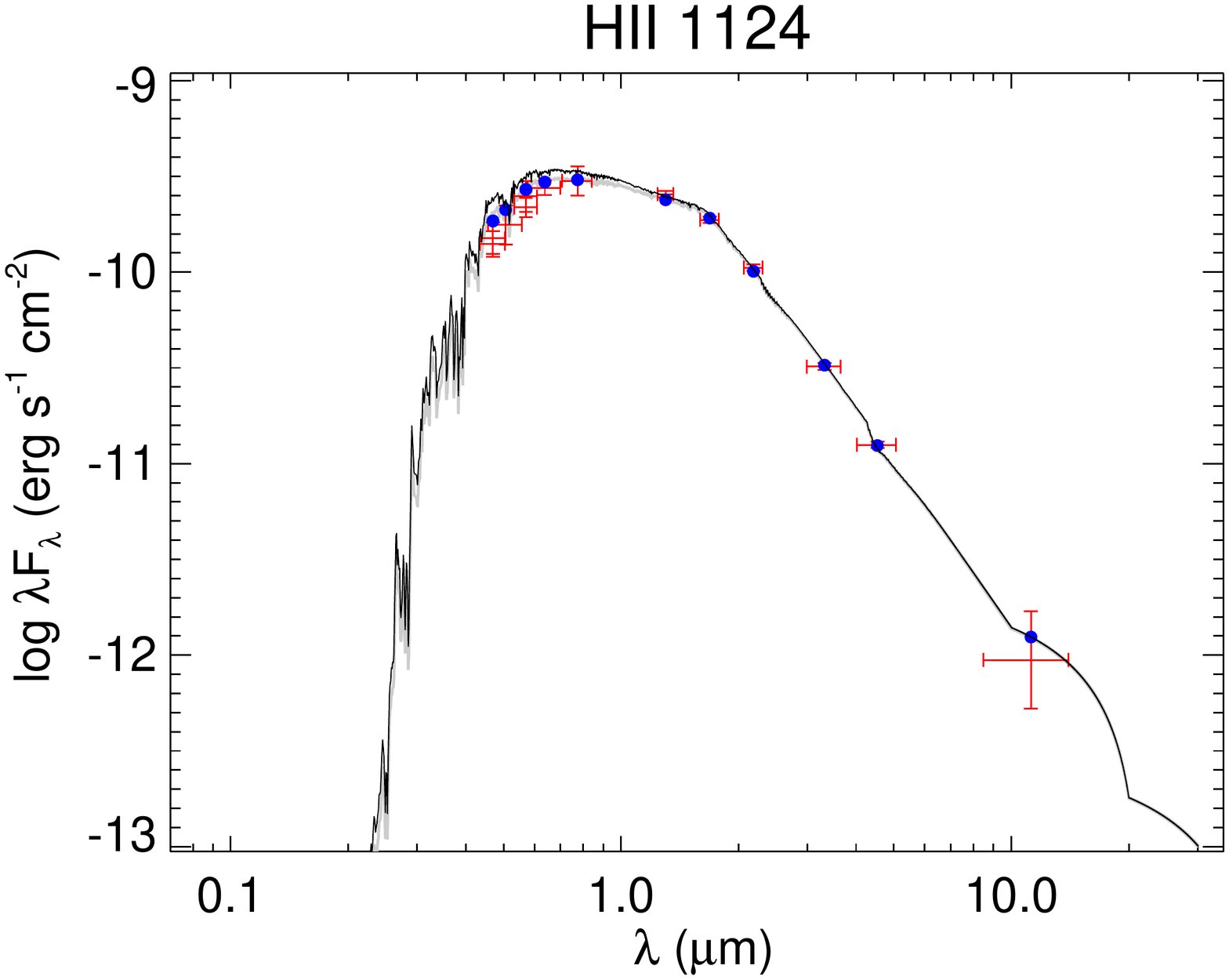}
  \includegraphics[trim=60 60 60 60,clip,width=0.49\linewidth]{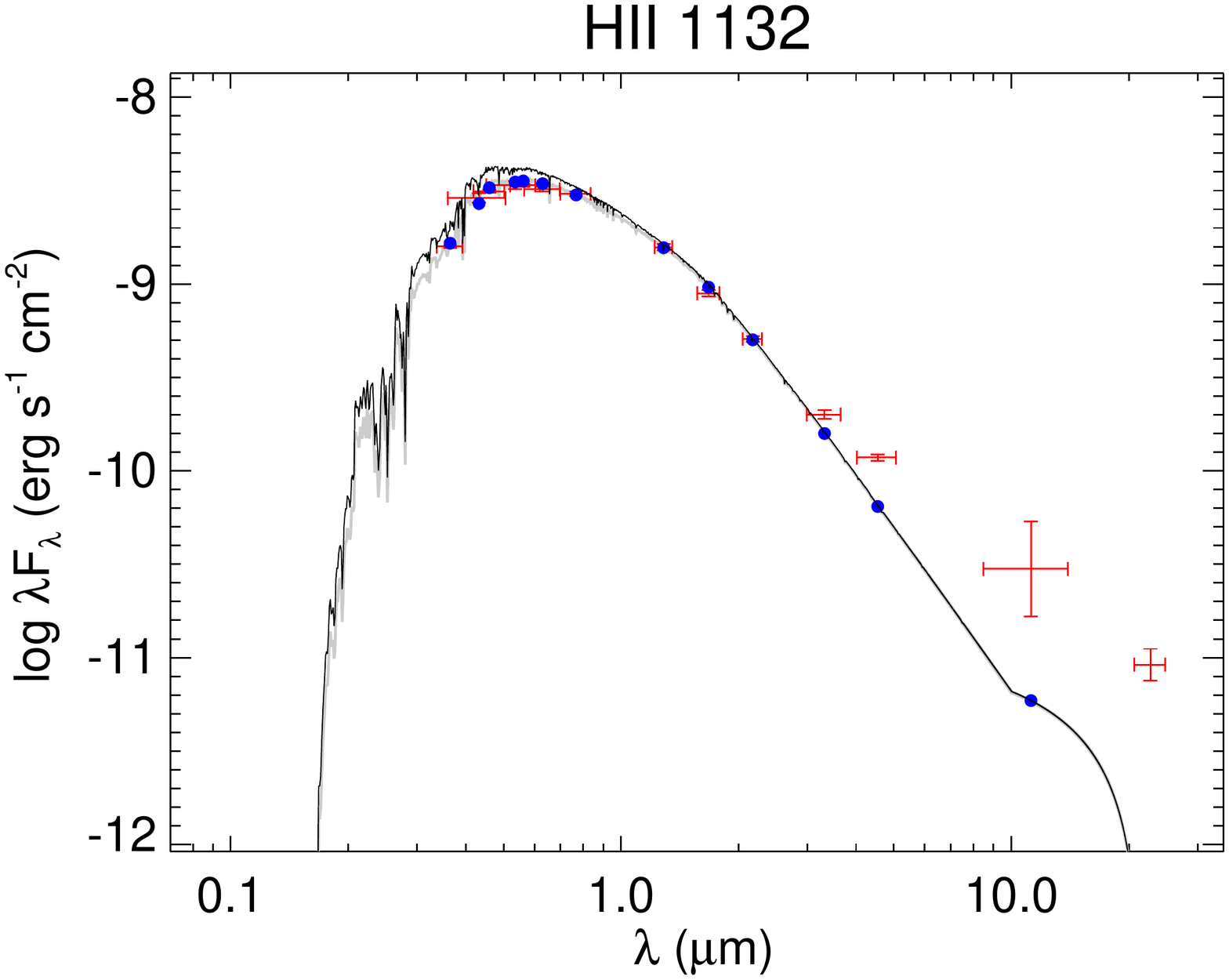}
  \includegraphics[trim=60 60 60 60,clip,width=0.49\linewidth]{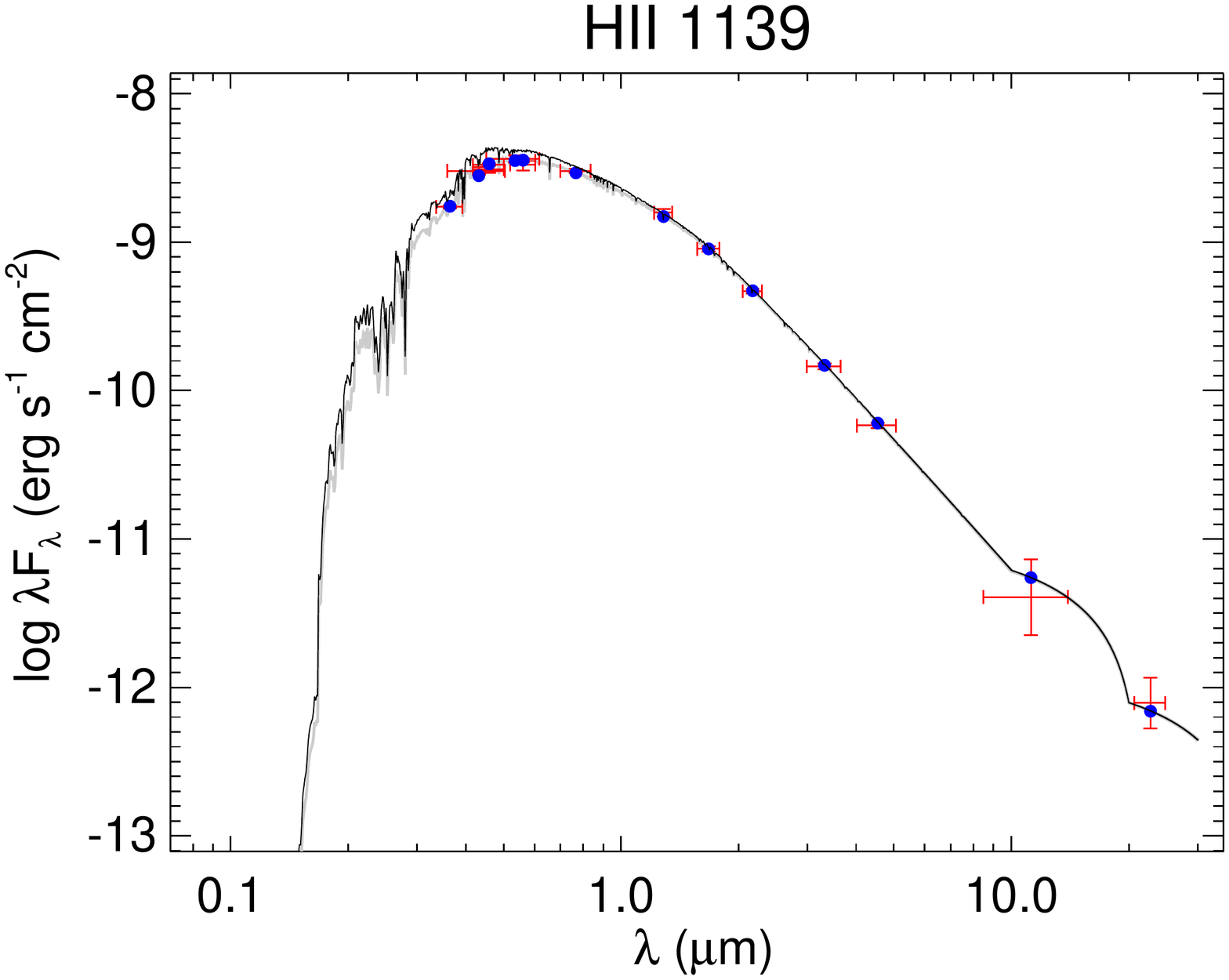}
  \includegraphics[trim=60 60 60 60,clip,width=0.49\linewidth]{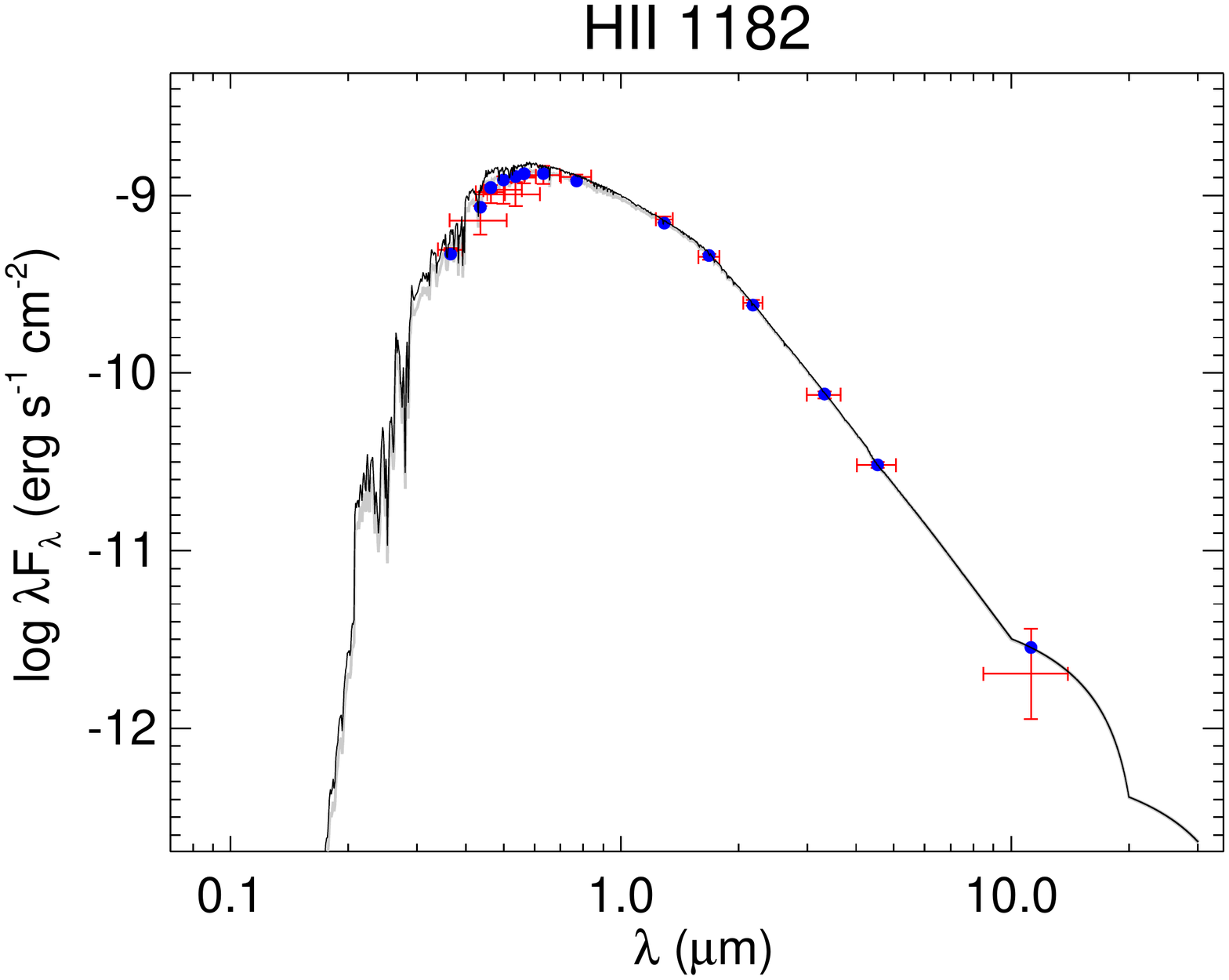}
  \includegraphics[trim=60 60 60 60,clip,width=0.49\linewidth]{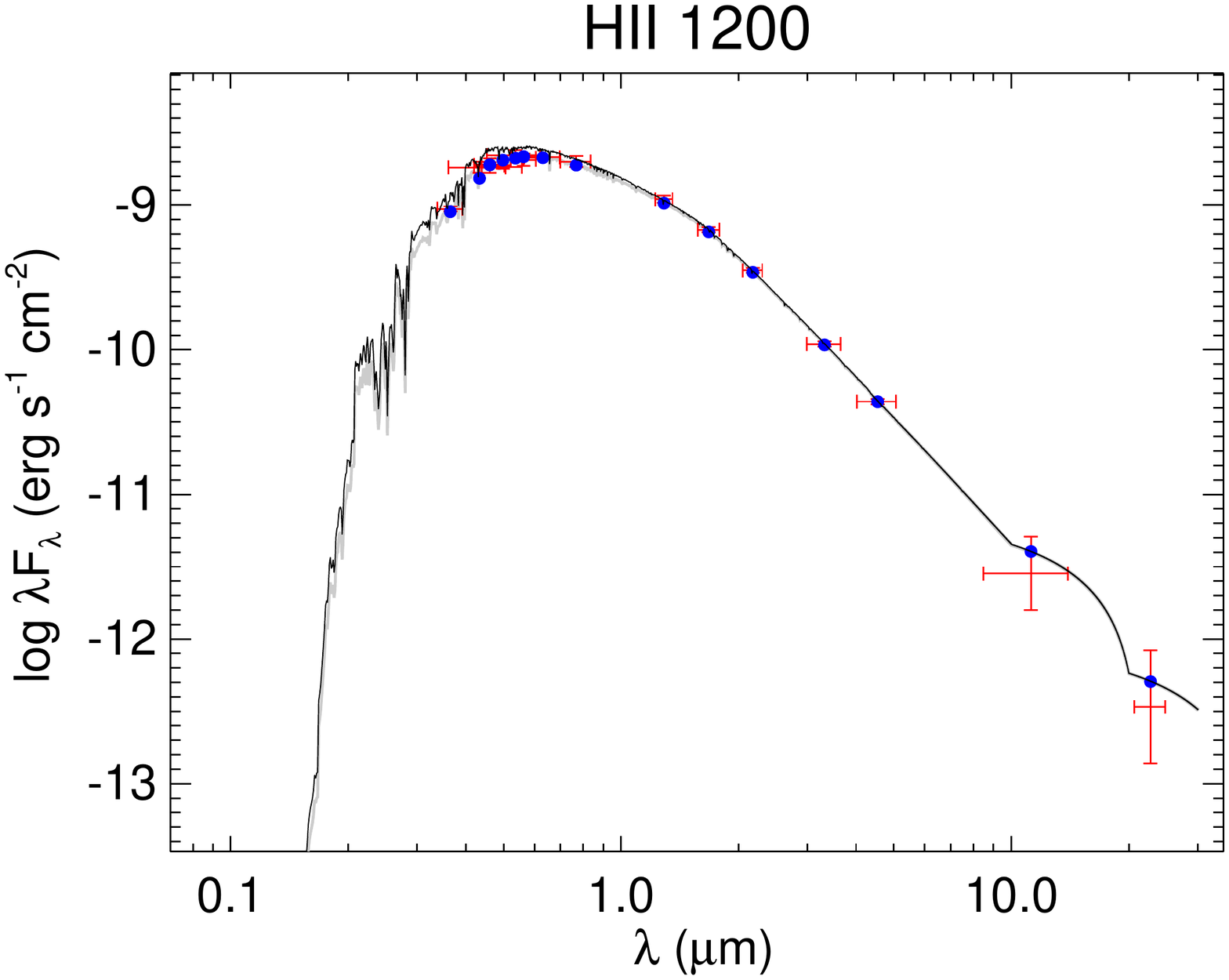}
  \includegraphics[trim=60 60 60 60,clip,width=0.49\linewidth]{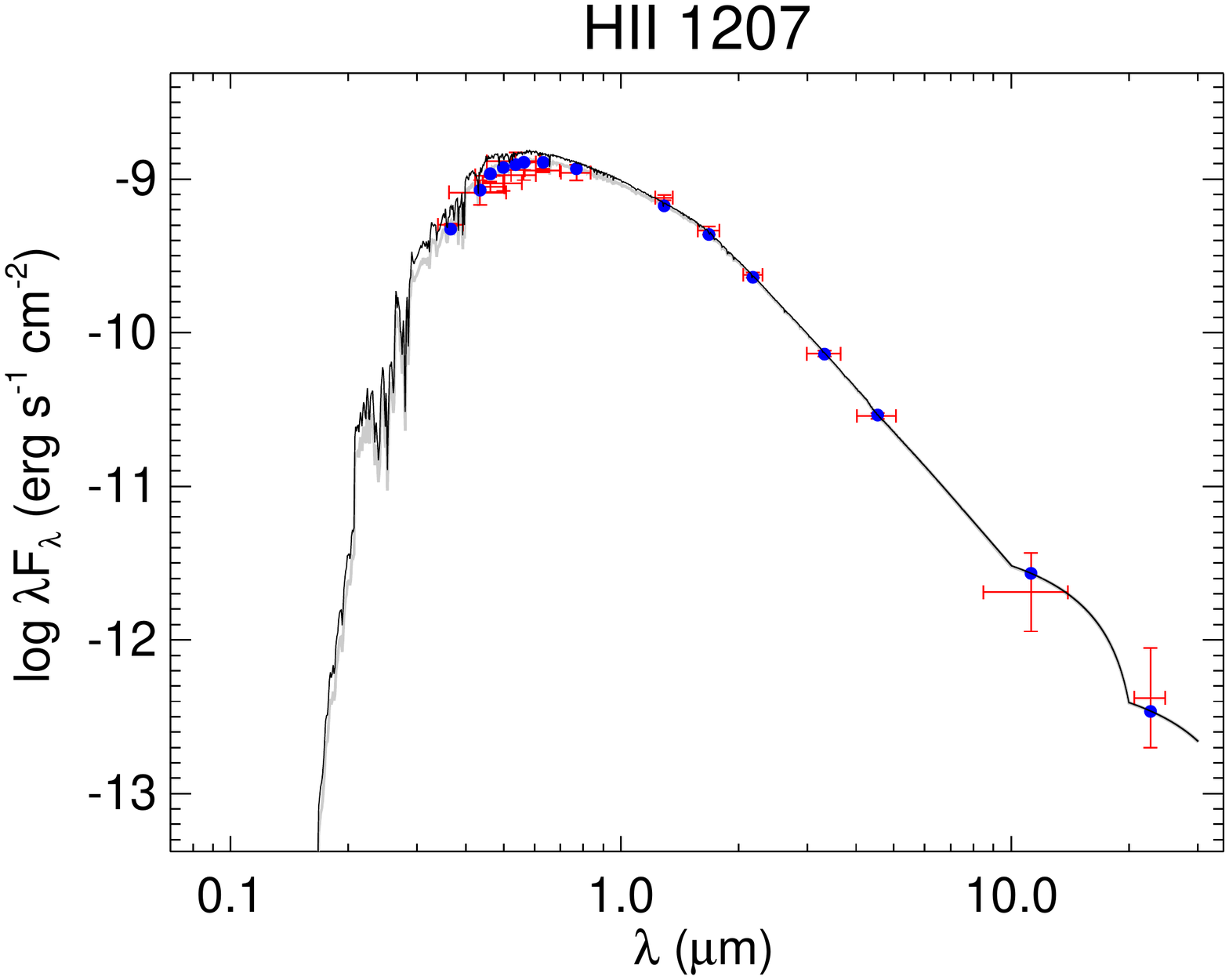}
  \caption{All labels, lines, symbols, and colors as in Figure \ref{fig:seds}.}
  \label{fig:seds_7}
\end{figure}

\begin{figure}[H]
  \centering
  \includegraphics[trim=60 60 60 60,clip,width=0.49\linewidth]{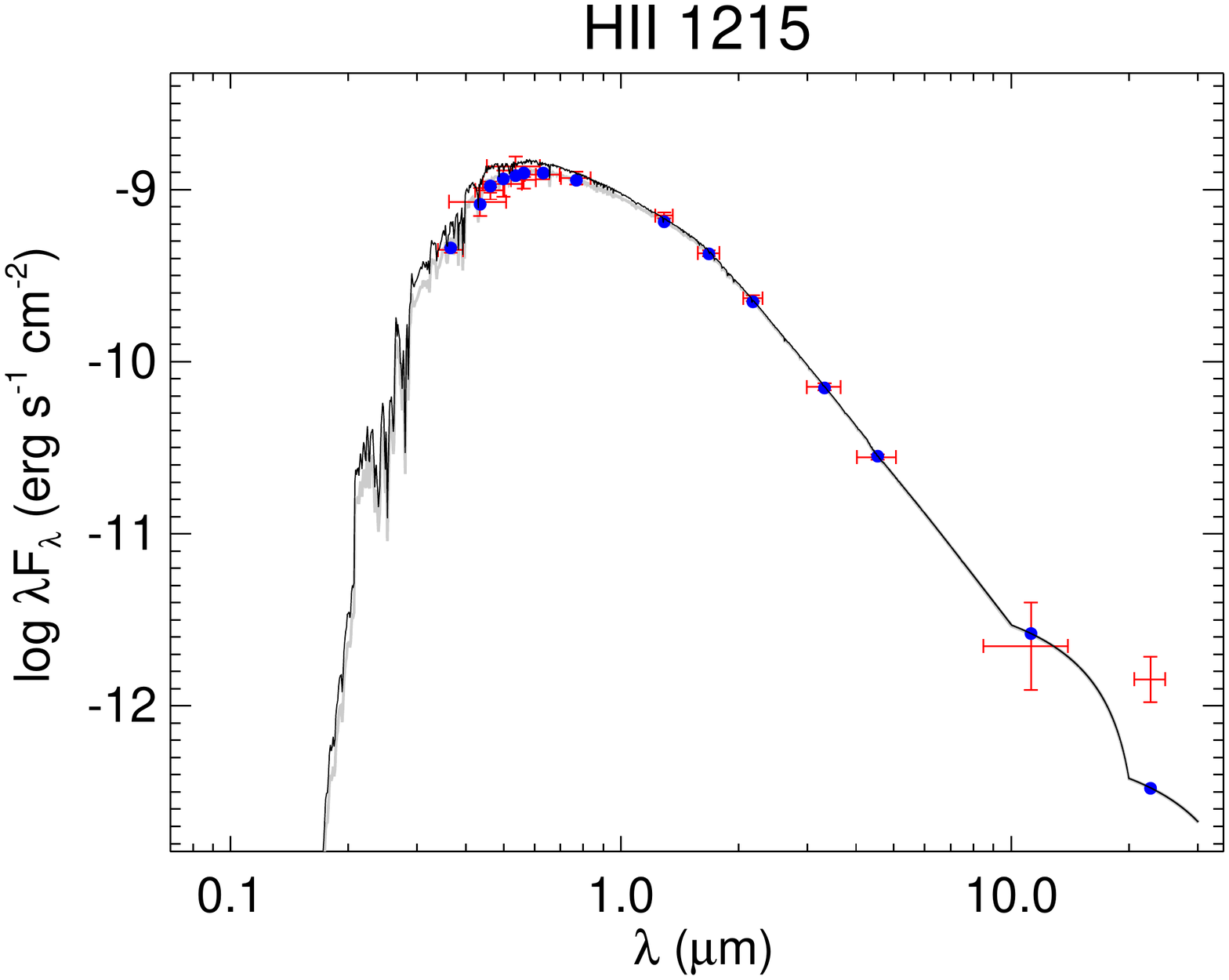}
  \includegraphics[trim=60 60 60 60,clip,width=0.49\linewidth]{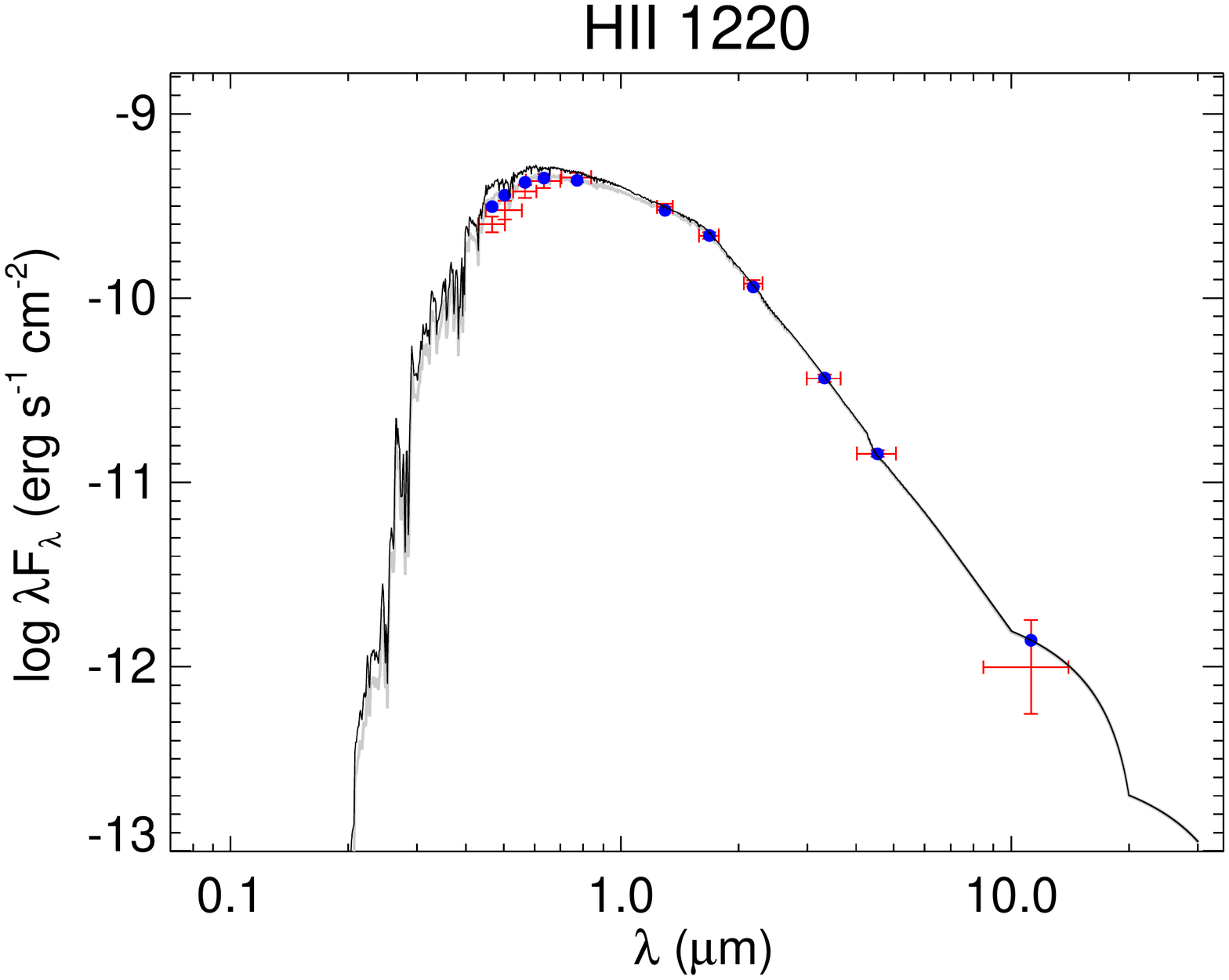}
  \includegraphics[trim=60 60 60 60,clip,width=0.49\linewidth]{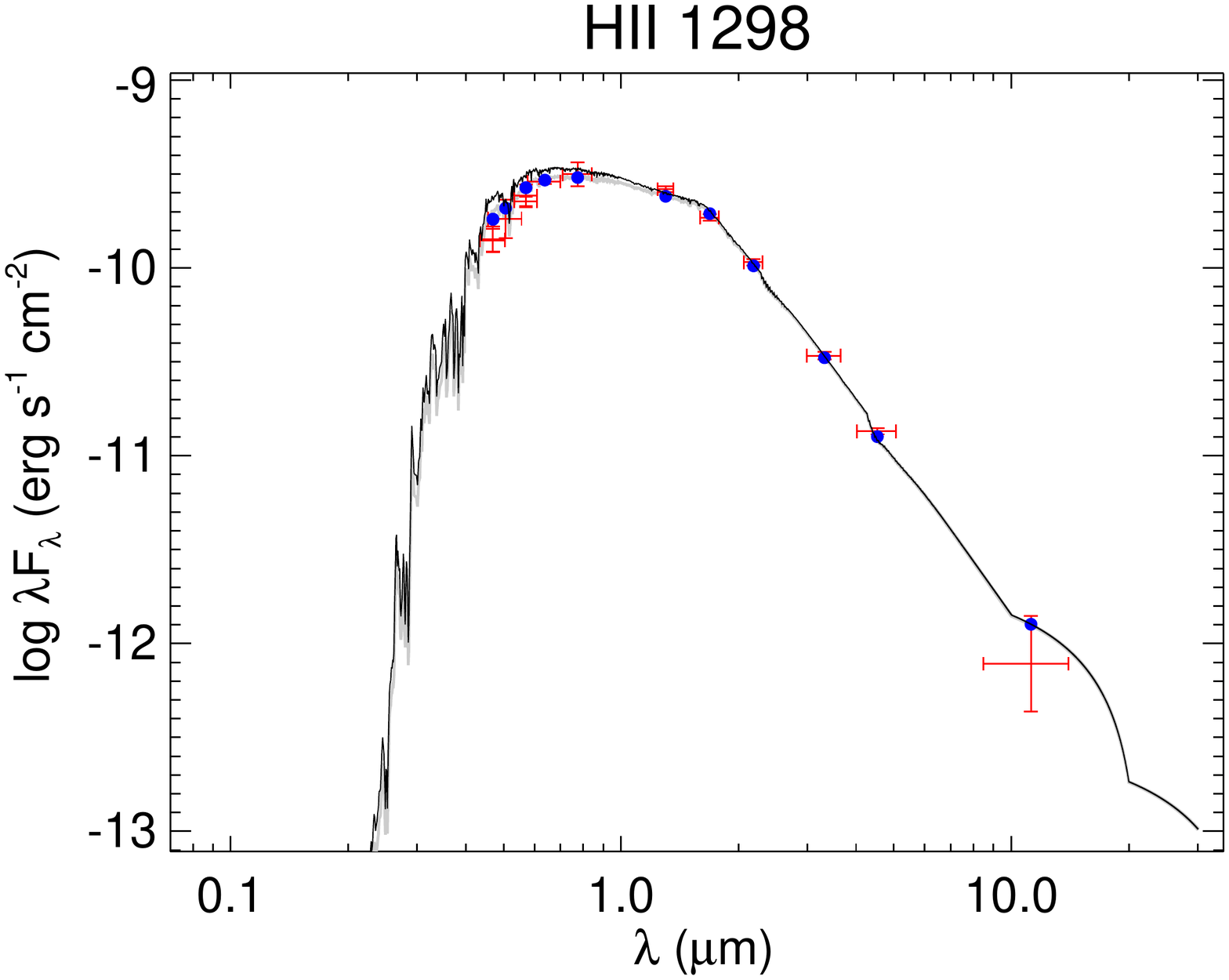}
  \includegraphics[trim=60 60 60 60,clip,width=0.49\linewidth]{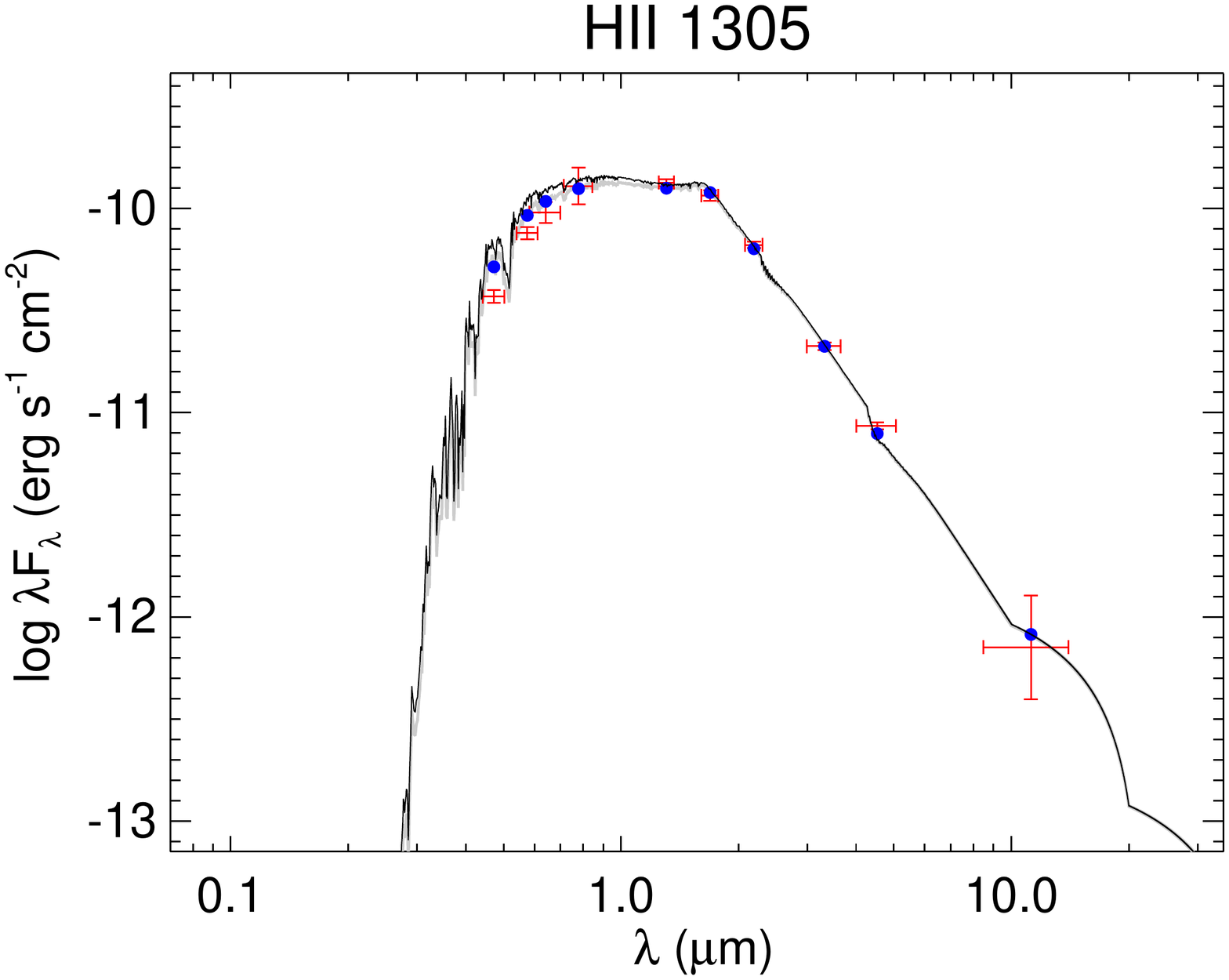}
  \includegraphics[trim=60 60 60 60,clip,width=0.49\linewidth]{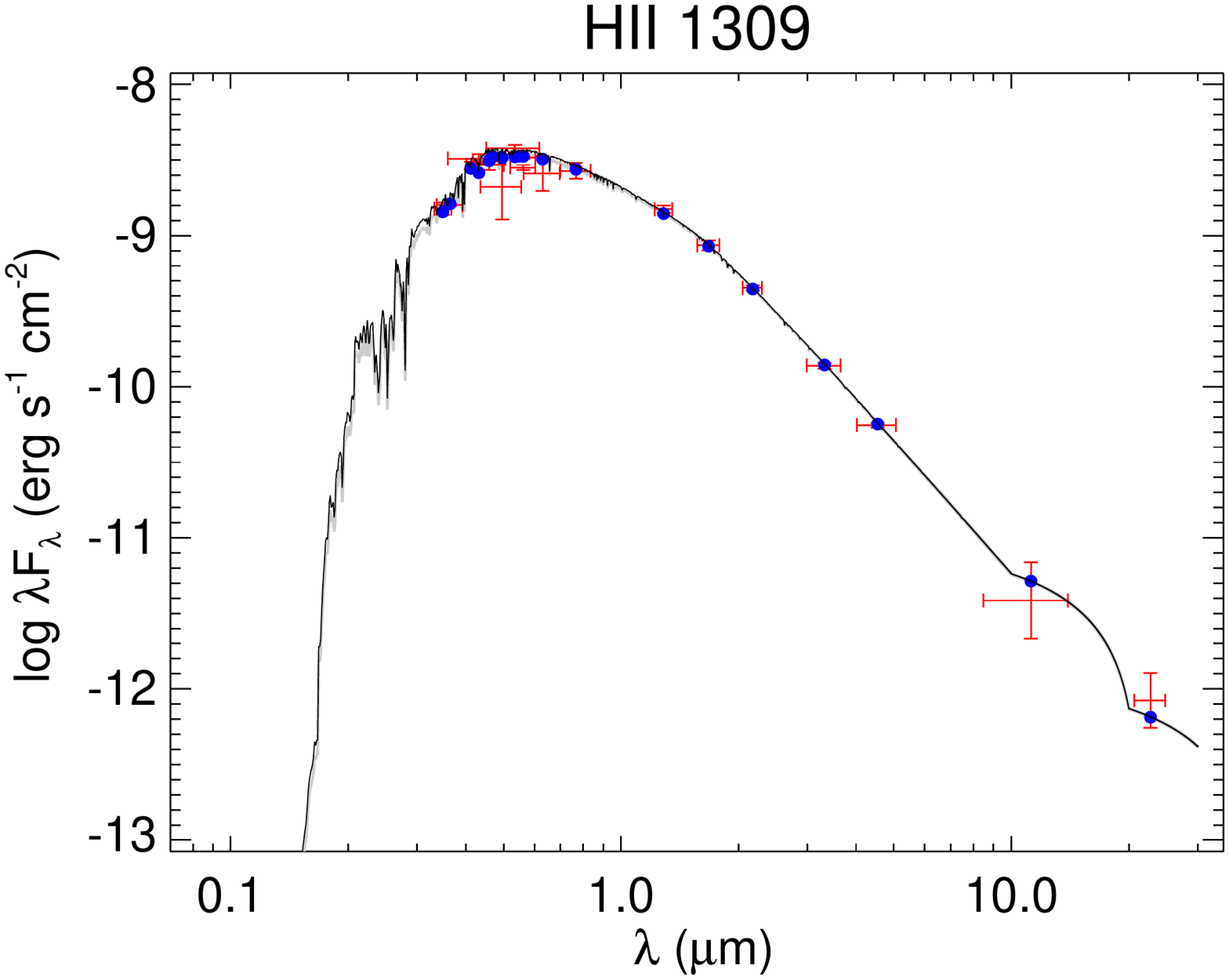}
  \includegraphics[trim=60 60 60 60,clip,width=0.49\linewidth]{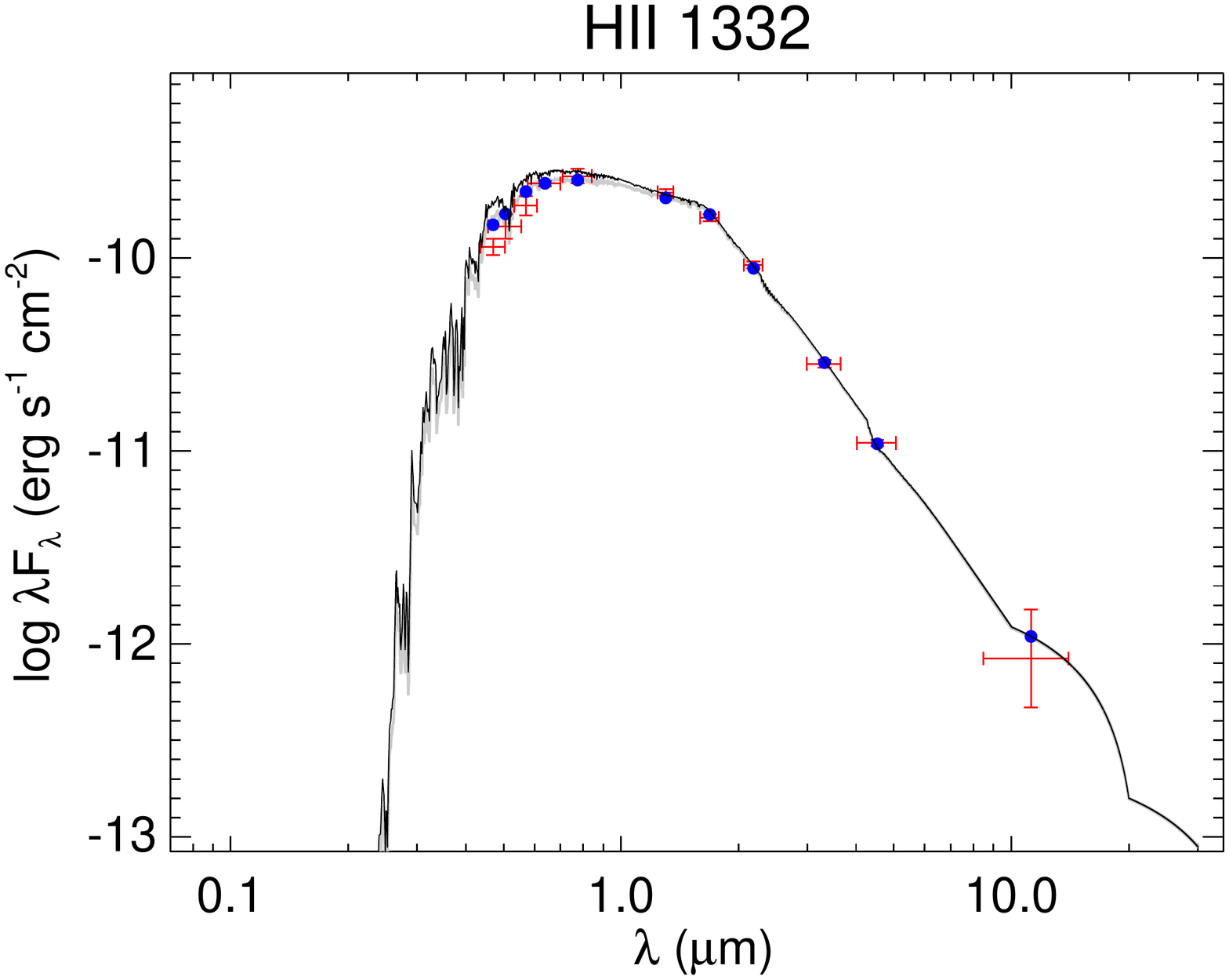}
  \caption{All labels, lines, symbols, and colors as in Figure \ref{fig:seds}.}
  \label{fig:seds_8}
\end{figure}

\begin{figure}[H]
  \centering
  \includegraphics[trim=60 60 60 60,clip,width=0.49\linewidth]{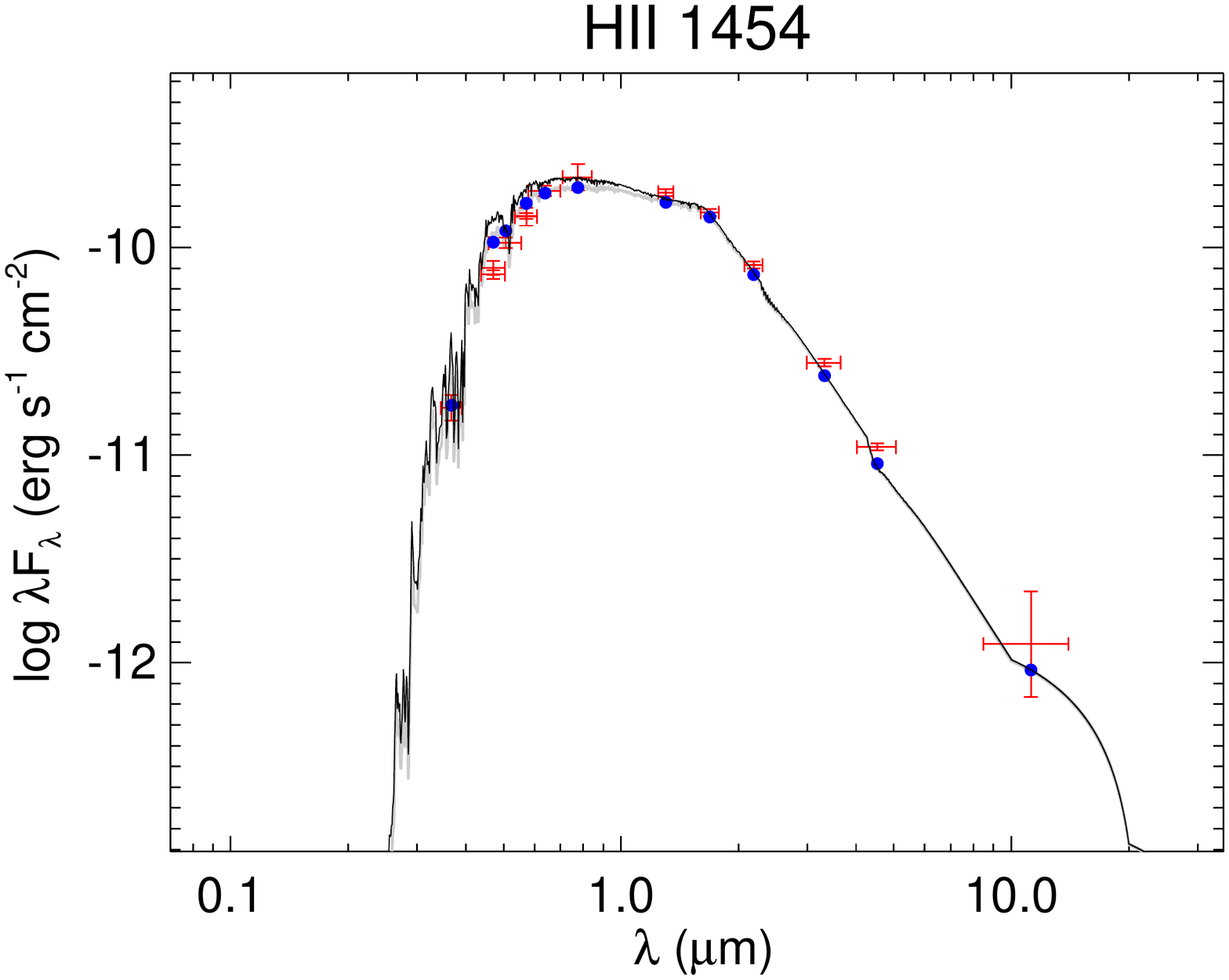}
  \includegraphics[trim=60 60 60 60,clip,width=0.49\linewidth]{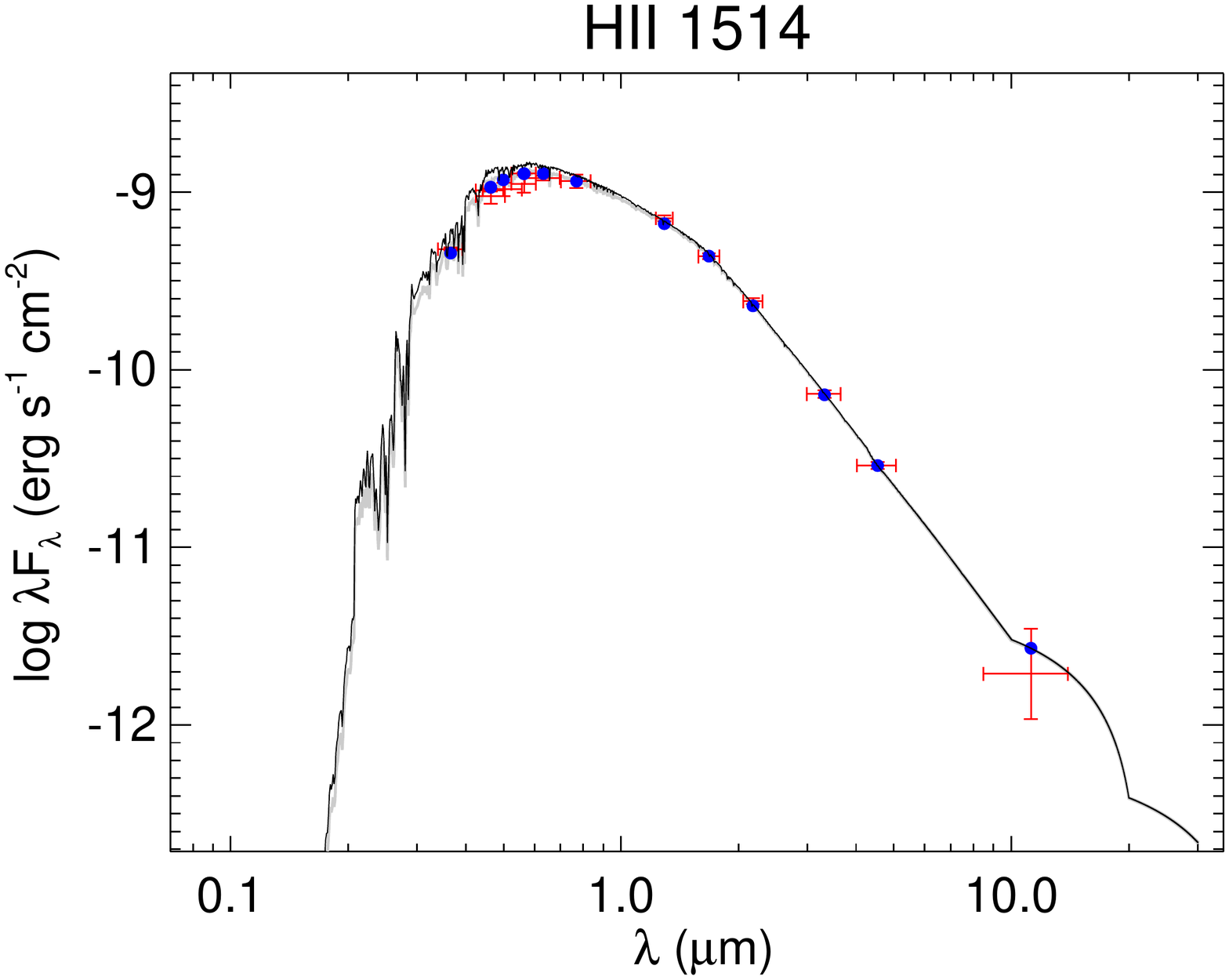}
  \includegraphics[trim=60 60 60 60,clip,width=0.49\linewidth]{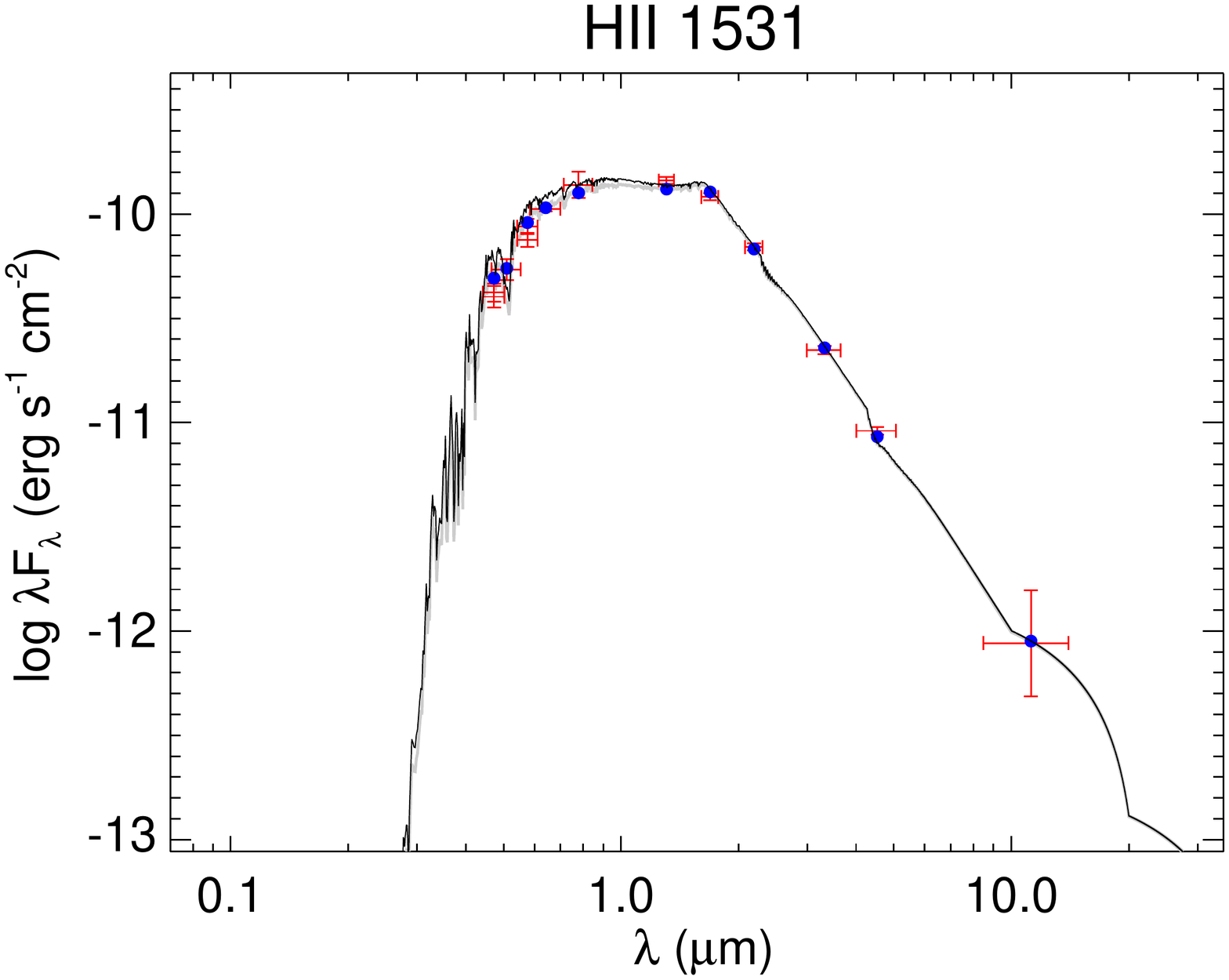}
  \includegraphics[trim=60 60 60 60,clip,width=0.49\linewidth]{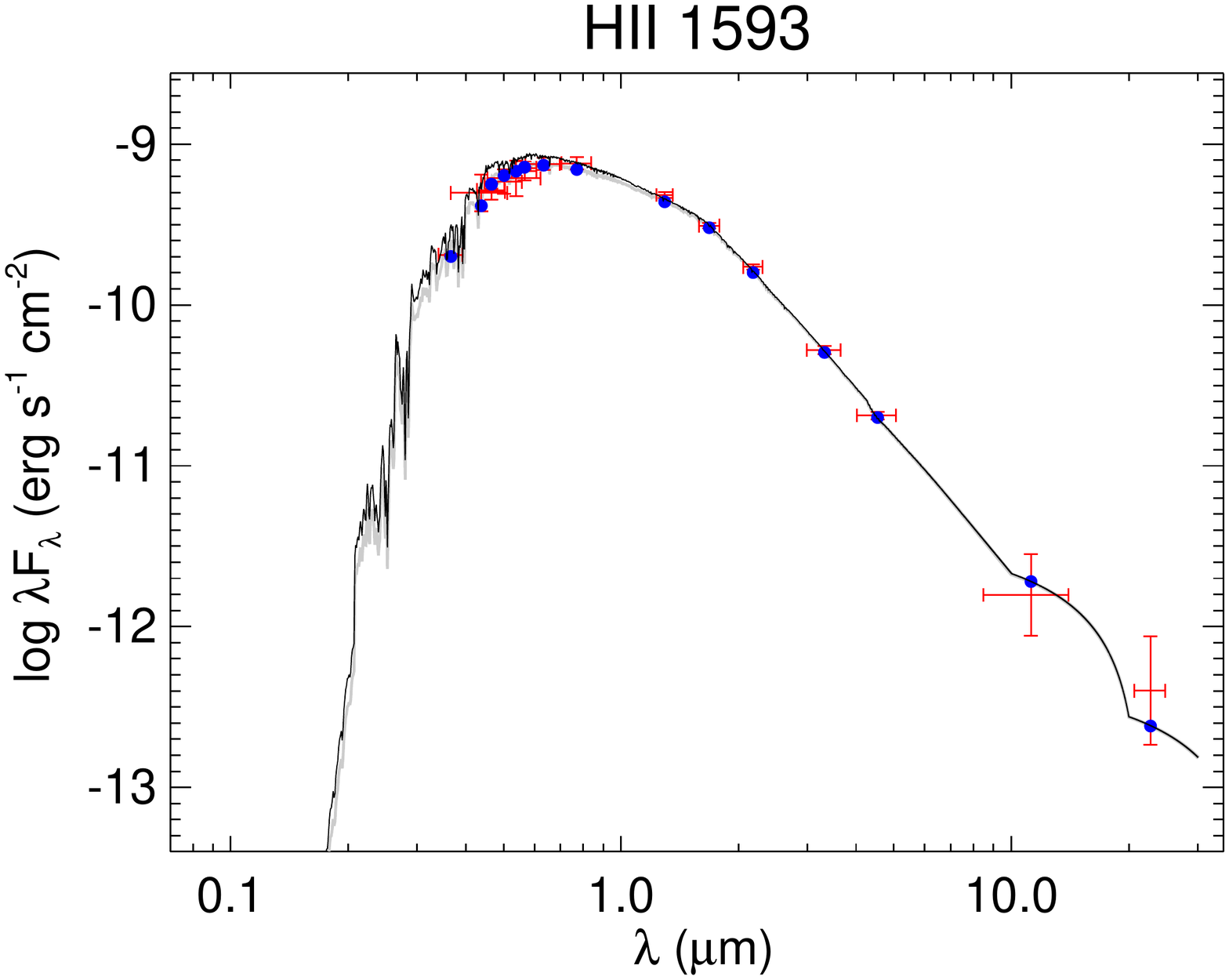}
  \includegraphics[trim=60 60 60 60,clip,width=0.49\linewidth]{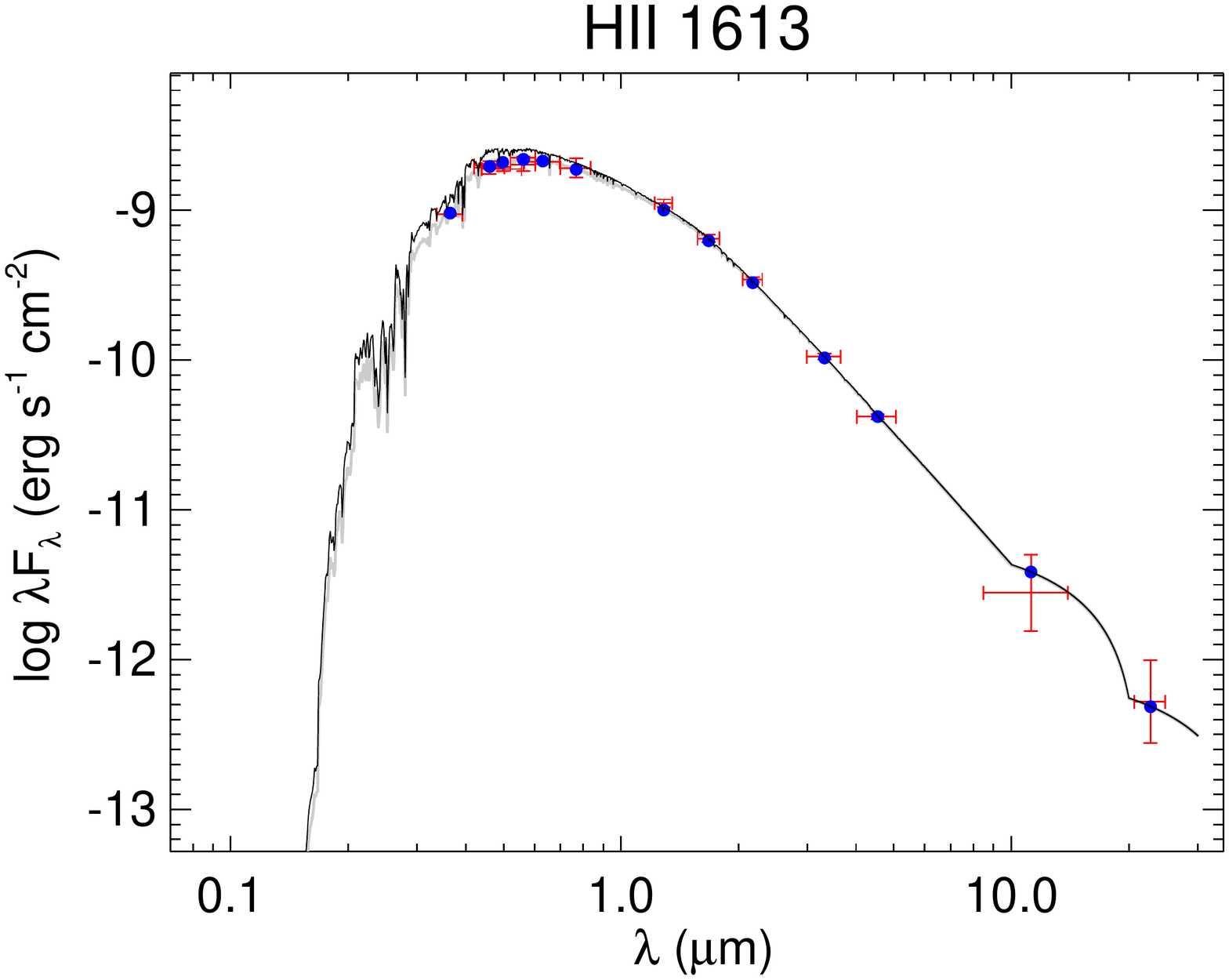}
  \includegraphics[trim=60 60 60 60,clip,width=0.49\linewidth]{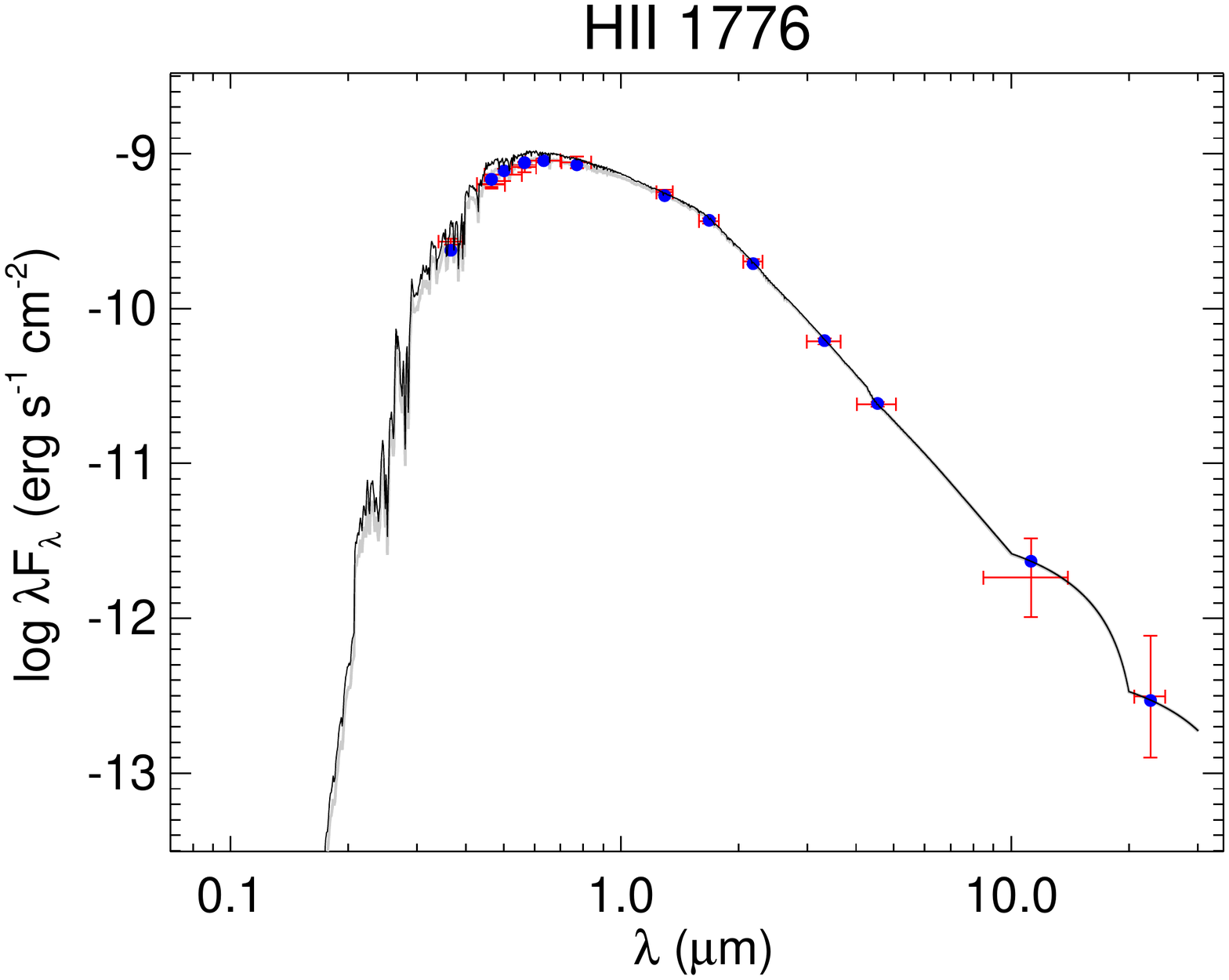}
  \caption{All labels, lines, symbols, and colors as in Figure \ref{fig:seds}.}
  \label{fig:seds_9}
\end{figure}

\begin{figure}[H]
  \centering
  \includegraphics[trim=60 60 60 60,clip,width=0.49\linewidth]{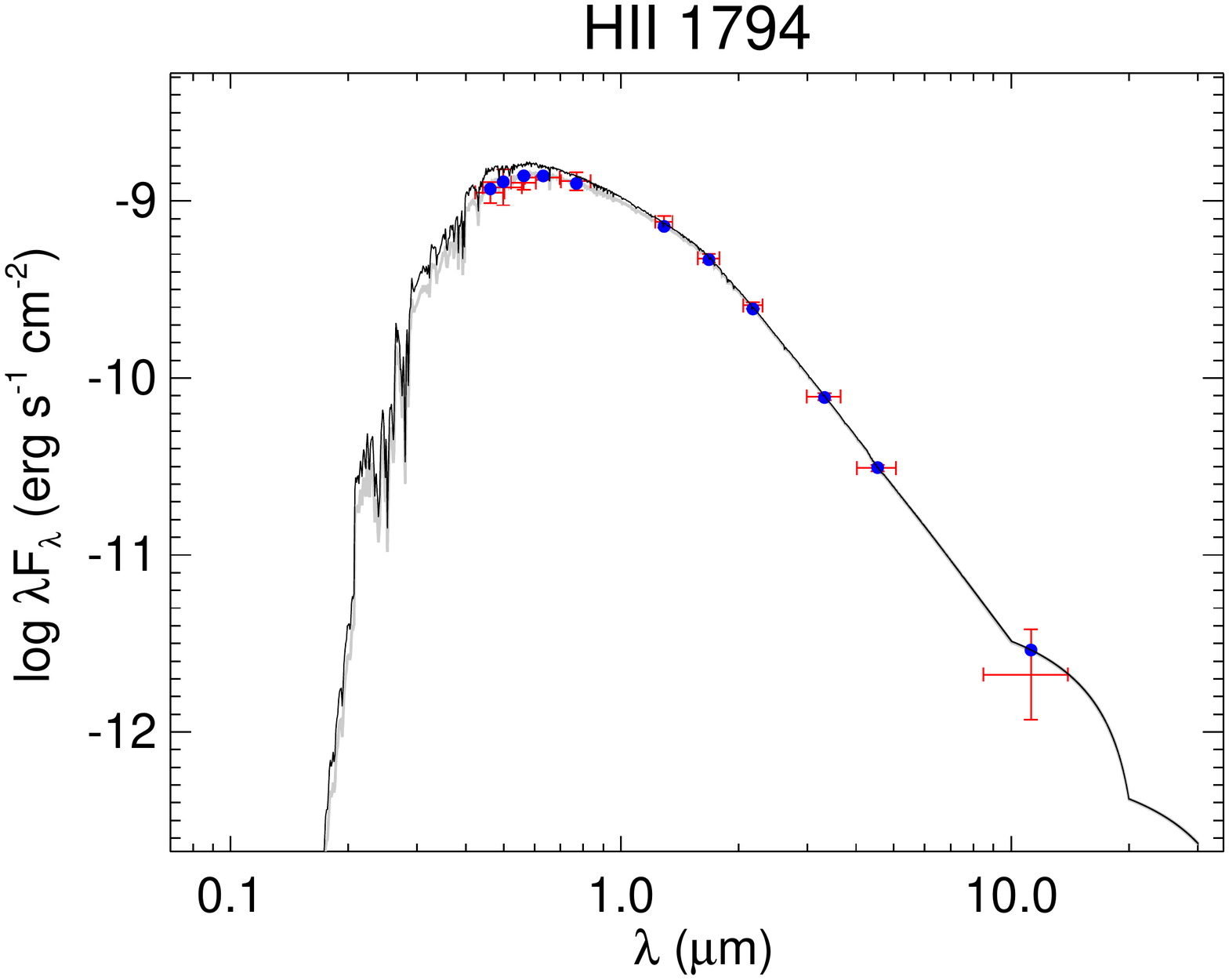}
  \includegraphics[trim=60 60 60 60,clip,width=0.49\linewidth]{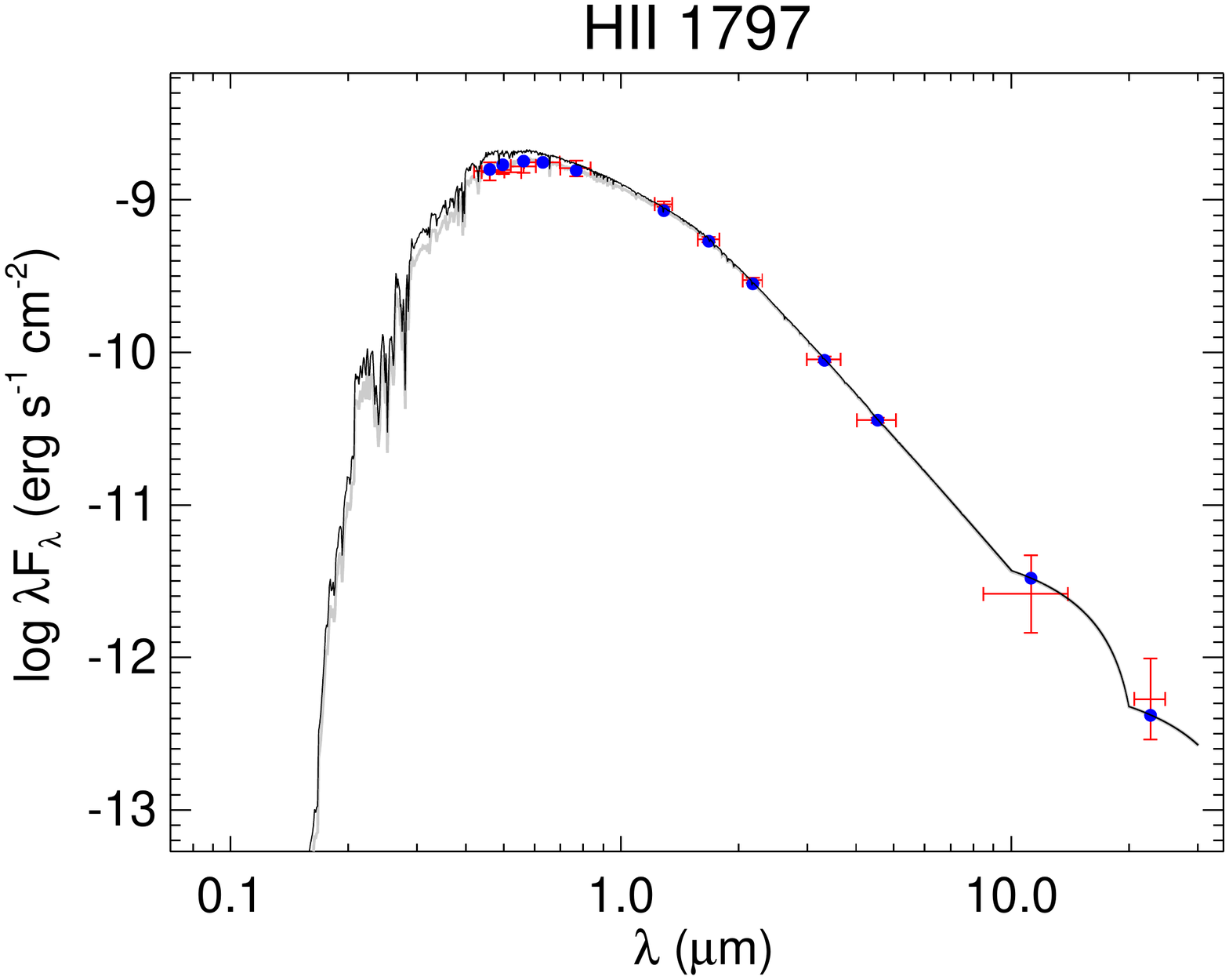}
  \includegraphics[trim=60 60 60 60,clip,width=0.49\linewidth]{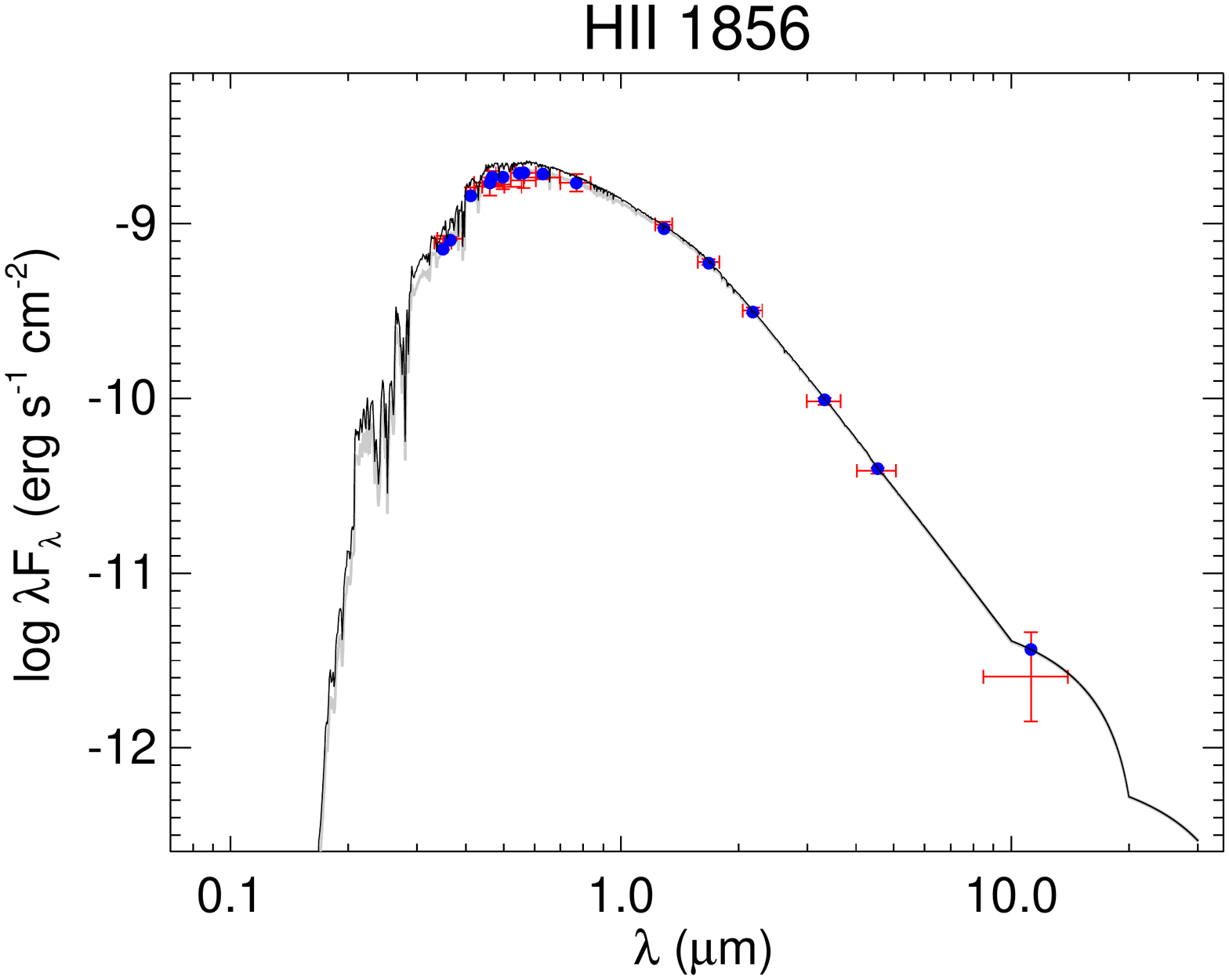}
  \includegraphics[trim=60 60 60 60,clip,width=0.49\linewidth]{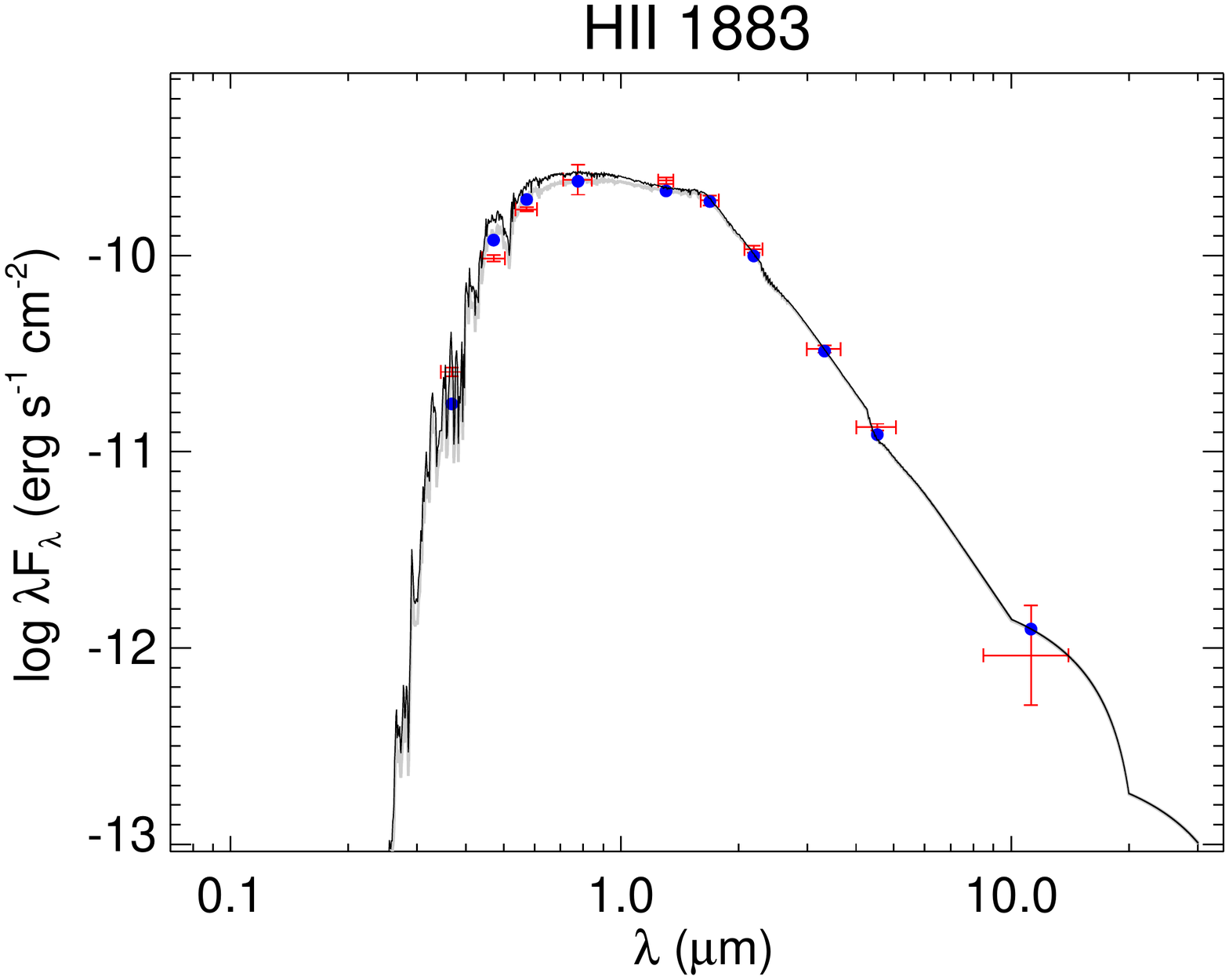}
  \includegraphics[trim=60 60 60 60,clip,width=0.49\linewidth]{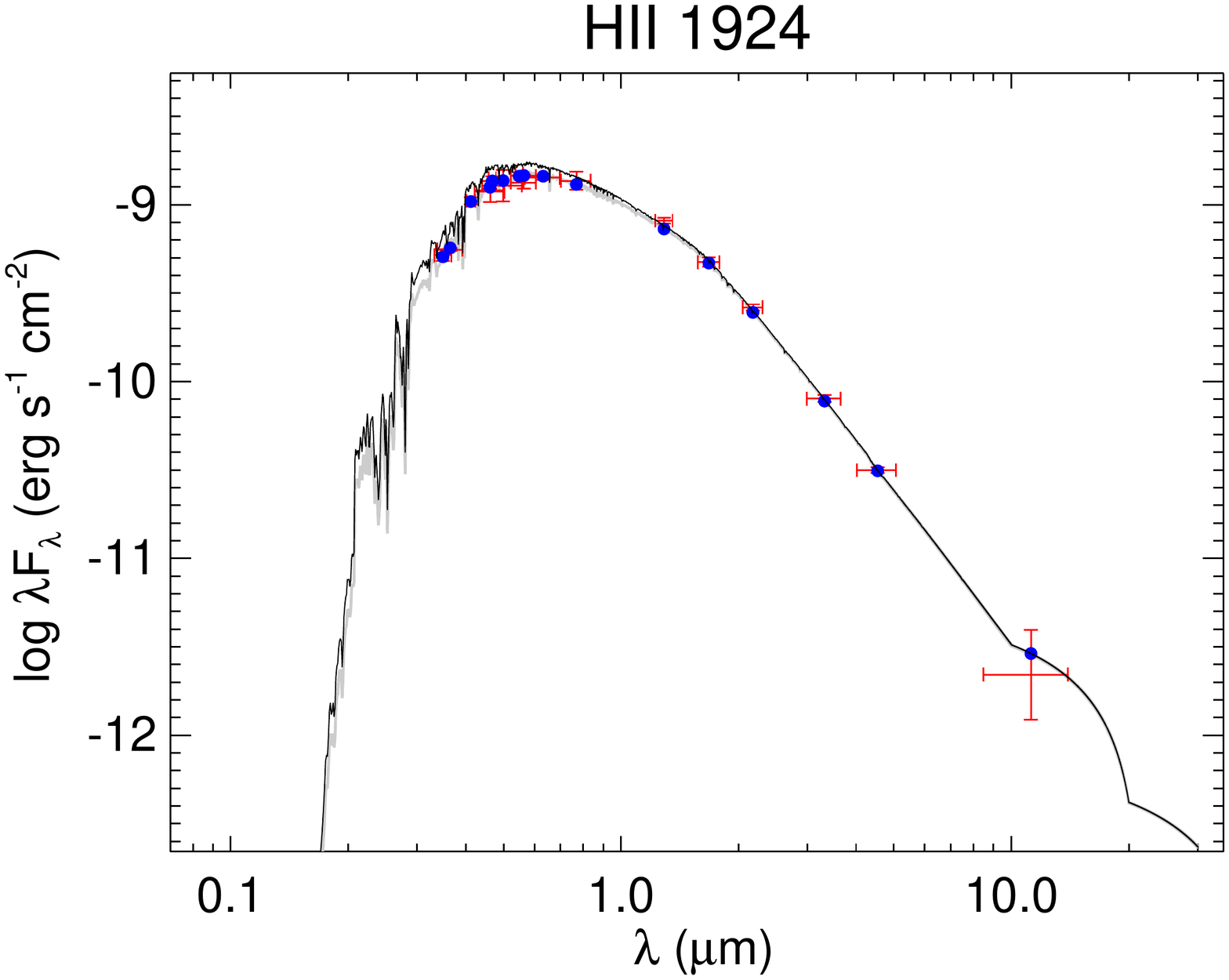}
  \includegraphics[trim=60 60 60 60,clip,width=0.49\linewidth]{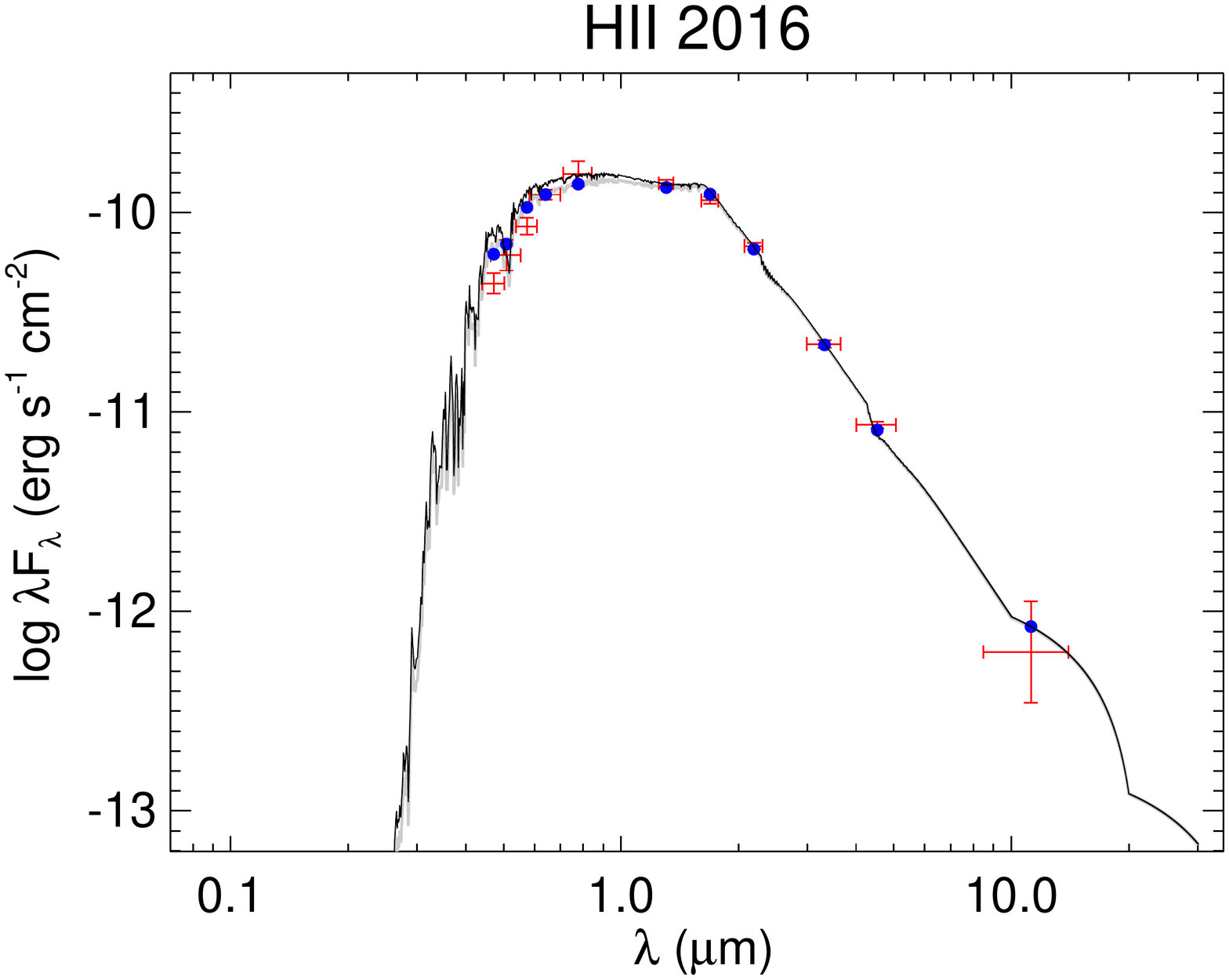}
  \caption{All labels, lines, symbols, and colors as in Figure \ref{fig:seds}.}
  \label{fig:seds_10}
\end{figure}

\begin{figure}[H]
  \centering
  \includegraphics[trim=60 60 60 60,clip,width=0.49\linewidth]{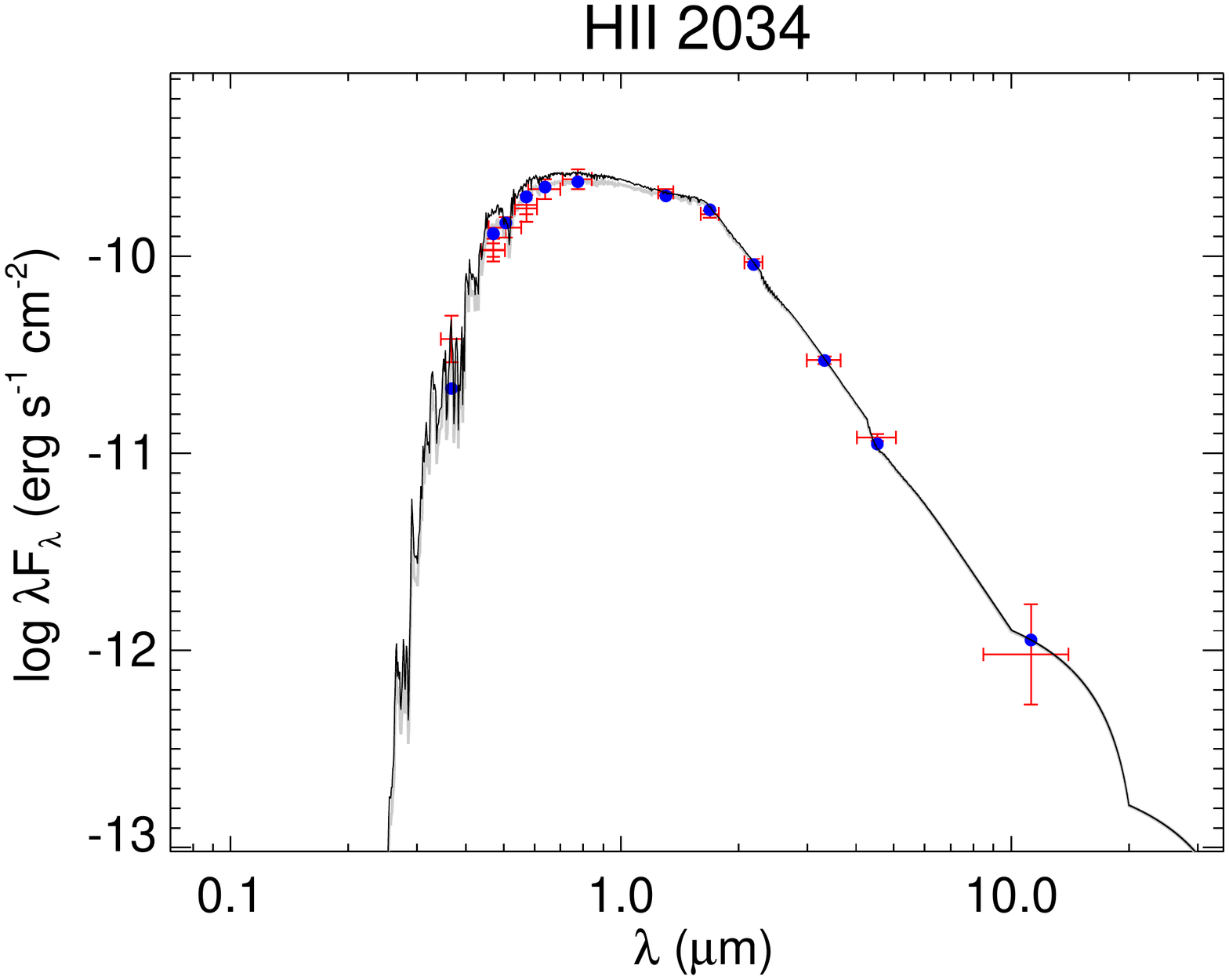}
  \includegraphics[trim=60 60 60 60,clip,width=0.49\linewidth]{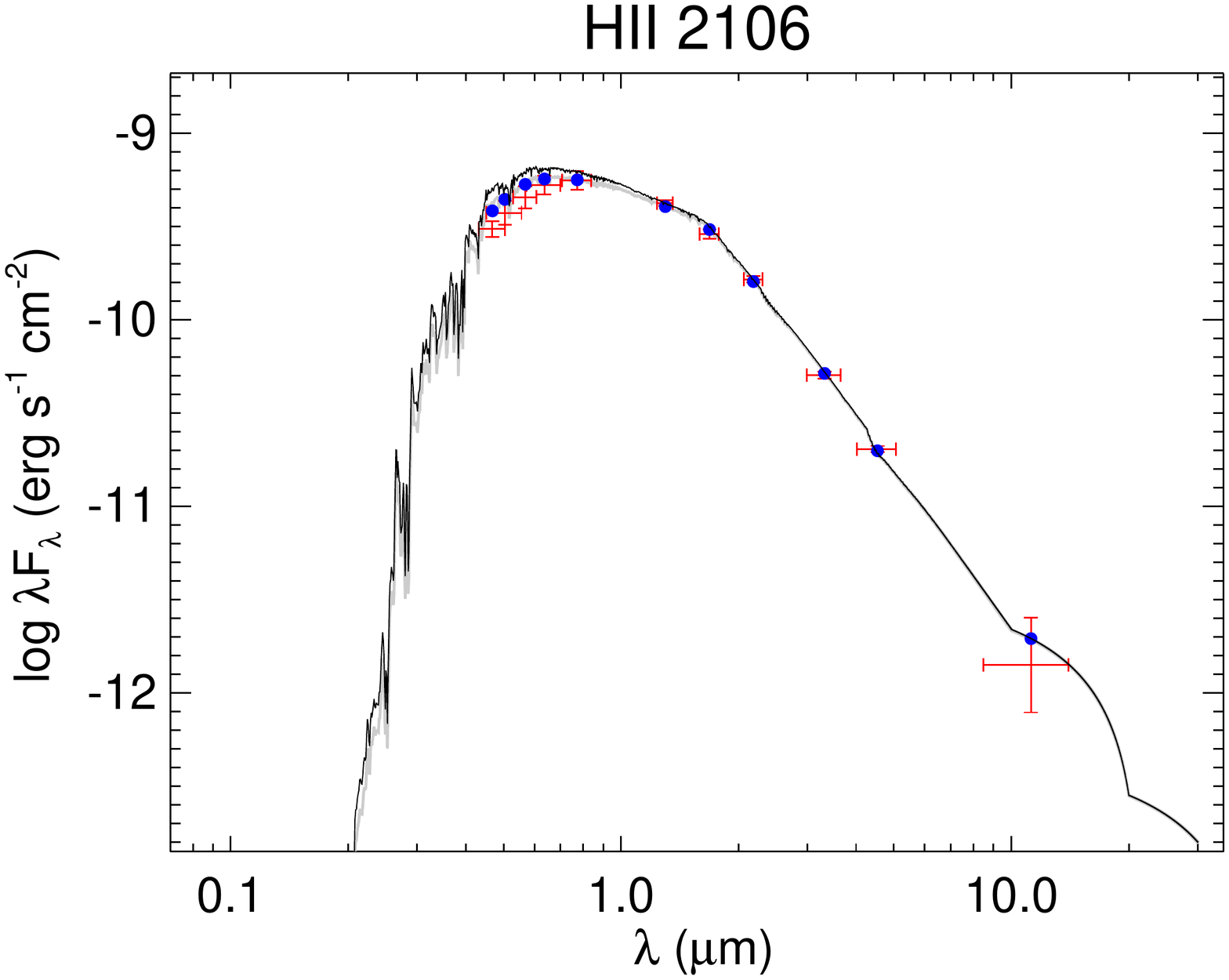}
  \includegraphics[trim=60 60 60 60,clip,width=0.49\linewidth]{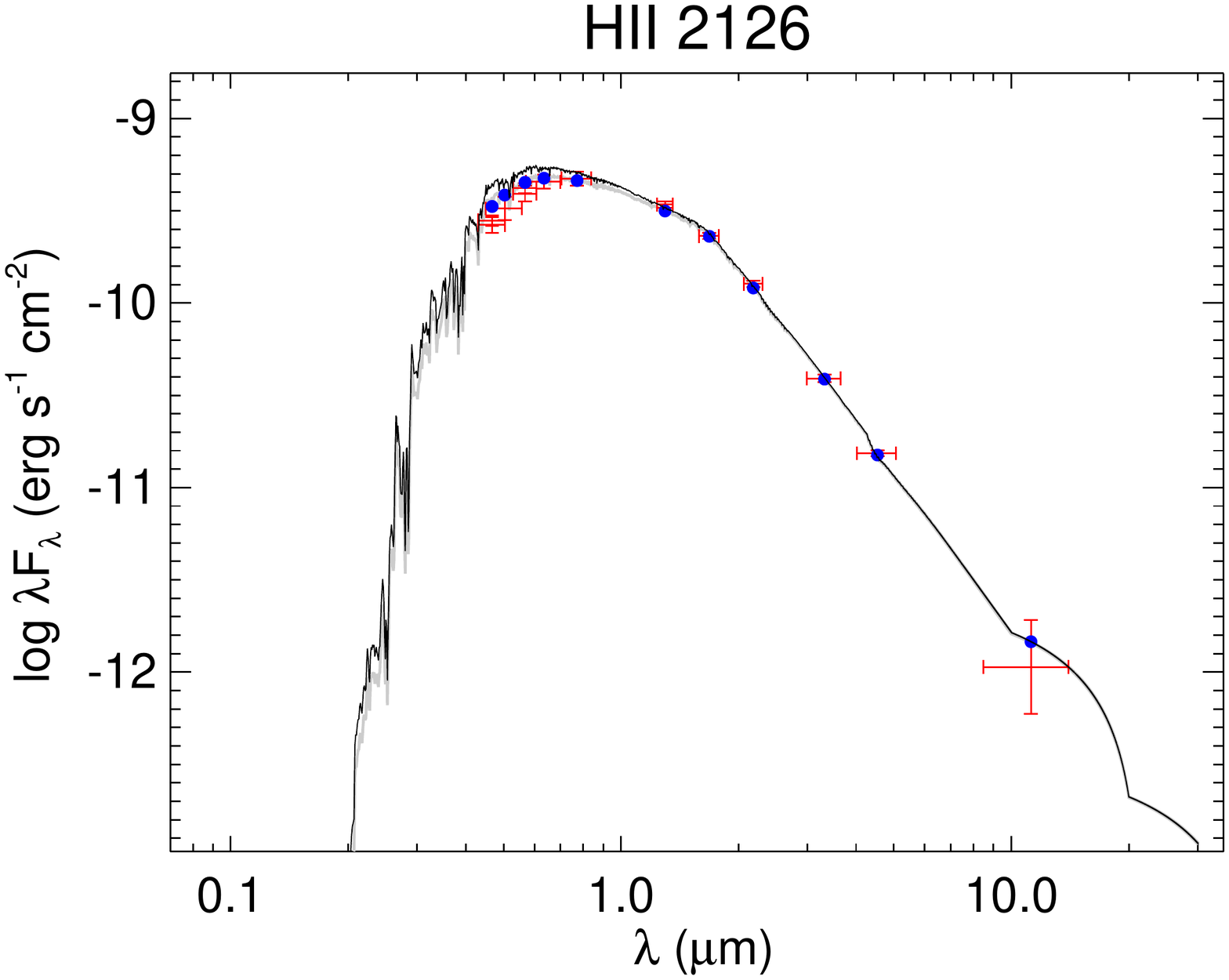}
  \includegraphics[trim=60 60 60 60,clip,width=0.49\linewidth]{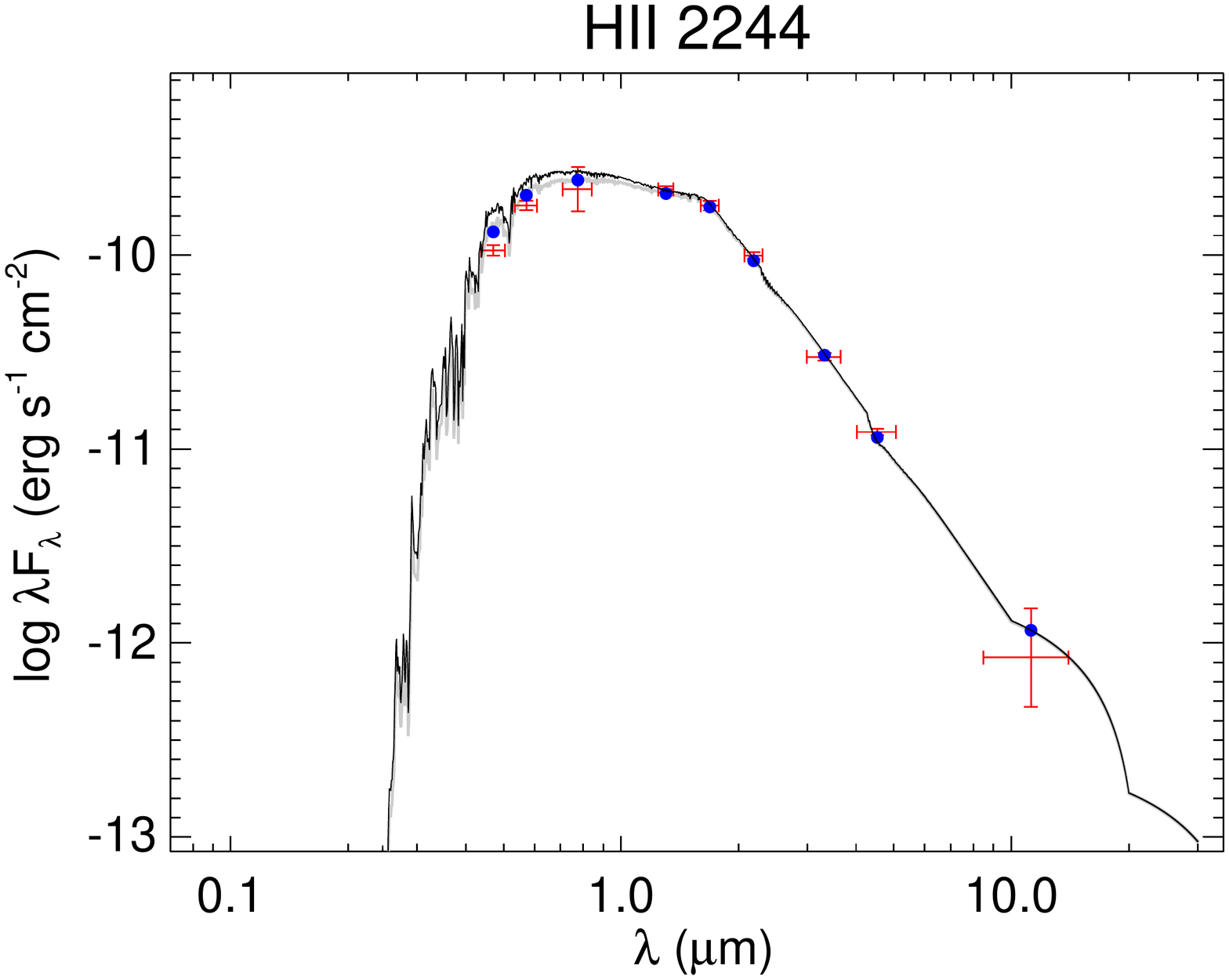}
  \includegraphics[trim=60 60 60 60,clip,width=0.49\linewidth]{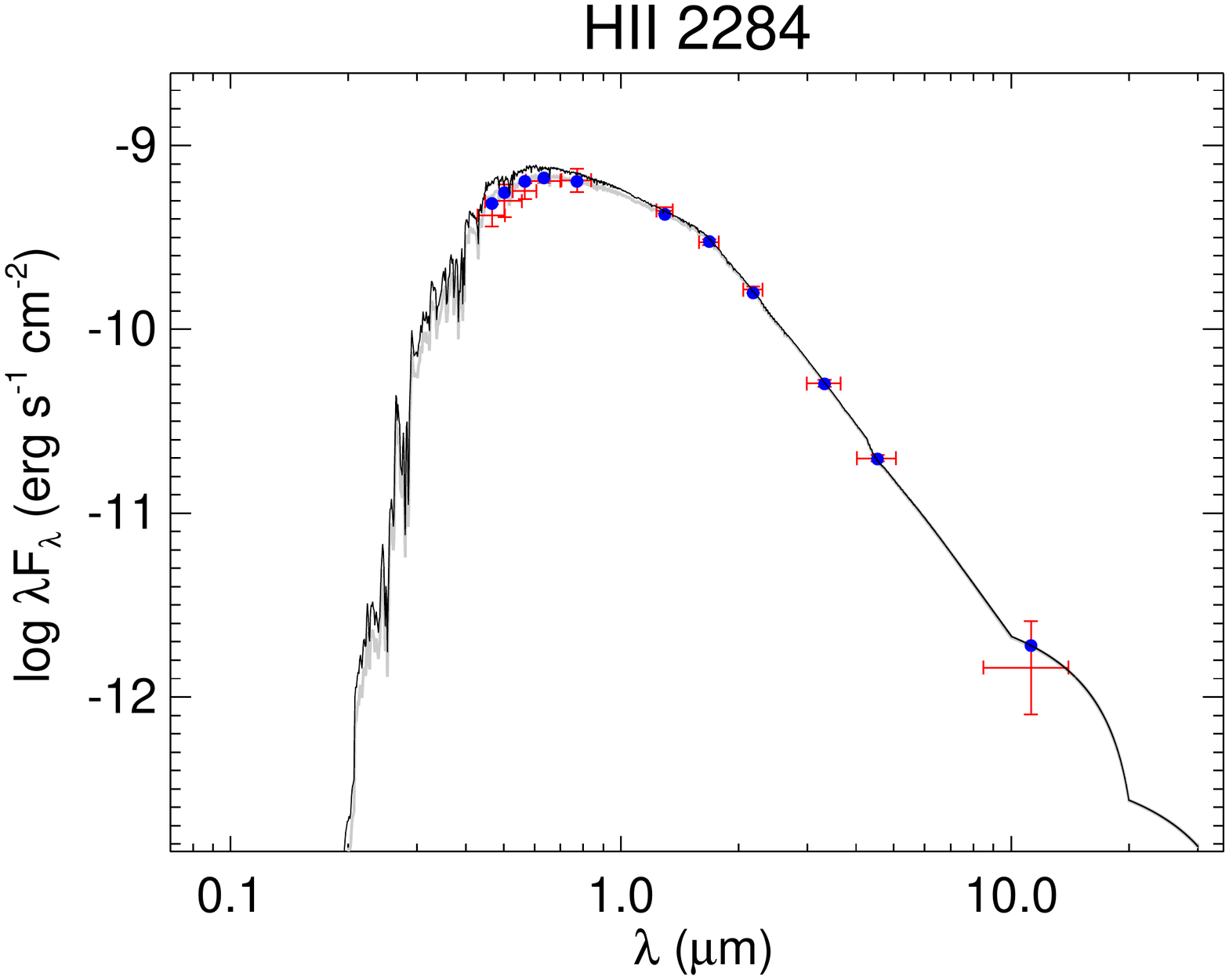}
  \includegraphics[trim=60 60 60 60,clip,width=0.49\linewidth]{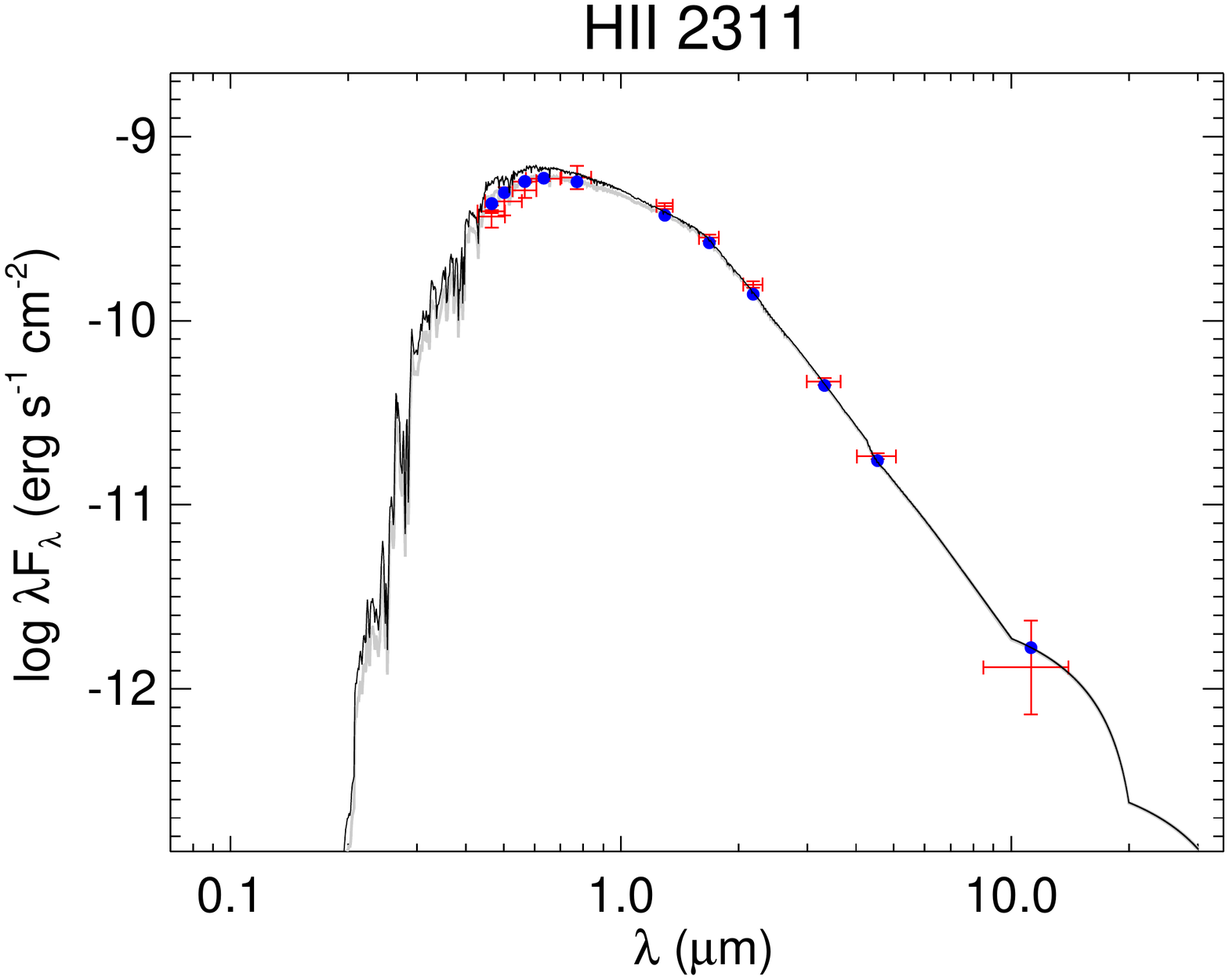}
  \caption{All labels, lines, symbols, and colors as in Figure \ref{fig:seds}.}
  \label{fig:seds_11}
\end{figure}

\begin{figure}[H]
  \centering
  \includegraphics[trim=60 60 60 60,clip,width=0.49\linewidth]{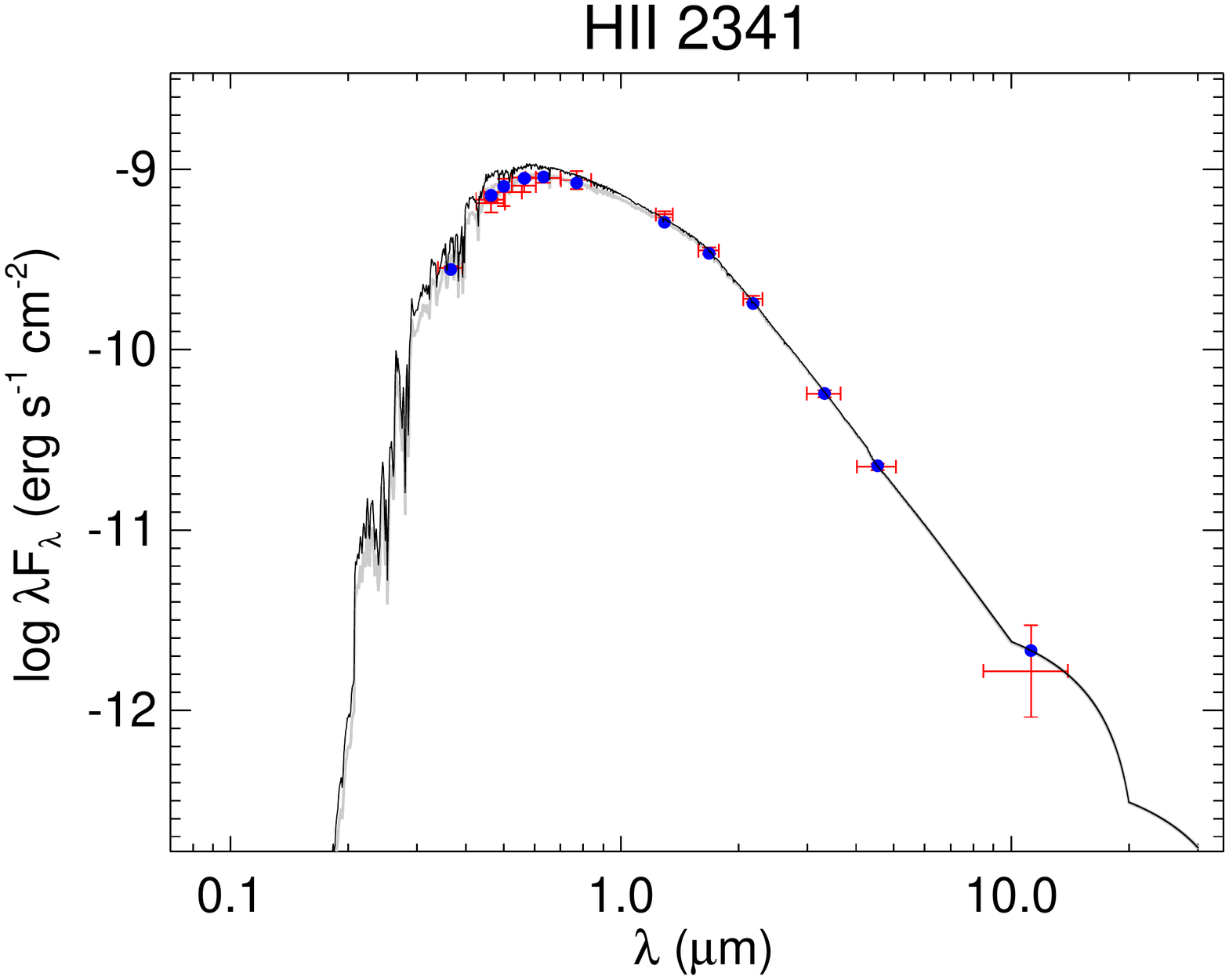}
  \includegraphics[trim=60 60 60 60,clip,width=0.49\linewidth]{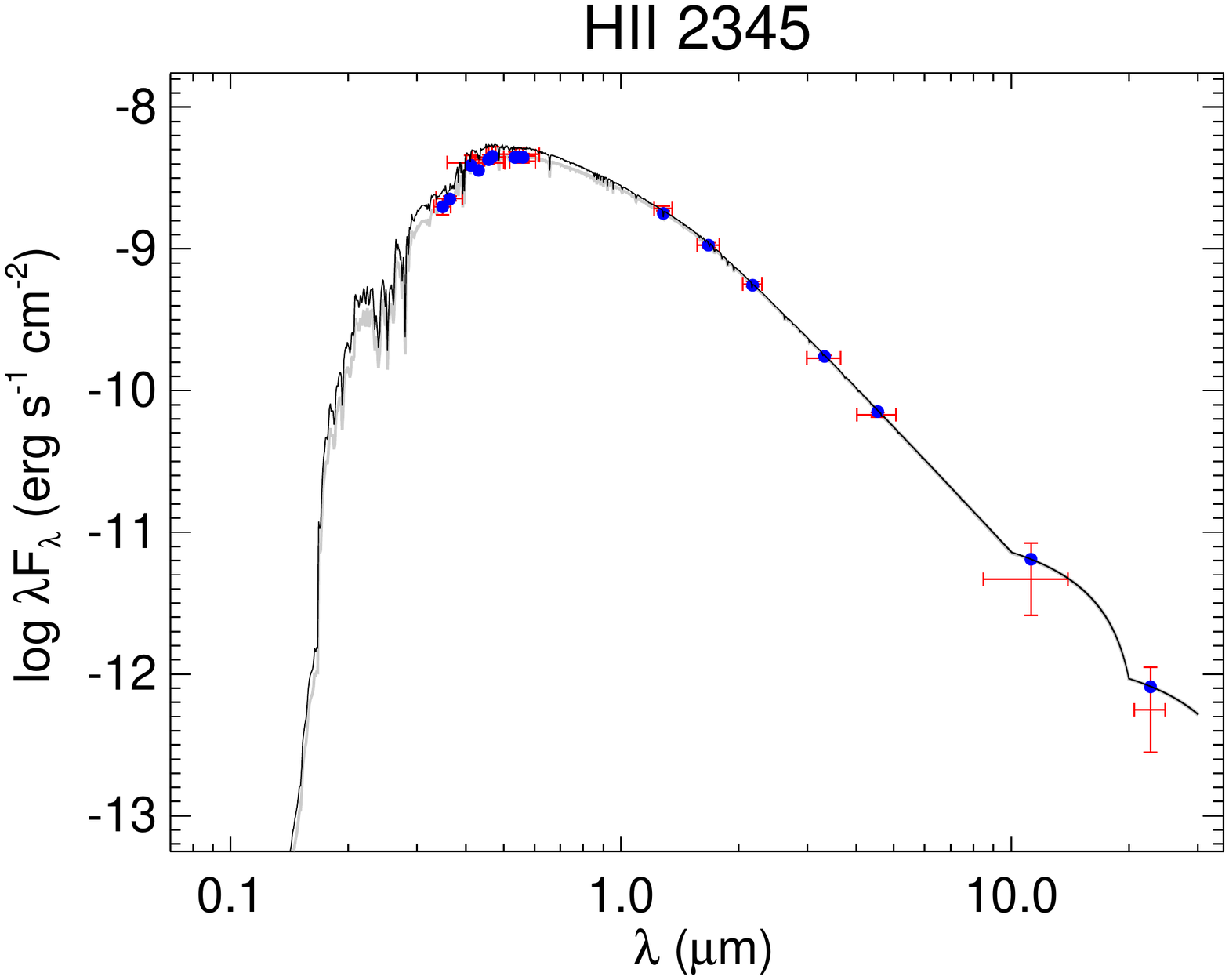}
  \includegraphics[trim=60 60 60 60,clip,width=0.49\linewidth]{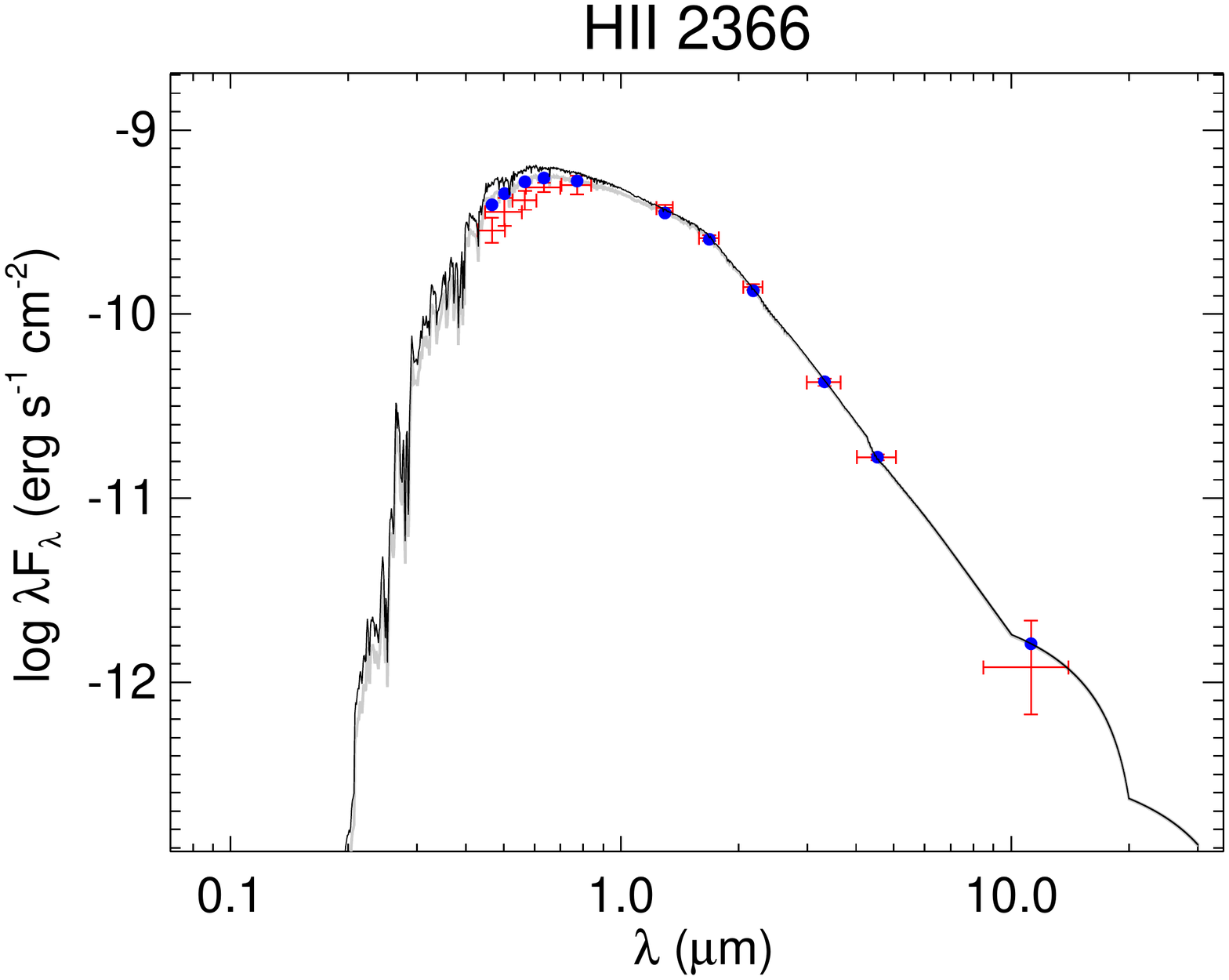}
  \includegraphics[trim=60 60 60 60,clip,width=0.49\linewidth]{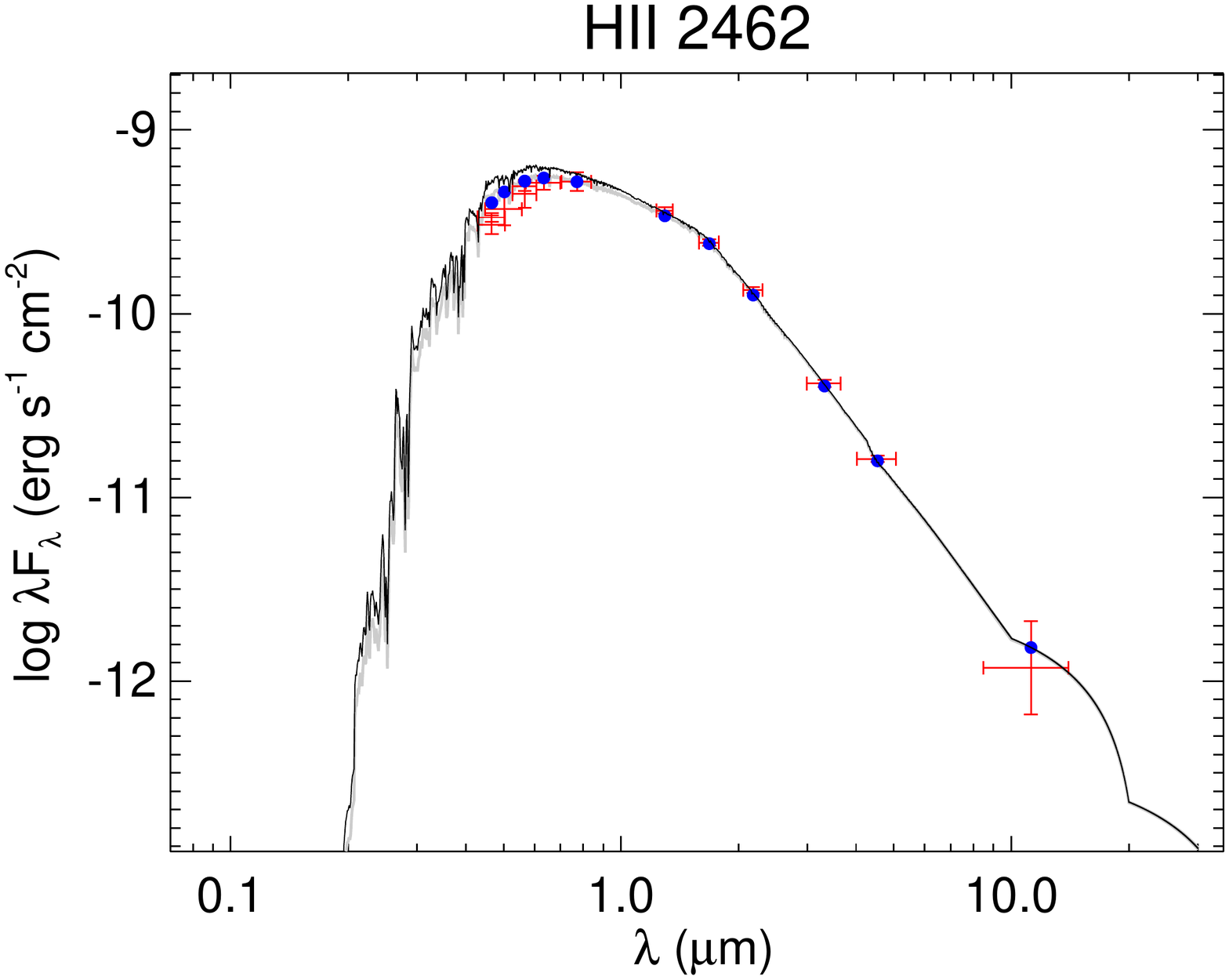}
  \includegraphics[trim=60 60 60 60,clip,width=0.49\linewidth]{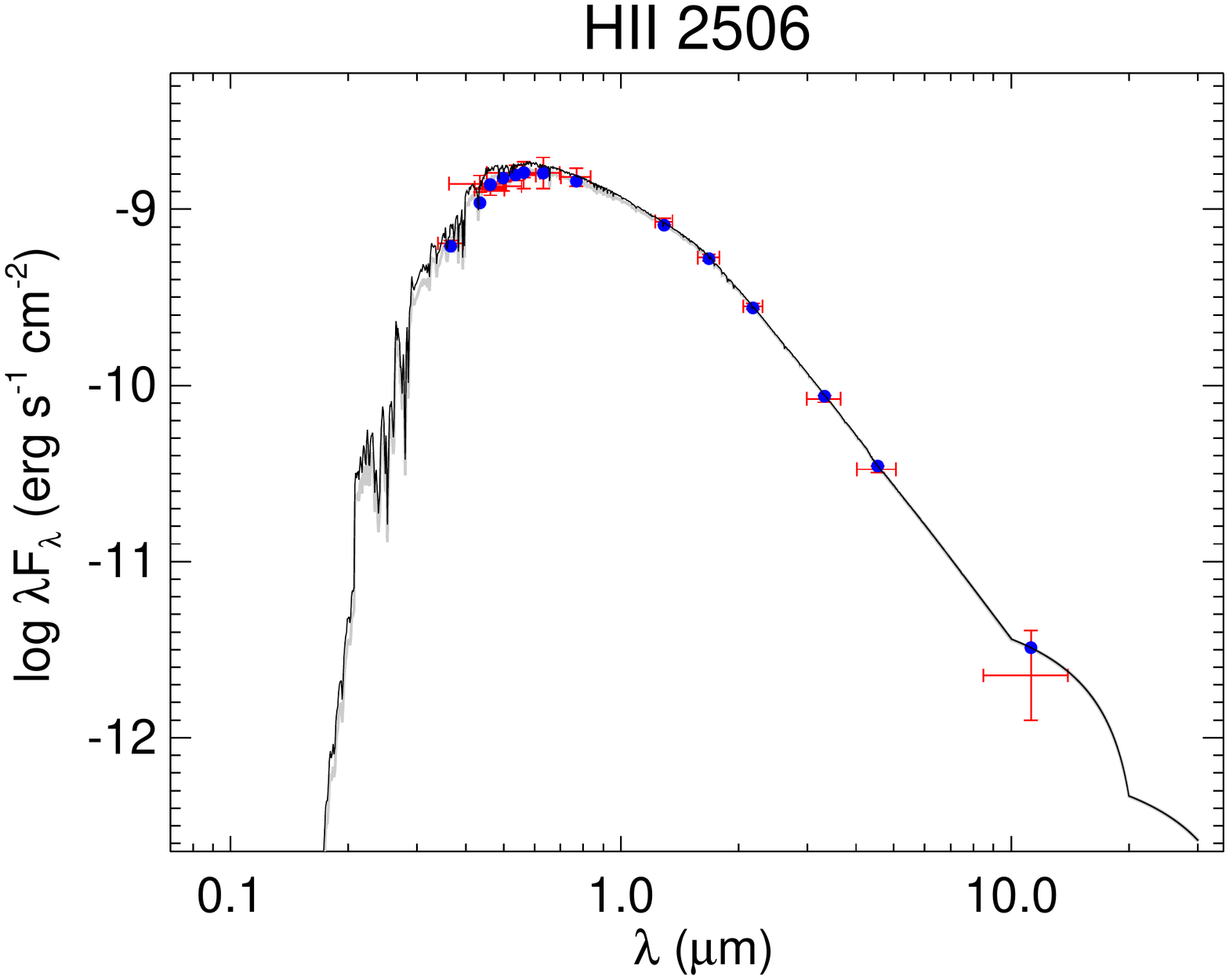}
  \includegraphics[trim=60 60 60 60,clip,width=0.49\linewidth]{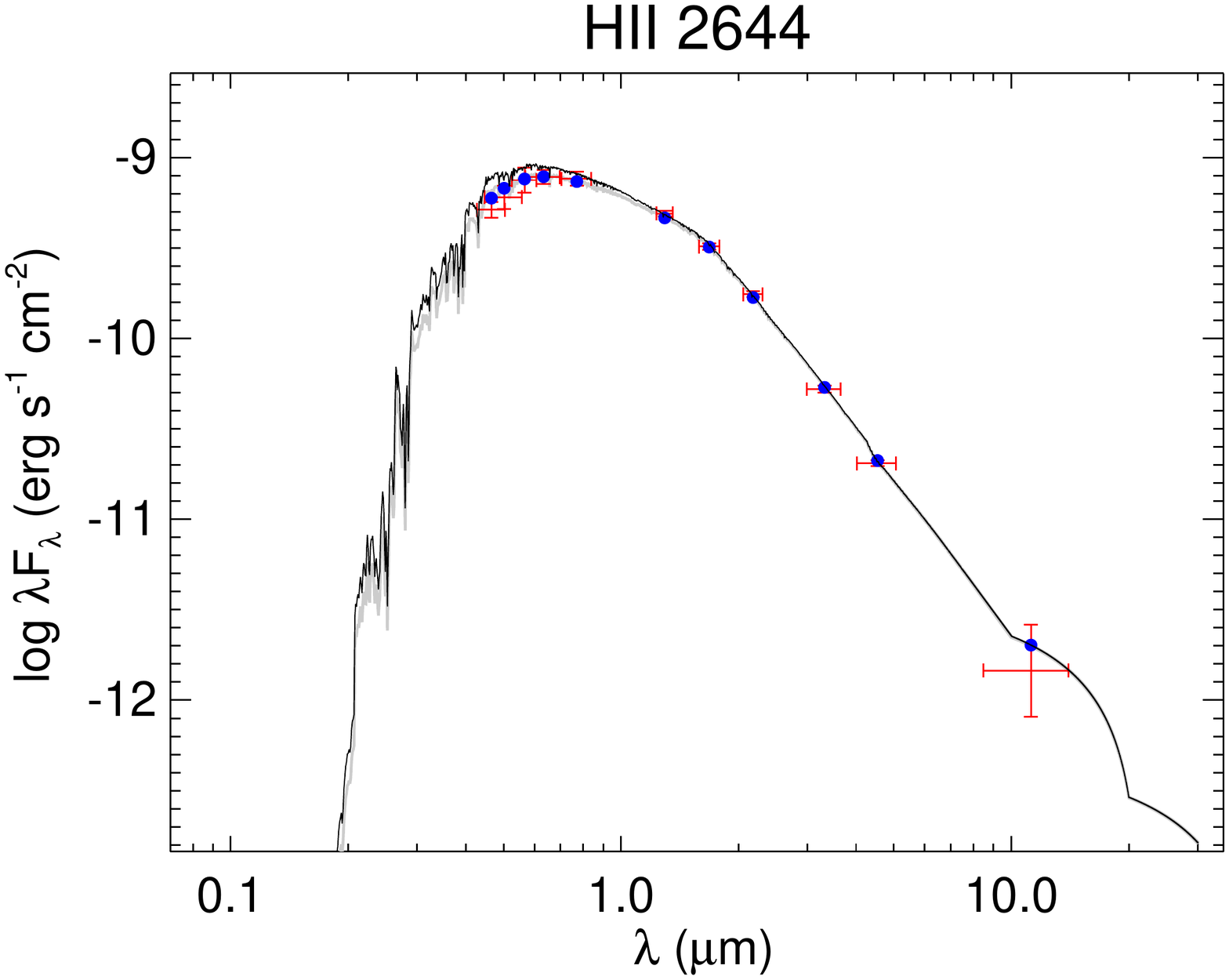}
  \caption{All labels, lines, symbols, and colors as in Figure \ref{fig:seds}.}
  \label{fig:seds_12}
\end{figure}

\begin{figure}[H]
  \centering
  \includegraphics[trim=60 60 60 60,clip,width=0.49\linewidth]{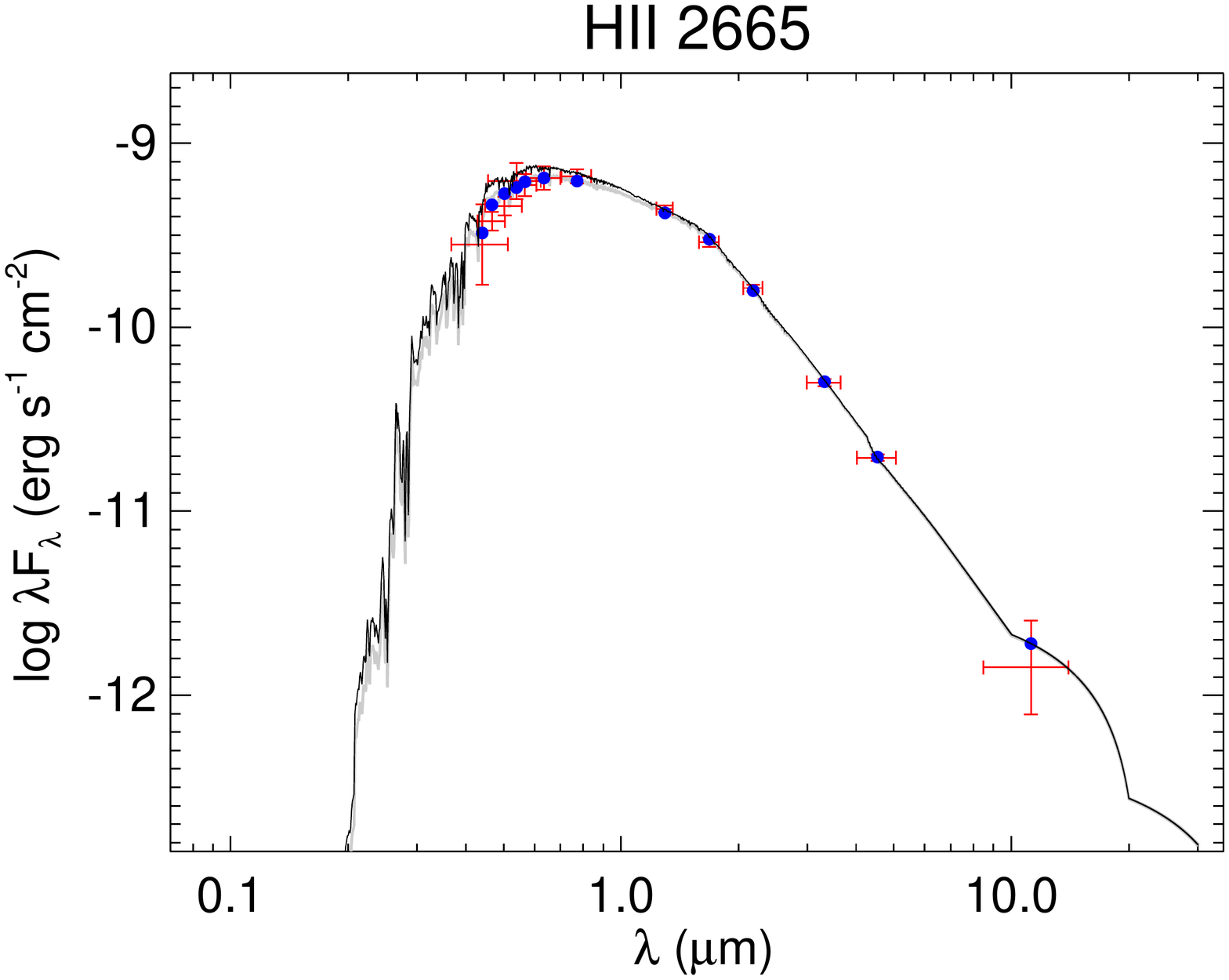}
  \includegraphics[trim=60 60 60 60,clip,width=0.49\linewidth]{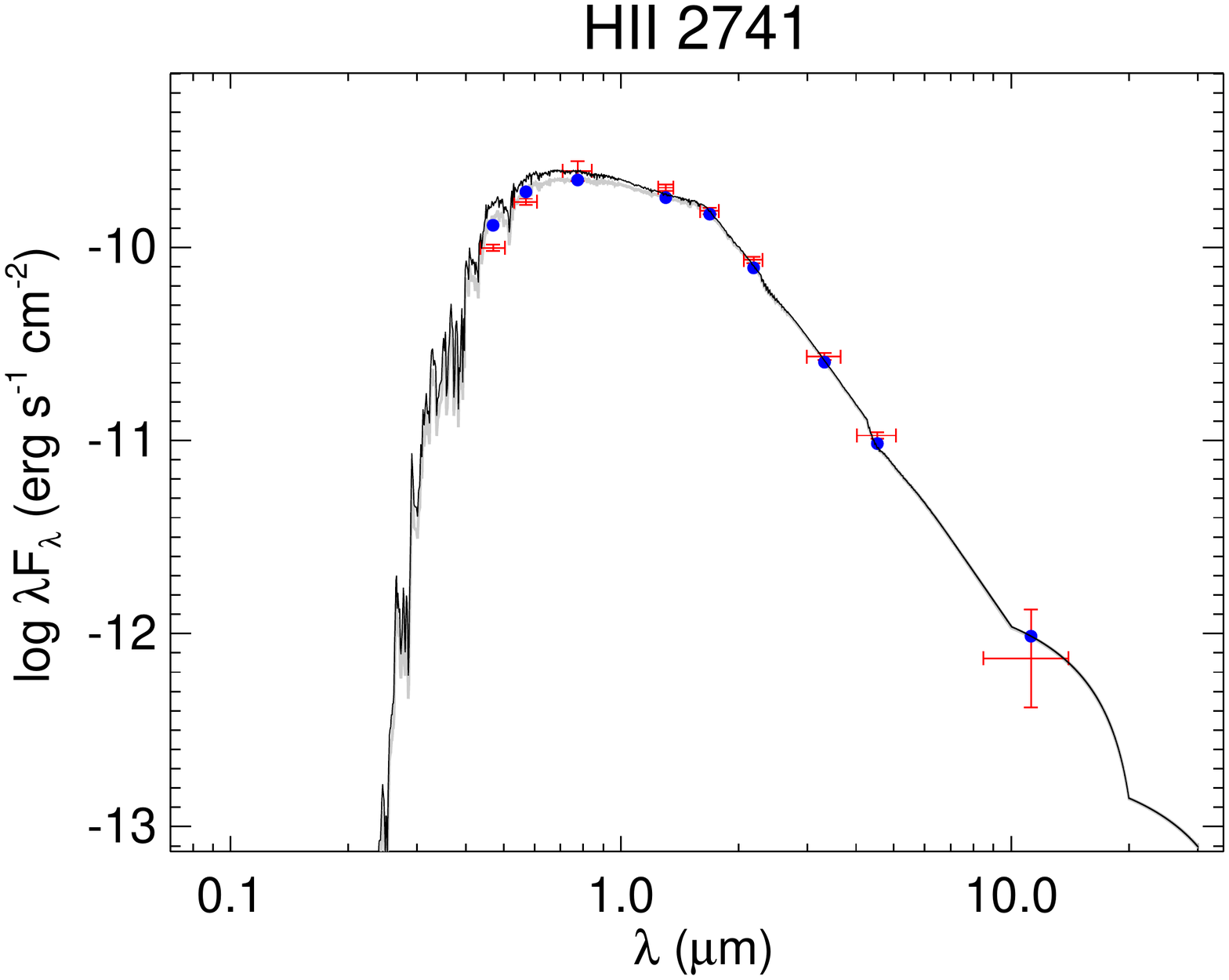}
  \includegraphics[trim=60 60 60 60,clip,width=0.49\linewidth]{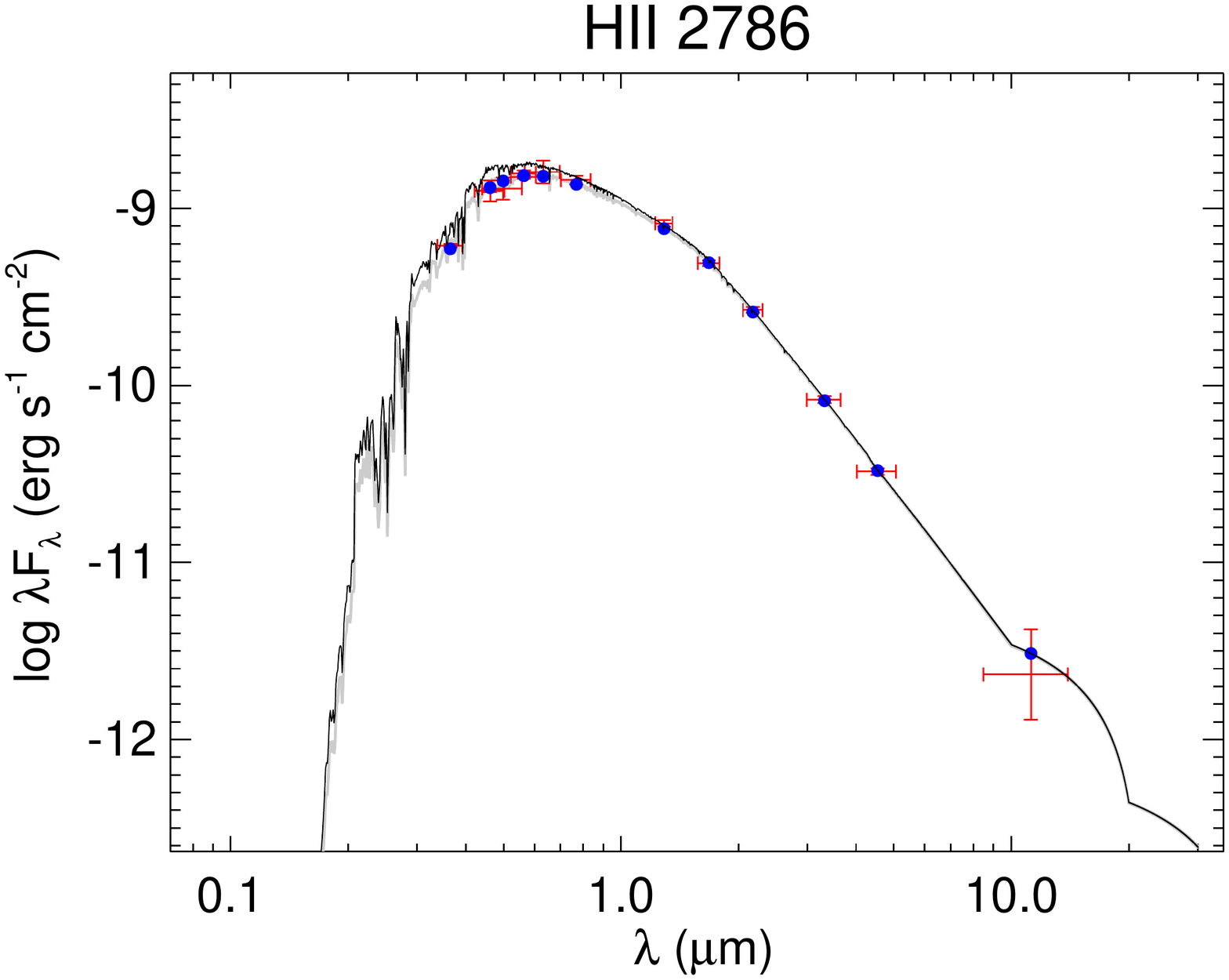}
  \includegraphics[trim=60 60 60 60,clip,width=0.49\linewidth]{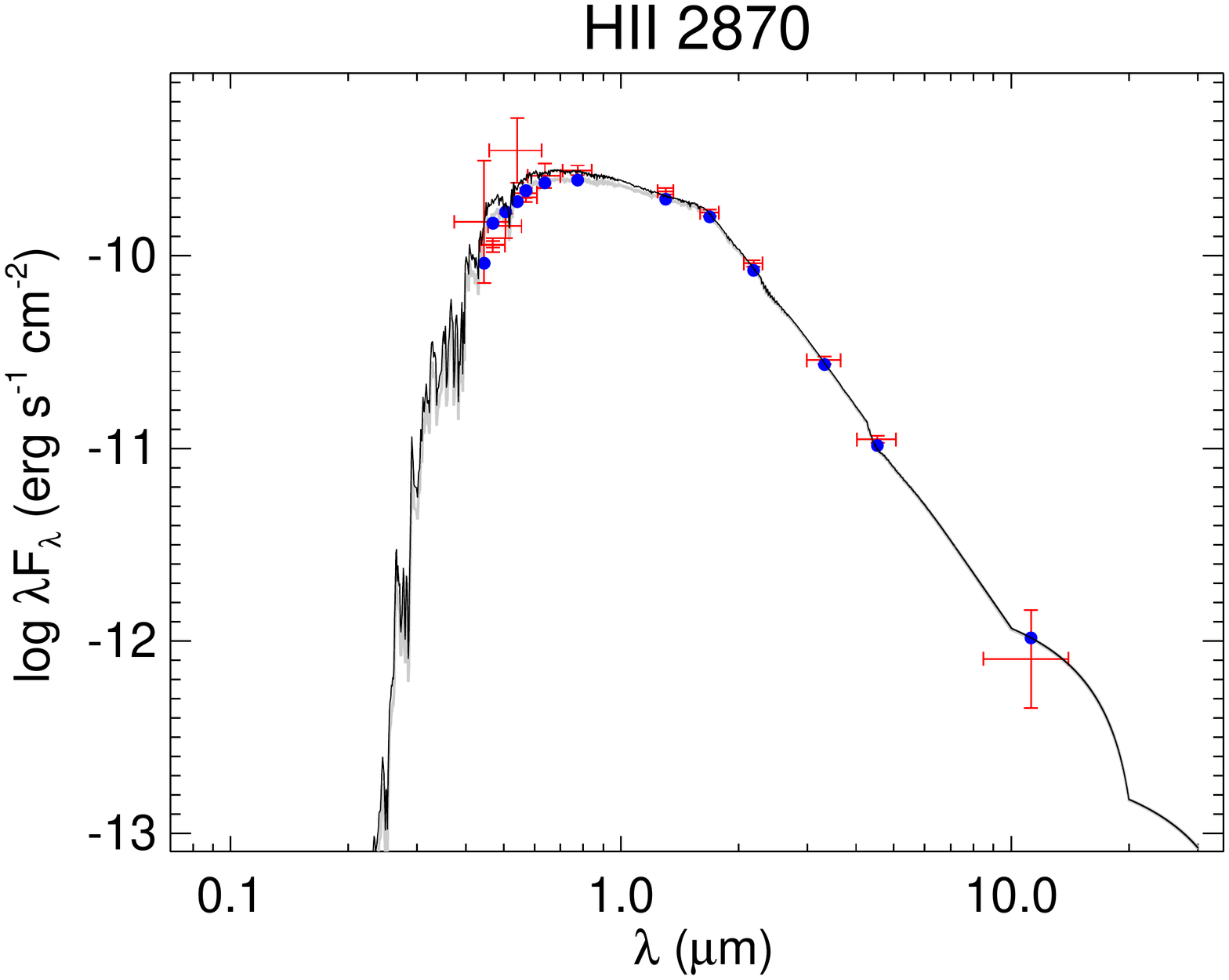}
  \includegraphics[trim=60 60 60 60,clip,width=0.49\linewidth]{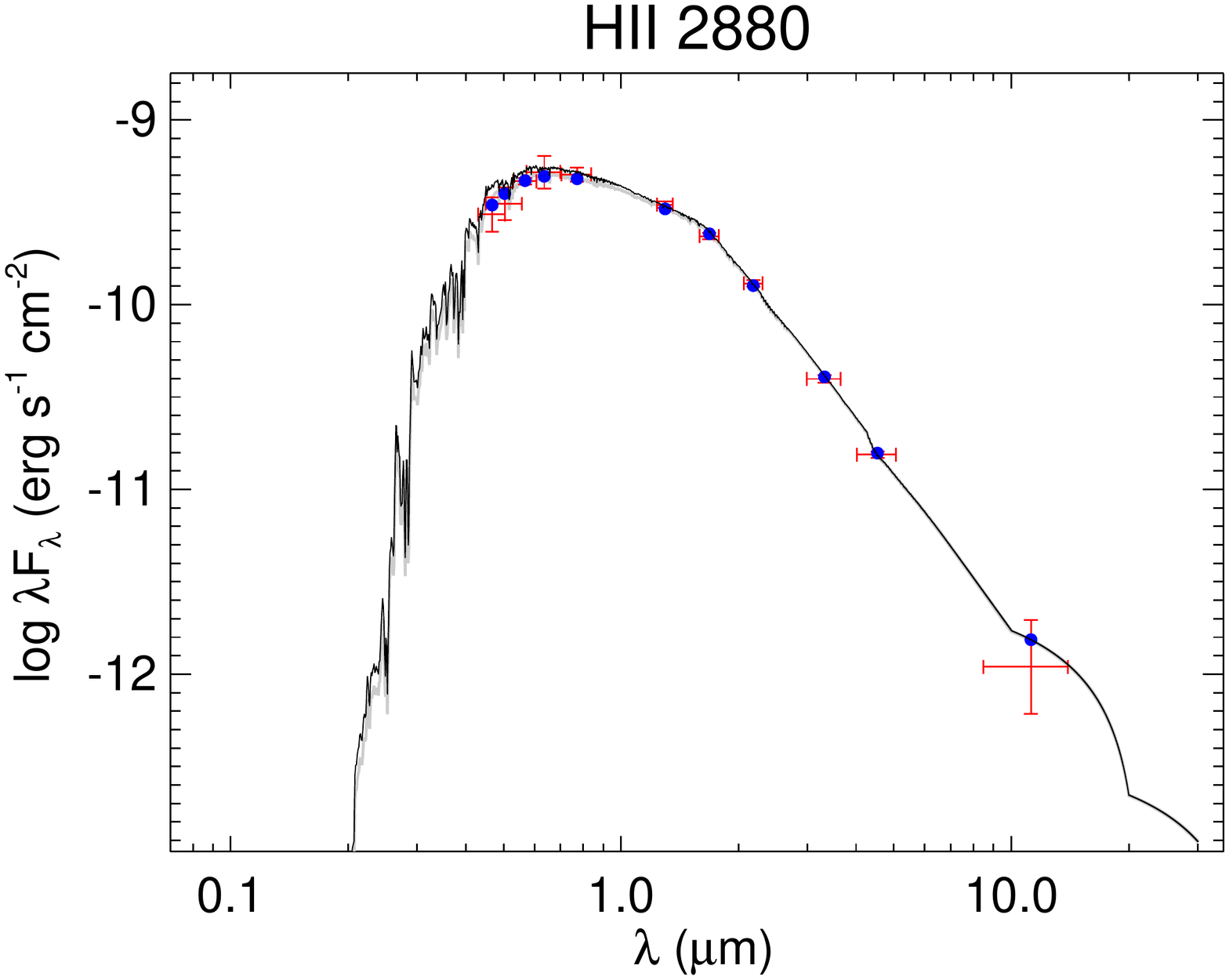}
  \includegraphics[trim=60 60 60 60,clip,width=0.49\linewidth]{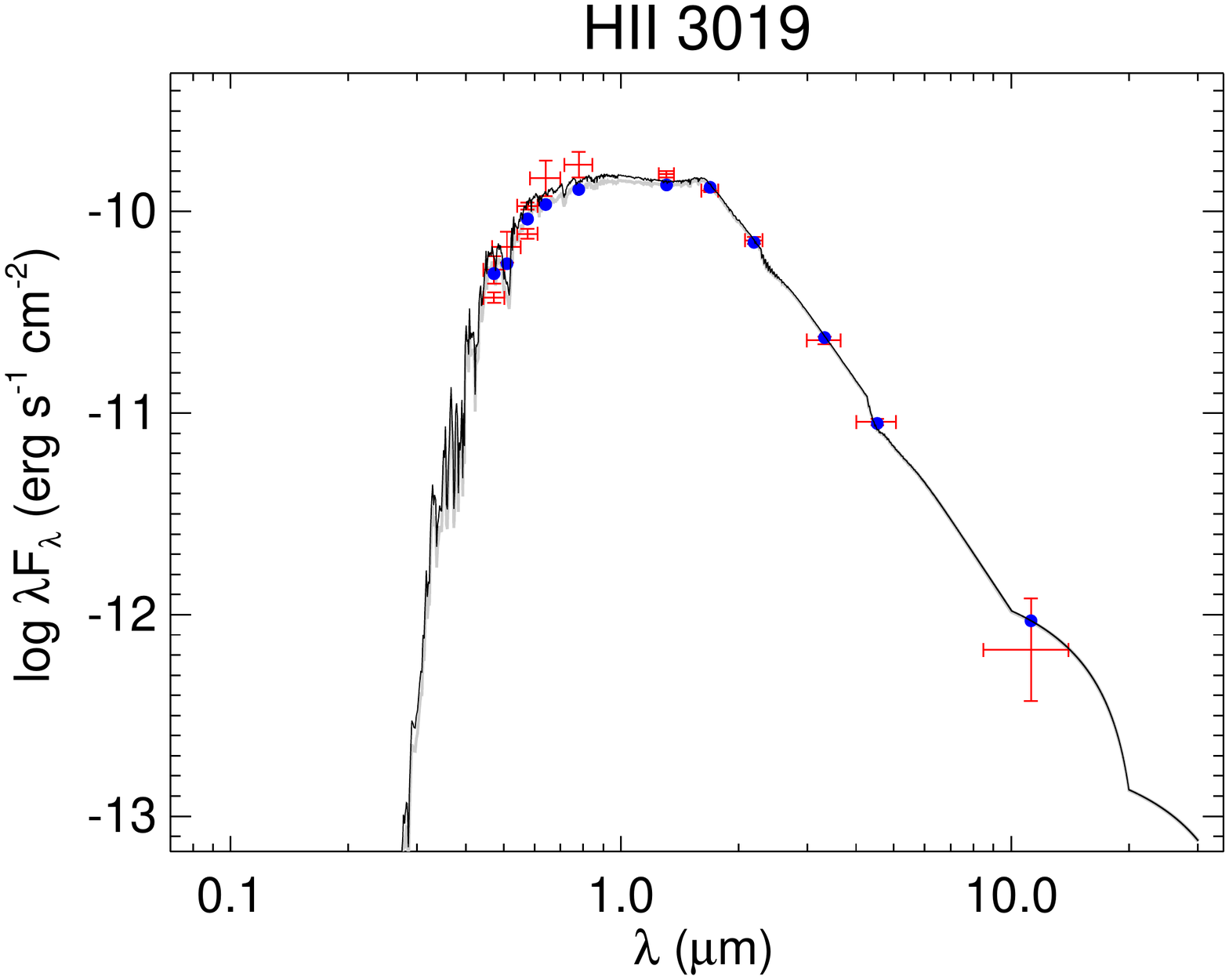}
  \caption{All labels, lines, symbols, and colors as in Figure \ref{fig:seds}.}
  \label{fig:seds_13}
\end{figure}

\begin{figure}[H]
  \centering
  \includegraphics[trim=60 60 60 60,clip,width=0.49\linewidth]{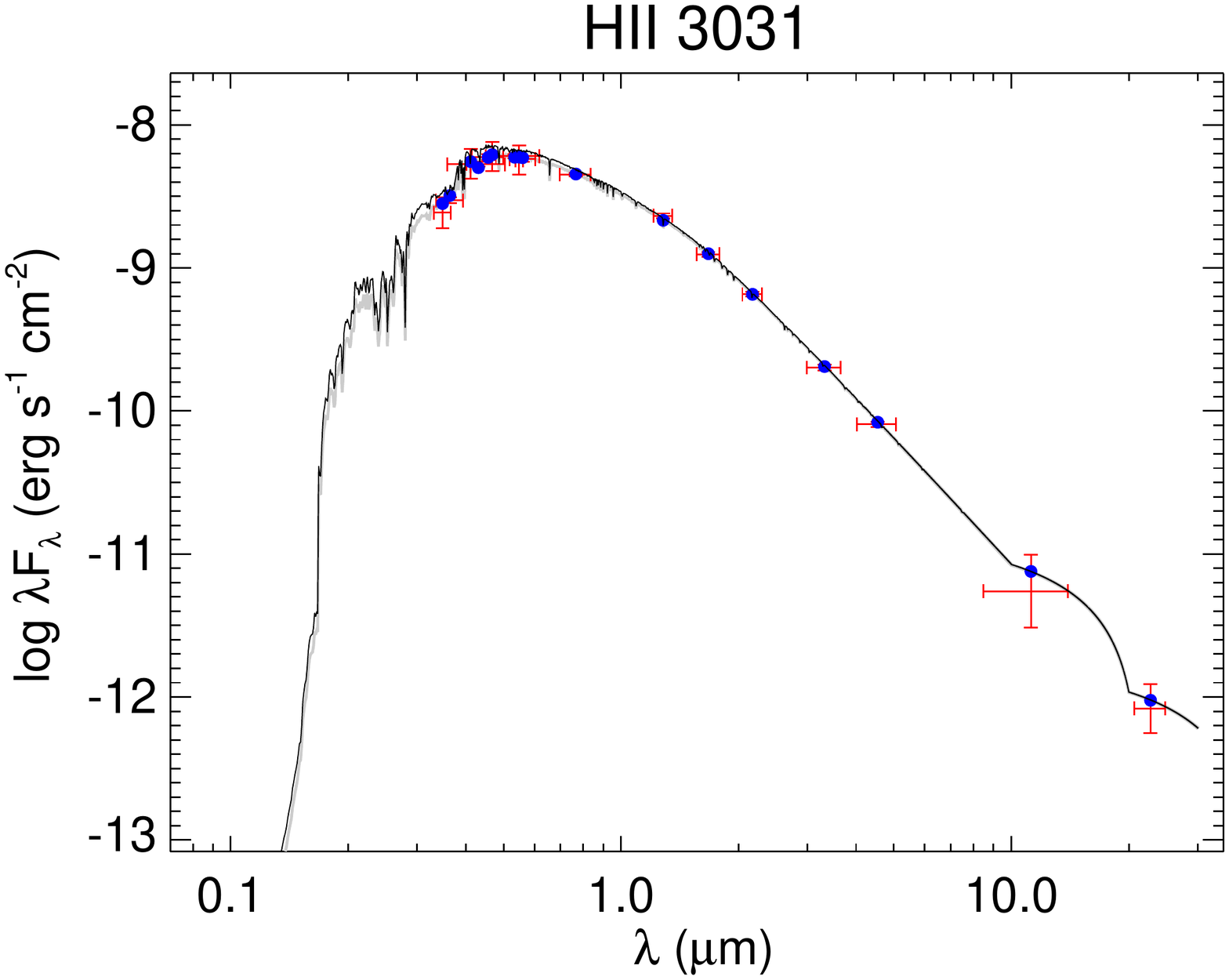}
  \includegraphics[trim=60 60 60 60,clip,width=0.49\linewidth]{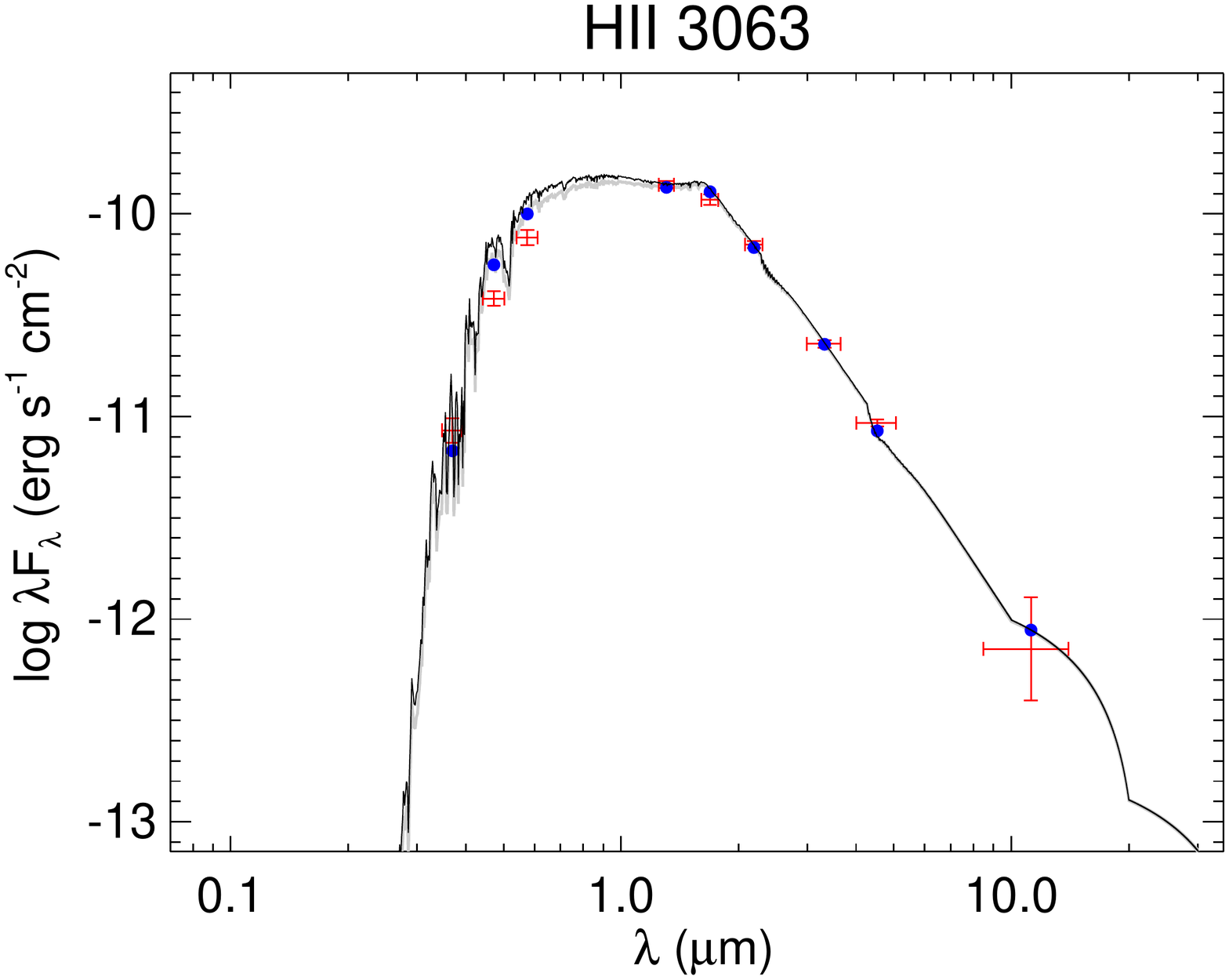}
  \includegraphics[trim=60 60 60 60,clip,width=0.49\linewidth]{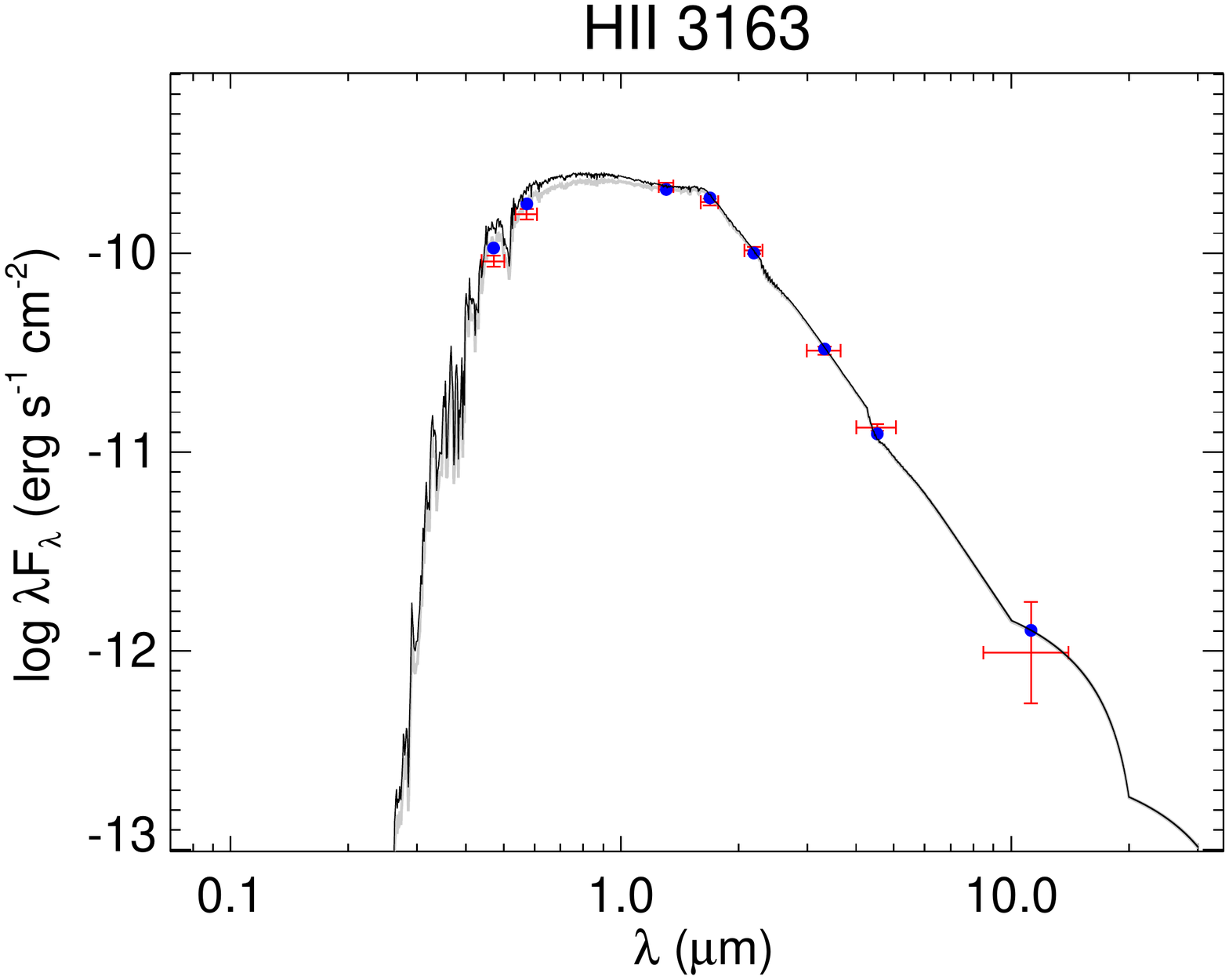}
  \includegraphics[trim=60 60 60 60,clip,width=0.49\linewidth]{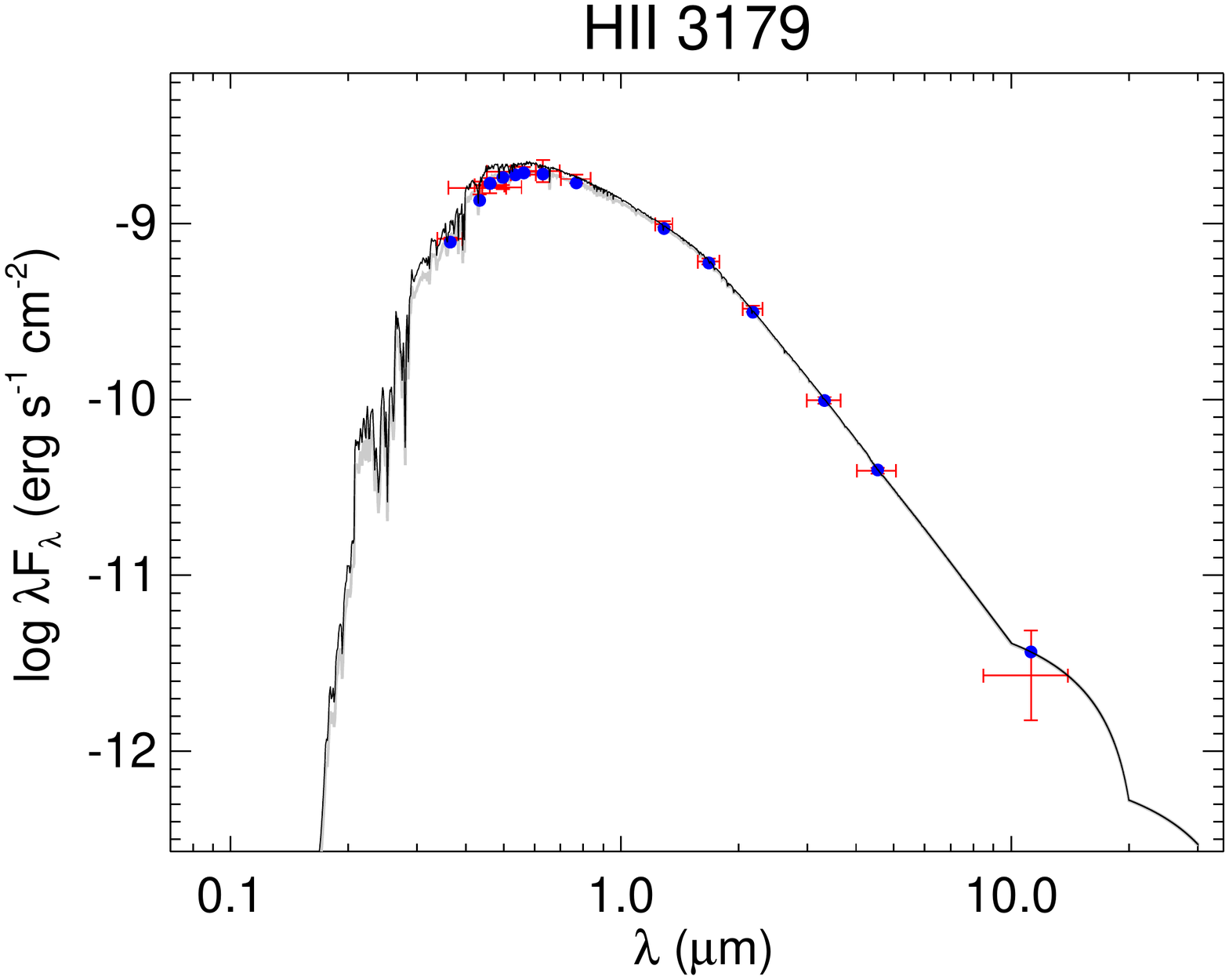}
  \includegraphics[trim=60 60 60 60,clip,width=0.49\linewidth]{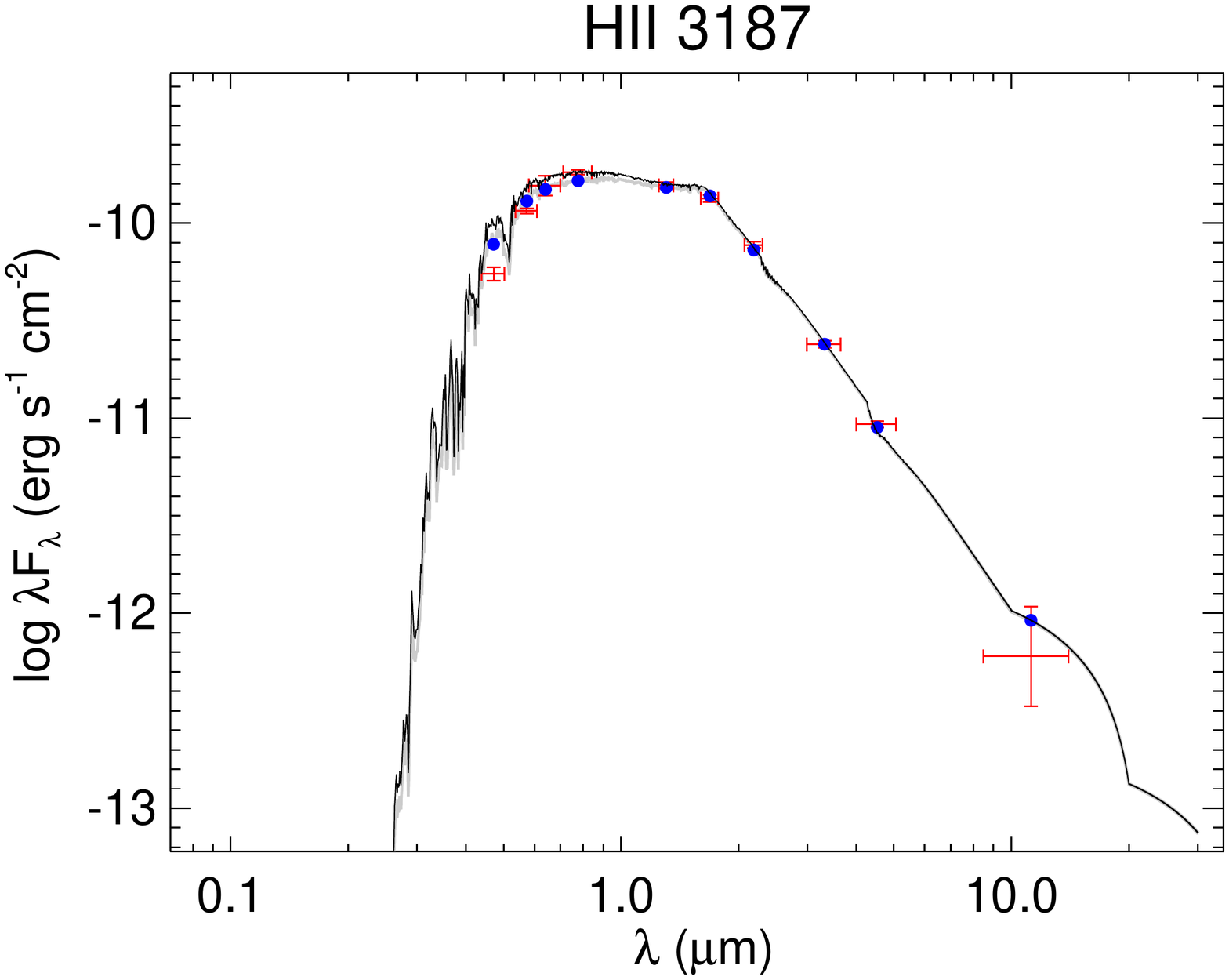}
  \caption{All labels, lines, symbols, and colors as in Figure \ref{fig:seds}.}
  \label{fig:seds_14}
\end{figure}

\end{document}